\documentclass[ aps, twocolumn, floatfix, letterpaper, nofootinbib ]{revtex4-2}

\usepackage{graphicx}
\usepackage{dcolumn}
\usepackage{bm}
\usepackage{multirow}

\usepackage{centernot}

\usepackage[normalem]{ulem}

\usepackage{amsmath}
\usepackage{amsthm}
\usepackage{amstext}
\usepackage{amssymb}
\usepackage{mathrsfs}
\usepackage{amsfonts}
\usepackage{amsbsy} 
\usepackage{tensor}
\usepackage{physics}

\usepackage[all,matrix,cmtip]{xy}

\usepackage{csquotes}
\MakeOuterQuote{"}

\usepackage[dvipsnames]{xcolor}

\definecolor{red}{rgb}{1,0,0}
\definecolor{blue}{rgb}{0,0,1}
\definecolor{dblue}{rgb}{0,0,0.4}
\definecolor{green}{rgb}{0,1,0}
\definecolor{black}{rgb}{0,0,0}
\definecolor{white}{rgb}{1,1,1}
\definecolor{niceBlue}{RGB}{20,10,237}

\usepackage[colorlinks,citecolor=niceBlue,linkcolor=niceBlue,urlcolor=niceBlue,hyperindex]{hyperref}

\definecolor{brn}{rgb}{.8,.4,.0}
\definecolor{redo}{rgb}{1,.5,.0}
\definecolor{ddgrn}{rgb}{0,0.4,0}
\definecolor{dgrn}{rgb}{0,0.55,0}
\definecolor{dbl}{rgb}{0,0,0.5}

\usepackage[bbgreekl]{mathbbol}
\newcommand{\one}{\mathbf{1}}

\newcommand{\Z}{\mathbb{Z}}

\newcommand{\R}{\mathbb{R}}

\newcommand{\U}[1]{U(1)^{(#1)}}
\newcommand{\ZN}[1]{\Z_N^{(#1)}}

\renewcommand{\t}[1]{\widetilde{#1}} 
\newcommand{\h}[1]{\hat{#1}} 

\newcommand{\ii}{\hspace{1pt}\mathrm{i}\hspace{1pt}}
\newcommand{\ee}{\hspace{1pt}\mathrm{e}}
\renewcommand{\dd}{\hspace{1pt}\mathrm{d}}
\newcommand{\da}{\dagger}
\newcommand{\<}{\langle}
\renewcommand{\>}{\rangle}

\newcommand{\Rf}[1]{Ref.~\onlinecite{#1}}
\newcommand{\Rfs}[1]{Refs.~\onlinecite{#1}}

\newcommand{\pp}{\partial}

\newcommand{\ie}{{\it i.e.~}} 
 
\newcommand{\etc}{{\it etc}}

\newcommand{\bpm}{\begin{pmatrix}}
\newcommand{\epm}{\end{pmatrix}}
\newcommand{\bmm}{\begin{matrix}}
\newcommand{\emm}{\end{matrix}}

\newcommand{\cA}{ {\cal A} } 
\newcommand{\cB}{ {\cal B} }
 
\newcommand{\cD}{ {\cal D} }

\newcommand{\cL}{ {\cal L} }

\newcommand{\cO}{ {\cal O} } 
\newcommand{\cP}{ {\cal P} } 
\newcommand{\cR}{ {\cal R} } 
\newcommand{\cS}{ {\cal S} } 
\newcommand{\cT}{ {\cal T} } 
\newcommand{\cV}{ {\cal V} }

\newcommand{\cZ}{ {\cal Z} } 

\usepackage{euscript}

\newcommand\eB           {\EuScript{B}}

\newcommand\eM          {\EuScript{M}}

\newcommand\eV        {\EuScript{V}}

\newcommand\eZ         {\EuScript{Z}}

\newcommand{\al}{\alpha} 
\newcommand{\bt}{\beta} 
\newcommand{\del}{\delta} 
\newcommand{\Del}{\Delta} 
\newcommand{\eps}{\epsilon} 
 
\newcommand{\ga}{\gamma} 
\newcommand{\Ga}{\Gamma} 
\newcommand{\ka}{\kappa} 
\newcommand{\la}{\lambda} 
 
\newcommand{\om}{\omega}

\newcommand{\Th}{\Theta} 
 
\newcommand{\Si}{\Sigma}

\newcommand{\txti}[1]{\textit{#1}}
\renewcommand{\txt}[1]{\text{#1}}

\newcommand{\hstar}{\mathop{*}} 
\def\wdg{{\mathchoice{\,{\scriptstyle\wedge}\,}{{\scriptstyle\wedge}}{{\scriptscriptstyle\wedge}}{{\scriptscriptstyle\wedge}}}} 

\newcommand{\gs}{\ket{\txt{vac}}}

\newcommand\toZ[1]{\left\lfloor #1 \right\rceil}

\renewcommand{\mod}{\mathrm{mod}}

\newcommand\cVec{\cV\mathrm{ec}}
\newcommand\eVec{\eV\mathrm{ec}}

\begin{document}

\title{Exact emergent higher-form symmetries in bosonic lattice models}

\author{Salvatore D. Pace}
\affiliation{Department of Physics, Massachusetts Institute of Technology, Cambridge, Massachusetts 02139, USA}

\author{Xiao-Gang Wen} 
\affiliation{Department of Physics, Massachusetts Institute of Technology,
Cambridge, Massachusetts 02139, USA}

\date{\today}

\begin{abstract}

Although condensed matter systems usually do not have higher-form symmetries,
we show that, unlike 0-form symmetry, higher-form symmetries can emerge as
exact symmetries at low energies and long distances. In particular, emergent
higher-form symmetries at zero temperature are robust to arbitrary local UV
perturbations in the thermodynamic limit. This result is true for both
invertible and non-invertible higher-form symmetries. Therefore, emergent
higher-form symmetries are \textit{exact emergent symmetries}: they are not UV
symmetries but constrain low-energy dynamics as if they were. Since phases of
matter are defined in the thermodynamic limit, this implies that a UV theory
without higher-form symmetries can have phases characterized by exact emergent
higher-form symmetries. We demonstrate this in three lattice models, the
quantum clock model and emergent ${\mathbb{Z}_N}$ and ${U(1)}$ ${p}$-gauge
theory, finding regions of parameter space with exact emergent (anomalous)
higher-form symmetries. Furthermore, we perform a generalized Landau analysis
of a 2+1D lattice model that gives rise to $\mathbb{Z}_2$ gauge theory. Using
exact emergent 1-form symmetries accompanied by their own energy/length scales,
we show that the transition between the deconfined and Higgs/confined phases is
continuous and equivalent to the spontaneous symmetry-breaking transition of a
$\mathbb{Z}_2$ symmetry, even though the lattice model has no symmetry. Also,
we show that this transition line must \txti{always} contain two parts
separated by multi-critical points or other phase transitions. We discuss the
physical consequences of exact emergent higher-form symmetries and contrast
them to emergent ${0}$-form symmetries.  Lastly, we show that emergent 1-form
symmetries are no longer exact at finite temperatures, but emergent $p$-form
symmetries with ${p\geq 2}$ are.

\end{abstract}

\maketitle
\makeatletter 
\def\l@subsubsection#1#2{}
\makeatother 

\tableofcontents

\section{Introduction}

A longstanding pillar for understanding strongly interacting quantum many-body
systems is to identify and understand their symmetries. Indeed, symmetries
provide powerful constraints and universal characterizations of a system's
dynamics and phases. This point of view has become increasingly fruitful with
modern generalizations of symmetry~\cite{NOc0605316,NOc0702377, GW14125148, KT13094721,S150804770, BH180309336, CI180204790, BCH220807367, BCH221111764, GLS220110589,
KS220809056, QRH201002254,OPH230104706, BT170402330,TW191202817,KZ200514178,
LV200811187, MT220710712,FT220907471} (see \Rfs{M220403045,CD220509545} for
recent reviews). For instance, topological order~\cite{W9039}, which provided
the first indication that conventional symmetries~\cite{L3726, GL5064} are not
all-powerful, can now be understood in a symmetry
framework~\cite{NOc0702377,W181202517,HS181204716,HT190411550, QRH201002254, KKO210713091,M220403045, BCH220807367, P230805730}. These generalizations open up an exciting frontier for the
discovery of new phases of quantum matter and the conceptual
organization/systematic understanding~\cite{KZ200514178} of quantum phases.

One of the simplest generalizations of symmetry is called higher-form symmetry.
For ordinary symmetries, charged operators act on a point in space and the
unitary operator that generates the symmetry transformation acts on all of
space. Higher-form symmetries generalize this by allowing charged operators to
be extended~\cite{NOc0605316, NOc0702377, GW14125148}. For a ${p}$-form
symmetry, the charged operators act on ${p}$-dimensional subspaces and the
unitary generating the transformation acts on a closed ${(d-p)}$-dimensional
subspace of ${d}$-dimensional space. So, an ordinary symmetry is just a
${0}$-form symmetry.

Most things 0-form symmetries can do, higher-form symmetries can also do. For example, higher-form symmetries can spontaneously break, giving rise to a topological ground state degeneracy (gapless Goldstone bosons) when discrete (continuous)~\cite{GW14125148, KR9210154, L180207747, HI180209512, KH190504617, HHY200715901, YY200907621, IM210612610}. Indeed, abelian topological orders reflect anomalous discrete 1-form symmetries spontaneously breaking, and photons in a Coulomb phase arise from ${U(1)}$ 1-form symmetries spontaneously breaking. A higher-form symmetry can also have a 't Hooft anomaly, providing powerful constraints on the IR through generalized Lieb-Schultz-Mattis-Oshikawa-Hastings theorems and introducing higher-form symmetry-protected topological phases~\cite{KT13094721, TK151102929, Y150803468, GKK170300501, KR180505367, HS181204716, W181202517, KR180505367, WW181211967, W181202517, TW190802613, JW200900023, MF220607725, PW220703544, VBV221101376, TRV230308136}.

These applications of higher-form symmetries make them a powerful tool in
studying quantum many-body systems. However, unfortunately, models with exact
higher-form symmetries are rather special and, in a sense, fine-tuned. So, it
is natural to wonder if they play a role in more typical, physically relevant
models. 

One possibility is that while they may not be exact microscopic symmetries,
they could still arise as emergent symmetries. However, experience with emergent ordinary (0-form) symmetries causes apprehension since their consequences are typically approximate since they can be violated by irrelevant operators.\footnote{\label{footnoteemanant}A counterexample to this usual rule are emanant symmetries~\cite{CS221112543}.} In other words, explicitly
breaking 0-form symmetries creates $\cO(E^\ga)$ errors at energy scale $E$,
even in the thermodynamic limit. Amazingly, common folklore suggests that this
0-form symmetry-based intuition does not carry over to higher-form symmetries.
They can constrain a system exactly even as emergent symmetries, as discussed
in the context of gauge theories~\cite{FNN8035, HW0541, PS08010587} in early
days, and in the context of higher-form symmetry~\cite{W181202517, SSN201215845, IM210612610, COR220205866, M220403045, CHB220914302, HK221011492, CS221112543, CJ230413751}.

Here we investigate this robustness of higher-form symmetries in detail and
from a lattice perspective, considering bosonic lattice Hamiltonian models without
higher-form symmetries. These UV-complete theories are simple,
well-defined, and relevant to condensed matter physics. 

For this class of
models, we demonstrate how higher-form symmetries can emerge and, when they do, why they constrain the IR exactly. We find that at energies with
the emergent higher-form symmetry, the dynamics of states are affected by the
emergent higher-form symmetry as if it were an exact UV symmetry. More
precisely, any errors coming from the higher-form symmetries being emergent below a finite
energy scale $E$ are of order $\cO(\ee^{-L^{\ga}})$, where $L$ is the
system size measured by lattice constant, and thus vanish in the thermodynamic limit. Our arguments apply to both invertible and non-invertible higher-form symmetries.

Therefore, phases of microscopic models without exact higher-form symmetries
can be exactly characterized by emergent higher-form symmetries. To
emphasize this, we refer to emergent higher-form symmetries as exact emergent
symmetries. Consequently, to understand how emergent higher-form
symmetries characterize phases, one should first partition parameter space
by the theory's exact \txti{and} exact emergent symmetries. These partitions then lay a foundation for the system's phases to be labeled and
characterized using generalized symmetries.

The rest of this paper goes as follows: In section~\ref{sec:midIRandQuasiAdiCont}, we show that emergent higher-form symmetries in lattice models are exact. We use the point of view that
symmetries are described by algebras of local symmetric
operators~\cite{CW220303596}, but also develop low-energy effective Hamiltonians with the emergent symmetries. In section~\ref{exactEmEx}, we consider three examples with exact emergent higher-form symmetries: the ${\Z_N}$ quantum clock model and models of emergent ${\Z_N}$ and ${U(1)}$
${p}$-gauge theory. In section~\ref{landauSec}, we use the concept of exact emergent symmetry to perform a complete generalized Landau analysis of the Fradkin-Shenker model with periodic boundary conditions. We recover known results regarding the universality classes of the phase transitions and make new predictions about the phase diagram's general structure. In section~\ref{sec:physCons}, we discuss general physical consequences of emergent higher-form symmetries being exact. In particular, how exact emergent higher-form
symmetries can characterize phases of systems, both when they are, and are not, spontaneously broken in the bulk. In section~\ref{sec:finiteT}, we consider emergent higher-form symmetries at finite temperature, showing that only emergent ${p}$-form symmetries with ${p>1}$ are exact. Then, in section~\ref{sec:conclusion}, we conclude and discuss some open questions arising from this work.

\section{Scales hierarchies and emergent symmetries}\label{sec:midIRandQuasiAdiCont}

Consider a lattice bosonic quantum system described by the local Hamiltonian
${H}$ and whose total Hilbert space ${\cV }$ is tensor product decomposable:
${\cV = \bigotimes_i \cV_i}$. Since ${H}$ includes the exact interactions at
the microscopic scale and describes the system at all energies throughout the
entire parameter space, we refer to it as the UV Hamiltonian, adopting the
language used in field theory. While, in theory, the system's physical
properties can be extracted from ${H}$, this proves much too difficult in
practice~\cite{LP0028}. 

\begin{figure}[t!]
\centering
    \includegraphics[width=.48\textwidth]{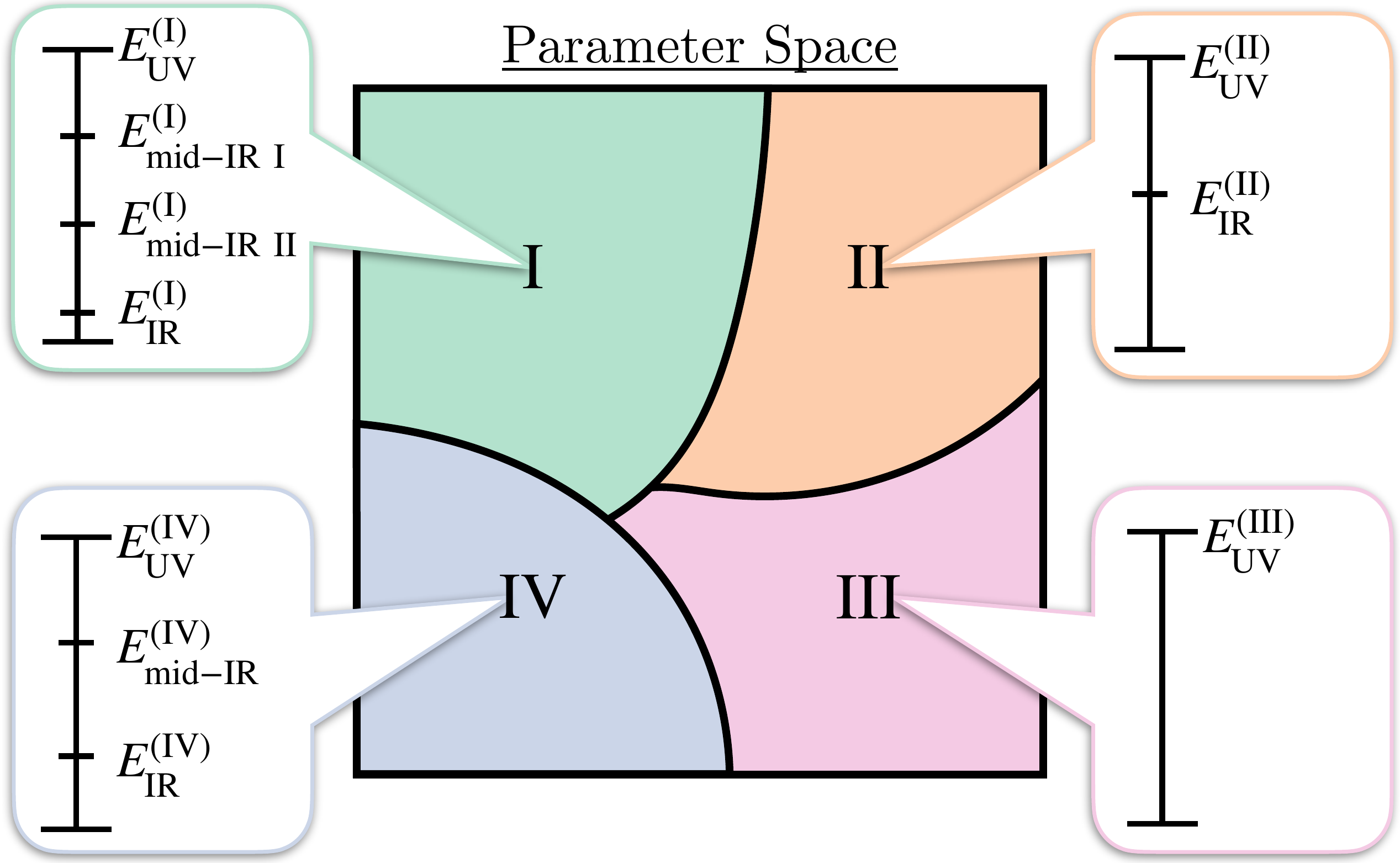}
    \caption{%
   The parameter space of a many-body Hamiltonian can be partitioned by its differing hierarchies of energy scales. A schematic depiction of this is shown here. The parameter space is partitioned into four regions, labeled I, II, III, and IV, with their differing energy scale hierarchies shown.
    }
    \label{fig:HierScales}
\end{figure}

A guiding principle to overcome this daunting problem is the separation of
energy scales (assuming no UV/IR mixing~\cite{SS200310466,
GLS210800020, GLS220110589, OJE211002658, PW220407111}). We will always denote the lowest
energy scale as ${E_{\txt{IR}}}$ and refer to the sub-Hilbert space
${\cV_{\txt{IR}} = \txt{span}\{\ket{E_n}~|~E_n < E_{\txt{IR}}\}}$, where
${\ket{E_n}}$ is an energy eigenstate, as the IR. Furthermore, we will always
denote the largest possible energy value as ${E_{\txt{UV}}}$. However, there
can be other interesting energy scales between the IR and the UV scales, which
we will call mid-IR energies ${E_{\txt{mid-IR}}}$ and refer to the sub-Hilbert
space ${\cV_{\txt{mid-IR}} = \txt{span}\{\ket{E_n}~|~E_n < E_{\txt{mid-IR}}\}}$
as the mid-IR. Generally, there can be multiple of these mid-IR scales, and as
demonstrated in Fig.~\ref{fig:HierScales}, different regions of parameter space
will have a different hierarchy of energy scales.

\subsection{Exact emergent generalized symmetries}\label{effHamSec111}
\label{lowexact}

\begin{figure}[t!]
\centering
    \includegraphics[width=.48\textwidth]{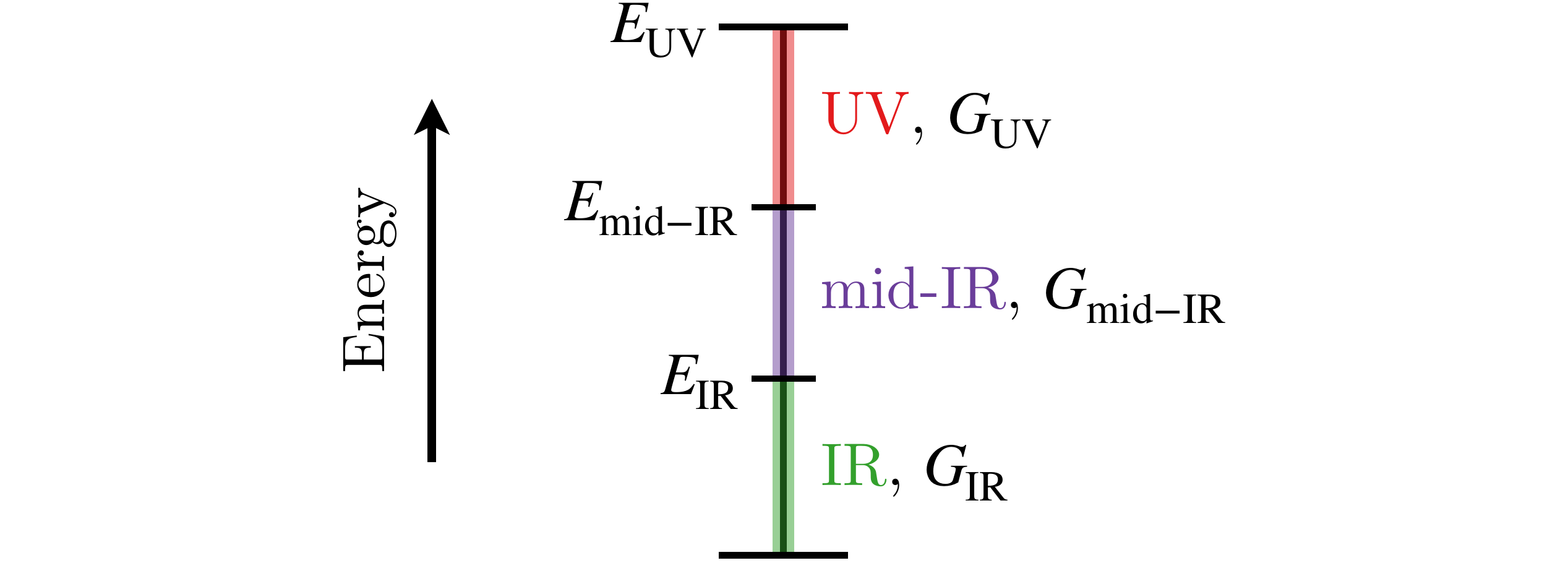}
    \caption{ The symmetries of a quantum many-body system generally depend on
the energy scale of an observer. In particular, there can be emergent symmetries at low energies absent from the microscopic (UV) symmetries
${G_{\txt{UV}}}$. These can be generalized symmetries and can be anomalous.  } \label{fig:HierScalesEx}
\end{figure}

Low-energy eigenstates will often have additional structures absent from high-energy eigenstates. For example, these additional structures could reflect the presence of new, emergent symmetries, as depicted in Fig.~\ref{fig:HierScalesEx}.

It is nontrivial to systematically identify the scale hierarchies of a general
UV Hamiltonian. Here we will specialize to a typical situation where the UV
Hamiltonian can be written as
\begin{equation}\label{genH}
H = H_0 + H_1,
\end{equation}
where a mid-IR scale $E_{\txt{mid-IR}}$ of ${H_0}$ is known (e.g., an energy
gap of a quasiparticle) and both ${H_0}$ and ${H_1}$ are translation-invariant. We assume $H_0$ is not pathological in the mid-IR and that the qualitative features of $H$ resemble those of $H_0$. For example, $H_0$ cannot have perfectly flat bands in the mid-IR since they'd introduce exponential amounts of degeneracies in the spectra, which will not arise in $H$ since a generic $H_1$ will lift the degeneracy. Furthermore, we assume there is a collection of
mutually commuting local projection operators ${\{\cP_i\}}$ acting only on degrees of freedom near site $i$ such that ${\cV^{(H_0)}_{\txt{mid-IR}}}$ is spanned by energy eigenstates of ${H_0}$ satisfying ${\<\cP_i\> = 0}$. In other words, ${\ket{\psi_{\txt{mid-IR}}}\in\cV^{(H_0)}_{\txt{mid-IR}}}$ satisfy
${\cP_i\ket{\psi_{\txt{mid-IR}}} = 0}$, while ${\cP_i\ket{\psi}\neq 0}$ for
${\ket{\psi}\not\in\cV^{(H_0)}_{\txt{mid-IR}}}$. Features of ${H_0}$ that emerge at ${E<E_{\txt{mid-IR}}}$ are determined by the constraint ${\cP_i = 0}$. In fact, as we
will soon argue, these local projection operators also determine the emergent symmetries.

We will assume
${H_1}$ includes terms that mix states with ${\<\cP_i\> = 0}$ and states with
${\<\cP_i\> \neq 0}$. Because of ${H_1}$, energy eigenstates of
${H}$ are a superposition of states with ${\<\cP_i\> = 0}$ and states with
${\<\cP_i\> \neq 0}$ and, therefore, the sub-Hilbert space spanned by states satisfying
${\<\cP_i\> = 0}$ is not a mid-IR of ${H}$.  Consequently, it appears that any emergent
structures arising from ${\cP_i = 0}$ (such as symmetry) are destroyed
by the $H_1$ term.

On the other hand, if all parameters in ${H_1}$ are much smaller than those in
${H_0}$, it is tempting to think that the mid-IR of ${H}$ is typically closely related to
the ${\<\cP_i\> = 0}$ states. This intuition motivates one to introduce the
parameter ${s\in[0,1]}$ and family of Hamiltonians
\begin{equation}
H(s) = H_0 + s\, H_1,
\end{equation}
from which the mid-IR of ${H}$ can be constructed from the mid-IR of ${H_0}$ by
slowly tuning ${s=0}$ to ${s=1}$~\cite{HW0541}. Indeed, let us denote the
${n^{\txt{th}}}$ many-body energy eigenstate of ${H(s)}$ as ${|\psi^{(s)}_n\>}$
and define the unitary operator ${V_s = \sum_n |\psi^{(s)}_n\>\<\psi^{(0)}_n|}$
which satisfies ${V_s|\psi^{(0)}_n\> = |\psi^{(s)}_n\>}$ and
\begin{equation}
 \<\psi^{(0)}_n|A |\psi^{(0)}_n\> = \<\psi^{(s)}_n|V_s A V_s^\da|\psi^{(s)}_n\>
\end{equation}
for any operator ${A}$. Therefore the ${\<\cP_i \>=0}$ sector of ${H_0}$ is related to the ${\<V_1\cP_i V_1^\da\> =
0}$ sector of ${H}$. However, this definition of ${V_s}$ is unphysical since
${V_s\cP V_s^\da}$ is likely to be nonlocal even if ${\cP }$ is local.
\Rf{HW0541} found a local unitary operator ${U_{\txt{LU}}}$ that approximates
${V_s}$ very well while ensuring local operators remain local when dressed. An
explicit form of ${U_{\txt{LU}}}$ is~\cite{BHS10010344}
\begin{equation}\label{eqn:Ulu}
\begin{aligned}
U_{\txt{LU}} &= \cS'\left\{\exp[\ii\int_0^1 \dd s'~\cD_{s'}]\right\},\\
\cD_s &\equiv \ii \int \dd t~ F(t) \ee^{\ii H(s) t} \partial_s H(s) \ee^{-\ii H(s) t} ,
\end{aligned}
\end{equation}
where ${\cS'}$ denotes ${s'}$-ordering and ${F(t)}$ is a function satisfying a
particular set of requirements, such as ${F(t)=-F(-t)}$ which ensures ${\cD_s}$ is anti-Hermitian.  

Motivated by those results, here we assume that there exists a proper local
unitary operator $U_{\txt{LU}}$ with the following properties: 
\begin{enumerate}
\item it maps a local operator ${O_i}$ to a local operator ${\t{O}_i \equiv
U_{\txt{LU}} O_i U_{\txt{LU}}^\da}$ that acts on degrees of freedom near ${i}$ (the operator is fattened); 
\item it maps the $n$th eigenstate of ${H(0)}$ to a superposition of some eigenstates ${|
\psi^{(1)}_{n'}\>}$ of $H(1)$ with energy ${E_{n}^{(1)}-\del < E_{n'}^{(1)}
< E_{n}^{(1)}+\del}$, where ${E_{n}^{(1)}}$ is
the energy of the $n$th eigenstate of ${H(1)}$ and ${\del \ll E_\txt{min-IR}}$;
\item it does not break the symmetries of $H_0$ and $H_1$.
\end{enumerate}
If such a unitary operator satisfying these properties does not exist for a particular $H$ in parameter space, it means that the mid-IR does not exist at that point of parameter space (e.g., due to the gapped quasiparticles defining $E_{\txt{mid-IR}}$ condensing). The existence of ${U_{\txt{LU}}}$ is a conjecture, and we will obtain our results based on this conjecture. It would be interesting to see if one could apply the mathematical techniques and proofs developed in \Rf{YL220911242} to construct ${U_{\txt{LU}}}$ rigorously.

Consider an eigenstate $|\psi^{(0)}_n\>$ of $H(0)$ with energy
${E_{n}^{(0)} \ll E_\txt{mid-IR}}$, thus satisfying $\cP_i
|\psi^{(0)}_n\>=0$.  $U_{\txt{LU}}$ will map it to some eigenstates of $H(1)$
with eigenvalues much less then $ E_\txt{mid-IR}$, which satisfy $\t\cP_i
|\psi^{(1)}_n\>=0$, where ${\t\cP_i =U_{\txt{LU}} \cP_i U_{\txt{LU}}^\dag}$ is also a set of mutually
commuting local projectors. This is true for all mid-IR eigenstates of $H(0)$, so the mid-IR of ${H}$ can be identified as the sub-Hilbert space spanned by the mid-IR
eigenstates of ${H(0)}$ transformed by $U_{\txt{LU}}$. In other words, the mid-IR states of ${H}$ which span ${\cV^{(H)}_{\txt{mid-IR}}}$
satisfy $\t \cP_i = 0$. Therefore, any emergent low-energy structures of $H_0$ specified by the projectors $\cP_i$ become emergent low-energy structures of $H$ specified by the projectors $\t\cP_i$.
Since $\{ \cP_i\}$ and $\{\t \cP_i\}$ are related by a local unitary
transformation $U_\txt{LU}$, we expect the two exact low-energy structures to be
equivalent.

This result was obtained in \Rf{HW0541} for emergent $\Z_2$ and
$U(1)$ gauge symmetry, and proved rigorously for the ${\Z_2}$ case. In that case, the low energy subspace satisfies the modified Gauss law
${\t\rho_i= U_{\txt{LU}}\rho_i U_{\txt{LU}}^\da =  0}$ exactly and is exactly gauge invariant. We believe such a result
remains valid for more general situations.

Having identified a mid-IR of ${H}$, we now identify its emergent symmetries at ${E < E_{\txt{mid-IR}}}$. It is useful to adopt
the perspective that a symmetry is described/defined by an algebra of local
symmetric operators~\cite{CW220303596,MM220903370, MM230203028}. For instance, if the UV symmetries are
generated by the unitaries ${\{U_g\}}$, \ie ${[U_g,H_0] =[U_g,H_1] =0}$, then
the associated algebra of local symmetric operators is
\begin{equation}\label{UValgSymOp}
\cA_{\txt{UV}} = \{O_{\txt{UV}}~|~O_{\txt{UV}}U_g = U_g O_{\txt{UV}}~\forall~g\},
\end{equation} 
where ${O_{\txt{UV}}}$ is a local operator acting on the full Hilbert space
${\cV}$. Indeed, given ${\cA}$, one can recover the symmetry transformation
operators by finding all operators that commute with its elements.

In the mid-IR, operators that violate the ${\<\t\cP_i\>=0}$ constraint are not allowed. Therefore, the mid-IR symmetries are described by the algebra of local
symmetric operators
\begin{equation}\label{midIRalgSymOp2}
\begin{aligned}
\hspace{-10pt}\cA_{\txt{mid-IR}} &=\{
O_{\txt{mid-IR}}~|
~O_{\txt{mid-IR}} \t\cP_i=  \t\cP_i O_{\txt{mid-IR}}~\forall ~i,\\
&\hspace{60pt}
~\forall~O_{\txt{mid-IR}}\in \cA_{\txt{UV}}\}.
\end{aligned}
\end{equation}
The symmetry transformations are then all operators that commute with the elements of ${\cA_{\txt{mid-IR}}}$ as well as the projectors ${\t\cP_i}$. These will include the UV symmetry operators but could include
additional emergent symmetries, reflecting the possibility depicted in
Fig.~\ref{fig:HierScalesEx}. We refer to emergent symmetries identified by ${\cA_{\txt{mid-IR}}}$ as \txti{exact emergent symmetries} to emphasize that at ${E < E_{\txt{mid-IR}}}$, they are equally impactful as UV symmetries. Indeed, given only ${\cA_{\txt{mid-IR}}}$, one cannot distinguish between UV symmetries and exact emergent symmetries. Importantly, exact emergent symmetries are not approximate symmetries.

The exact emergent symmetries found using $\cA_{\txt{mid-IR}}$ arise from the commuting projectors $\t \cP_i$. Since these projectors are local, ${0}$-form symmetries will typically not appear as exact emergent symmetries. Indeed, since the charged operators of 0-form symmetries are local, these projectors would likely have to be nonlocal to forbid them from appearing in $\cA_{\txt{mid-IR}}$. In other words, even weakly breaking a ${0}$-form symmetry in the UV theory, ${\cA_{\txt{mid-IR}}}$ will typically include terms charged under the symmetry, explicitly breaking the symmetry at all scales. This recovers how emergent 0-form symmetries generally are not exact and instead approximate symmetries. 

However, this description of symmetry is capable of describing all generalizations of symmetries. So, the exact emergent symmetries can be higher-form symmetries. 

The charged operators of higher-form symmetries are nonlocal, winding around
nontrivial cycles of the lattice, and thus will never appear in
$\cA_{\txt{mid-IR}}$. Therefore, emergent higher-form symmetries are always
exact symmetries in the thermodynamic limit since they cannot be broken by
low-energy local operators. In other words, emergent higher-form symmetries are
robust against translation-invariant local perturbations\footnote{More
precisely, any translation-invariant ${k}$-local perturbation where ${k}$ is
much smaller than the linear system size measured in units of lattice spacing.}
of the UV theory and, thus, are topologically robust. All of this is true for
both invertible and non-invertible higher-form symmetries.\footnote{An exact emergent ${p}$-form symmetry below an energy scale implies ${q}$-form symmetries (${q<p}$) below that same energy scale generated by ``condensation defects''~\cite{RSS220402407,CCH220409025, BSW220805973, LRS220805982, BCH220807367, DT230101259}. While these are ${q}$-form symmetries, they transform ${p}$-dimensional operators. Therefore they too are exact emergent symmetries even when ${q=0}$. These ${p}$-dimensional objects carry generalized charges of the ${q}$-form symmetry~\cite{BS230402660, BBG230403789, BS230517159}.
} 

To build some intuition for why this is true, we consider a spacetime
picture~\cite{LLc0507191, WP12016409}. Suppose ${H_0}$ has a 1-form symmetry,
so loops carrying the symmetry charge cannot be cut open by
unitary time evolution, in ${(3+1)D}$. In spacetime, it means that its worldsheet will not
have any holes. Turning on ${H_1}$ and explicitly breaking the 1-form symmetry,
these loops can now be cut open so their worldsheets will have holes. When the
perturbation ${H_1}$ is small, these holes are also small and can be
coarse-grained away to yield a low-energy subspace with only closed worldsheets
and an exact emergent 1-form symmetry. However, when the perturbation is large,
these holes are larger than the worldsheets themselves, disintegrating them by the Higgs mechanism and
preventing a 1-form symmetry from emerging. It is straightforward to generalize this to a general higher-form symmetry, and it would be interesting to study this spacetime picture further using the renormalization group.

Furthermore, an exact emergent ${p}$-form symmetry below an energy scale $E$ in
${d}$ dimensional space implies the exact emergence of its dual symmetry below
the same energy scale $E$, which is a ${(d-p-1)}$-form symmetry when the ${p}$-form
symmetry is discrete. When ${d-p-1=0}$, this leads to an exact emergent dual ${0}$-form symmetry.

For lattices with vacancy defects, there can be nontrivial ${p}$-cycles involving a finite number of ${p}$-cells. Then, operators charged under an emergent ${p}$-form symmetry could appear in ${\cA_{\txt{mid-IR}}}$, making it an approximate symmetry. However, if the vacancy defect density is small, these nontrivial ${p}$-cycles are also small. Therefore they will disappear under coarse-graining, and the emergent ${p}$-form symmetry will become exact.

Emergent higher-form symmetries, therefore, have an associated energy \txti{and} length scale. The former is ${E_{\txt{mid-IR}}}$, designating at which energies the symmetry emerges. The latter is the length scale that the symmetry's operators are fattened by ${U_{\txt{LU}}}$.  If this energy scale goes to zero (this length scale goes to infinity), the symmetry is explicitly broken at all scales (is no longer well defined) and cannot emerge. 

One can possibly discover all of the emergent symmetries of a system in a
Hamiltonian-independent way by constructing ${\cA}$ at all energy scales for
each energy hierarchy in parameter space. However, it is desirable to have a
Hamiltonian description reflecting the emergent symmetries. The symmetries that
emerge at ${E < E_{\txt{mid-IR}}}$ are hidden from the UV Hamiltonian ${H}$
since it describes the dynamics of both states with ${\t\cP_i = 0}$ and
${\t\cP_i =1}$. Therefore, to make the emergent symmetries manifest, we will develop an effective mid-IR theory
${H_{\txt{mid-IR}}}$ that describes only the dynamics of states with
${\t\cP_i = 0}$. Since ${H}$ is a sum of only operators in
${\cA_{\txt{UV}}}$, the effective mid-IR theory ${H_{\txt{mid-IR}}}$ should be a sum of only
operators in ${\cA_{\txt{mid-IR}}}$. Therefore, symmetries of $H_\txt{mid-IR}$ will include the
UV symmetries but also include additional ones, which we will identify as
exact emergent symmetries. 

The effective mid-IR Hamiltonian should act only on the mid-IR Hilbert space
${\cV_{\txt{mid-IR}}}$. The most general form for it is 
\begin{equation}\label{eqn:genMidIRHam}
H_{\txt{mid-IR}} = \hspace{-7pt}\sum_{\hspace{5pt}\cO \in \cA_{\txt{mid-IR}}}  \hspace{-10pt}\left(C_{\cO}~\cO + \txt{h.c.}\right).
\end{equation}
The constants ${\{C_{O}\}}$ are renormalized versions of the UV parameters. One can in principle determine ${\{C_{\cO}\}}$ by requiring the spectra and correlation functions of ${H_{\txt{mid-IR}}}$ to match with the UV theory's for ${\ket{\psi}\in\cV_{\txt{mid-IR}}}$. Since UV theory is local, we require the effective theory to be a local Hamiltonian. Therefore, the greater number of operators involved in
${\cO}$ or the larger the region of the lattice ${\cO}$ acts on, the smaller
${C_{\cO}}$ is. We view the ability to define a local effective mid-IR Hamiltonian as a requirement for the mid-IR itself to be well defined. 

The mid-IR having an exact emergent symmetry means there is a transformation that leaves the ${H_{\txt{mid-IR}}}$ unchanged. This implies the existence of an emergent conservation law obeyed at ${E<E_{\txt{mid-IR}}}$ which cannot be broken by local operators. The existence of this exact emergent conservation law can be used as a definition of the existence of the exact emergent symmetry. For ${p}$-form symmetries, this conservation law means that at ${E<E_{\txt{mid-IR}}}$, there are only closed ${p}$-branes excitations.

It is important to note that this effective Hamiltonian is different than those found using, for instance, Brillouin-Wigner perturbation theory~\cite{HW1037} or Schrieffer-Wolff transformations~\cite{BDL11050675}. Indeed, these effective Hamiltonians describe the mid-IR of ${H_0}$ (the ${\<\cP_i\> = 0}$ subspace), where as the effective Hamiltonian Eq.~\eqref{eqn:genMidIRHam} describes the genuine mid-IR of ${H}$ (the ${\<\t\cP_i\> = 0}$ subspace).

The philosophy behind ${H_{\txt{mid-IR}}}$ is similar to that of effective field theory, where one writes down an effective action that includes all allowed terms. Its definition is physically reasonable but not rigorously derived. We will
not present a rigorous justification or proof of ${H_{\txt{mid-IR}}}$. Here, we state Eq.~\eqref{eqn:genMidIRHam} with the
restrictions on ${C_{\cO}}$ as the conjectured form of ${H_{\txt{mid-IR}}}$,
and we will examine the consequences of this conjecture.

\subsection{A holographic picture}\label{holoSubSec}

An algebra of local symmetric operators (e.g., Eqs.~\eqref{UValgSymOp}
and~\eqref{midIRalgSymOp2}) determines a non-degenerate braided fusion (higher)
category $\eM$~\cite{KZ220105726,CW220303596}. To emphasize its utility in
describing a symmetry, ${\eM}$ is called a \emph{symmetry category} or
\emph{symmetry topological-order}.\footnote{Symmetry category was called
\emph{categorical symmetry} in \Rfs{JW191213492, KZ220105726}.  Since the term
categorical symmetry is now commonly used to mean non-invertible/algebraic
symmetry, we rename \emph{categorical symmetry} to \emph{symmetry category} to
avoid confusion.} For a finite symmetry, its symmetry category $\eM$ describes
a topological order in one higher dimension, which we also denote by $\eM$.
\Rfs{KZ200308898, KZ200514178} proposed a unified description of all types of
symmetries (including their higher-form, higher-group, anomaly, and
non-invertibility properties) using such topological order in one higher
dimension, which leads to a symmetry/topological order (Symm/TO)
correspondence~\cite{JW191213492,KZ220105726}.  The bulk topological order (\ie
the symmetry topological-order) can be realized by a topological field theory
(TFT).  To emphasize its utility in describing symmetry, this bulk TFT is
called symmetry TFT~\cite{AS211202092}.

\begin{figure}[t]
\centering
    \includegraphics[width=.48\textwidth]{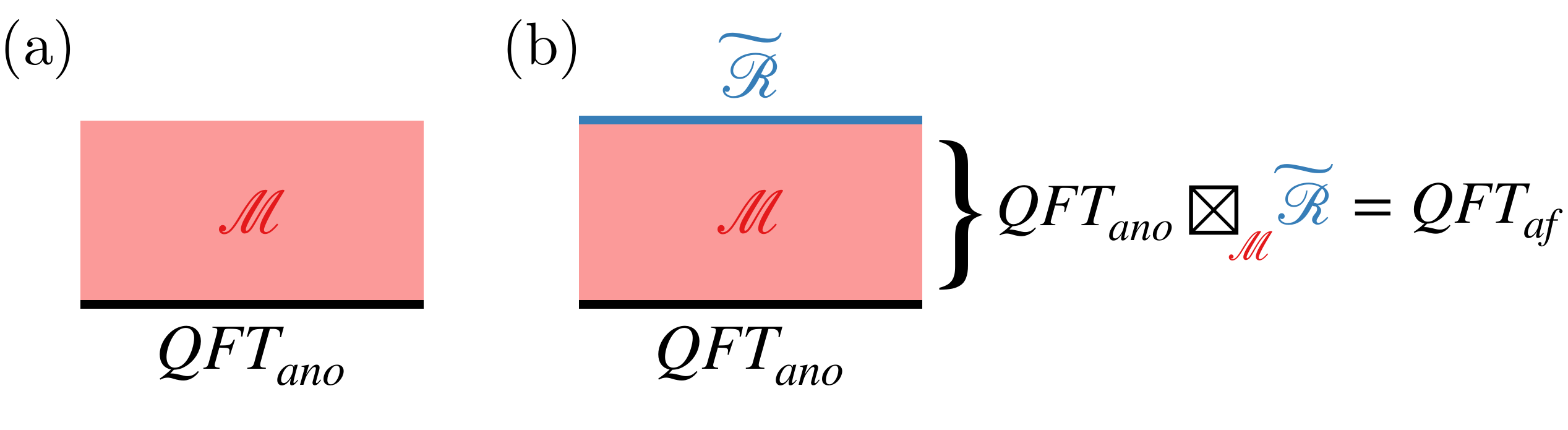}
    \caption{(a) The low
energy properties of $QFT_{ano}$ (${QFT_{af}}$ restricted to the ${\cR}$-symmetric sub-Hilbert space) can be exactly simulated by a boundary of a
topological order $\eM$ in one higher dimension.  $QFT_{ano}$ uniquely
determines $\eM$. (b) The low energy properties of $QFT_{af}$ throughout all ${\cR}$ sectors can be simulated by including the boundary $\t\cR$.} \label{fig:symTO}
\end{figure}

More precisely, given an anomaly-free system\footnote{An anomaly-free system is
one with a lattice UV completion.} $QFT_{af}$ with a symmetry, its low
energy properties within the symmetric sub-Hilbert space are exactly simulated
by a boundary of a topological order $\eM$ in one higher dimension. Indeed,
$QFT_{af}$ restricted within the symmetric sub-Hilbert space can be viewed as a
new system $QFT_{ano}$ with a non-invertible gravitational
anomaly~\cite{JW190513279,JW191213492}, which corresponds to $\eM$ in one
higher dimension~\cite{W1313,KW1458} (see Fig.~\ref{fig:symTO}a).

Fully simulating $QFT_{af}$ requires the boundary of ${\eM}$ to also capture
states outside the symmetric subspace. This can be achieved by adding an
additional gapped boundary $\t \cR$ of  $\eM$~\cite{KZ200514178, FT220907471},
as shown in Fig.~\ref{fig:symTO}b. Provided that the topological order $\eM$
and the boundary $\t\cR$ have an infinite energy gap, the low-energy properties
of ${QFT_{af}}$ are described by the composition of the topological order $\eM$
with two boundaries ${QFT_{ano}}$ and $\t\cR$, which we denote as ${QFT_{ano}
\boxtimes_{\eM} \t\cR}$.  

If the spatial dimension of the boundary $\t\cR$ is ${d}$, its gapped
excitations are described by a fusion ${d}$-category, which we will also denote
as ${\t\cR}$. On the other hand, the excitations of $\eM$ are described by a
braided fusion $d$-category denoted as $\eM$.  The boundary fusion $d$-category
$\t\cR$ uniquely determines the bulk braided fusion $d$-category $\eM$, and are
related to one another by
\begin{align}\label{BounBulkRel}
 \eM =\eZ(\t\cR),
\end{align}
where ${\eZ(\t\cR)}$ is the center of $\t\cR$. When ${d=1}$, ${\eZ(\t\cR)}$ is
the Drinfeld center of ${\t\cR}$.  

We say $QFT_{af}$ is described by the symmetry category ${\eM}$ if admits a
decomposition ${QFT_{af} = QFT_{ano} \boxtimes_{\eM} \t\cR}$.  The
decomposition also implies that $QFT_{af}$ has a symmetry whose
symmetry defects (\ie symmetry transformations) are described by fusion
$d$-category ${\t\cR}$.  We will call such a symmetry as $\t\cR$-symmetry.  For
example, ${d\txt{-}\cVec_G}$-symmetry is the ordinary global symmetry described by a
group $G$.

If $\t\cR$ is a local\footnote{A fusion $d$-category $\t\cR$ is \emph{local} if
there exists a fusion $d$-category $\cR$ such that ${\eZ(\cR)=\eZ(\t\cR)}$ and
${\cR \boxtimes_\eM \t\cR = n\eVec}$, where $d\eVec$ is the braided fusion
$d$-category describing excitations in a trivial topological order (i.e., above
a trivial product state).  Such $\cR$ is called the dual of $\t\cR$.} fusion
$d$-category, then the $\t\cR$-symmetry is an anomaly-free  symmetry.  The
symmetry charges of an anomaly-free $\t\cR$-symmetry are described by a fusion
$d$-category $\cR$, which is the dual of $\t\cR$.  However, if $\t\cR$ is not
local, the $\t\cR$-symmetry is anomalous.  We note that in \Rf{FT220907471},
the pair ${(\t\cR, \eM)}$ is regarded as a generalized symmetry regardless if
$\t\cR$ is local or not. Thus, such a description includes both anomaly-free and
anomalous symmetries.

Using ${QFT_{ano} \boxtimes_{\eM} \t\cR}$ to describe the symmetries of
${QFT_{af}}$ is very general and provides a unifying formalism capable of
describing all generalizations of symmetry. As we will now argue,
${QFT_{ano} \boxtimes_{\eM} \t\cR}$ is also able to describe the exact emergent
symmetries of ${QFT_{af}}$ discussed in the previous subsection.

Recall from the previous subsection the general Hamiltonian~\eqref{genH}, where a known mid-IR of ${H_0}$ was spanned by states satisfying ${\cP_i\ket{\psi} = 0}$ for local commuting projectors ${\{\cP_i\}}$. Any exact emergent symmetries in the mid-IR are determined entirely by ${\{\cP_i\}}$, and the exact emergent symmetries of ${H}$ could be found by dressing ${\{\cP_i\}}$ with ${U_{\txt{LU}}}$. Here for simplicity, we will just consider ${H_0}$ since our results will hold even after we include
an arbitrary translation-invariant perturbation ${H_1}$, as we discussed in the last subsection. Without a loss of generality, we will assume that ${H_0}$ has the form
\begin{align}
\label{H0OP}
 H_0 = \sum_i O_i + E_\text{mid-IR} \sum_i \cP_i,
\end{align}
where $O_i$ are local operators and ${[O_i, \cP_j]=0}$ is required since $O_i$
is assumed not to mix the ${\cP_i = 0}$ and ${\cP_i = 1}$ states. Thus, the local
energy dynamics controlled by $O_i$ are constrained by $\cP_i$, and consequently, the local projectors $\{\cP_i\}$ can give rise to an emergent symmetry.

It is believed that a topological order with a gapped boundary can be realized
by a commuting projector model.  Therefore, the $\t\cR$-boundary and the
$\eM$-bulk in Fig.~\ref{fig:symTO}b can be realized by a commuting projector model.
To have $H_0$ as the
Hamiltonian described by the slab in Fig.~\ref{fig:symTO}b, we take $\cP_i$ in $H_0$ to be those commuting projectors. $O_i$'s in $H_0$ are the boundary Hamiltonian
terms describing the $QFT_{ano}$ boundary in Fig.~\ref{fig:symTO}b.

We also enlarge ${\{O_i\}}$ to include local operators in the bulk $\eM$, and require that they still commute with ${\cP_i}$. Since the thickness of the bulk is finite, ${\{O_i\}}$ can include operators that connect the $QFT_{ano}$
and $\t\cR$ boundaries, which we will refer to as inter-boundary operators. There are also intra-boundary operators, which are those that act only on the degrees of freedom near the boundary $QFT_{ano}$. To explicitly break a symmetry on ${QFT_{ano}}$, ${\{O_i\}}$ must include an inter-boundary operator that transfers symmetry charge from ${\t\cR}$ to ${QFT_{ano}}$. Doing this for a ${p}$-form symmetry requires a ${(p+1)}$-dimensional operator.

For a ${0}$-form symmetry, such an operator would include a finite number of local operators acting in a line from ${\t\cR}$ to ${QFT_{ano}}$. This is a local operator and thus allowed in ${H_0}$. It is unlikely local projectors $\cP_i$ can forbid such an operator, and therefore they cannot produce exact emergent 0-form symmetries. For emergent higher-form symmetries, any inter-boundary operators that transfer symmetry charge are non-contractible extended operators, acting on the whole system. These operators are not local and are not included in the set of local operators $\{O_i\}$. Thus emergent higher-form symmetries are exact.

\begin{figure}[t]
\centering
    \includegraphics[width=.48\textwidth]{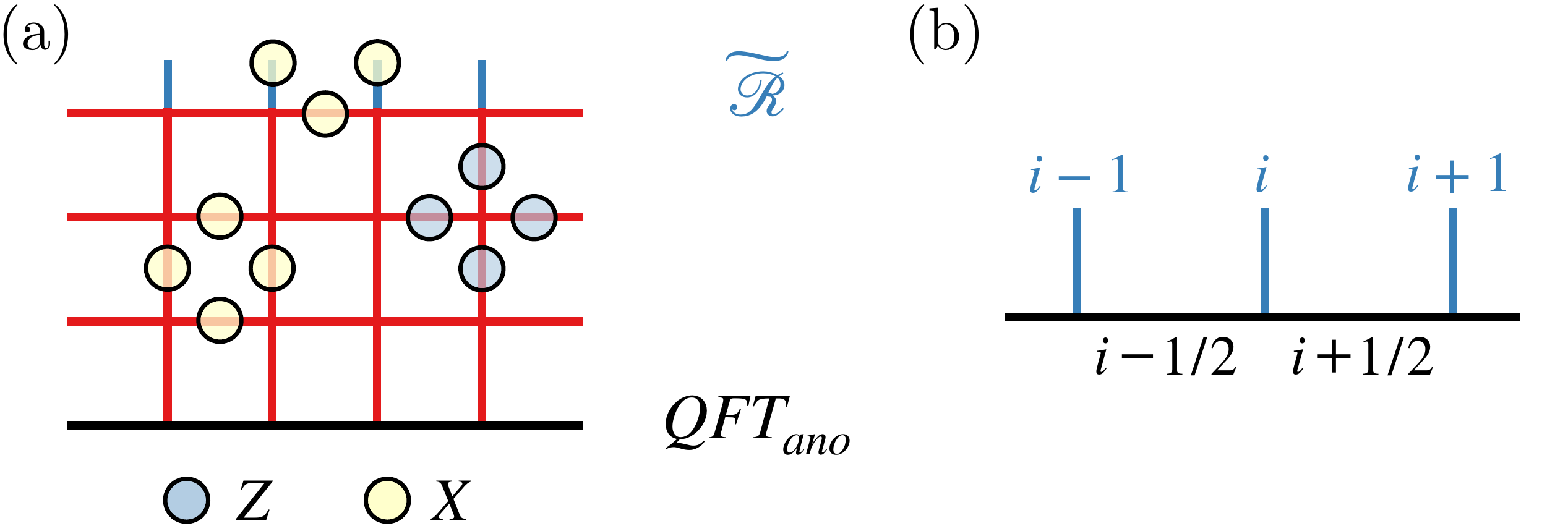}
    \caption{(a) A lattice realization of Fig.~\ref{fig:symTO}b, where qubits live on
the links. The star and plaquette terms of the toric code model are shown both in the bulk and on the ${\t\cR}$ boundary. (b) Since the bulk is topological, one can take the thin slab limit.}
\label{fig:latSymTO}
\end{figure}

The discussion so far has been pretty general, so let us consider an example of a model with an exact emergent $\Z^{(1)}_2$ symmetry in 1+1D. We consider ${\eM = \cZ(\cVec_{\Z_2})}$, which corresponds to the ${\Z_2}$ toric code model, and $\t\cR$ the $\Z_2$-charge condensed boundary (the rough boundary~\cite{KK11045047}). The slab ${QFT_{ano} \boxtimes_{\eM} \t\cR}$ in Fig.~\ref{fig:symTO}b becomes the lattice shown in Fig.~\ref{fig:latSymTO}a, with qubits residing on the links.

The toric code model contains two types of projectors: star terms ${\cP_i^{vert}= (1-Z_1Z_2Z_3Z_4)/2}$ acting on the qubits of the four links touching each vertex and plaquette terms ${\cP_i^{plaq}=(1-X_1X_2X_3X_4)/2}$ acting on the qubits of the four links around each square.  The $\t\cR$ boundary has truncated plaquette terms ${\cP_i^{bdry}=\frac 12 (1 - X_1X_{2}X_{3})}$, as shown in Fig.~\ref{fig:latSymTO}a.

Since the bulk is topological, let us take a thin slab limit of Fig.~\ref{fig:latSymTO}a, which gives us Fig.~\ref{fig:latSymTO}b where we label vertical (horizontal) links by $i$ (${i+\frac12}$). The only remain projector surviving this limit is ${\cP^{bdry}_i = \frac 12 (1 - X_iX_{i+\frac12}X_{i+1})}$, and thus all allowed $O_i$'s are generated by products of $X_i$,
$X_{i+\frac12}$, and ${Z_{i-\frac12}Z_{i}Z_{i+\frac12}}$. Notice that while $X_{i+\frac12}$, ${X_iX_{i+\frac12}X_{i+1}}$, ${Z_{i-\frac12}Z_{i}Z_{i+\frac12}}$ are intra-boundary operators, the $X_i$'s are inter-boundary operators.

Let us first restrict ourselves to only the
intra-boundary operators. We will consider the full setup, where ${O_i}$ includes intra and inter-boundary operators, after. The intra-boundary operators form an algebra of the local symmetric operators generated by
\begin{align}
\hspace{-2pt}\cA^{(\txt{intra-only})}_\txt{mid-IR} \hspace{-2pt}=\{ X_{i+\frac12}, \hspace{1pt} X_iX_{i+\frac12}X_{i+1}, \hspace{1pt} Z_{i-\frac12}Z_{i}Z_{i+\frac12} \}.  
\end{align}
The operators (local or non-local) that commute with $\cA^{(\txt{intra-only})}_\txt{mid-IR}$ are generated by
\begin{align}
\cT^{(\txt{intra-only})}_\txt{mid-IR} =\{ X_iX_{i+\frac12}X_{i+1},~ \prod_i (Z_{i-\frac12}Z_iZ_{i+\frac12}) \},
\end{align}
and give rise to all symmetry transformations. Therefore, when restricted to the intra-boundary operators, there are two mid-IR symmetries arising from $\cP_i$. ${X_iX_{i+\frac12}X_{i+1}}$ acts on loops and corresponds to a ${\Z_2^{(1)}}$ symmetry, while ${\prod_i (Z_{i-\frac12}Z_iZ_{i+\frac12})}$ acts on the entire lattice and corresponds to a ${\Z_2^{(0)}}$ symmetry. When the ${(2+1)}$D system ${QFT_{ano} \boxtimes_{\eM} \t\cR}$ is mapped to the ${(1+1)}$D system ${QFT_{af}}$, the ${\Z_2^{(0)}}$ symmetry still acts on the entire lattice while the ${\Z_2^{(1)}}$ symmetry now acts on a single lattice site.

Let us now include the inter-boundary operators. Doing so, the algebra of local symmetric operators becomes
\begin{align}
\cA_\txt{mid-IR} =\{ X_{i+\frac12},~ X_i,~ Z_{i-\frac12}Z_{i}Z_{i+\frac12} \},
\end{align}
and the symmetry transformations are now generated by
\begin{align}
\cT_\txt{mid-IR} =\{ X_iX_{i+\frac12}X_{i+1} \}.
\end{align}
The ${\Z_2^{(1)}}$ symmetry is still present, but the ${\Z_2^{(0)}}$ symmetry is now gone. This is because the allowed inter-boundary operators $X_i$ transfer the charges of the
$\Z_2^{(0)}$ symmetry between the two boundaries, breaking the $\Z^{(0)}_2$ symmetry. The operators that could transform the ${\Z_2^{(1)}}$ symmetry charges between the two boundaries were not included in ${\cA_\txt{mid-IR}}$ since they are not local operators, and as a result, the ${\Z_2^{(1)}}$ symmetry is still a mid-IR symmetry.

\begin{figure*}[t!]
\centering
    \includegraphics[width=.98\textwidth]{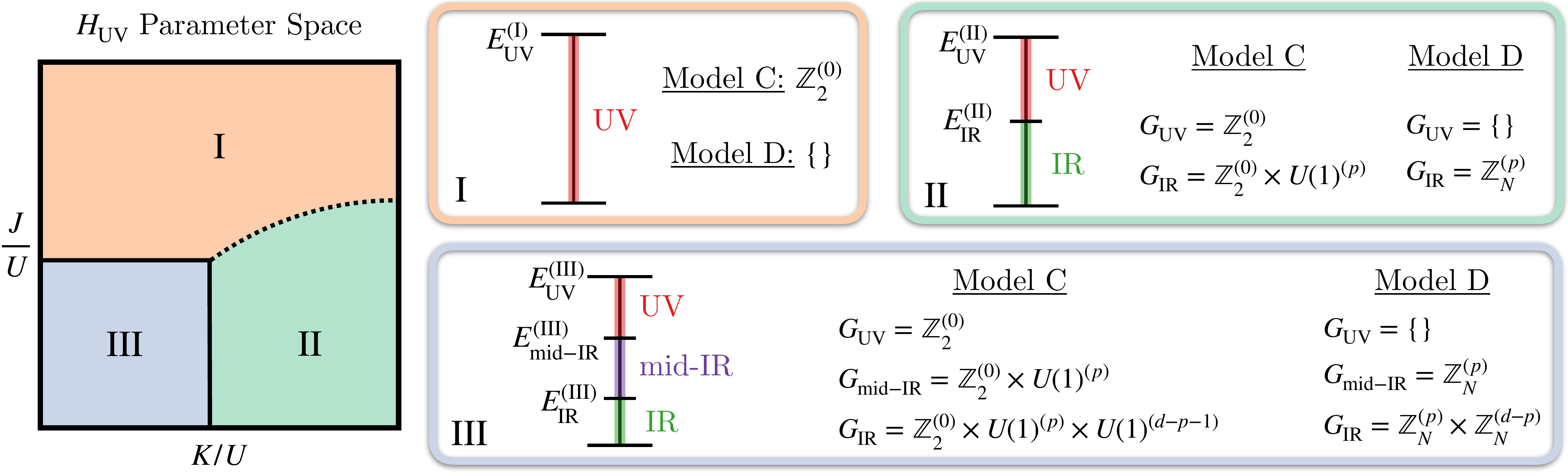}
    \caption{(left) The parameter space of models C~\eqref{2bodyU1Ham}
and D~\eqref{ZNuvHam} can be partitioned into three regions, I, II, and
III, corresponding to its different energy scale hierarchies. These regions are
not necessarily distinct phases of the models. Here, solid lines indicate a
phase transition while dashed lines do not. (right) In these regions, we
identify the exact emergent symmetries at each energy scale. Here ${p}$ is a
positive integer and $d$ is the dimension of space such that ${d > p + 1}$ (${d
> p}$) for model C (D). The emergent $\U{d-p-1}$ symmetry in region III for
model C is only nontrivial in the continuum limit.  When ${J/U = 0}$ (${K/U =
0}$), the exact emergent ${\U{p}}$ and ${\ZN{p}}$ (${\U{d-p-1}}$ and
${\ZN{d-p}}$) symmetries of models C and D, respectively, are exact UV
symmetries.  } \label{fig:phaseDia} 
\end{figure*}

\section{Examples of exact emergent higher-form symmetries}\label{exactEmEx}

In this section, using the point of view discussed in
section~\ref{effHamSec111}, we go through examples of lattice models without higher-form symmetries that have exact emergent higher-form symmetries. These models are all described by Hamiltonians governing degrees
of freedom on a ${d}$-dimensional cubic spatial lattice.
We extensively use discrete differential geometry notation, which we review in appendix section~\ref{sec:diffgeoLat}. The remainder of this section is organized as follows, with each subsection dedicated to a single model and entirely self-contained.

In
subsection~\ref{mainSec:qClockModel}, we consider the quantum
clock model (Eq.~\eqref{qClockModelHam}). When its UV ${\Z_N}$ symmetry is
spontaneously broken, we find there is an exact emergent ${\ZN{d}}$ symmetry at
energies below the domain wall gap. The IR symmetry operators form a projective representation of ${\Z_N\times\ZN{d}}$, signaling the presence of a mixed 't Hooft anomaly that protects the ground state degeneracy in the SSB phase.

In subsection~\ref{sec:modelDEES}, we consider a model of emergent ${\Z_N}$
${p}$-gauge theory called model D (Eq.~\eqref{ZNuvHam}), and in subsection~\ref{sec:modelCEES} a model of emergent ${U(1)}$
${p}$-gauge theory called model C (Eq.~\eqref{2bodyU1Ham}). The exact
emergent symmetries and energy scale hierarchies of these models are summarized
in Fig.~\ref{fig:phaseDia}. The left panel of
Fig.~\ref{fig:phaseDia} is a schematic depiction, and we do not investigate the precise shapes of the regions nor the nature of their boundaries. Region $\mathrm{III}$ corresponds to the deconfined
phase of the emergent gauge theory. Region $\mathrm{II}$ corresponds to the confined ``phase,''
where the gauge charges are still well-defined gapped excitations and confined. Fig.~\ref{fig:phaseDia} and our discussion throughout subsections~\ref{sec:modelDEES} and~\ref{sec:modelCEES} portray the possibility
that region II exists for ${J\neq 0}$. Lastly, region $\mathrm{I}$ is where the gauge charges are condensed/are no longer well-defined gapped excitations.

Model D with ${p=1}$, ${d=2}$, and ${N=2}$ is emergent
$(2+1)$D $\Z_2$ lattice gauge theory. In Eq.~\eqref{ZNuvHam}, the ${J}$ term creates $\Z_2$ charge fluctuations and the ${K}$ term creates $\Z_2$ flux fluctuations, both breaking UV $\Z^{(1)}_2$ symmetries. Region III corresponds to the deconfined phase of the $\Z_2$ gauge theory, and in the IR, there is a spontaneously broken exact emergent anomalous $\Z_2^{(1)}\times \Z_2^{(1)}$ symmetry. This mixed anomaly is characterized by the SPT
(see Eq. \eqref{ZinvZN})
\begin{align}
\label{Z2Z2}
\cZ[\cA,\h\cA] = \ee^{\ii \pi \int_{M^{4}} \cA \cup \hat \cA},
\end{align}
where $\cA$ and $\h\cA$ are $\Z_2$-valued 2-cocycles, and protects the ground
state degeneracy on a torus. Large $J$
drives a $\Z_2$-charge condensation transition (\ie a Higgs transition) and
large $K$ drives a $\Z_2$-flux condensation transition (\ie a confinement
transition). When the $\Z_2$ gauge charges have an energy gap, the exact emergent $\Z_2^{(1)}$ symmetry is present on both sides of the confinement transition, ${\mathrm{II}\leftrightarrow \mathrm{III}}$, controlling the transition, and its
unphysical part corresponds to the exact emergent $\Z_2$ gauge
redundancy~\cite{HW0541}.

Model C with ${p=1}$ and ${d=3}$ is emergent ${(3+1)}$D $U(1)$ lattice gauge theory. In Eq.~\eqref{2bodyU1Ham}, the ${J}$ creates $U(1)$-charge fluctuations and the ${K}$ term creates magnetic monopole fluctuations. Region III corresponds to the deconfined
phase of the $U(1)$ gauge theory, and in the continuum limit of the IR, there is a spontaneously broken exact emergent ${U(1)^{(1)}\times U(1)^{(1)}}$ symmetry. This mixed 't Hooft anomaly is characterized by the SPT
(see Eq. \eqref{U1U1})
\begin{align}
\label{U1U1a}
\cZ[\cA,\h\cA] = \ee^{\ii 2 \pi \int_{M^5} \cA \wedge \dd \hat \cA},
\end{align}
where $\cA$ and $\hat \cA$ are $\R/\Z$-valued 2-form fields, and protects the gaplessness of the photon~\cite{W9125,DM190806977}. Large $J$ drives a $U(1)$-charge condensation transition (\ie a Higgs
transition) and large $K$ drives a magnetic monopole condensation transition
(i.e., a confinement transition). When the ${U(1)}$ charges have an energy gap, the exact emergent $U(1)^{(1)}$ symmetry is present on both sides of confinement transition II
$\leftrightarrow$ III, controlling the transition, and its unphysical part corresponds to the exact emergent $U(1)$ gauge
redundancy~\cite{HW0541}.

\subsection{Quantum clock model}\label{mainSec:qClockModel}

Consider
${\Z_N}$ quantum rotors residing on the 0-cells (sites) of the spatial
${d}$-dimensional cubic lattice at zero temperature. These are described by the clock operators ${X_{c_0}}$ and ${Z_{c_0}}$, which are unitary
operators satisfying 
\begin{equation}\label{ZNclockAlg}
\begin{aligned}
Z_{\t{c}_0}X_{c_0} &= \om^{\delta_{c_0,\t{c}_0}}X_{c_0}Z_{\t{c}_0},\quad\quad\quad X_{c_0}^{N}  &= Z_{c_0}^{N} = \one,
\end{aligned}
\end{equation}
where ${\om\equiv \ee^{\ii 2\pi/N}}$. They are ${N}$-dimensional generalizations of the Pauli matrices and have eigenvalues ${\{1,\om,\om^2,\cdots,\om^{N-1}\}}$. The Hamiltonian of the quantum clock model is
\begin{equation}\label{qClockModelHam}
H = -\frac{J}{2}\sum_{c_1} \prod_{c_0\in\pp c_1}X_{c_0}  - \frac{K}{2}\sum_{c_0} Z_{c_0} + \txt{h.c.},
\end{equation}
where ${X_{-c_0}\equiv X^\da_{c_0}}$, the first sum is over 1-cells,  and the second sum is over all 0-cells. This theory has an exact ${\Z_N}$ 0-form---${\ZN{0}}$---symmetry, which is generated by
\begin{equation}\label{clockModelSym}
U = \prod_{c_0} Z_{c_0}.
\end{equation}
The charged operator of this symmetry is ${X_{c_0}}$, which from the clock operator algebra transform as ${X_{c_0} \to \ee^{2\pi \ii/N} X_{c_0}}$. Therefore, the algebra of local symmetric operators is generated by
\begin{equation}\label{cloclLocAlgUV}
\cA_{\txt{UV}} = \{Z_{c_0}, \prod_{c_0\in\pp c_1}X_{c_0}\}.
\end{equation}

\subsubsection{An exact emergent ${\ZN{d}}$ symmetry and mixed 't Hooft anomaly}\label{qClockModel}

When ${K/J \ll 1}$, the quantum clock model lies in a ${\ZN{0}}$ spontaneous symmetry broken (SSB) phase. Indeed, in the tractable ${K/J\to 0}$ limit, the ground state satisfies ${X^\da_{c_0}X_{\t{c}_0} \gs = \gs}$ for all neighboring 0-cells, and thus ${\< X^\da_{c_0}X_{c'_0} \> = 1}$ for any 0-cells ${c_0}$ and ${c'_0}$. In this phase, there are gapped domain walls carrying ${\Z_N}$ topological charge. Indeed, in the ${K/J\to 0}$ limit, the domain-wall density ${\h\rho}$ for a state ${\ket{\psi}}$ is defined by
\begin{equation}\label{0formZnTopDefDen}
\prod_{c_0\in\pp c_{1}} X_{c_0}\ket{\psi} = \ee^{\frac{2\pi\ii}{N} (\hstar\h{\rho})_{c_{1}}}\ket{\psi},
\end{equation}
where ${(\hstar\h{\rho})_{c_{1}}\equiv \h{\rho}_{\hstar c_{1}}}$. Therefore, the operator ${(\hstar Z)_{\h{c}_{d}}}$ excites a domain wall on ${\pp \h{c}_{d}}$.

The domain wall gap ${J}$ provides a candidate energy scale below which new symmetries may emerge. Let us call this energy scale the IR. However, when ${K/J\neq 0}$, there no longer exists a low-energy sub-Hilbert space spanned by states satisfying ${\<\h\rho_{\h c_{d-1}}\> = 0}$ mod ${N}$. This is because the ${K}$ term in ${H}$ causes the ${\<\h \rho_{\h c_{d-1}}\> = 0}$ and ${\<\h \rho_{\h c_{d-1}}\> \neq 0}$ states to mix. This does not necessarily mean the domain walls no longer exist when ${K>0}$, just that their operators depend on ${K}$. Indeed, a corresponding low-energy sub-Hilbert space can be identified using ${U_{\txt{LU}}}$ from section~\ref{effHamSec111}. Therefore, there exists a low-energy sub-Hilbert space for ${K/J\ll 1}$ in the SSB phase spanned by states satisfying ${\<\t{\h\rho}_{\h c_{d-1}}\> \equiv \<U_{\txt{LU}} \h\rho_{\h c_{d-1}}U_{\txt{LU}}^\da\> = 0}$. By the definition of ${U_{\txt{LU}}}$, the ${\ZN{0}}$ symmetry operator satisfies
\begin{equation}\label{dressclockModelSym}
U = \t U = \prod_{c_0} \t{Z}_0.
\end{equation}

Having identified the IR of the SSB phase, we would now like to find an effective IR theory. This IR satisfies the constraint ${\t{\h{\rho}}_{\h{c}_{d-1}} = 0}$, or equivalently ${\prod_{c_0\in\pp c_{1}} \t X_{c_0} = 1}$. Due to the constraint, the ${\Z_N^{(0)}}$ symmetric IR operators must be constructed from only ${\t{Z}_0}$. Only one such operator commutes with the constraint: Eq.~\eqref{dressclockModelSym}. Therefore, for a finite-size system, the algebra of local symmetric IR operators is
\begin{equation}
\cA_{\txt{IR}}^{\txt{finite-}L} = \{ \prod_{c_0} \t{Z}_0\} = \{U\}.
\end{equation}
The corresponding effective IR Hamiltonian is 
\begin{equation}
H^{\txt{finite-}L}_{\txt{IR}} =-J\ka^{N_0}~U,
\end{equation}
where ${\ka\sim K/J}$ and ${N_0}$ is the total number of ${0}$-cells. The ${K/J}$ dependence of ${\ka}$ comes from the fact that ${H=0}$ in the IR when ${K/J\to 0}$. 

In the thermodynamic limit, ${U}$ is a nonlocal operator, so the algebra of local symmetric IR operators becomes
\begin{equation}
\cA_{\txt{IR}} = \{\}.
\end{equation}
Indeed, since ${K/J \ll 1}$, in the thermodynamic limit the effective IR Hamiltonian becomes ${H_{\txt{IR}} = 0}$. Since ${\cA_{\txt{IR}}}$ is the empty set (or equivalently, since ${H_{\txt{IR}}}$ is zero), any IR operator commutes with the local symmetric IR operators and thus corresponds to a symmetry. This includes ${U}$, and thus the UV ${\Z_N^{(0)}}$ symmetry, as expected. However, the operator ${\t X_{c_0}}$ is also allowed and it generates the transformation
\begin{equation}\label{0ZNlatticeMagSym}
U\to \ee^{\frac{2\pi\ii}N } U.
\end{equation}
Since the charged object is supported on ${d}$-cycles and transforms by an element of ${\Z_N}$, the operator ${\t X_{c_0}}$ generates a ${\ZN{d}}$ symmetry (which is always a higher-form symmetry). 

This emergent ${\ZN{d}}$ symmetry has been noted previously throughout the literature~\cite{KS14010740, IM200304349, VBV221101376}. Indeed, it is an emergent symmetry since it does not commute with ${H}$. Here we find it is an exact emergent symmetry since it exactly commutes with the IR effective Hamiltonian. Therefore, the ground state subspace of the ${\Z_N}$ SSB phase has an exact ${\ZN{0}\times\ZN{d}}$ symmetry. Furthermore, this IR symmetry is anomalous, which can be noticed from the fact the symmetry operator of the ${\ZN{0}}$ symmetry is charged under the ${\ZN{d}}$ symmetry. This mixed 't Hooft anomaly protects the ground state degeneracy of the SSB phase. The only way to eliminate it is to prevent the ${\ZN{d}}$ symmetry from emerging by condensing domain walls.

\subsection{Emergent ${\Z_N}$ ${p}$-gauge theory}\label{sec:modelDEES}

\begin{figure}[t!]
\centering
    \includegraphics[width=.48\textwidth]{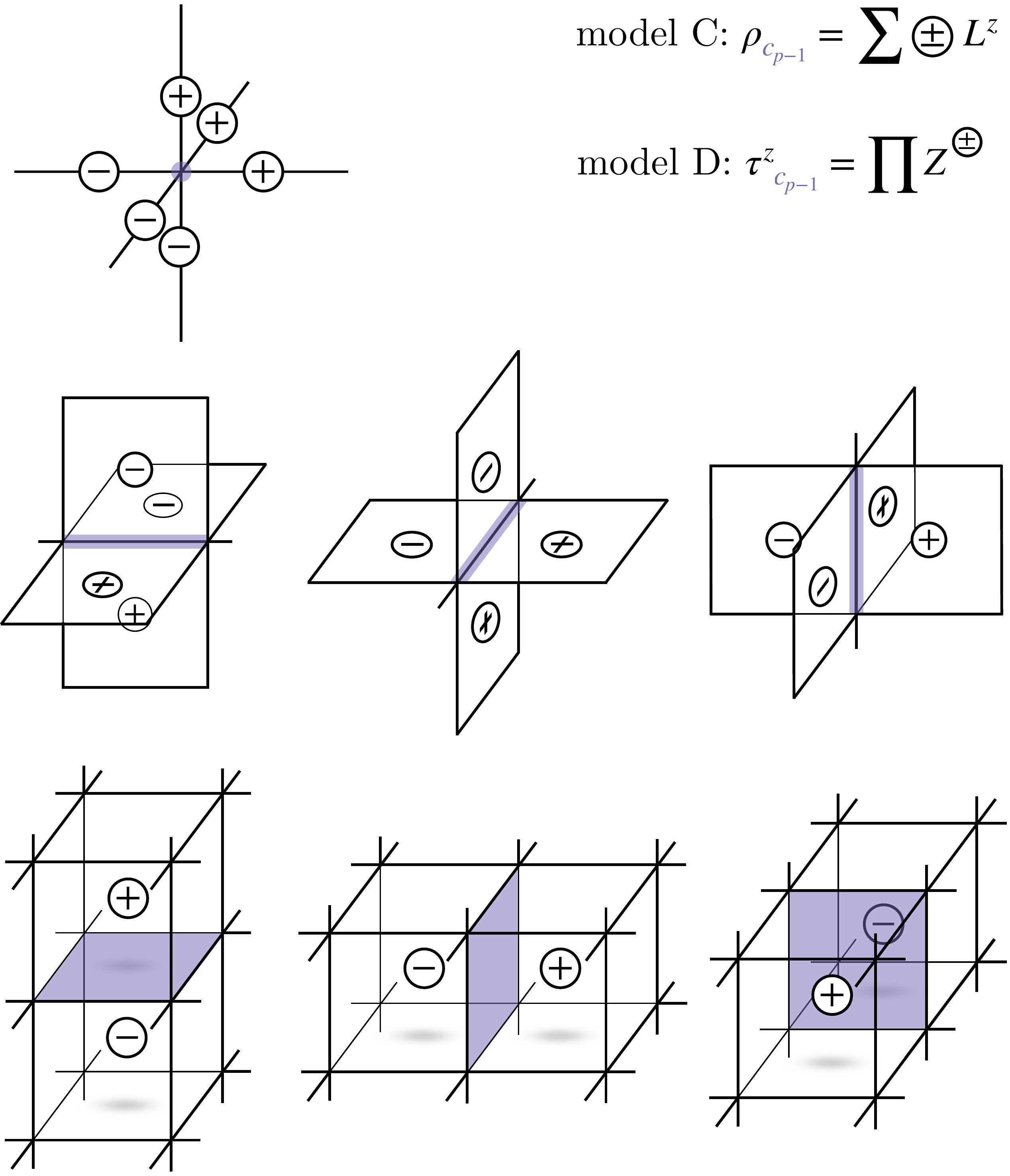}
    \caption{%
    Graphical representation of ${\rho_{c_{p-1}}}$ in model C~\eqref{eqn:rhoOp} and ${\tau^z_{c_{p-1}}}$ in model D~\eqref{ZNuvHam} in 3d space for (1st row) ${p=1}$, (2nd row) ${p=2}$, and (3rd row) ${p=3}$. The ${(p-1)}$-cell ${c_{p-1}}$ is colored purple. The ${\pm}$ disks on the ${p}$-cells denote the sign in front of ${L^z_{c_p}}$ in the sum for ${\rho_{c_{p-1}}}$, or whether ${Z_{c_p}}$ is ${Z^+\equiv Z}$ or ${Z^-\equiv Z^\da}$  in the product for ${\tau^z_{c_{p-1}}}$.
    }
    \label{fig:rhoOpp}
\end{figure}

In this section, we consider a model for emergent ${\Z_N}$ ${p}$-gauge theory, which we call model D. When ${p=1}$, this is just typical ${\Z_N}$ gauge theory. Consider ${\Z_N}$ quantum rotors residing on the ${p}$-cells of the spatial ${d}$-dimensional cubic lattice with ${p>0}$. A ${\Z_N}$ quantum rotor is an ${N}$-level system described by clock operators ${X_{c_p}}$ and ${Z_{c_p}}$ which satisfy Eq.~\eqref{ZNclockAlg}. Model D is described by the Hamiltonian
\begin{align}\label{ZNuvHam}
&H_{\txt{UV}} \hspace{-2pt}= -\frac{U}2\sum_{c_{p-1}}\tau^z_{c_{p-1}}  - U \sum_{c_{p+1}}W^{\da}_{c_{p+1}}\nonumber\\&\hspace{40pt}- \frac{K}2 \sum_{c_p} Z_{c_p}  - \frac{J}2\sum_{c_p} X_{c_p} + \txt{h.c.},\\
&\tau^z_{c_{p-1}} \equiv \prod_{c_p\in\del c_{p-1}}Z_{c_p},\quad\quad,W^{\da}_{c_{p+1}} = \prod_{c_p\in \pp c_{p+1}}X_{c_p}\nonumber
\end{align}
where ${\sum_{c_{p}}}$ is over all ${p}$-cells, ${\del c_{p-1}}$ is the coboundary of ${c_{p-1}}$ (see Eq.~\eqref{coboundaryDef}), and${Z_{-c_p} \equiv
Z_{c_p}^\da}$. ${\tau^z}$ is generally a product of ${2(d-p+1)}$ operators,
examples of which are shown in Fig.~\ref{fig:rhoOpp}.

Since ${H_{\txt{UV}}}$ has terms
linear in ${Z_{c_p}}$ and ${X_{c_p}}$, the algebra of local symmetric UV operators is generated by
\begin{equation}
\cA_{\txt{UV}} = \{Z_{c_p}, X_{c_p}\}.
\end{equation}
Nothing commutes with both
${Z_{c_p}}$ and ${X_{c_p}}$, and thus there are no UV
symmetries in this theory.

\begin{figure}[t!]
\centering
    \includegraphics[width=.48\textwidth]{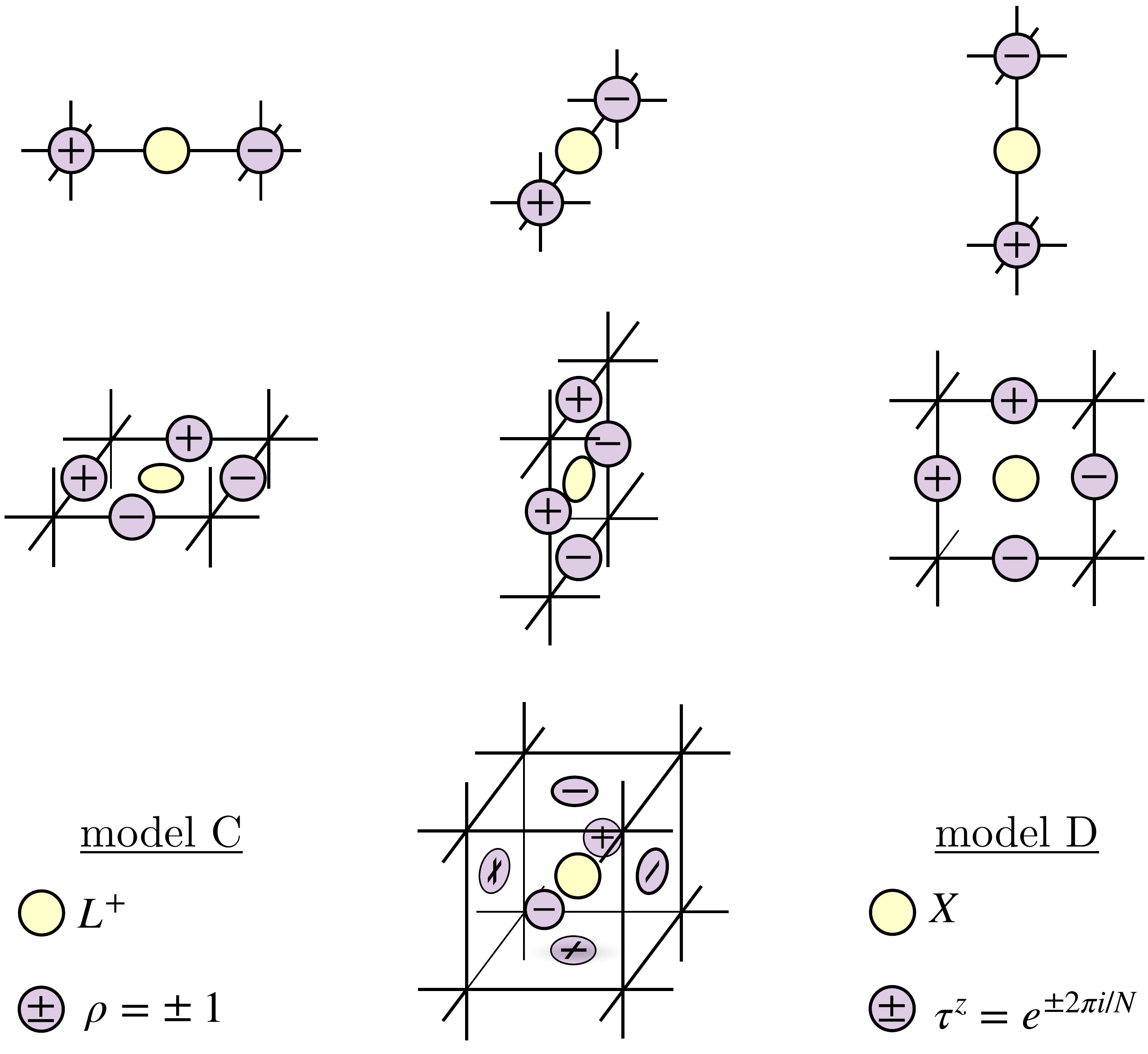}
    \caption{%
    Graphical representation of the excitation created by ${L^+_{c_p}}$ [${X_{c_p}}$] in model C [D] in 3d space for (1st row) ${p=1}$, (2nd row) ${p=2}$, and (3rd row) ${p=3}$. The yellow disk represents ${L^+_{c_{p}}}$ [${X_{c_{p}}}$]. For model C, the purple ${\pm}$ disk denotes the sign in ${\rho_{c_{p-1}}(L_{c_p}^+\ket{0})=\pm( L_{c_p}^+\ket{0})}$ for that ${(p-1)}$-cell. For model D, the ${\pm}$ disk instead denotes the sign in ${\tau^z_{c_{p-1}}(X_{c_p}\ket{0})=\om^{\pm 1}( X_{c_p}\ket{0})}$ for that ${(p-1)}$-cell.}
    \label{fig:charges}
\end{figure}

\subsubsection{An exact emergent $\ZN{p}$ symmetry}

In the limit ${J/U\to0}$, there exists a low energy sub-Hilbert space spanned by states satisfying ${\<\tau^z_{c_{p-1}}\> = 1}$. Violating this constraint costs energy ${U}$, and we interpret states that do so in this limit as having a gapped excitation, a segment of which residing on ${c_{p-1}}$. We'll refer to these bosonic ${(p-1)}$-dimensional (in space) excitations as ``charges'' since they are the gauge charges of the emergent ${\Z_N}$ ${p}$-gauge theory. From the clock operators' algebra, ${X_{c_p}}$ excites a charge on ${\pp c_p}$, examples of which are shown in Fig.~\ref{fig:charges}. Since ${X_{c_p}^N = \one}$, exciting ${N}$ charges is the same as not exciting any. Thus, the charge number takes values in ${\Z_N}$.

The charge gap ${U}$ provides a candidate energy scale below which new symmetries may emerge. However, when ${J/U\neq 0}$, there no longer exists a low-energy sub-Hilbert space spanned by states satisfying ${\<\tau^z_{c_{p-1}}\> = 1}$. This is because the ${J}$ term in ${H_{\txt{UV}}}$ causes the ${\<\tau^z_{c_{p-1}}\> = 1}$ and ${\<\tau^z_{c_{p-1}}\> \neq 1}$ states to mix. This does not necessarily mean the charges no longer exist when ${J>0}$, just that their operators depend on ${J}$. Indeed, a corresponding low-energy sub-Hilbert space can be identified using the local unitary from section~\ref{effHamSec111}, which we denote as ${U^{(1)}_{\txt{LU}}}$. Therefore, there exists a low-energy sub-Hilbert space spanned by states satisfying ${\<\t\tau^z_{c_{p-1}}\> \equiv \<U^{(1)}_{\txt{LU}}\tau^z_{c_{p-1}}U_{\txt{LU}}^{(1)\da}\> = 1}$.

We will not find an explicit form for ${U^{(1)}_{\txt{LU}}}$ and thus will not precisely know throughout how much of parameter space the dressed (fattened) operators can be defined without violating the assumptions of ${U^{(1)}_{\txt{LU}}}$. Instead, we will assume that such an operator exists and can access a greater than measure-zero part of parameter space and will investigate the consequences of this conjecture.

At this point, we cannot tell if the charge gap ${\Del}$ is an IR scale or mid-IR I scale or mid-II scale, etc. In section~\ref{section:znSSB}, we find it is a mid-IR scale in region III, but an IR scale in region II of parameter space (see Fig.~\ref{fig:phaseDia}). For the rest of this section, however, we will adopt the language from the perspective of region III and call the charge gap a mid-IR scale.

\begin{figure}[t!]
\centering
    \includegraphics[width=.48\textwidth]{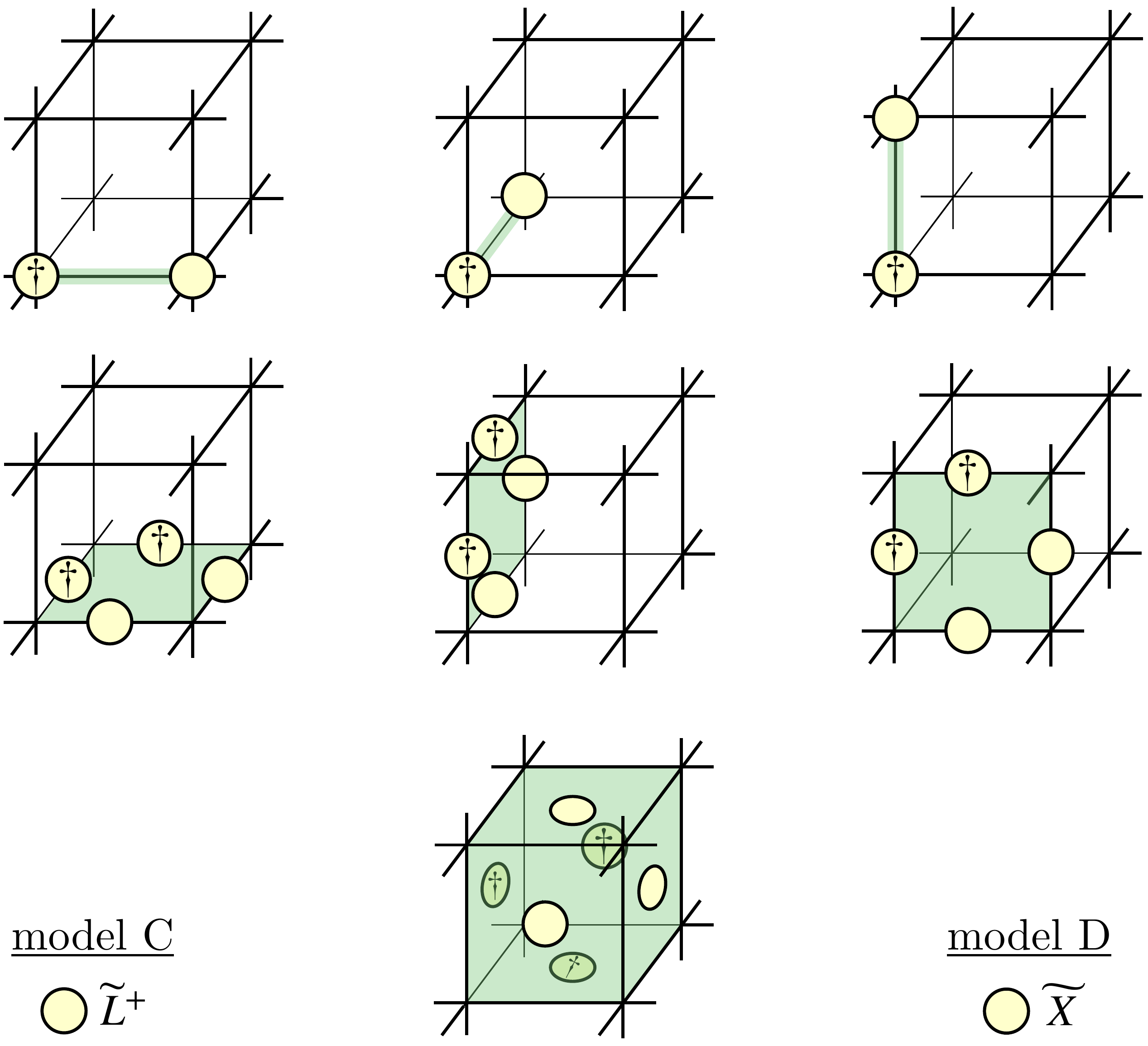}
    \caption{%
    Graphical representation of ${\t W^\da(\pp c_{p+1})}$ for model C (Eq.~\eqref{eqn:wOp}) and D (Eq.~\eqref{eqn:ZNwOp}) in ${3}$d space for (1st row) ${p=0}$, (2nd row) ${p=1}$, and (3rd row) ${p=2}$. For model C (D), the yellow colored disks represent ${\t L^+}$ (${\t X}$) operators, the product of which yields ${\t W^\da(\pp c_{p+1})}$. Discs labeled by ${\da}$ denote the hermitian conjugate and ${c_{p+1}}$ is colored green.
    }
    \label{fig:wOpp}
\end{figure}

Given the mid-IR scale ${E_{\txt{mid-IR}} \equiv \Del}$, we would now like to find an effective mid-IR theory describing states at energies ${E < E_{\txt{mid-IR}}}$. Since ${\t{Z}_{c_{p}}}$ commutes with ${\t{\tau}^z_{c_{p-1}}}$, it does not excite any charges and is an allowed mid-IR operator. The operators ${\t{X}_{c_p}}$ are not allowed as they excite charges. the allowed operators constructed from ${\t{X}_{c_p}}$ are
\begin{equation}\label{eqn:ZNwOp}
\t{W}^{\da}[C_p] = \prod_{c_p\in C_p}\t{X}_{c_p},
\end{equation}
which we call the Wilson operator (see Fig.~\ref{fig:wOpp}). It has the interpretation of exciting a charge, transporting it along a ${p}$-cycle, and then annihilating it.

While ${\t{W}^{\da}[C_p]}$ does not excite charges, not all ${\t{W}^{\da}[C_p]}$ are mid-IR operators. Indeed, when ${K/U\gg 1}$ the $p$-brane excitation created by ${\t{W}^{\da}[C_p]}$ costs energy ${\sim|C_p| K}$, where $|C_p|$ is the number of $p$-cells $C_p$ is made of. So, roughly, ${\t{W}^{\da}[C_p]}$ is allowed in this limit only if ${|C_p| \ll \Del/K}$. On the other hand, when ${K/U\ll 1}$ this $p$-brane's gap does not increase linearly with $|C_p|$ and all ${\t{W}^{\da}[C_p]}$ are mid-IR allowed operators. We will denote the set of ${C_p}$ for which ${\t{W}^{\da}[C_p]}$ is an allowed Mid-IR operator when acting on ${\gs}$ by ${\eZ_p}$. This set of $p$-cycles depends on the value of ${K/U}$.

For a finite size system, the algebra of local symmetric mid-IR operators is generated by 
\begin{equation}
\cA^{\txt{finite-}L}_{\txt{mid-IR}} = \{\t{Z}_{c_p},\t{W}^{\da}[C_p ]:C_p \in \eZ_p\}.
\end{equation}
Strictly speaking, this is only approximate since ${\t W[C_p\in \eZ_p]}$ is only a mid-IR operator when acting on low-energy eigenstates in the mid-IR, not mid-IR states with $E$ close to $E_{\txt{mid-IR}}$. Nevertheless, the symmetries of this should be the same as the exact form of the effective mid-IR theory. The mid-IR Hamiltonian under this approximation is
\begin{equation}\label{ZNfullHeff}
\begin{aligned}
H^{\txt{finite-}L}_{\txt{mid-IR}} = &- \ka U \sum_{c_p}\t{Z}_{c_p}  -U\sum_{c_{p+1}}\t{W}[\pp c_{p+1}] \\
&\hspace{20pt}-U  \sum_{C_{p}\in \eZ_p} \varepsilon_{C_p}\t{W}[C_{p}]+\cdots
,
\end{aligned}
\end{equation}
where ${\ka \sim K/U}$ and ${\varepsilon_{C_p}\sim (J/U)^{|C_p|}}$.

In the thermodynamic limit, ${\t{W}^{\da}}$ acting on non-contractible ${p}$-cycles is a non-local operator. Denoting the subset of ${\eZ_p}$ with only contractible ${p}$-cycles as ${\eB_p}$, the algebra of local symmetric mid-IR operators is now generated by
\begin{equation}
\cA_{\txt{mid-IR}} = \{\t{Z}_{c_p},\t{W}^{\da}[C_p]: C_p \in \eB_p\}.
\end{equation}
From the effective Hamiltonian point of view, ${H^{\txt{finite-}L}_{\txt{mid-IR}}}$ must be a local Hamiltonian, so the mid-IR theory is only well-defined provided ${J/U \ll 1}$. Therefore, in the thermodynamic limit, terms with Wilson operators supported on nontrivial ${p}$-cycles vanish, and the mid-IR theory becomes
\begin{equation}\label{ZNfullHeffThermoLim}
\begin{aligned}
H_{\txt{mid-IR}} = &- \ka U \sum_{c_p}\t{Z}_{c_p}  - U\sum_{c_{p+1}}\t{W}[\pp c_{p+1}] \\
&\hspace{20pt}-U  \sum_{C_{p}\in \eB_p} \varepsilon_{C_p}\t{W}[C_{p}]+\cdots
,
\end{aligned}
\end{equation}

\begin{figure}[t!]
\centering
    \includegraphics[width=.48\textwidth]{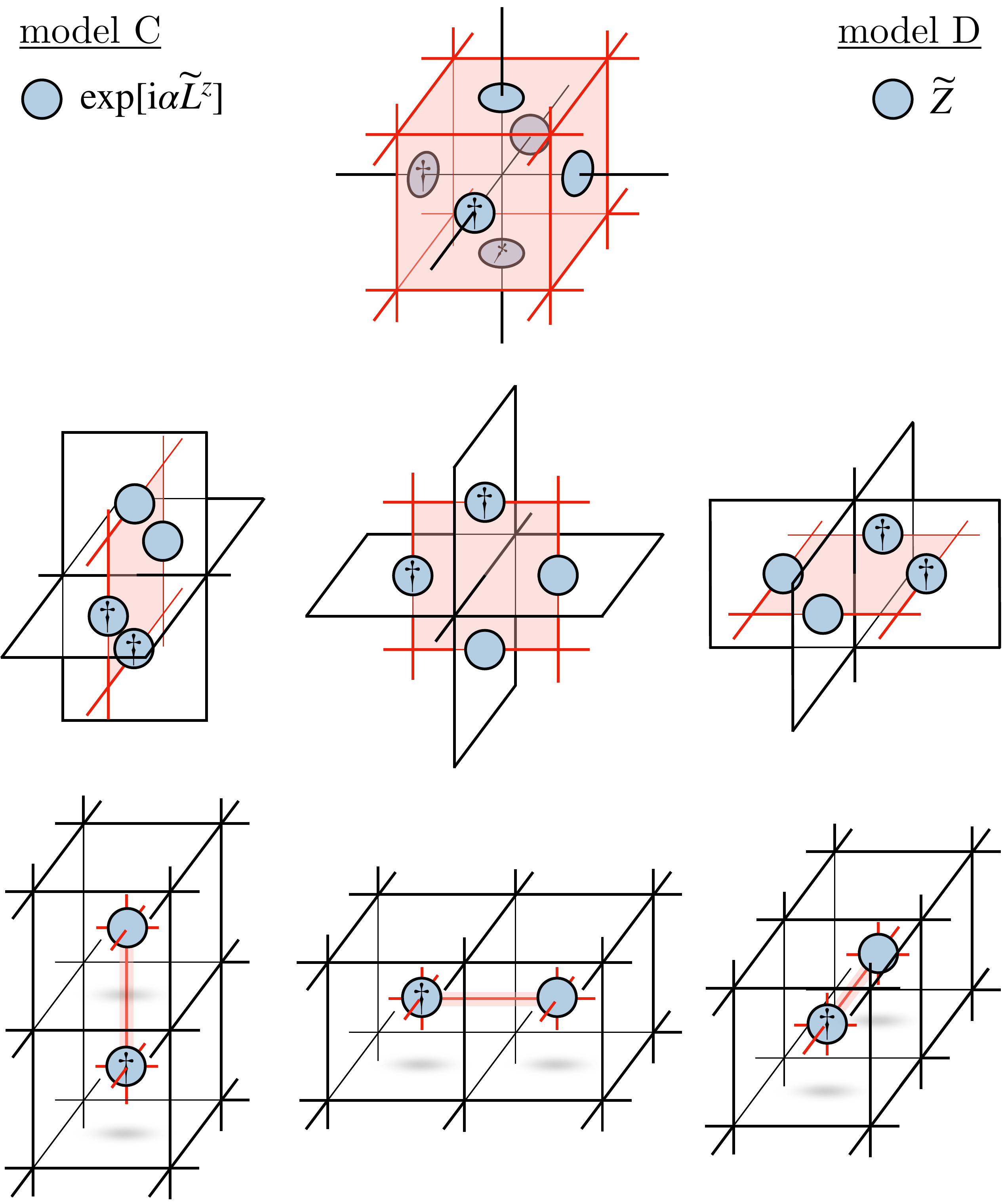}
    \caption{%
    Graphical representation of the emergent ${\U{p}}$ [${\ZN{p}}$] symmetry operator Eq.~\eqref{Upsymop} [\eqref{ZNsymop}] in model C [D] acting on ${\pp \h{c}_{d-p+1}}$ in 3d space for (1st row) ${p=1}$, (2nd row) ${p=2}$, and (3rd row) ${p=3}$. The blue-colored disks denote the operator ${\t L^+_{c_p}}$ [${\t X_{c_p}}$], the product of which yields the symmetry operator. Discs labeled by ${\da}$ denote the hermitian conjugate of the operator and ${\h{c}_{d-p+1}}$ is colored red.
    }
    \label{fig:symOp}
\end{figure}

${\cA_{\txt{mid-IR}}}$ includes an new symmetry absent from ${\cA_{\txt{UV}}}$. Indeed, the mid-IR theory is invariant under the transformation
\begin{equation}\label{ZnWsymTrans}
\t X_{c_p} \hspace{-3pt}\to \ee^{\ii\Ga_{c_p}}\t X_{c_p}\implies \t W[C_p]\to \ee^{\ii \sum_{c_p  \in C_p}\Ga_{c_p}} \t W[C_p],
\end{equation}
where ${(\dd \Ga)_{c_{p+1}} \equiv \sum_{c_p\in\pp c_{p+1}}\Ga_{c_p}=0}$ and ${\Ga_{c_p}\in 2\pi \Z/N}$ for ${(X_{c_p})^N = \one}$ to be invariant. This is a symmetry because ${\sum_{c_p  \in C_p}\Ga_{c_p} = 0}$ for ${C_p\in \eB_p}$ since ${(\dd\Ga)_{c_{p+1}}=0}$. It is not a gauge symmetry as it transforms Wilson operators on non-contractible ${p}$-cycles by a nontrivial element of ${\Z_N}$. Therefore, since the charged operators are supported on a ${p}$-cycle, the mid-IR has an exact emergent ${\ZN{p}}$ symmetry. The operator generating this ${\ZN{p}}$ symmetry is
\begin{equation}\label{ZNsymop}
\t U(\h\Sigma_{d-p}) = \hspace{-5pt}\prod_{\h{c}_{d-p}\in\h\Sigma_{d-p}}\hspace{-5pt}(\hstar \t Z)_{\h{c}_{d-p}},
\end{equation}
where ${\h{\Si}_{d-p}}$ is a ${(d-p)}$-cycle of the dual lattice and ${(\hstar \t Z)_{\h{c}_{d-p}}\equiv \t Z_{\hstar \h{c}_{d-p}}}$ (see Fig.~\ref{fig:symOp}).

This ${\ZN{p}}$ symmetry only emerges in parameter space where the mid-IR---the low-energy regime without gapped charges---exists. Here we associate the mid-IR's existence to the existence of a well-defined effective mid-IR Hamiltonian, so the exact emergent ${\ZN{p}}$ symmetry exists only when ${H_{\txt{mid-IR}}}$ converges. In the approximation scheme used, this requires ${(\varepsilon_{C_p})^{1/|C_p|}  < 1}$ for all ${C_p\in\eB_p}$. The constant of proportionality in ${(\varepsilon_{C_p})^{1/|C_p|} \propto J/U}$ increases with ${|C_p|}$ since there are more ways ${\t{W}[C_p]}$ can be generated from ${\t{X}_{c_p}}$ for larger ${|C_p|}$. Therefore, it is sufficient to consider only the largest ${p}$-cycle in ${\eB_p}$. For small enough ${K/U}$ where ${\eB_p}$ includes all trivial $p$-cycles, the largest ${C_p}$ does not depend on ${K/U}$, and so the largest value of ${J/U}$ with the exact emergent ${\ZN{p}}$ symmetry is independent of ${K/U}$. This value of ${J/U}$ defines the boundary between regions I and III in Fig.~\ref{fig:phaseDia}. For large enough ${K/U}$, ${\eB_p}$ does not include all trivial ${p}$-cycles. The larger ${K/U}$ is, the smaller the maximum value of ${|C_p|}$ is, and thus the larger the maximum value of ${J/U}$ with the exact emergent ${\ZN{p}}$ symmetry is. This value of ${J/U}$ increasing with ${K/U}$ defines the boundary between regions I and II shown schematically in Fig.~\ref{fig:phaseDia}.

\subsubsection{An exact emergent anomalous $\ZN{p}\times\ZN{d-p}$ symmetry}\label{section:znSSB}

The exact emergent ${\ZN{p}}$ symmetry can be spontaneously broken, and the SSB phase corresponds to the deconfined phase of ${\Z_N}$ ${p}$-gauge theory. To gain some intuition, we consider two tractable limits of the effective mid-IR theory Eq.~\eqref{ZNfullHeffThermoLim}. When ${J/U = 0}$ but ${K/U\neq 0}$ (which is in region II), the ground state satisfies ${\t Z_{c_{p}} \gs = \gs}$, and therefore ${\<\t W^\da[C_p]\> = 0}$ for all ${C_p}$. Consequently, this limit lies in a ${\ZN{p}}$ symmetric phase. On the other hand, when ${K/U = 0}$ but ${J/U\neq 0}$ (which is in region III), the ground state satisfies ${\t{W}[C_{p}\in\eB_p]\gs = \gs}$, and consequently ${\<\t W^\da[C_p]\> = 1}$ for all trivial ${p}$-cycles. Therefore, the ${\ZN{p}}$ symmetry is spontaneously broken in this limit. 

A ${\ZN{p}}$ symmetry at zero temperature can spontaneously break when ${d>p}$~\cite{GW14125148, L180207747}. Therefore, when ${d>p}$, we expect the symmetry to be broken even for ${K/U \neq 0}$ and a stable SSB phase to exist. For small ${\ka}$ and ${\varepsilon_{C_p}}$, a reasonable expectation from Eq.~\eqref{ZNfullHeffThermoLim} is the SSB phase occurs when ${\ka \lesssim  1}$. This determines the boundary between the ${\ZN{p}}$ symmetric and ${\ZN{p}}$ SSB phases and regions II and III shown in Fig.~\ref{fig:phaseDia}. We leave a more detailed investigation of this phase transition to future work.

\begin{figure}[t!]
\centering
    \includegraphics[width=.48\textwidth]{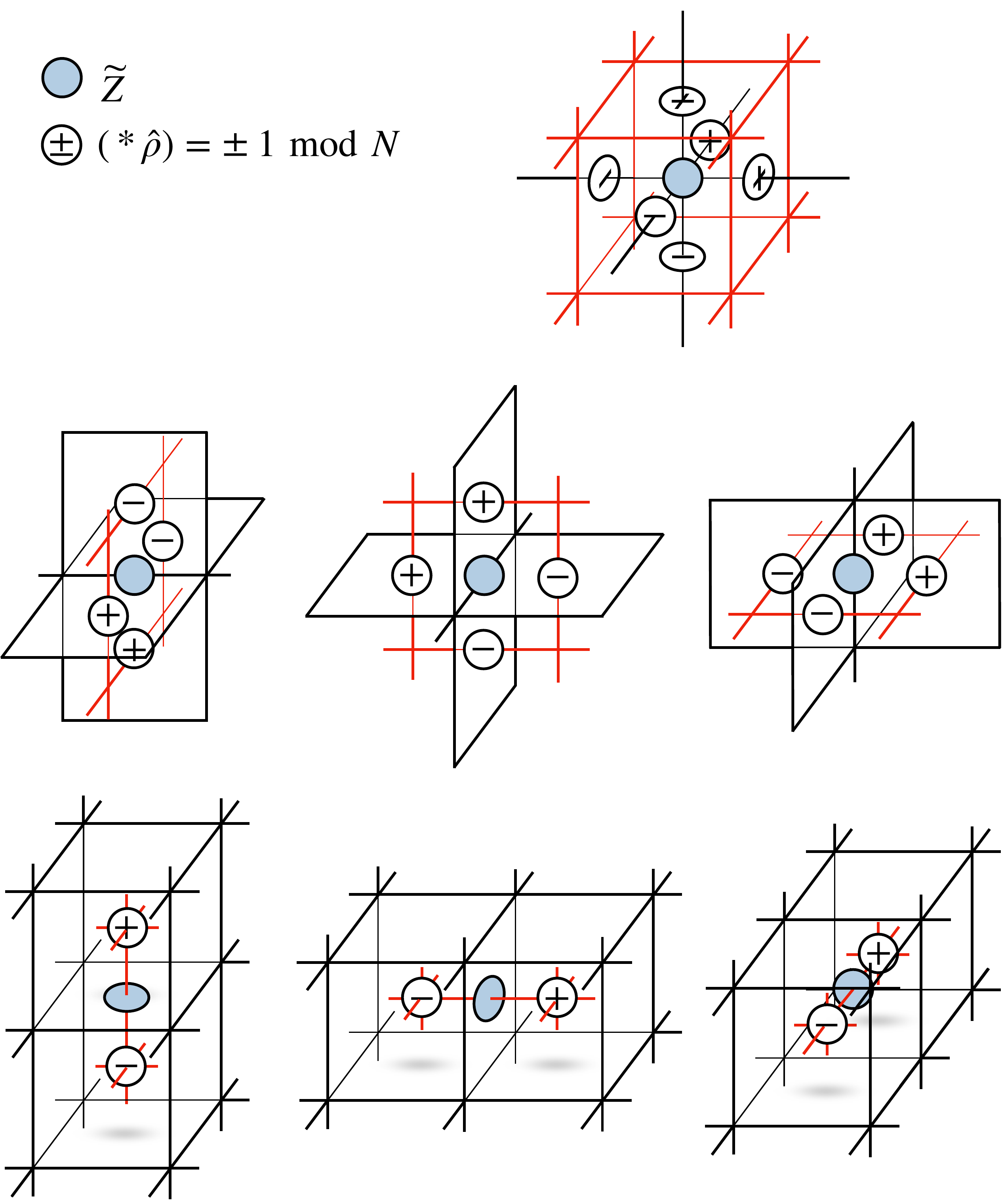}
    \caption{%
    Graphical representation of the topological defects created by ${(\hstar \t Z)_{\h{c}_{d-p}}}$ in model D for for 3d space and (1st row) ${p=0}$, (2nd row) ${p=1}$, and (3rd row) ${p=2}$. The blue disk represents the ${(\hstar \t Z)_{\h{c}_{d-p}}}$ operator. The disks labeled by ${\pm}$ represent ${(\hstar\h{\rho})_{c_{p+1}}=\pm 1}$ for that ${c_{p+1}}$ (see Eq.~\eqref{ZnTopDefDen}).}
    \label{fig:ZNtopDefect}
\end{figure}

Let us now restrict our considerations to the SSB phase. Like 0-form symmetries, breaking higher-form symmetries gives rise to gapped topological defects and, in this case, arise from the nontrivial mappings ${Z_p(M_d;\Z_N)\to \Z_N}$. In the ${K/U\to 0}$ limit, the topological defect density ${\h\rho}$ for a state ${\ket{\psi}}$ is defined by\footnote{This is a natural generalization of the ${p=0}$ case, where the topological defects are domain walls (see Eq.~\eqref{0formZnTopDefDen}).}
\begin{equation}\label{ZnTopDefDen}
\prod_{c_p\in\pp c_{p+1}}\t X_{c_p}\ket{\psi} = \ee^{\frac{2\pi\ii}{N} (\hstar\h{\rho})_{c_{p+1}}}\ket{\psi}.
\end{equation}
The topological defects are ${(d-p-1)}$-dimensional excitations in space, residing on the dual lattice, and carry ${\Z_N}$ charge. In the familiar ${p=1}$ case, they correspond to the ${\Z_N}$ flux excitations of ${\Z_N}$ gauge theory. Furthermore, from Eq.~\eqref{ZNclockAlg}, the operator ${(\hstar \t Z)_{\h{c}_{d-p}}}$ excites a topological defect on ${\pp \h{c}_{d-p}}$ (see Fig.~\ref{fig:ZNtopDefect}).

The topological defect gap ${\Del_{\txt{defect}}}$ provides a candidate energy scale below which new symmetries may emerge. Let us call this energy scale the IR. However, when ${K/U\neq 0}$, there no longer exists a low-energy sub-Hilbert space satisfying ${\<\h\rho_{\h c_{d-p-1}}\> = 0}$ mod ${N}$. This is because the ${\ka U}$ term in ${H_{\txt{mid-IR}} }$ causes the ${\<\h\rho_{\h c_{d-p-1}}\> = 0}$ and ${\<\h\rho_{\h c_{d-p-1}}\> \neq 0}$ states to mix. This does not necessarily mean the topological defects no longer exist when ${K/U>0}$, just that their operators depend on ${K/U}$. Indeed, we can use the local unitary discussed in section~\ref{effHamSec111}, which we will denote as ${U^{(2)}_{\txt{LU}}}$, to identify the corresponding low-energy sub-Hilbert space. In general, ${U^{(2)}_{\txt{LU}}}$ is different than the local unitary ${U^{(1)}_{\txt{LU}}}$ used in the previous subsection. Therefore, there exists a low-energy sub-Hilbert space in region III spanned by states satisfying ${\< \h\rho'_{\h c_{d-p-1}}\> \equiv \< U^{(2)}_{\txt{LU}}\h\rho_{\h c_{d-p-1}}U^{(2)\da}_{\txt{LU}}\> = 0}$. By the definition of ${U^{(2)}_{\txt{LU}}}$, the emergent ${\ZN{p}}$ symmetry operator satisfies
\begin{equation}\label{dressZNsymop}
\t U(\h\Sigma_{d-p}) = \t U'(\h\Sigma_{d-p}) = \hspace{-5pt}\prod_{\h{c}_{d-p}\in\h\Sigma_{d-p}}\hspace{-5pt}(\hstar \t Z')_{\h{c}_{d-p}}.
\end{equation}

Having identified the IR of region III, we would now like to find an effective IR theory. This IR satifies ${\h{\rho}'_{\h{c}_{d-p-1}} = 0}$, or equivalently ${\t W'[\pp c_{p+1}] = 1}$. Due to the constraint, the ${\ZN{p}}$ symmetric IR operators must be constructed from only ${\t{Z}_{c_p}}$. Only one such type of operator commutes with the constraint: Eq.~\eqref{dressZNsymop}. Therefore, for finite-size systems, the algebra of local symmetric IR operators is
\begin{equation}
\cA_{\txt{IR}}^{\txt{finite-}L} =\{\t U(\h\Sigma_{d-p}) : \h\Sigma_{d-p}\in Z_{d-p}(\h{M}_d;\Z_N)\},
\end{equation}
where ${Z_{d-p}(\h{M}_d;\Z_N)}$ is the set of all dual ${(d-p)}$-cycles with ${\Z_N}$ coefficients. The corresponding effective IR Hamiltonian is 
\begin{equation}\label{ZNIReffHam}
\hspace{-3pt}H^{\txt{finite-}L}_{\txt{IR}} \hspace{-1pt}=-U\hspace{-36pt}\sum_{\hspace{24pt}\h\Sigma_{d-p}\in Z_{d-p}(\hspace{-1pt}\h{M}_d;\Z_N)}\hspace{-22pt}\ka^{|\h\Sigma_{d-p}|}~
\t U(\h\Sigma_{d-p}),
\end{equation}
where ${\ka\sim K/U}$ and ${|\h\Sigma_{d-p}|}$ is the number of ${(d-p)}$-cells in ${\h\Sigma_{d-p}}$.

In the thermodynamic limit, ${\t U}$ acting on non-contractible ${(d-p)}$ cycles are non-local operators and Eq.~\eqref{dressZNsymop} includes only contractible ${(d-p)}$-cycles. However, such ${\t U}$ can be written as ${\t U(\pp O) = \prod_{\h{c}\in O}(\hstar \t{\tau}^z)_{\h c}}$, and are thus trivial in the IR since there are no charges. So, in the thermodynamic limit
\begin{equation}
\cA_{\txt{IR}} =\{\},
\end{equation}
and ${H_{\txt{IR}} = 0}$.

Since ${\cA_{\txt{IR}}}$ is the empty set, all nontrivial IR operators commute with it and correspond to symmetries. This includes ${\t U}$ supported on nontrivial dual ${(d-p)}$ cycles, and thus the mid-IR ${\ZN{p}}$ symmetry, as expected. However, ${\t W'}$ supported on nontrivial ${p}$-cycles is also allowed and from Eq.~\eqref{ZnWsymTrans}, they generate the transformation
\begin{equation}\label{ZNlatticeMagSym}
\t U(\h\Sigma_{d-p}) \to \ee^{\ii \sum_{\h{c}_{d-p}  \in \h\Sigma_{d-p}}\h{\Ga}_{\h{c}_{d-p}}} \t U(\h\Sigma_{d-p}),
\end{equation}
where ${(\dd\h\Ga)_{\h{c}_{d-p+1}}\hspace{-3pt}=0}$ and ${\h\Ga_{\h{c}_{d-p}}\in 2\pi\Z/N}$. Since the charged operator is supported on ${(d-p)}$ cycles and transforms by an element of ${\Z_N}$, ${\t W'}$ generates a ${\ZN{d-p}}$ symmetry. This is an exact emergent symmetry because it is not an exact symmetry of the UV but is an exact symmetry of the IR. 

So, in the IR of region III---the ground state subspace of the deconfined phase of emergent ${\Z_N}$-${p}$ gauge theory---there is an exact emergent ${\ZN{p}\times\ZN{d-p}}$ symmetry. Furthermore, this IR symmetry is anomalous, which can be noticed from the fact that the symmetry operator of the ${\ZN{p}}$ is charged under the ${\ZN{d-p}}$ symmetry. This mixed 't Hooft anomaly protects the deconfined phase's topological degeneracy and topological order. The only way to get rid of it is to prevent the ${\ZN{p}\times\ZN{d-p}}$ symmetries from emergent by either condensing the topological defects (to destroy ${\ZN{d-p}}$) or the charges (to destroy the entire ${\ZN{p}\times\ZN{d-p}}$).

This emergent anomalous ${\ZN{p}\times \ZN{d-p}}$ symmetry is the same as the exact symmetry of ${p}$-form ${BF}$ theory and ${p}$-form toric code. This is no accident. In Appendix~\ref{subSec:ZNcontLim}, we show how the ground states in region III are also the ground states of the ${p}$-form toric code and that their topological quantum field theory description is ${p}$-form ${BF}$ theory.

\subsection{Emergent ${U(1)}$ ${p}$-gauge theory}\label{sec:modelCEES}

In this section, we consider a model for emergent ${U(1)}$ ${p}$-gauge theory, which we call model C. When ${p=1}$, this is just typical ${U(1)}$ gauge theory. Consider ${U(1)}$ quantum rotors residing on the ${p}$-cells of the spatial ${d}$-dimensional cubic lattice with ${p>0}$. Each rotor can be viewed as a particle on an infinitesimal circle, whose position we denote as the angle ${\Th}$, carrying angular momentum ${L^z}$. The operators ${L^z}$ and ${\Th}$ are hermitian and satisfy ${[\Th_{c_p},L^z_{\t{c}_p}]=\ii\del_{c_p,\t{c}_p}}$, so ${L_z = -\ii \frac{\pp}{\pp \Th}}$. Since the eigenvalue of ${\Th}$ is an angle, the eigenvalues of ${L^z}$ are integers. 

Model C is described by the Hamiltonian
\begin{align}\label{2bodyU1Ham}
&H_{\txt{UV}} = \frac{U}2\sum_{c_{p-1}} \rho_{c_{p-1}}^2 -U\sum_{c_{p+1}}W^{\da}_{c_{p+1}}\nonumber\\
& \hspace{40pt}\frac{K}2\sum_{c_p}(L_{c_p}^{z})^2 + \frac{J}2 \sum_{c_p}\left(L^+_{c_p}+ \txt{h.c.}\right),\\
&\rho_{c_{p-1}}=\sum_{c_p\in\del c_{p-1}}L_{c_p}^{z},\quad\quad W^{\da}_{c_{p+1}} = \prod_{c_p\in \pp c_{p+1}} L_{c_p}^+\nonumber
\end{align}
where ${\sum_{c_{p}}}$ is over all ${p}$-cells, ${\del c_{p-1}}$ is the coboundary of ${c_{p-1}}$ (see Eq.~\eqref{coboundaryDef}), and ${L^+_{c_p} = (L^-_{c_p})^\da = \exp[\ii \Th_{c_p}]}$ is the raising operator for ${L^z_{c_p}}$. Using the definition of ${\del c_p}$, ${\rho_{c_{p-1}}}$ can be written as (see Fig.~\ref{fig:rhoOpp})
\begin{equation}\label{eqn:rhoOp}
\rho(\bm x)_{\mu_1\cdots\mu_{p-1}}\hspace{-3pt}=\hspace{-1pt}\sum_{\nu}L^z(\bm x)_{\nu\mu_{1} \ldots \mu_{p-1} }\hspace{-1pt}-L^z(\bm x-\bm{\h\nu})_{\nu\mu_{1} \ldots \mu_{p-1} }.
\end{equation}

The algebra of local symmetric UV operators is generated by 
\begin{equation}
\cA_{\txt{UV}} = \{\rho_{c_{p-1}}^2, (L_{c_p}^{z})^2, L^+_{c_p}\}.
\end{equation}
This is invariant under the transformation ${L^z_{c_p}\to -L^z_{c_p}}$, and thus there is a UV ${\Z^{(0)}_2}$ symmetry.

\subsubsection{An exact emergent $\U{p}$ symmetry}

In the limit ${J/U \to 0}$, there exists a low energy sub-Hilbert space spanned by states satisfying ${\<\rho_{c_{p-1}}\> = 0}$. Violating this constraint costs energy ${U}$, and we interpret states that do so in this limit as having a gapped excitation, a segment of which resides on ${c_{p-1}}$. We'll refer to these bosonic ${(p-1)}$-dimensional (in space) excitations as ``charges'' since they are the gauge charges of the emergent ${U(1)}$ gauge theory. From the commutation relation satisfied by ${L^z}$ and ${\Th}$, ${L^+_{c_p}}$ excites a charge on ${\pp c_p}$, examples of which are shown in Fig.~\ref{fig:charges}. 

The charge gap ${U}$ provides a candidate energy scale below which new symmetries may emerge. However, when ${J/U\neq 0}$, there no longer exists a low-energy sub-Hilbert space spanned by states satisfying ${\<\rho_{c_{p-1}}\> = 0}$. This is because the ${J}$ term in ${H_{\txt{UV}}}$ causes the ${\<\rho_{c_{p-1}}\> = 0}$ and ${\<\rho_{c_{p-1}}\> \neq 0}$ states to mix. This does not necessarily mean the charges no longer exist when ${J>0}$, just that their operators depend on ${J}$. Indeed, a corresponding low-energy sub-Hilbert space can be identified using the local unitary from section~\ref{effHamSec111}, which we denote as ${U_{\txt{LU}}}$. Therefore, there exists a low-energy sub-Hilbert space spanned by states satisfying ${\<\t{\rho}_{c_{p-1}}\> \equiv \<U_{\txt{LU}}\rho_{c_{p-1}}U_{\txt{LU}}^\da\> = 0}$.

We will not find an explicit form for ${U_{\txt{LU}}}$ and thus will not precisely know throughout how much of parameter space the dressed (fattened) operators can be defined without violating the assumptions of ${U_{\txt{LU}}}$. Instead, we will assume that such an operator exists and can access a greater than measure-zero part of parameter space and will investigate the consequences of this conjecture.

At this point, we cannot tell if the charge gap ${\Del}$ is an IR scale or mid-IR I scale or mid-II scale, etc. In section~\ref{section:SSB}, we find it is a mid-IR scale in region III but an IR scale in region II of parameter space (see Fig.~\ref{fig:phaseDia}). For the rest of this section, however, we will adopt the language from the perspective of region III and call the charge gap a mid-IR scale.

Given the mid-IR scale ${E_{\txt{mid-IR}} \equiv \Del}$, we would now like to find an effective mid-IR theory describing states at energies ${E < E_{\txt{mid-IR}}}$. Since the UV-symmetric operator ${(\t{L}^z_{c_{p}})^2}$ commutes with ${\t{\rho}_{c_{p-1}}}$, it does not excite any charges and is an allowed mid-IR operator. The operators ${\t{L}^+_{c_p}}$ are not allowed as they excite charges. the allowed operators constructed from ${\t{L}^+_{c_p}}$ are
\begin{equation}\label{eqn:wOp}
\t{W}^{\da}[C_p] = \prod_{c_p\in C_p}\t{L}^+_{c_p}.
\end{equation}
which we call the Wilson operator (see Fig.~\ref{fig:wOpp}). It has the interpretation of exciting a charge, transporting it along a ${p}$-cycle, and then annihilating it.

While ${\t{W}^{\da}[C_p]}$ does not excite charges, not all ${\t{W}^{\da}[C_p]}$ are mid-IR operators. Indeed, when ${K/U\gg 1}$ the $p$-brane excitation created by ${\t{W}^{\da}[C_p]}$ costs energy ${\sim|C_p| K}$, where $|C_p|$ is the number of $p$-cells $C_p$ is made of. So, roughly, ${\t{W}^{\da}[C_p]}$ is allowed in this limit only if ${|C_p| \ll \Del/K}$. On the other hand, when ${K/U\ll 1}$ this $p$-brane's gap does not increase linearly with $|C_p|$ and all ${\t{W}^{\da}[C_p]}$ are mid-IR allowed operators. We will denote the set of ${C_p}$ for which ${\t{W}^{\da}[C_p]}$ is an allowed Mid-IR operator when acting on ${\gs}$ by ${\eZ_p}$. This set of $p$-cycles depends on the value of ${K/U}$.

For a finite size system, the algebra of local symmetric mid-IR operators is generated by 
\begin{equation}
\cA^{\txt{finite-}L}_{\txt{mid-IR}} = \{(\t{L}^z_{c_p})^2,\t{W}^{\da}[C_p ]:C_p \in \eZ_p\}.
\end{equation}
Strictly speaking, this is only approximate since ${\t W[C_p\in \eZ_p]}$ is only a mid-IR operator when acting on low-energy eigenstates in the mid-IR, not mid-IR states with $E$ close to $E_{\txt{mid-IR}}$. Nevertheless, the symmetries of this should be the same as the exact form of the effective mid-IR theory. The mid-IR Hamiltonian under this approximation is
\begin{equation}\label{fullHeff}
\begin{aligned}
H^{\txt{finite-}L}_{\txt{mid-IR}} =&\ka U\sum_{c_p}(\t{L}_{c_p}^{z})^2 -U\sum_{c_{p+1}}\t{W}[\pp c_{p+1}] 
\\&-U\sum_{C_{p}\in \eZ_p}\varepsilon_{C_p} \t{W}[C_{p}] +\cdots,
\end{aligned}
\end{equation}
where ${\ka \sim K/U}$ and ${\varepsilon_{C_p}\sim (J/U)^{|C_p|}}$.

In the thermodynamic limit, ${\t{W}^{\da}}$ acting on non-contractible ${p}$-cycles is a non-local operator. Denoting the subset of ${\eZ_p}$ with only contractible ${p}$-cycles as ${\eB_p}$, the algebra of local symmetric mid-IR operators is now generated by
\begin{equation}
\cA_{\txt{mid-IR}} = \{(\t{L}^z_{c_p})^2,\t{W}^{\da}[C_p ]:C_p \in \eB_p\}.
\end{equation}
From the effective Hamiltonian point of view, ${H^{\txt{finite-}L}_{\txt{mid-IR}}}$ must be a local Hamiltonian, so the mid-IR theory is only well-defined provided ${J/U \ll 1}$.\footnote{When ${p=1}$, Eq.~\eqref{fullHeff} can be thought of as a lattice regularization of the string field theory in \Rf{IM210612610}. Here, the suppression of large loops automatically arises from the locality of the UV theory.} Therefore, in the thermodynamic limit, terms with Wilson operators supported on nontrivial ${p}$-cycles vanish, and the mid-IR theory becomes
\begin{equation}\label{fullHeffThermoLim}
\begin{aligned}
H_{\txt{mid-IR}} =&\ka U\sum_{c_p}(\t{L}_{c_p}^{z})^2 -U\sum_{c_{p+1}}\t{W}[\pp c_{p+1}] 
\\&-U\sum_{C_{p}\in \eB_p}\varepsilon_{C_p} \t{W}[C_{p}] +\cdots,
\end{aligned}
\end{equation}

${\cA_{\txt{mid-IR}}}$ includes an new symmetry absent from ${\cA_{\txt{UV}}}$. Indeed, the mid-IR theory is invariant under the transformation
\begin{equation}\label{WsymTrans}
\t{L}^+_{c_p} \hspace{-3pt}\to \ee^{\ii\Ga_{c_p}}\t{L}^+_{c_p}\implies \t W[C_p]\to \ee^{\ii \sum_{c_p  \in C_p}\Ga_{c_p}} \t W[C_p],
\end{equation}
where ${(\dd \Ga)_{c_{p+1}} \equiv \sum_{c_p\in\pp c_{p+1}}\Ga_{c_p}=0}$. This is a symmetry because ${\sum_{c_p  \in C_p}\Ga_{c_p} = 0}$ for ${C_p\in \eB_p}$ since ${(\dd\Ga)_{c_{p+1}}=0}$. It is not a gauge symmetry as it transforms Wilson operators on non-contractible ${p}$-cycles by a nontrivial element of ${U(1)}$. Therefore, since the charged operators are supported on a ${p}$-cycle, the mid-IR has an exact emergent ${\U{p}}$ symmetry. The symmetry operator of this ${\U{p}}$ symmetry is
\begin{equation}\label{Upsymop}
\t U_\al(\h\Sigma_{d-p}) = \hspace{-5pt}\prod_{\h{c}_{d-p}\in\h\Sigma_{d-p}}\hspace{-5pt}\exp[\ii\alpha~(\hstar \t L^z)_{\h{c}_{d-p}} ],
\end{equation}
where ${\al\in [0,2\pi)}$, ${\h{\Si}_{d-p}}$ is a ${(d-p)}$-cycle of the dual lattice, and ${(\hstar \t L^z)_{\h{c}_{d-p}}\equiv \t L^z_{\hstar \h{c}_{d-p}}}$ (see Fig.~\ref{fig:symOp}). 

This ${\U{p}}$ symmetry only emerges in parameter space where the mid-IR---the low-energy regime without gapped charges---exists. Here we associate the mid-IR's existence to the existence of a well-defined effective mid-IR Hamiltonian, so the exact emergent ${\U{p}}$ symmetry exists only when ${H_{\txt{mid-IR}}}$ converges. In the approximation scheme used, this requires ${(\varepsilon_{C_p})^{1/|C_p|}  < 1}$ for all ${C_p\in\eB_p}$. The constant of proportionality in ${(\varepsilon_{C_p})^{1/|C_p|} \propto J/U}$ increases with ${|C_p|}$ since there are more ways ${\t{W}[C_p]}$ can be generated from ${\t{L}^+_{c_p}}$ for larger ${|C_p|}$. Therefore, it is sufficient to consider only the largest ${p}$-cycle in ${\eB_p}$. For small enough ${K/U}$ where ${\eB_p}$ includes all trivial $p$-cycles, the largest ${C_p}$ does not depend on ${K/U}$, and so the largest value of ${J/U}$ with the exact emergent ${\U{p}}$ symmetry is independent of ${K/U}$. This value of ${J/U}$ defines the boundary between regions I and III in Fig.~\ref{fig:phaseDia}. For large enough ${K/U}$, ${\eB_p}$ does not include all trivial ${p}$-cycles. The larger ${K/U}$ is, the smaller the maximum value of ${|C_p|}$ is, and thus the larger the maximum value of ${J/U}$ with the exact emergent ${\U{p}}$ symmetry is. This value of ${J/U}$ increasing with ${K/U}$ defines the boundary between regions I and II shown schematically in Fig.~\ref{fig:phaseDia}.

\subsubsection{An exact emergent anomalous $\U{p}\times\U{d-p-1}$ symmetry in the continuum}\label{section:SSB}

The exact emergent ${\U{p}}$ symmetry can be spontaneously broken, and the SSB phase corresponds to the deconfined phase of ${U(1)}$ ${p}$-gauge theory. To gain some intuition, we consider two tractable limits of the effective mid-IR theory Eq.~\eqref{fullHeffThermoLim}. When ${J/U = 0}$ but ${K/U\neq 0}$ (which is in region II), the ground state satisfies ${\t{L}^z_{c_{p}} \gs = 0}$, and therefore ${\<\t W^\da[C_p]\> = 0}$ for all ${C_p}$. Consequently, this limit lies in a ${\U{p}}$ symmetric phase. On the other hand, when ${K/U = 0}$ but ${J/U\neq 0}$ (which is in region III), the ground state satisfies ${\t{W}[C_{p}\in\eB_p]\gs = \gs}$, and consequently ${\<\t W^\da[C_p]\> = 1}$ for all trivial ${p}$-cycles. Therefore, the ${\U{p}}$ symmetry is spontaneously broken in this limit. 

A ${\U{p}}$ symmetry at zero temperature can spontaneously break when ${d>p+1}$~\cite{GW14125148, L180207747}. Therefore, when ${d>p+1}$, we expect the symmetry to be broken even for ${K/U \neq 0}$ and a stable SSB phase to exist. For small ${\ka}$ and ${\varepsilon_{C_p}}$, a reasonable expectation from Eq.~\eqref{fullHeffThermoLim} is the SSB phase occurs when ${\ka \lesssim  1}$. This determines the boundary between the ${\U{p}}$ symmetric and ${\U{p}}$ SSB phases and regions II and III shown in Fig.~\ref{fig:phaseDia}. We leave a more detailed investigation of this phase transition to future work.

Let us now restrict our considerations to the SSB phase. Like 0-form symmetries, breaking higher-form symmetries gives rise to gapped topological defects and, in this case, arise from the nontrivial mappings ${Z_p(M_d;\Z)\to U(1)}$. The topological defects excited in a state ${\ket{\psi}}$ are probed by repeatedly acting the Wilson operator over a trivial ${(p+1)}$-cycle ${C_{p+1}}$:\footnote{This is a natural generalization of the ${p=0}$ case, where the topological defects are vortices.}
\begin{equation}
\prod_{c_{p+1}\in C_{p+1}}\t W^\da[\pp c_{p+1}]\ket{\psi} = \ee^{2\pi\ii \h{Q}(C_{p+1})}\ket{\psi}.
\end{equation}
The eigenvalue ${\h Q(C_{p+1})}$ is the winding number and yields the net number of topological defects enclosed by ${C_{p+1}}$. It is given by
\begin{equation}\label{winder}
\h Q(C_{p+1}) = \frac{1}{2\pi} \hspace{-3pt}\sum_{c_{p+1}\in C_{p+1}} \hspace{-4pt}F_{c_{p+1}},
\end{equation}
where ${F_{c_{p+1}} = (\dd\t\Th)_{c_{p+1}}~\mod~2\pi}$. Using the identity ${x~\mod~n = x - n\toZ{x/n}}$, where ${\toZ{\bm{\cdot}}}$ rounds its input to the nearest integer, ${F_{c_{p+1}}}$ can be written as
\begin{equation}
F_{c_{p+1}} \equiv (\dd\t\Th)_{c_{p+1}} + \om_{c_{p+1}},
\end{equation}
where ${\om_{c_{p+1}}\equiv -2\pi\toZ{(\dd\t\Th)_{c_{p+1}}/(2\pi)}}$.

\begin{figure}[t!]
\centering
    \includegraphics[width=.48\textwidth]{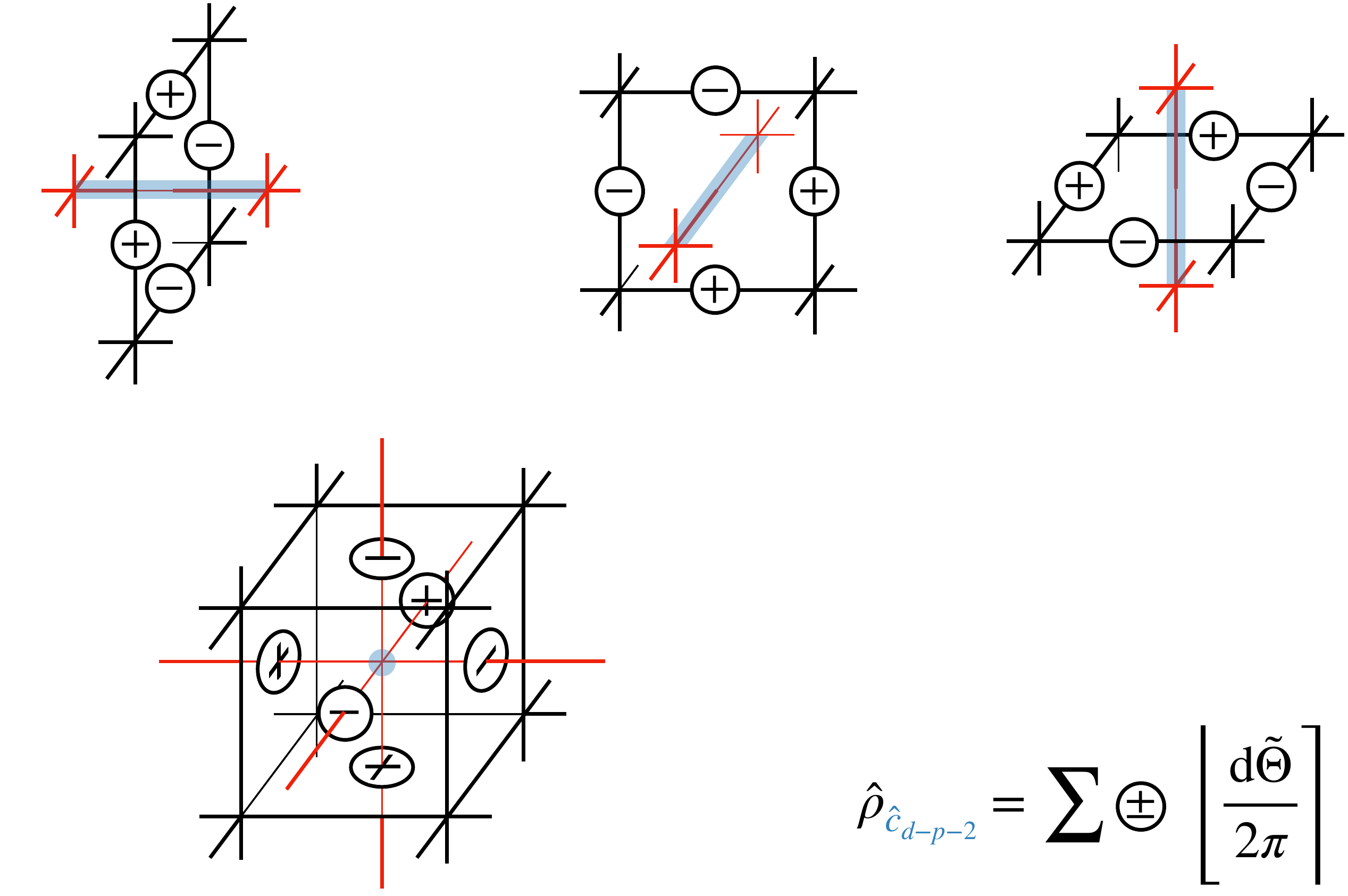}
    \caption{%
    Graphical representation of ${\h{\rho}_{\h{c}_{d-p-2}}}$ (see Eq.~\eqref{topdefectrhoop}) in 3d space for (1st row) ${p=0}$ and (2nd row) ${p=1}$. The ${\pm}$ disks denote the sign in front of that ${\toZ{\dd\t\Th/(2\pi)}}$ in the sum for ${\h{\rho}_{\h{c}_{d-p-2}}}$. The direct lattice is colored in black, the dual lattice is in red, and ${\h{\rho}_{\h{c}_{d-p-2}}}$ is in blue.
}
    \label{fig:topDefect}
\end{figure}

The topological defects can be characterized locally by paramerizing ${\h Q(C_{p+1} = \pp O_{p+2}) \equiv \sum_{c_{p+2}\in O_{p+2}}(\hstar \h\rho)_{c_{p+2}}}$, where the topological defect density ${\h\rho}$ is
\begin{equation}\label{topdefectrhoop}
(\hstar \h\rho)_{c_{p+2}}  =  \frac{1}{2\pi}(\dd F)_{c_{p+2}}.
\end{equation}
Therefore, they are ${(d-p-2)}$-dimensional excitations in space, residing on the dual lattice, and carry ${\Z}$ charge (see Fig.~\ref{fig:topDefect}). In the familiar ${p=1}$ case, they correspond to the magnetic monopole excitations of ${U(1)}$ gauge theory.

However, these topological defects cannot be observed directly in the lattice model.\footnote{One could instead consider a Villain type Hamiltonian model for which these topological defects are observable even in the UV/mid-IR~\cite{GLS210301257, CS221112543, FS221113047}. Nevertheless, these different UV lattice models should have the same IR effective field theory.} Indeed since ${\Th_{c_{p}}}$ always appears as ${L^+_{c_p} = \ee^{\ii\Th_{c_p}}}$, ${(\hstar \h\rho)_{c_{p+2}}}$ too always appears as ${\ee^{\ii 2\pi (\hstar \h\rho)_{c_{p+2}}}}$ and so ${\h\rho_{\h{c}_{d-p-2}} \sim  \h\rho_{\h{c}_{d-p-2}} + 1}$ on the lattice.

While the topological defects are unobservable on the lattice, their effects emerge in the continuum limit. The general paradigm for lattice models (without UV/IR mixing) is that the effective IR theory deep into a phase of matter is a continuum quantum field theory reflecting that phase's universal properties. Finding the IR effective field theory involves going deep into the ${\U{p}}$ SSB phase and taking the continuum limit. Deep into the SSB phase, the effective IR hamiltonian Eq.~\eqref{fullHeffThermoLim} includes only the leading order in ${\ka}$ and ${\varepsilon_{C_p}}$ terms:
\begin{equation}\label{deepIRU1Ham}
H_{\txt{deep~IR}} \hspace{-1pt}\approx\hspace{-1pt}\frac{\ka U}2 \hspace{-1pt}\sum_{c_p}\hspace{-1pt}\left(\t L_{c_p}^{z}\right)^2 \hspace{-2pt}+ \frac{U }2\sum_{c_{p+1}}\hspace{-1pt}\left( F_{c_{p+1}}\hspace{-1pt}\right)^2 \hspace{-3pt}.
\end{equation}
In the field theory, these higher-order terms could contribute as higher-derivative terms but do not affect the deep IR.

Appendix~\ref{subSec:contLim} shows how we take the continuum limit of ${H_{\txt{deep~IR}}}$, doing so carefully to capture the topologically nontrivial parts of the quantum fields from the lattice operators. We find that the IR effective field theory is compact ${p}$-form Maxwell theory, described by the path integral
\begin{equation}
\cZ_{\txt{deep~IR}} = \hspace{-4pt}\int\cD [a]\hspace{-5pt}\sum_{\om_a\in 2\pi H^{p+1}(X;\Z)}\hspace{-5pt}\ee^{-\frac{\ii}{2g^2}\int_{X}  F_a\wdg\hstar F_a },
\end{equation}
where ${F_a = \dd a + \om_a}$, ${a}$ is a ${p}$-form in Minkowski spacetime ${X}$, and ${H^{p+1}(X;\Z)}$ is the ${(p+1)}$th de Rham cohomology group with integral periods. This field theory describes the dynamical fluctuations of the ${\frac{(d-1)!}{p!(d-p-1)!}}$ ${p}$-form Goldstone bosons of the ${\U{p}}$ SSB phase~\cite{HHY200715901} traveling at the ``speed of light'' ${c = U\sqrt{\ka }}$. Furthermore, as reviewed in appendix section~\ref{pformMaxReviewSec}, it has an anomalous ${\U{p}}\times\U{d-p-1}$ symmetry~\cite{GW14125148}. Therefore, deep into the ${\U{p}}$ SSB phase of the lattice model, a new symmetry emerges in the continuum. So, the IR of region III has an exact emergent ${\U{p}}\times\U{d-p-1}$ symmetry in the continuum.

\section{Generalized Landau paradigm in practice}
\label{landauSec}

As mentioned in the introduction, an exciting prospect of generalized
symmetries is the expansion of Landau's original symmetry paradigm~\cite{L3726,
GL5064}. Since, as we have argued, emergent higher-form symmetries are exact
symmetries, to fully utilize the power of a generalized Landau symmetry
paradigm, we must consider both a system's microscopic and exact emergent
symmetries.  In particular, every emergent symmetry is accompanied by an
energy scale or a length scale. As we change parameters, those energy scales
may become zero, or the length scales diverge. This modified generalized
Landau paradigm can give new results, which is one of the key results of the
paper.  In this section, we demonstrate how this can be done in practice by
studying how the exact emergent 1-form symmetries affect the structure of the
phases and phase transitions of a simple concrete model. 

\subsection{Fradkin-Shenker model}

\begin{figure}[t!]
\centering
    \includegraphics[width=.48\textwidth]{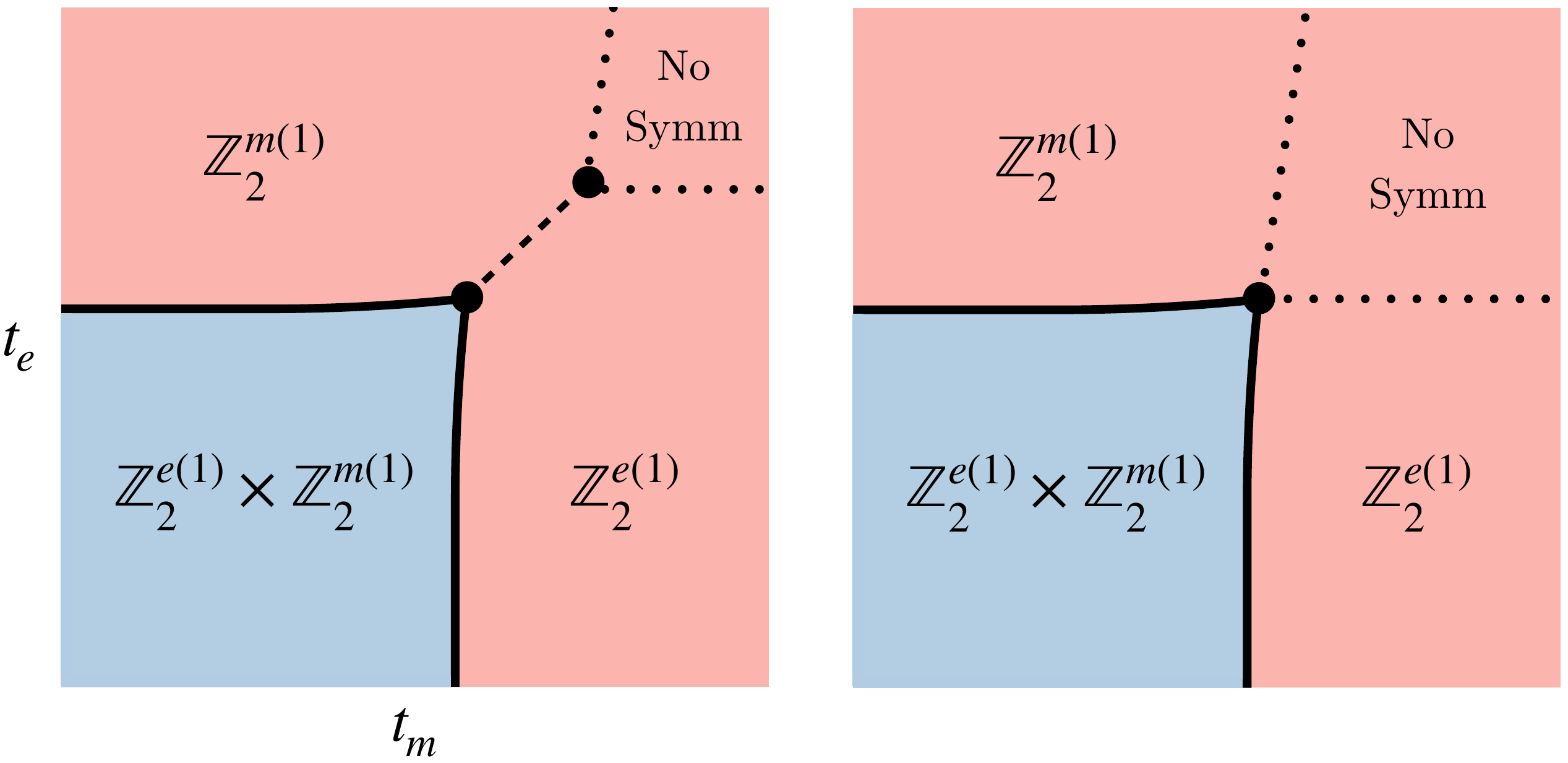}
	\caption{
(Left) The phase diagram of Fradkin-Shenker model~\eqref{Ham} labeled by its exact emergent symmetries. The topological phase is colored blue while the trivial phase is red. The solid line corresponds to a continuous transition, the dashed line corresponds to a first-order transition, and the dotted line is not a phase transition.
(Right) Another possible phase diagram. The phase transition curve is not smooth and has a singularity, which is a consequence of exact emergent 1-form symmetries.
} \label{TCphaseSymm} 
\end{figure}

We will study ${(2+1)}$D ${\Z_2}$ lattice gauge theory with matter. Let us consider the square lattice with periodic boundary conditions and a qubit residing on each link ${l}$ acted on by the Pauli matrices ${X_l}$ and ${Z_l}$. The Hamiltonian is
\begin{align}
\label{Ham}
 H &= - \sum_{s}  Q_{s} -\sum_{p}  F_{p}
 - t_e \sum_{l} X_{l}  - t_m \sum_{l} Z_{l}, 
\nonumber\\
& Q_s = \prod_{l\supset s} Z_{l},\ \ \ \ \ F_{p} = \prod_{l\subset p} X_{l},
\end{align}
where ${t_e, t_m\geq 0}$, ${Q_s}$ is a product over the four links meeting at site ${s}$, and ${F_p}$ is a product over the four links surrounding plaquette ${p}$. ${H}$ is the toric code~\cite{K032} in a magnetic field ${(t_e, t_m)}$, which is equivalent to the Fradkin-Shenker model~\cite{FS7982}. It has been intensely studied~\cite{TKP08043175, VDS08070487,DKO10121740, WP12016409, IS201106611, SSN201215845, BPV211201824, CHB220914302, VBV221101376} and famous for its phase at ${t_e,t_m\ll 1}$ with ${\Z_2}$ topological order~\cite{RS9173,W9164} (see Fig.
\ref{TCphaseSymm}).

This model has exact 1-form symmetries when ${t_e}$ and/or ${t_m}$ vanishes. When ${t_e = 0}$, it enjoys an anomaly-free ${\Z_2}$ 1-form symmetry generated by
\begin{align}
U_1^{(e)}(\t\ga) =\prod_{l\perp\t\ga} Z_{l},
\end{align}
where the product is over all links crossing the loop ${\t\ga\in Z_1(\t M;\Z_2)}$ of the dual lattice ${\t M}$. When ${t_m = 0}$, ${H}$ has a different anomaly-free ${\Z_2}$ 1-form symmetry generated by
\begin{align}
U_1^{(m)}(\ga) =\prod_{l\subset \ga} X_{l},
\end{align}
where the product is over all links in the loop ${\ga\in Z_1(M;\Z_2)}$ of the lattice ${M}$. We distinguish these by calling them the electric and magnetic symmetries, respectively, and denoting them as ${\Z_2^{e(1)}}$ and ${\Z_2^{m(1)}}$. When ${t_e = t_m = 0}$ and ${H}$ is the toric code model, both symmetries are present and act anomalously. This anomaly ensures that both symmetries are spontaneously broken in any gapped phase~\cite{CO191004962}.

There are additional symmetries besides these two 1-form symmetries. For
instance, when ${t_e = t_m}$,  
${H}$ has a $e$-$m$-exchange symmetry ${\Z_2^{em}}$
whose action exchanges ${X_l\leftrightarrow Z_{l+\h x /2 + \h y /
2}}$. There are other ${0}$-form symmetries at the toric code point as well,
which are generated by condensation defects of the two 1-form symmetries.
However, these additional symmetries do not appear to play a role in
characterizing the model's phase diagram, and we will therefore focus on the
two 1-form symmetries from here on

These microscopic symmetries are present in a small subspace of parameter space and thus are inadequate in classifying the model's phases and phase transitions. However, since they include 1-form symmetries, after explicitly breaking them they will survive as exact emergent symmetries at low energies.

For instance, the ${\Z_2^{e(1)}}$ symmetry at ${t_e = 0}$ exists as an exact emergent symmetry whenever there exists an ${e}$ anyon string operator. Thus, it emerges at energy scales below twice the ${e}$ anyon gap when ${t_e \neq 0}$. When ${t_e = 0}$, the ${e}$ anyon's string operator is simply ${X_l}$, while for ${t_e\neq 0}$ they are fattened and dressed by a particular local unitary ${U_{\txt{LU}}}$. The emergent symmetry operators will be generated by the fattened loop operators, which is ${U_1^{(e)}}$ dressed by ${U_{\txt{LU}}}$. As an exact emergent symmetry, ${\Z_2^{e(1)}}$ has a length scale $l_\text{symm}$ and an energy scale ${E_\text{symm} \sim \cO(1-t_e)}$. If $t_e$ is too large then ${l_\text{symm}=\infty}$ and/or ${E_\text{symm}=0}$ causing the exact emergent higher-form symmetry to disappear.

When ${t_e = 0}$ and the ${\Z_2^{e(1)}}$ symmetry is a microscopic symmetry that is spontaneously broken for small ${t_m}$. The phase transition at ${t_m\approx 0.34}$~\cite{TKP08043175} is controlled by ${\Z_2^{e(1)}}$. Since the dual symmetry of ${\Z_2^{e(1)}}$ is a ${\Z_2}$ ${0}$-form symmetry, the transition is described by the gauged Ising---Ising*---conformal field theory. Since ${\Z_2^{e(1)}}$ is an exact emergent symmetry when ${t_e\neq 0}$, the symmetry broken phase and phase transition persist away from ${t_e=0}$, as shown in Fig.~\ref{TCphaseSymm}, despite the lattice model no longer having a microsopic ${\Z_2^{e(1)}}$ symmetry.

This discussion applies to ${\Z_2^{m(1)}}$ as well, just with all electric and magnetic variables exchanged. Therefore, the topological phase has an exact emergent anomalous ${\Z_2^{e(1)}\times \Z_2^{m(1)}}$ symmetry spontaneously broken and a phase transition into the Higgs (confined) regime is driven by restoring the magnetic (electric) 1-form symmetry. This means that the Higgs (confined) regime has an exact emergent unbroken $\Z_2^{m(1)}$ ($\Z_2^{e(1)}$) symmetry. We propose that across the first-order transition line, the exact emergent $\Z_2^{m(1)}$ symmetry switches to the exact emergent $\Z_2^{e(1)}$ symmetry, and vice versa. 

There should be no region outside of the topological phase with both emergent $\Z_2^{m(1)}$ and $\Z_2^{e(1)}$ symmetries since otherwise, their mixed anomaly would require the gapped phase at the upper right corner of Fig. \ref{TCphaseSymm} to be a non-product state. This implies that there is a region in the upper right corner of the phase diagram with no emergent symmetry. 

Since both the Higgs and confined regimes lie in the trivial
phase~\cite{FS7982, BR1979349}, one may wonder how they can have different
symmetries. Indeed, Higgs and confined phases can be distinguished when they
have different realizations of symmetries (see recent work \Rfs{CJS200708539,
HK221011492, VBV221101376, TRV230308136} on the subject). However, with
periodic boundary conditions, both symmetries are realized trivially since the
charged states cost infinite energy in the thermodynamic limit. This is why
there is no phase transition when the exact emergent unbroken 1-form symmetries
disappear across the dotted line in Fig.~\ref{TCphaseSymm}. While the emergent
symmetry is not represented faithfully, it is still important to keep track of
it since it controls the universality class of the phase transition out of the
topological phase.

Since the two continuous transitions out of the topological phase (the Higgsing
and confining transitions) have different 1-form symmetries, there must be a singularity (i.e., a multi-critical point) where the transition lines meet since their symmetries will switch
${\Z_2^{m(1)}\leftrightarrow\Z_2^{e(1)}}$. In model \eqref{Ham}, this
singularity happens to be at the end of the first-order transition line.
It would be interesting to study a generalization of the model that has
no $e$-$m$-exchange symmetry even along the diagonal line, to see if there are
other forms of singularities along the continuous transition line, in
particular, if the first-order transition line can shrink to a point [see
Fig.~\ref{TCphaseSymm}~(right)]. We predict the existence of singularities
along the continuous transition line even for general models.

When transitioning from the topological phase to the Higgs regime, the energy
scale for the exact emergent ${\Z_2^{e(1)}}$ symmetry vanishes which causes the
electric symmetry to no longer emerge. We therefore say that the critical point
of the transition has a marginal emergent ${\Z_2^{e(1)}}$ symmetry.  As a
definition, a system has a \emph{marginal emergent symmetry} if there exists an
infinite sequence of systems approaching the original system, such that each
system in the sequence has the exact emergent symmetry. This appears to be a
concept that is unique to exact emergent higher-form symmetries. Similarly, the
multi-critical point does not have exact emergent ${\Z_2^{m(1)}\times
\Z_2^{e(1)}}$ symmetry, but instead a marginal emergent $\Z_2^{m(1)}\times
\Z_2^{e(1)}$ symmetry. An interesting future direction is to investigate the
role marginal emergent symmetries play in characterizing phase transitions.  

\section{Physical consequences}\label{sec:physCons}

Emergent ${0}$-form symmetries are typically not exact, so their
consequences are approximate. However, as we have shown, emergent higher-form
symmetries are exact emergent symmetries. Therefore, their low-energies consequences are exact and equivalently powerful as UV symmetries.
Furthermore, since emergent higher-form symmetries are robust against translation-invariant local
perturbations, physical properties arising from their existence are also
robust. In this section, we summarize the physical consequences of emergent
higher-form symmetries being exact. We emphasize their role in characterizing
phases of matter, fitting exact emergent symmetries into the generalized Landau
classification scheme (see \Rf{M220403045}).

\subsection{Spontaneous symmetry breaking}\label{SEC:physConsSSB}

Since spontaneous symmetry breaking (SSB) is diagnosed using the ground state, an emergent higher-form symmetry can be spontaneously broken in the same way a UV symmetry can be spontaneously broken. A consequence of this is that a phase with an emergent discrete higher-form symmetry spontaneously broken has an exact ground state degeneracy (GSD) which depends on spacetime's topology. Similarly, a phase with an emergent continuous higher-form symmetry spontaneously broken has Goldstone bosons. If the continuous higher-form symmetry emerges at ${E < E_{\txt{mid-IR}}}$, these Goldstone bosons are exactly gapless for mid-IR states. However, for states in the sub-Hilbert space spanned by energy eigenstates with ${E \geq E_{\txt{mid-IR}}}$, the Goldstone bosons acquire a gap.\footnote{This is a familiar concept in the ${p=1}$ case where electric screening causes the photon to acquire a gap.} 

Since emergent higher-form symmetries are topologically robust, a local translation-invariant UV perturbation does not gap out their Goldstone bosons nor lift the topological GSD. This is very different from ${0}$-form symmetries where even weakly breaking the symmetry in the UV gaps out the Goldstone boson~\cite{W19721698} or lifts the GSD.

The SSB phase of an emergent higher ${p}$-form symmetry has gapped topological defect excitations. When an anomaly-free ${\U{p}}$ (${\ZN{p}}$) symmetry spontaneously breaks in ${d}$-dimensional space, there are ${d-p-2}$ (${d-p-1}$) dimensional topological defects carrying ${\Z}$ (${\Z_N}$) topological charge~\cite{S150804770, IM210612610, PL231109293}. For ${p=1}$ and ${d=3}$, this is the magnetic monopole of ${U(1)}$ gauge theory (the flux loop of ${\Z_N}$ gauge theory). In the trivial symmetric phase, the topological defects are condensed. 

As we saw in section~\ref{exactEmEx}, when the ${\U{p}}$ (${\ZN{p}}$) topological defect has a gap, there is a low-energy regime with an exact emergent ${\U{d-p-1}}$ (${\ZN{d-p}}$) symmetry. We can flip this around and define the existence of gapped topological defects by the presence of these exact emergent symmetries. Therefore, a ${\U{p}}$ (${\ZN{p}}$) symmetry can spontaneously break in ${d}$-dimensional space only when ${\U{d-p-1}}$ (${\ZN{d-p}}$) can be an exact emergent symmetry. Since only emergent higher-form symmetries are exact at zero temperature (see section~\ref{sec:finiteT}), a ${\U{p}}$ (${\ZN{p}}$) symmetry can only spontaneously break when ${d>p+1}$ (${d>p}$), agreeing with \Rfs{GW14125148, L180207747}.

In the SSB phase of an emergent higher-form symmetry, the low-energy states and observables are sometimes organized into the symmetry's representations. This is not generically true since emergent ${p}$-form symmetries have nontrivial charged operators only when there are nontrivial ${p}$-cycles in space. When nontrivial ${p}$-cycles exist, charged operators create ${p}$-brane excitations costing finite energy in the SSB phase, so low-energy states fall into representations of the emergent symmetry. This then gives rise to selection rules on the correlation functions of low-energy operators. 

When space has no nontrivial ${p}$-cycles, the emergent ${p}$-form symmetry is trivialized, and one may be tempted to say there is no emergent symmetry. Nevertheless, it still has a corresponding exact emergent conservation law, and thus there is a transformation that leaves the low-energy effective theory unchanged. Furthermore, the SSB phase still has neutral charges condensed and, in the ${U(1)}$ case, gapless Goldstone bosons. Therefore, the emergent symmetry in this case still has many nontrivial consequences of a symmetry, and we thus interpret it as a symmetry.

\subsection{'t Hooft anomalies}

Exact emergent anomaly-free symmetries can be gauged at the energy scales they exist.  Since every symmetry implies a dual symmetry~\cite{JW191213492} found by
gauging~\cite{BT170402330}, this implies that exact emergent higher-form symmetries also have
dual symmetries. However, it also implies there can be obstructions to gauging and thus 't Hooft anomalies

An emergent higher-form symmetry can be anomalous with or without spontaneous symmetry breaking and has consequences regardless of space's topology and boundaries (see also \Rf{CJ230413751}). Such an anomaly can include only exact emergent symmetries or both exact emergent and exact symmetries.

A 't Hooft anomaly prevents the ground state from being a trivial product state due to anomaly matching, providing IR constraints from UV data. For an exact emergent anomalous symmetry, all energy scales below which the emergent anomalous symmetry is present must also respect anomaly matching. Therefore, exact emergent anomalous symmetries also obstruct a trivial ground state, thus providing useful IR constraints using mid-IR data.

In the examples from section~\ref{exactEmEx}, the SSB phases of the models had exact emergent anomalous higher-form symmetries below the topological defect's gap. One can view the ground state degeneracies and gaplessness of Goldstone bosons in these phases as being protected by the 't Hooft anomaly. Another example is a bosonic superfluid. There is an exact emergent higher-form ${U(1)}$ symmetry below the vortex gap that is not spontaneously broken but whose existence contributes to a 't Hooft anomaly protecting superflow~\cite{ES210615623}. Thus, if the topological defect's gaps were held at infinity in these examples, the ground state could never become a trivial product state.

\subsection{Without spontaneous symmetry breaking}

When a ${p}$-form symmetry (${p>0}$) is unbroken, its ${p}$-brane symmetry
excitations are gapped, and their gap grows with their size. Therefore, in the
thermodynamic limit of a compact space, charged states cost infinite energy, and all finite-energy states are in the symmetric sector of the emergent
higher-form symmetry. Since the emergent symmetry is trivial, one may again be tempted to say there is no emergent symmetry at all. However, even exact
symmetries trivialize at low energies when unbroken, and they still have
physical consequences, although very subtle. Therefore, as we will discuss,
unbroken emergent higher-form symmetries can still have the nontrivial
effects of a symmetry, and we thus interpret it as a symmetry.

If an exact
emergent higher-form symmetry is anomaly-free and not spontaneously broken in the absence of a boundary, it can characterize nontrivial
symmetry-protected topological (SPT)
phases~\cite{PW220703544,MF220607725,VBV221101376, TRV230308136}. Indeed, the existence of an emergent higher-form symmetry implies that there
are boundaries with the emergent symmetry. Arbitrary perturbations of such boundaries also have the emergent higher-form symmetry. Therefore, a corresponding nontrivial SPT order could exist in the bulk if the emergent higher-form symmetry is realized anomalously on such boundaries. The bulk SPT order cancels the 't Hooft anomaly by anomaly in-flow~\cite{CH8536}, ensuring the theory remains gauge invariant when background gauge fields are turned on.

Emergent SPT orders have direct physical
consequences in the presence of a spatial boundary. Indeed, since the emergent higher-form symmetry is realized anomalously on the boundary, all the physical consequences discussed in the previous two subsections, like symmetry breaking and obstructions to trivial ground states, apply on this boundary.

In the absence of a boundary, the effective IR theory of the SPT will be an invertible topological field theory in terms of the background fields. From a low-energy point of view, this is no different from an SPT protected by a UV symmetry. Indeed, the UV symmetry is trivial in the IR since it is unbroken, but an invertible topological field theory in terms of its background gauge fields characterizes the SPT order~\cite{K1459, WGW1489}. Moreover, this invertible theory has physical meaning: it is the effective response theory of the SPT.

This emphasizes an important distinction between SPTs protected by 0-form and higher-form symmetries. 0-form SPTs cannot occur in regions of parameter space where the 0-form symmetry is explicitly broken. However, this is untrue of higher-form SPTs since emergent higher-form symmetries are exact. Therefore, to identify higher-form SPT phases, instead of partitioning parameter space by exact higher-form symmetries, one should partition it by the emergent higher-form symmetries.

We note that this perspective of SPTs privileges the characteristic that the protecting symmetry is realized anomalously on a boundary. It then uses anomaly inflow to relate this boundary feature to a bulk property, which can be detected through its topological response at low energy. Whether this is enough to sharply define a bulk property/observable that characterizes a phase of matter is important to further investigate.

\begin{figure}[t!]
\centering
    \includegraphics[width=.48\textwidth]{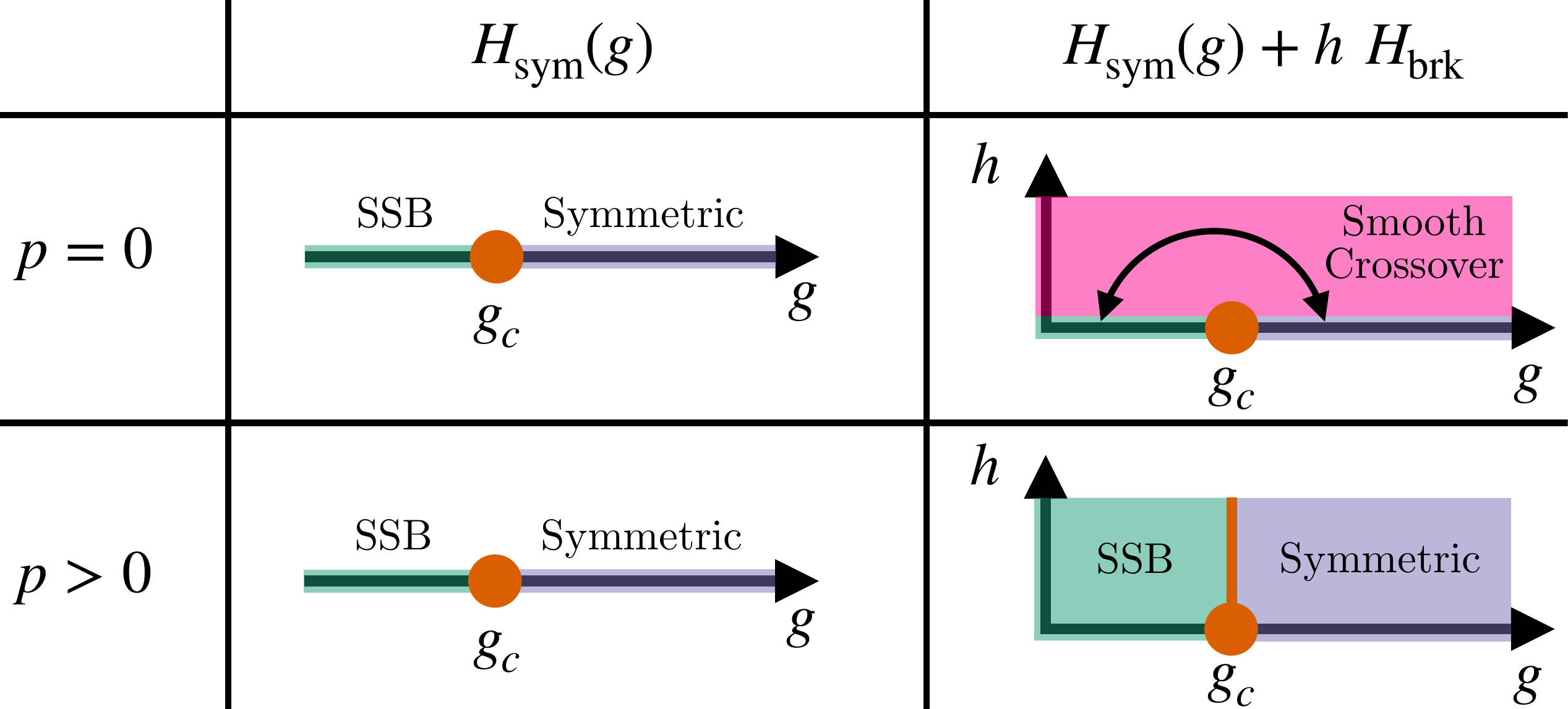}
    \caption{A Hamiltonian ${H_{\txt{sym}}(g)}$ with a ${p}$-form
symmetry spontaneously broken at ${g<g_c}$ has a  qualitatively similar phase diagram whether ${p=0}$ or ${p>0}$. However, explicitly breaking the
symmetry with ${h H_{\txt{brk}}}$ ${(|h|\ll 1)}$, the phase diagram of
${H_{\txt{sym}}(g)+h H_{\txt{brk}}}$ differs between ${p=0}$
and ${p>0}$. For ${p=0}$, there is now a smooth crossover between what used to be the SSB
phase and the symmetric phase. Shown here is the case where the symmetry-breaking perturbation is relevant. For ${p>0}$, ${H_{\txt{sym}}(g) + h H_{\txt{brk}}}$ has an exact emergent ${p}$-form symmetry, so there is not a smooth crossover between the phases. Starting in the SSB phase and increasing ${h}$ would lead to an eventual phase transition out of the SSB phase.
}
    \label{fig:excplitbrkCP}
\end{figure}

Emergent higher-form symmetries can also exactly characterize their SSB phase transitions (with or without spatial boundaries)~\cite{SSN201215845, IM210612610}, as we saw in Sec.~\ref{landauSec}. Indeed, transitioning from the SSB phase of an exact emergent symmetry into its symmetric phase, the critical point will have the exact emergent symmetry and can be in its symmetry-breaking pattern universality class (see Fig.~\ref{fig:excplitbrkCP}). An example is the confinement transition of
${(2+1)}$D ${\mathbb{Z}_2}$ lattice gauge theory with dynamical matter. The matter fluctuations explicitly break a ${\mathbb{Z}_2^{(1)}}$ symmetry, but the transition is still in the Ising universality class since there is an exact emergent ${\mathbb{Z}_2^{(1)}}$
symmetry~\cite{SSN201215845}.

Emergent invertible higher-form symmetries can interact nontrivially with exact symmetries, forming an emergent higher-group symmetry~\cite{BH180309336, BCH221111764}. For example, when an ${A^{(1)}}$ symmetry emerges in the presence of an exact 2-group symmetry, the total low-energy symmetry is described by a 2-group ${\mathbb{G}^{(2)} = (G, A,\rho,[\beta])}$, where ${G}$ is the IR 0-form symmetry, ${\rho: G \to \txt{Aut}(A)}$, and ${[\beta]\in H_\rho^3(\cB G; A)}$. The 1-form symmetry, even without spontaneous symmetry breaking, can be nontrivial if the Postnikov class ${[\beta]}$ is nontrivial.

\subsubsection{A dynamical effect}

An exact emergent higher-form symmetry constrains the ${p}$-brane symmetry excitations’ dynamics. For simplicity, let us set ${p=1}$ and consider a state with symmetry excitation excited on a contractible ${1}$-cycle ${C_1}$. 

\begin{figure}[t!]
\centering
    \includegraphics[width=.48\textwidth]{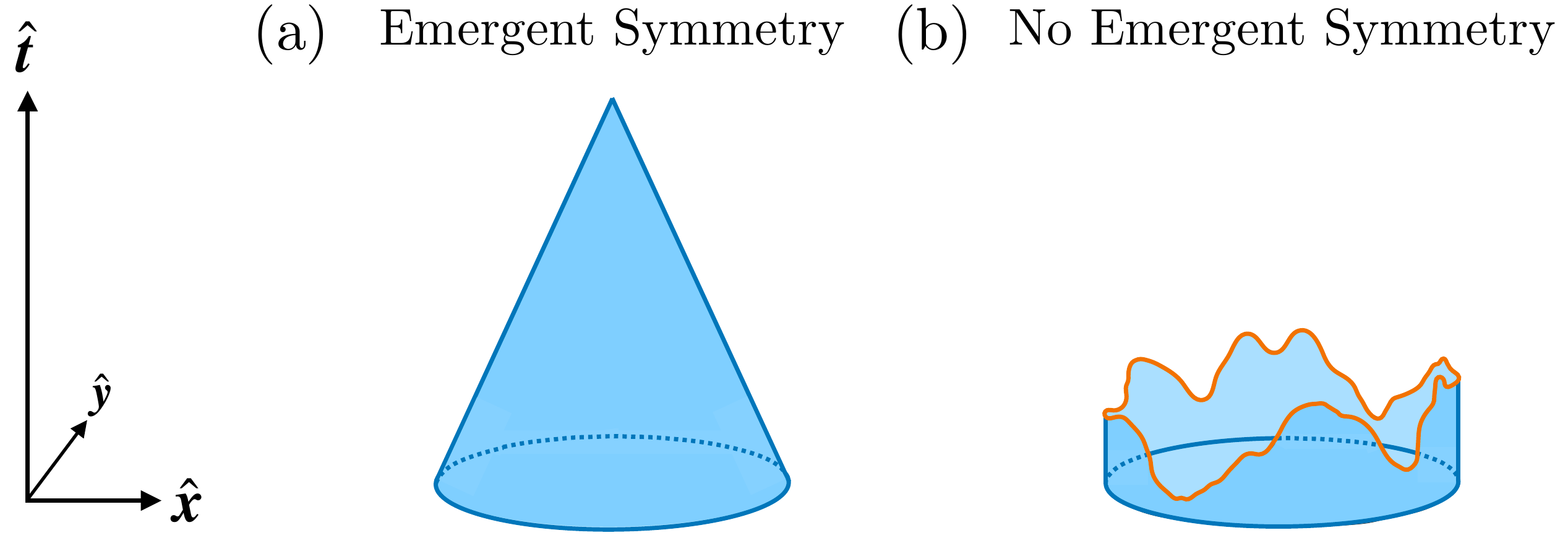}
    \caption{Shows a cartoon of the different decay processes of 1-form symmetry excitations in ${2+1}$d spacetime. The world sheet of the ${1}$-brane excitation is colored blue, while the worldline of gauge charge excitations is orange.}
    \label{fig:dynam}
\end{figure}

We first assume that symmetry excitations have a fixed energy $\eps$ per lattice edge and that open string ends (e.g., gauge charges) have an energy gap $\Del$. The 1-form symmetry exists only at energies ${E<\Del}$ and affects symmetry excitations with ${|C_1|\eps \ll \Del}$. For symmetry excitations with ${|C_1| \ll \Del/\eps}$, due to the emergent 1-form symmetry, the only way for them to decay is by contracting to a point (see Fig.~\ref{fig:dynam}a). So, their life time ${\tau}$ grows with their size ${|C_1|}$. Symmetry excitations with ${|C_1| > \Del/\eps}$ are not affected by the 1-form symmetry and can, therefore, decay by quantum tunneling to states with open string ends (see Fig.~\ref{fig:dynam}b). Therefore, their lifetime is independent of their size, instead going like ${\tau \sim \ee^{\Del/\eps}}$.

Let us now assume that a symmetry excitation is trapped by a trap potential~\cite{W150605768} and has a fixed total energy ${|C_1|\eps}$. If the trapped symmetry excitation has ${|C_1| \ll \Del/\eps}$, it can no longer decay since the trap potential prevents it from contracting to a point. Such a symmetry excitation is an exact quantum many-body scar (QMBS) state~\cite{SAP201109486}. If the trapped symmetry excitation has ${|C_1| > \Del/\eps}$, it will still decay with a finite lifetime ${\tau \sim \ee^{\Del/\eps}}$. However, it will have a long lifetime when ${\eps}$ is small, making large symmetry excitation loops approximate QMBS states. The larger the trap, the better the QMBS state with a given energy, so infinite-sized loop excitations are exact QMBS states. Furthermore, any small perturbation of the trap would still lead to an approximate QMBS state, but with lifetime $\tau \sim \ee^{\Del/|\eps+\del \eps|}$ where $\del \eps$ is the strength of perturbation. Thus, the existence of the emergent 1-form symmetry at ${E<\Del}$ implies that there exists a large potential trap leading to exact QMBS states at ${E<\Del}$ and approximate QMBS states at ${E>\Del}$.

Both of these scenarios apply to ${p}$-form symmetries with ${p>1}$ under a straightforward generalization. In the latter, the trap potential can trap ${p>1}$ dimensional symmetry excitations which lead to QMBS states. However, it can also trap topologically ordered states which can lead to QMBS states. So, to be precise, we say that an exact emergent ${p}$-form symmetry implies the existence of ${p}$-dimensional trap potentials leading to QMBS states besides those corresponding to topologically ordered states.

\section{Finite temperature effects}\label{sec:finiteT}

Here we discuss how our results are modified at finite temperature ${T}$. When ${T\neq 0}$, the imaginary time direction becomes ${S^1}$ with radius ${1/T}$. Thus ${S^1}$ is a small dimension of spacetime when the linear system size ${L \gg 1/T}$, and spacetime can be dimensionally reduced from ${(d+1)}$-dimensional ${M_d\times S^1}$ to ${d}$-dimensional ${M_d}$ in the thermodynamic limit.

Let us warm up by first considering how the SSB critical dimension for higher-form symmetries changes.\footnote{We thank Carolyn Zhang for helpful discussions about this} At ${T=0}$, a ${\ZN{p}}$ (${\U{p}}$) symmetry can spontaneously break when ${d+1>p+1}$ (${d+1>p+2}$)~\cite{GW14125148, L180207747}. Using dimensional reduction, we can understand the finite temperature case by replacing ${d+1}$ with ${d}$. Thus, at ${T\neq 0}$, a ${\ZN{p}}$ (${\U{p}}$) symmetry can spontaneously break when ${d>p+1}$ (${d>p+2}$).

So, are emergent higher-form symmetries still exact when ${T\neq 0}$? At finite temperature, ${p}$-form symmetry charged operators can act on nontrivial ${p}$-cycles winding around the compactified imaginary time direction. When dimensionally reducing spacetime, this direction becomes negligible, and any such ${p}$-cycles become ${(p-1)}$-cycles. So, it is as if ${p}$-form symmetries acts like ${(p-1)}$-form symmetries. Since emergent ${0}$-form symmetries are not exact, emergent ${1}$-form symmetries are no longer exact at finite temperature. However, emergent ${p}$-form symmetries with ${p>1}$ are still exact at ${T\neq 0}$.

This can be seen without having to dimensionally reduce. Indeed, at ${T\neq 0}$, ${1}$-cycles winding around imaginary time are nontrivial 1-cycles of finite length even in the thermodynamic limit. Assuming the UV theory is local, the low-energy effective Euclidean action includes terms suppressed by ${\ee^{-1/T}}$ with the emergent 1-form symmetry's charged operators~\cite{IM210612610}. Hence, emergent 1-form symmetries at ${T\neq 0}$ are approximate, not exact. For ${p}$-form symmetries with $p>1$, there are no finite sized nontrivial ${p}$-cycles at ${T \neq 0}$. Indeed, while a ${p}$-cycle could wind around the finite-sized imaginary time direction, it must also wind around ${(p-1)}$ spatial directions, which are not finite-sized in the thermodynamic limit. Therefore, emergent ${p}$-form symmetries with ${p>1}$ are still exact at ${T\neq 0}$.

Let us now apply the above discussion to known examples. 

The ${p}$-form toric code model (see Eq.~\eqref{dressedpTC}) lies in the SSB phase of an anomalous ${\ZN{p}\times\ZN{d-p}}$ symmetry, where ${0<p<d}$. The ${e}$ excitations are ${(p-1)}$ branes while the ${m}$ excitations are ${(d-p-1)}$ branes. Whether or not its topological order is robust at finite temperature depends on if this symmetry remains spontaneously broken at ${T\neq 0}$. This requires ${d>p+1}$ for ${\ZN{p}}$ and ${d>d-p+1}$ for ${\ZN{d-p}}$. Therefore, for ${\ZN{p}\times\ZN{d-p}}$ to spontaneously break at ${T\neq 0}$, ${p}$ and ${d}$ must satisfy
\begin{equation}\label{pformfiniteTto}
2 \leq p \leq d-2.
\end{equation}

This is never satisfied for ${d=2}$ or ${3}$, recovering that the toric code in ${d=2}$ and ${3}$ at finite temperature does not have topological order~\cite{DKL0252,CC08043591,AFH08104584,AHH08110033}. In the ${d=3}$ case, the exact symmetry is ${\ZN{1}\times\ZN{2}}$, and there is no topological order at finite temperature since ${\ZN{2}}$ cannot spontaneously break. However, the ${\ZN{1}}$ still can, giving rise to the ``classical topological order'' discussed in \Rf{CC08043591}. 

${d=4}$ is the smallest spatial dimension for which Eq.~\eqref{pformfiniteTto} is satisfied, which recovers that the 2-form toric code's topological order is robust at finite temperature~\cite{DKL0252, AHH08110033}. In this case, ${p=2}$, so both ${e}$ and ${m}$ excitations are loops.

Weakly perturbing the ${p}$-form toric code explicitly breaks its ${\ZN{p}\times\ZN{d-p}}$ symmetry. When ${T\neq 0}$, its can only emergent exactly if ${p>1}$ and ${d-p>1}$, which is precisely Eq.~\eqref{pformfiniteTto}. The ``classical topological order'' when ${d=3}$ is not robust to these perturbations at ${T\neq 0}$ since the emergent ${\ZN{1}}$ symmetry will be approximate.

Next, consider ${U(1)}$ quantum spin liquids, a class of spin liquid phases whose effective description is the deconfined phase of pure ${U(1)}$ gauge theory~\cite{SB160103742}. The prototypical example in ${d=3}$ is quantum spin ice~\cite{GM13111817, HFB0404, BSS12041325, PMM200904499}. More generally, a ${p}$-form ${U(1)}$ quantum spin liquid's effective IR theory is ${p}$-form Maxwell theory, which lies in the SSB phase of a ${\U{p}\times \U{d-p-1}}$ symmetry. If such an SSB phase is robust at finite temperature, ${d>p+2}$ and ${d>d-p-1+2}$ for ${\U{p}}$ and ${\U{d-p-1}}$, respectively, to spontaneously break. Therefore, for ${\U{p}\times\U{d-p-1}}$ to spontaneously break at ${T\neq 0}$, ${p}$ and ${d}$ must satisfy
\begin{equation}\label{UpfiniteT}
2 \leq p \leq d-3.
\end{equation}

This is never satisfied for ${d=2}$, ${3}$, or ${4}$. ${d=5}$ is the smallest spatial dimension where a ${U(1)}$ quantum spin liquid phase is stable at finite temperature. The low-energy description of such a phase is ${(5+1)}D$ ${U(1)}$ ${2}$-form Maxwell theory, where both electric and magnetic excitations are loops. 

When ${d=4}$ and ${p=1}$, the emergent symmetry is ${\U{1}\times\U{2}}$. While the ${\U{2}}$ symmetry cannot spontaneously break at ${T\neq 0}$, the ${\U{1}}$ symmetry can, giving rise to a ``classical ${U(1)}$ topological order.'' However, this emergent ${\U{1}}$ symmetry is not exact. Indeed, the emergent higher-form ${U(1)}$ symmetries are only exact at ${T\neq 0}$ when Eq.~\eqref{UpfiniteT} is satisfied.

\section{Conclusion and discussion}\label{sec:conclusion}

In this paper, we have investigated the robustness of emergent higher-form symmetries from a UV perspective, considering bosonic lattice Hamiltonian models. In section~\ref{sec:midIRandQuasiAdiCont}, we showed how emergent higher-form symmetries in lattice models are generally exact symmetries and not approximate symmetries. To emphasize this robustness, we referred to emergent higher-form symmetries as \txti{exact emergent symmetries}. This means that lattice models without exact higher-form symmetries can have emergent higher-form symmetries whose effects at low energy are the same as if they were exact symmetries. Therefore, emergent higher-form symmetries can exactly characterize phases of systems without exact higher-form symmetries. We considered three examples of this in section~\ref{exactEmEx}, discussed the general physical consequences in section~\ref{sec:physCons}, and discussed finite temperature effects in section~\ref{sec:finiteT}.

As discussed in section~\ref{sec:midIRandQuasiAdiCont}, when higher-form symmetries are emergent, their symmetry and charged operators are ``fattened''~\cite{HW0541}. The exact expression of these operators depends on the UV parameters, and finding such a closed form requires an exact expression for the local unitary $U_{\txt{LU}}$. Not only is this highly nontrivial, but likely analytically intractable for generic models. That said, a promising approach to find exact expressions is using numerical schemes. This was done for untwisted and twisted $\Z_2$ lattice gauge theory in \Rf{CHB220914302}, which developed a novel unbiased numerical optimization scheme to systemically find the dressed symmetry operators. It would be interesting to extend these machine-learning approaches to other models.

An important follow-up to our paper is an in-depth study of the boundary between regions I and II in Fig.~\ref{fig:phaseDia}. One possibility is that a boundary phase transition separates the two regions, as in~\Rfs{VBV221101376, TRV230308136}. What about a bulk point of view? Recall from section~\ref{sec:midIRandQuasiAdiCont} that emergent higher-form symmetries have an associated energy and length scale. For instance, the emergent symmetry is destroyed if the energy scale vanishes, which is precisely what happens when going from region III to I in Fig.~\ref{fig:phaseDia}. Perhaps when going from region II to I, it is the length scale that blows up. This would prevent the symmetry and charged operators from being well-defined, destroying the emergent symmetry. It would be interesting to further investigate this possibility.

\section{Acknowledgements}

We are grateful for fun and helpful discussions with
Arkya Chatterjee,
Xie Chen,
Hart Goldman,
Ethan Lake,
Ho Tat Lam,
Yu Leon Liu,
Zijian Xiong,
and Carolyn Zhang.
S.D.P. is supported by the National Science Foundation Graduate Research Fellowship under Grant No. 2141064 and by the Henry W. Kendall Fellowship.
This work is partially supported by NSF DMR-2022428
and by the Simons Collaboration on Ultra-Quantum Matter, which is a grant from
the Simons Foundation (651446, XGW).

\appendix

\section{Discrete differential geometry for hypercubic lattices}\label{sec:diffgeoLat}

\begin{figure}[t!]
\centering
    \includegraphics[width=.48\textwidth]{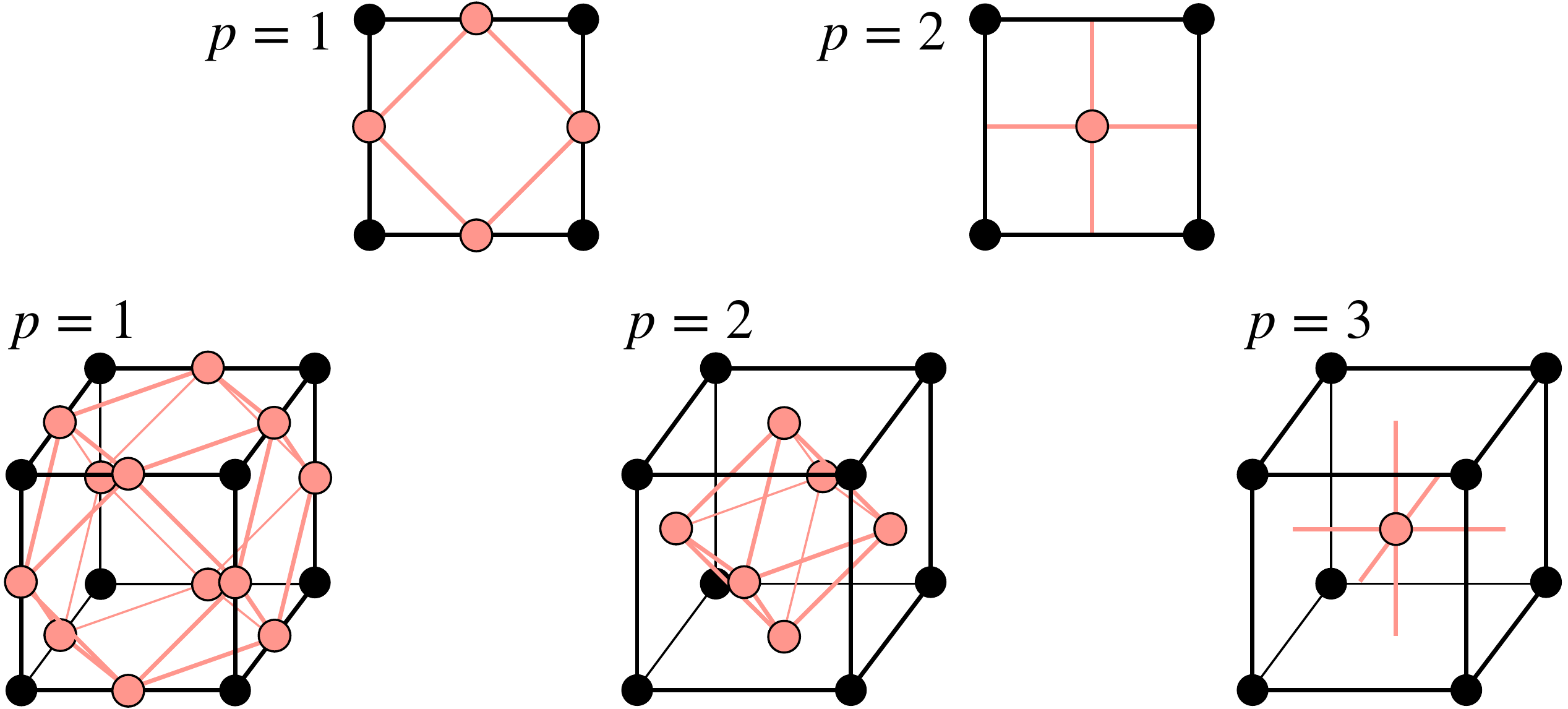}
    \caption{%
    The ${p}$-cells of the ${d}$-dimensional cubic lattice are equivalently the ${0}$-cells of another lattice. Shown here are examples of this equivalent lattice embedded in the conventional unit cell of the cubic lattice, drawn in pink and black, respectively, for (first row) ${d=2}$ and (second row) ${d=3}$.
    }
    \label{fig:lattices}
\end{figure}

This appendix reviews relevant parts of discrete differential geometry (in a non-rigorous fashion) used throughout the main text.\footnote{We adopt the notation and conventions used in \Rf{SG190102637}.} Consider a hypercubic lattice in ${d}$-dimensional space, denoted by ${M_d}$. While a Bravais lattice is a collection of lattice sites ${\bm x \in \Z^d}$, it is useful to view it as also formed by higher-dimensional objects, like links, plaquettes, cubes, \etc. We call a ${p}$-dimensional object a ${p}$-cell, with ${0\leq p \leq d}$. So, a ${0}$-cell is a lattice site, a ${1}$-cell is a link, a ${2}$-cell is a plaquette, \etc. The ${p}$-cells of the ${d}$-dimensional cubic lattice are equivalently viewed as the ${0}$-cells of some other lattice in ${d}$-dimensions, as demonstrated for ${d=2}$ and ${3}$ in Fig.~\ref{fig:lattices}.

${p}$-cells do not add additional structures to the lattice but are just a useful way of organizing the lattice sites. Indeed, denoting a ${p}$-cell associated with site ${\bm x}$ as ${c_p(\bm x)_{\mu_1\mu_2\cdots\mu_p}}$, where ${\mu_1 < \mu_2 < \cdots < \mu_p}$ and ${\mu_i\in\{1,2,\cdots,d\}}$, a ${p}$-cell of the cubic lattice is the set of ${2^p}$ lattice sites
\begin{equation}
\begin{aligned}
\hspace{-4pt}c_p(\bm x)_{\mu_1\mu_2\cdots\mu_p} \hspace{-4pt}= 
\{\bm x\} 
&\cup \{\bm x+\bm{\h\mu}_{i} ~|~ 1 \leq i \leq p\} \\
&\cup\{\bm x+\bm{\h\mu}_{i}+\bm{\h\mu}_{j} ~|~ 1 \leq i<j \leq p\} \\
&\cup \cdots\cup \{\bm x+\bm{\h\mu}_{1}+\ldots+\bm{\h\mu}_{p}\},
\end{aligned}
\end{equation}
where ${\bm{\h\mu}_i}$ is the unit vector in the ${\mu_i}$-direction. It is often convenient to drop the requirement that the indices are ordered (i.e., ${\mu_1<\mu_2<\cdots<\mu_p}$) and instead let ${c_p(\bm x)_{\mu_1\mu_2\cdots\mu_p}}$ obey ${c_p(\bm x)_{\cdots\mu_1\mu_2\cdots} = -c_p(\bm x)_{\cdots\mu_2\mu_1\cdots}}$.

\begin{figure}[t!]
\centering
    \includegraphics[width=.48\textwidth]{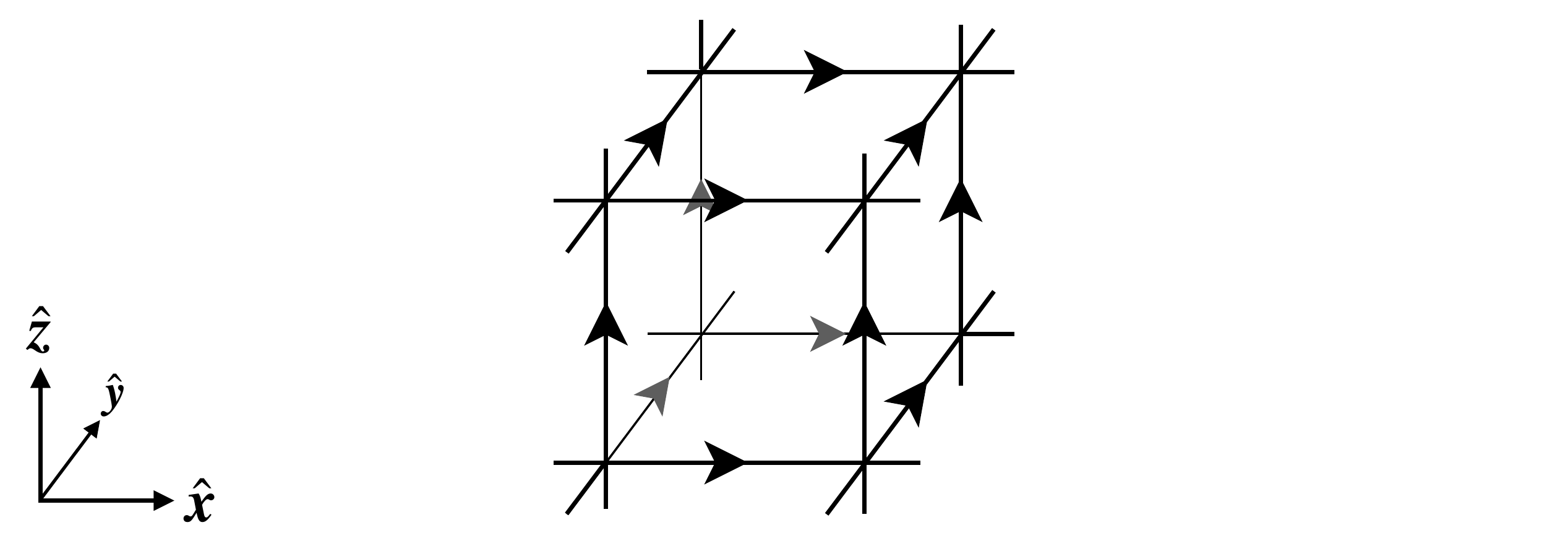}
    \caption{%
    Example of the branching structure used for a chunk of the cubic lattice in three-dimensional space.
    }
    \label{fig:branchingStructure}
\end{figure}

Introducing the concept of ${p}$-cells is convenient since ``sewing'' ${p}$-cells together gives a natural way to form ${p}$-dimensional subspaces of the lattice. Furthermore, these subspaces can be given an orientation by defining an orientation structure to the lattice. A nice local scheme for the lattice orientation is a branching structure, where the orientation on each ${1}$-cell is chosen such that a collection of ${1}$-cells cannot form an oriented closed loop. A canonical orientation on all other ${p}$-cells then follows from the branching structure. We use the branching structure where each ${1}$-cell ${c_1(\bm x)_\mu}$ has an arrow pointing in the ${\h{\bm{\mu}}}$ direction (see Fig.~\ref{fig:branchingStructure}). However, it is important to note that the choice of lattice orientation is a formal convention, and choosing different branching structures does not affect the physics.\footnote{\Rf{WWW211212148} conjectures that observables are branching structure independent only if the continuum effective field theory is framing anomaly free.}

A ${p}$-cell can be related to ${(p-1)}$ cells using the boundary operator ${\pp}$. The boundary operator acting on a ${p}$-cell---${\pp c_{p}}$---is the \txti{oriented} sum of $(p-1)$-cells on the boundary of ${c_{p}}$. For our branching structure, it is
\begin{equation}\label{boundaryDef}
\begin{aligned}
\hspace{-5pt}\partial c_p(\bm x)_{\mu_1 \cdots \mu_p}&\hspace{-4pt}=\hspace{-2pt}\sum_{k=1}^{p}(-1)^{k+1}\hspace{-2pt}\left[c_{p-1}\hspace{-2pt}\left(\bm x+\bm{\h\mu}_{k}\right)_{\mu_1 \cdots \stackrel{\txt{o}}{\mu}_k \cdots \mu_p }\right.\\
&\left.\hspace{70pt}-c_{p-1}(\bm x)_{\mu_1 \cdots \stackrel{\txt{o}}{\mu}_k \cdots \mu_p }\right],
\end{aligned}
\end{equation}
where the notation ${\stackrel{\txt{o}}{\mu}_k}$ indicates that the ${\mu_k}$ index is omitted. From its definition, the boundary operator satisfies ${\pp^2 c_p = 0}$ for any ${p}$-cell. Furthermore, as there are no ${(-1)}$-cells, the boundary operator acting on a ${0}$-cell is defined to be zero.

On the other hand, a ${p}$-cell can be related to ${(p+1)}$-cells using the coboundary operator ${\del}$. The coboundary operator acting on a ${p}$-cell---${\del c_{p}}$---is an \txti{oriented} sum of all ${(p+1)}$-cells whose boundary includes ${c_{p}}$. For our branching structure we use, it is
\begin{equation}\label{coboundaryDef}
\del c_p(\bm x)_{\mu_1 \cdots \mu_p}\hspace{-2pt}=\sum_{\nu}c_{p+1}(\bm x)_{\nu \mu_{1} \ldots \mu_{p} }-c_{p+1}(\bm x-\bm{\h\nu})_{\nu \mu_{1} \ldots \mu_{p} }.
\end{equation}
From its definition, the coboundary operator satisfies ${\del^2 c_p = 0}$ for any ${p}$-cell. Furthermore, as there are no ${(d+1)}$-cells, the coboundary operator acting on a ${d}$-cell is defined to be zero.

Lastly, the lattice has an associated dual lattice. The dual lattice has its lattice sites centered at the ${d}$-cells of the direct lattice. For the cubic lattice, one way to relate a dual lattice site ${\bm{\h x}}$ to a direct lattice site ${\bm x}$ is by ${\bm{\h x} = \bm x + \frac{1}{2}\bm{\h r}}$ with ${\bm{\h r} = \sum_i \bm{\h\mu}}_i$.

Each ${p}$-cell ${c_p}$ on the direct lattice corresponds to a ${(d-p)}$-cell ${\h{c}_{d-p}}$ on the dual lattice. Mapping between them is done using the dual operator ${\hstar}$. A ${p}$-cell ${c_{p}(\bm x)_{\mu_{1} \cdots \mu_{p}}}$ (with canonical ordering ${\mu_1<\cdots<\mu_p}$) and a ${(d-p)}$-cell of the dual lattice ${\h c_{d-p}(\bm{\h x})_{\mu_{1} \cdots \mu_{d-p}}}$ (${\mu_{1}<\cdots<\mu_{d-p}}$) are related by
\begin{align}
\hstar c_{p}(\bm x)_{\mu_{1} \cdots \mu_{p}}&= \eps_{\mu_1 \cdots \mu_p \mu_{p+1} \cdots \mu_{d}}\\
&\quad\quad\times\h{c}_{d-p}(\bm{\h x}-\bm{\h\mu}_{p+1} - \ldots-\bm{\h\mu}_{d})_{\mu_{p+1} \cdots \mu_{d}},\nonumber\\
\hstar \h{c}_{p}(\bm{\h x})_{\mu_1 \ldots \mu_p}&=\eps_{\mu_{1} \cdots \mu_{p} \mu_{p+1} \cdots \mu_{d}}\\
&\quad\quad\times c_{d-p}(\bm x+\bm{\h\mu}_{1} + \ldots+\bm{\h\mu}_p)_{\mu_{p+1} \cdots \mu_{d}} \nonumber,
\end{align}
where summation is \txti{not} implied on the right-hand side. Here ${\eps}$ is the Levi-Civita symbol, which takes into account the lattice's and dual lattice's relative orientations. From the definition of ${\hstar}$, acting ${\hstar}$ twice on a ${p}$-cell of the direct (dual) lattice yields ${\hstar\hstar c_p = (-1)^{p(d-p)}c_p}$ (${\hstar\hstar \h{c}_p = (-1)^{p(d-p)}\h{c}_p}$). Furthermore, from the definitions of the boundary, coboundary, and dual operators, they are related to one another by
\begin{equation}\label{cobdyBdy}
\del c_p = (-1)^{d(p+1)+1}\hstar\pp\hstar c_p,
\end{equation}
which, equivalently, is ${\hstar\del c_p = (-1)^{p}\pp\hstar c_p}$.

\section{TQFT of the ${p}$-form toric code ground states}\label{subSec:ZNcontLim}

In section~\ref{section:znSSB} of the main text, we found that the ground states of the ${\ZN{p}}$ SSB phase satisfy
\begin{equation}\label{pFormTCgs}
\hspace{-6pt}\prod_{c_p \in \delta c_{p-1}} \hspace{-6pt}\t Z'_{c_p} \hspace{-2pt}\gs \hspace{-1pt}=\hspace{-1pt} \gs,\quad \prod_{c_p\in\pp c_{p+1}}\hspace{-6pt}\t X'_{c_p}\hspace{-2pt}\gs\hspace{-1pt} = \hspace{-1pt}\gs,
\end{equation}
where ${\t{X}'}$ and ${\t{Z}'}$ are the ${\Z_N}$ clock operators dressed by unitaries. We note that these ground states are also the ground states of the ${p}$-form toric code Hamiltonian
\begin{equation}\label{dressedpTC}
\hspace{-3pt}H_{p\txt{TC}} = -\sum_{c_{p-1}}\hspace{-14pt}\prod_{\hspace{15pt}c_p \in \delta c_{p-1}} \hspace{-10pt}\t Z'_{c_p}\hspace{-2pt}  -\sum_{c_{p+1}}\hspace{-14pt}\prod_{\hspace{15pt}c_p\in\pp c_{p+1}}\hspace{-10pt}\t X'_{c_p}+\txt{hc}.
\end{equation}

In this section, we relate the lattice description of the ground states to an equivalent topological quantum field theory description. Doing so demonstrates the connection between exact emergent higher-form symmetries in lattice models and exact higher-form symmetries in Lagrangian quantum field theories, where higher-form symmetries are most commonly studied. 

To develop a field theory description of these ground states, we take inspiration from \Rf{SK170804619} and parametrize the dressed clock operators in the ${\ZN{p}}$ SSB phase by
\begin{align}
\t X'_{c_p} = \exp[\ii \t\Th'_{c_{p}}],\quad\quad\quad \t Z'_{c_p} = \exp[\ii(\hstar \t \Phi')_{c_{p}}].
\end{align}
Note that in order for ${(\t X'_{c_p})^N = (\t Z'_{c_p})^N = \one}$, it must be that the eigenvalues of ${\t\Th'_{c_p}}$ and ${(\hstar \t\Phi')_{c_p}}$ satisfy ${\t\Th'_{c_p}\in 2\pi\Z/N}$, and ${(\hstar \t\Phi')_{c_p}\in2\pi\Z/N}$. Furthermore, in order for the clock operators algebra Eq.~\eqref{ZNclockAlg} to be satisfied, ${\t\Th'_{c_p}}$ and ${(\hstar \t\Phi')_{c_p}}$ must obey the commutation relation ${[\t\Th'_{c_p},(\hstar \t\Phi')_{\t{c}_p}] = \frac{2\pi\ii}{N}\del_{c_p,\t{c}_p}}$. In terms of ${\t\Th'_{c_p}}$ and ${(\hstar \t\Phi')_{c_p}}$, the constraints Eq.~\eqref{pFormTCgs} defining the IR are
\begin{equation}\label{pFormTCgs2}
\frac{N}{2\pi}\del(\hstar \t\Phi')_{c_{p-1}} = \frac{N}{2\pi}(\dd \t\Th')_{c_{p+1}} = 0.
\end{equation}

The lattice Heisenberg operators ${\t\Th'_{c_p}(t)}$ and ${(\hstar \t\Phi')_{c_p}(t)}$ are related to their continuum counterparts ${\t\Th'(t,\bm x)}$ and ${\hstar\t\Phi'(t,\bm x)}$ by
\begin{equation}
\t\Th'_{c_p}  \hspace{-2pt}= \hspace{-2pt} \int_{c_p}  \hspace{-2pt} \t\Th',\hspace{40pt}(\hstar \t\Phi')_{c_p}\hspace{-2pt}= \hspace{-2pt} \int_{c_{p}} \hspace{-5pt} \hstar\t\Phi',
\end{equation}
where ${\int_{c_p}}$ denotes spatial integration over the ${p}$-cell ${c_p}$. 

For simplicity, we will work locally and treat the continuum quantum fields as differential forms in space (${\t\Th'}$ is a ${p}$-form while ${\t\Phi'}$ is a ${(d-p)}$-form) taking values in ${[-\pi,\pi)}$, ignoring that the holonomies of ${\t\Th'}$ and ${\hstar \t\Phi'}$ are restricted to values in ${2\pi \Z/N}$. In the continuum limit, the lattice operators simply become their continuum versions, so the constraint Eq.~\eqref{pFormTCgs2} in the continuum limit becomes
\begin{equation}\label{pFormTCgsCONT}
\frac{N}{2\pi}\dd^\da\hstar \t\Phi' = \frac{N}{2\pi}\dd \t\Th' = 0.
\end{equation}
Here, ${\dd^\da}$ is the adjoint of ${\dd}$, which is ${\dd^\da \equiv (-1)^{d(p+1)+1}\hstar\dd\hstar}$ when acting on a ${p}$-form.

The lattice Hamiltonian in the IR is just the ground state energy. Setting this to zero, in the continuum ${H^{(\mathrm{III})}_{\txt{IR}}  = 0}$. The continuum Lagrangian is thus
\begin{equation}
\cL^{(\mathrm{III})}_{\txt{IR}} = \frac{N}{2\pi p!} (\hstar \t\Phi')_{i_1\cdots i_p} \pp_t \t\Th'_{i_1\cdots i_p},
\end{equation}
which enforces the equal-time commutation relation
\begin{equation}
\hspace{-4pt}\frac{[\t\Th'_{i_1\cdots i_p}(\bm{x}),(\hstar\t{\Phi}^{\prime})_{j_1\cdots j_p}(\bm{y})]}{p!} \hspace{-1pt}=\hspace{-1pt}\frac{2\pi\ii}N \del\indices{^{i_1}_{[j_1}}\hspace{-3pt}\cdots\del\indices{^{i_p}_{j_p]}}\del^d(\bm{x}-\bm{y}).
\end{equation}
The low-energy path integral only integrates over field configurations satisfying Eq.~\eqref{pFormTCgsCONT}, and is
\begin{align}
\cZ^{(\mathrm{III})}_{\txt{IR}}  \hspace{-2pt}&= \hspace{-4pt}\int\hspace{-2pt} \cD[\t\Th']\cD[\t\Phi']~\del\left(\frac{N}{2\pi}\dd^\da \hstar \t\Phi'\hspace{-1pt}\right)\label{ZNPI1}\\
&\hspace{90pt}\times\del\left(\frac{N}{2\pi}\dd \t\Th'\hspace{-1pt}\right)~\ee^{\ii\hspace{-2pt}\int\hspace{-2pt}\dd t  \dd^{d}\bm x  \cL^{(\mathrm{III})}_{\txt{IR}} }.\nonumber
\end{align}

Let us now massage this path integral into a more familiar form. We can rewrite both functional delta functions by integrating in new fields acting as Lagrange multipliers and modifying the action. The first delta function is rewritten using a ${(p-1)}$-form Lagrange multiplier ${\la}$ as
\begin{equation}
\hspace{-4pt}\del\left(\frac{N}{2\pi}\dd^\da \hspace{-2pt}\hstar \t\Phi'\hspace{-2pt}\right) \hspace{-2pt}=\hspace{-4pt} \int\hspace{-2pt} \cD\la~\ee^{\frac{\ii N}{2\pi}\int \frac{\la_{i_1\cdots i_{p-1}}(\dd^\da\hstar \t\Phi')_{i_1\cdots i_{p-1}}}{(p-1)!}}\hspace{-2pt},
\end{equation}
and using a ${(p+1)}$-form Lagrange multiplier ${\hstar\eta}$, the second delta function is
\begin{equation}
\del\left(\frac{N}{2\pi}\dd \t\Th'\right) \hspace{-2pt}= \int\cD\eta~\ee^{\frac{\ii N}{2\pi}\int \frac{(\hstar\eta)_{i_1\cdots i_{p+1}} (\dd \t\Th')_{i_1\cdots i_{p+1}}}{(p+1)!}}.
\end{equation}
Plugging these expressions into the Eq.~\eqref{ZNPI1} and the components of ${\dd\t\Th'}$ and ${\dd^\da\hstar\t\Phi'}$ and simplifying, the path integral becomes
\begin{align}
&\cZ^{(\mathrm{III})}_{\txt{IR}}  = \int \cD[\t\Th']\cD[\t\Phi']\cD[\la]\cD[\eta]~\ee^{\ii\hspace{-2pt}\int\hspace{-2pt}\dd t  \dd^{d}\bm x  \cL^{(\mathrm{III})}_{\txt{IR}} },\\
&\cL^{(\mathrm{III})}_{\txt{IR}}  = \frac{N}{2\pi p!} \bigg( (\hstar \t\Phi')_{i_1\cdots i_p} \left[\pp_t \t\Th'_{i_1\cdots i_p} + p~\partial_{\left[i_1\right.} \la_{\left.i_2 \cdots i_{p}\right]} \right]\nonumber\\
& \hspace{100pt}+ (\hstar\eta)_{i_1i_2\cdots i_{p+1}} \partial_{\left[i_1\right.} \t\Th'_{\left.i_2 \cdots i_{p+1}\right]}\bigg).\nonumber
\end{align}

Let's now introduce the ${p}$-form ${a}$ and ${(d-p)}$-form ${b}$ in spacetime whose components are
\begin{align*}
a_{i_1\cdots i_p} &= \t\Th'_{i_1\cdots i_p} ,\hspace{72pt} a_{0i_2\cdots i_p} = -\la_{i_2\cdots i_p},\\
b_{i_1\cdots i_{d-p}} &= (-1)^{d-p}\t\Phi'_{i_1\cdots i_{d-p}},\quad\quad b_{0i_2\cdots i_{d-p}} = -\eta_{i_2\cdots i_{d-p}},
\end{align*}
where ${b}$ satisfies 
\begin{equation*}
(\hstar\eta)_{i_1\cdots i_{p+1}} = - (\hstar b )_{i_1\cdots i_{p+1}},\quad\quad (\hstar \t\Phi')_{i_1\cdots i_p} = (\hstar b)_{0i_1\cdots i_p}.
\end{equation*}
Using these, the path integral becomes
\begin{equation}
\begin{aligned}
\cZ^{(\mathrm{III})}_{\txt{IR}}  &= \int \cD[a]\cD[b]~\ee^{\ii\hspace{-2pt}\int\hspace{-2pt}\dd t  \dd^{d}\bm x  \cL^{(\mathrm{III})}_{\txt{IR}} },\\
\cL^{(\mathrm{III})}_{\txt{IR}}  &= \frac{N}{2\pi p!} \bigg( (\hstar b)_{0i_1\cdots i_p}\left[ \pp_t a_{i_1\cdots i_p} +(-1)^p p~ \partial_{\left[i_1\right.} a_{\left.i_2 \cdots i_{p}\right]0} \right]\\
&\hspace{70pt}- (\hstar b)_{i_1i_2\cdots i_{p+1}} \partial_{\left[i_1\right.} a_{\left.i_2 \cdots i_{p+1}\right]}\bigg).
\end{aligned}
\end{equation}
The term in square brackets can be rewritten as ${(p+1)\pp_{[0}a_{i_1\cdots i_p]}}$. Furthermore, working in flat spacetime, ${X}$ is equipped with Minkowski metric ${(-,+,\cdots +)}$. Using it and summing over spacetime indices ${\mu}$, ${\cL^{(\mathrm{III})}_{\txt{IR}} }$ can be rewritten as
\begin{equation}
\begin{aligned}
\cL^{(\mathrm{III})}_{\txt{IR}}  &= -\frac{N}{2\pi } \left( \frac{(\hstar b)^{\mu_1\mu_2\cdots \mu_{p+1}} \partial_{\left[\mu_1\right.} a_{\left.\mu_2 \cdots i_{p+1}\right]}}{p!}\right).
\end{aligned}
\end{equation}
Lasting, using differential forms notation, we arrive at our final expression for the path integral
\begin{equation}
\cZ^{(\mathrm{III})}_{\txt{IR}}  = \int \cD[a]\cD[b]~\ee^{\ii\int \frac{N}{2\pi }  b\wdg \dd a }.
\end{equation}

As anticipated, the low-energy effective field theory, which describes the ground states of the ${\ZN{p}}$ symmetry broken phase, is ${p}$-form ${\Z_N}$ gauge theory. This is arguably the simplest field theory with an anomalous ${\ZN{p}}\times\ZN{d-p}$ symmetry~\cite{KS14010740}.

\subsection{Review of ${p}$-form ${BF}$ theory}\label{pformBFReviewSec}

In the remainder of this section, we will review ${p}$-form ${BF}$ theory, focusing on its symmetries and anomalies, working in ${D=d+1}$ dimensional spacetime. From canonical quantization, the fields ${a}$ and $b$ satisfy the equal-time commutation relations
\begin{equation}\label{BFcomrel}
\hspace{-2pt}[a_{\mu_1\cdots\mu_{p}}(\bm{x}),b_{\mu_{p+1}\cdots\mu_{d}}(\bm{y})] = \frac{2\pi \ii}N  \eps_{0\mu_{1}\cdots\mu_{d}}\del^d(\bm{x}-\bm{y}).
\end{equation}

\subsubsection{${\Z_N}$ ${p}$-form gauge theory in the continuum}

Let us first review how ${p}$-form ${BF}$ theory can be obtained by condensing charge-$N$ gauge charges in ${p}$-form Maxwell theory~\cite{MMS0105,HOS0497, BS1119}. ${p}$-form Maxwell theory is reviewed in appendix~\ref{pformMaxReviewSec}. We will always assume that ${p>0}$. For the reader who would like to jump straight to $p$-form BF theory action, they should skip to Eq.~\eqref{HiggsModel2}.

We modify ${p}$-form Maxwell theory Eq.~\eqref{eqn:SIR} by introducing the dynamical ${(p-1)}$-form bosonic field $H$ and the gauge redundancy 
\begin{equation}\label{ZnGaugeRedundancy}
a \to a + \dd\chi,\quad\quad\quad\quad H \to H + N \chi,
\end{equation}
where ${N\in\Z}$. A gauge-invariant globally defined quantity in terms of only ${H}$ is ${F_H = \dd H + \om_H}$, where ${\om_H \in 2\pi H^p(X;\Z)}$. $F_H$ satisfies a Bianchi identity ${\frac{1}{2\pi}\hstar \dd F_H = 0}$ and its periods are quantized as ${\oint F_H \in 2\pi \Z}$.

With the additional degrees of freedom provided by $H$, we introduce the gauge invariant Wilson operator
\begin{equation}
W_{a,H}(O) = \ee^{\ii \int_O N a - \dd H},
\end{equation}
where ${O}$ is an open ${p}$-submanifold. Physically ${W_{a, H}(O)}$ is an operator that creates a charge excitation on $\pp O$, but one carrying $N$-units of ${a}$-charge. Minimally coupling $F_H$ to ${a}$, the partition function is
\begin{equation}\label{HiggsModel1}
\cZ = \int \cD[a]\cD[H]~e^{-\int_X \frac1{2g^2}|F_{a}|^2 + \frac{v^2}2|F_H - N a|^2}.
\end{equation}
The Lagrangian density now includes the term ${\cL \supset a\wdg\hstar Nv^2 F_H}$ and a mass term ${\cL\supset \frac{N^2 v^2}{2} |a|^2}$. Indeed, the new term added to ${p}$-form Maxwell theory is essentially a Higgs term with $H$ the phase of the Higgs field and ${v\in\R}$ the vev of the Higgs field. The gauge redundancy described by Eq.~\eqref{ZnGaugeRedundancy} is a ${\Z_N}$ gauge redundancy, reflecting how the initial ${U(1)}$ gauge redundancy has been Higgsed down to a ${\Z_N}$ gauge redundancy.

As discussed in appendix section~\ref{abelianDualAp}, these types of theories have a generalized ``particle-vortex'' like duality called abelian duality. For instance, since the action's dependency on ${H}$ is entirely in the form of ${F_H}$, we can dualize ${H\to\h H}$ using the same method shown in section~\ref{abelianDualAp}. Indeed, dualizing $H$ to the ${(D-p-1)}$-form ${\hat{H}}$ satisfying ${\oint F_{\hat{H}}}\in 2\pi\Z$ and ${\dd F_{\h H}=0}$ (where ${F_{\hat{H}} = \dd\hat{H} + \om_{\h H}}$), the Euclidean Lagrangian becomes
\begin{equation}\label{asdfjlkhlsdfh}
\cL =  \frac{ |F_{a}|^2}{2g^2} + \frac{|F_{\hat{H}}|^2}{8\pi^2 v^2}   - \frac{\ii N}{2\pi}a\wdg F_{\hat{H}}.
\end{equation}
The Euclidean path integral now integrates over the dynamical fields ${a}$ and $\hat{H}$ and sums over ${\om_{a}\in 2\pi H^{p+1}(X;\Z)}$ and ${\om_{\h H}\in 2\pi H^{D-p}(X;\Z)}$. We note that without changing the action amplitude, the Lagrangian density can be rewritten as
\begin{equation}\label{asdfjlkhlsdfh2}
\cL =  \frac{ |F_{a}|^2}{2g^2} + \frac{|F_{\hat{H}}|^2}{8\pi^2 v^2}  - \frac{\ii N}{2\pi}\h H\wdg F_{a}.
\end{equation}
Locally, we have just integrated by parts in the $BF$ term. However, keeping track of the globally nontrivial parts of $F_{\h H}$ and $F_{a}$ makes showing this difficult (it's most naturally seen using Deligne-Beilinson cohomology~\cite{MT150904236}).

Utilizing abelian duality, we have found two representations for the theory: the ${(a, H)}$ representation Eq.~\eqref{HiggsModel1} and the ${(a,\h H)}$ representation Eqs.~\eqref{asdfjlkhlsdfh} and~\eqref{asdfjlkhlsdfh2}, the latter being dual only locally.

In representation ${(a,\h H)}$, the deep IR is governed by ${p}$-form ${BF}$ theory. Here, the deep IR refers to energies below the gap of ${a}$ and ${\h H}$, which have a gap through topological mass generation. Indeed, to find their energy gaps, first note that in the ${(a,\h H)}$ representation, the Lorentzian action is
 \begin{equation}\label{LorBFFull}
S =  \int_X\left(-\frac{ |F_{a}|^2}{2g^2}  - \frac{|F_{\hat{H}}|^2}{8\pi^2 v^2}   + \frac{N}{2\pi}a\wdg F_{\hat{H}}\right).
\end{equation}
Since this theory is Gaussian, we can show that the ${a\wdg F_{\h H}}$ term causes all excitations to be gapped using the equations of motion. Minimizing the action and using that $\dd^\da \hstar F_{b,\h H} = 0$, we find that the classical equations of motion are
\begin{equation}
\begin{aligned}
\left(\del  +  N^2g^2v^2\right)\hstar F_{\hat{H}} &=0,\\
\left(\del+ N^2 g^2 v^2\right)\hstar F_{a} &= 0.
\end{aligned}.
\end{equation}
where ${\del = \dd^\da\dd + \dd\dd^\da}$ is the Hodge Laplacian. Therefore, we see that the ${p}$-form ${\hstar F_{\hat{H}}}$ and the ${(D-p-1)}$-form ${\hstar F_{a}}$ both have an energy gap ${N g v}$.

To go below the energy gap into the deep IR, we take the limit ${g\to\infty}$ and ${v\to\infty}$. In this limit, the Euclidean path integral becomes
\begin{equation}\label{HiggsModel2}
\cZ_{BF} = \int \cD[a]\cD[\hat{H}]~e^{ \frac{\ii N}{2\pi}\int_Xa\wdg F_{\hat{H}}}.
\end{equation}
This is $p$-form $BF$ theory, and it is in terms of the $p$-form bosonic field ${a}$, which is the ${U(1)}$ gauge field we started with and $\hat{H}$, which is the abelian dual of the Higgs field phase. Taking the deep IR limit using the Lagrangian density in this representation written as Eq.~\eqref{asdfjlkhlsdfh2}, the Lagrangian in the topological limit is equivalent to ${\cL_{BF}= -\frac{\ii N}{2\pi}\hat{H}\wdg F_{a}}$.

Plugging in ${F_{\h H} = \dd\h H+ \om_{\h H}}$ into Eq.~\eqref{HiggsModel2}, the path integral becomes
\begin{equation}
\cZ_{BF} = \hspace{-4pt}\int \cD [a] \cD[\h H]  \hspace{-15pt}\sum_{\frac{\om_{\h H}}{2\pi}\in H^{D-p}(X;\Z)}\hspace{-10pt} \ee^{\frac{\ii N}{2\pi}\int \left(a\wdg\dd\h H + a\wdg\om_{\h H}\right)}
\end{equation}
Integrating by parts on the first term and using Poincaré duality on the second term, we can rewrite this as
\begin{equation}\
\cZ_{BF} = \hspace{-4pt}\int \cD [a] \cD[\h H]  \hspace{-10pt}\sum_{\om\in H_{p}(X;\Z)}\hspace{-10pt} \ee^{\frac{\ii N}{2\pi}\int_X \h H\wdg\dd a + \ii N\int_\om a}.
\end{equation}
Integrating over $\h H$ and summing over $\om$, the path integral becomes
\begin{equation}\label{BFthy22}
\cZ_{BF} = \hspace{-4pt}\int \cD [a] ~ \del(\dd a)~\del\left(\oint a\in \frac{2\pi\Z}{N}\right).
\end{equation}
Notice that if we would have instead started with the Lagrangian density written as ${\cL_{BF}= -\frac{\ii N}{2\pi}\hat{H}\wdg F_{a}}$, upon integrating out ${a}$ and ${\om_a}$ we would get
\begin{equation}
\cZ_{BF} = \hspace{-4pt}\int \cD [\h H] ~ \del(\dd \h H)~\del\left(\oint \h H\in \frac{2\pi\Z}{N}\right).
\end{equation}

Having massaged ${p}$-form ${BF}$ theory into Eq.~\eqref{BFthy22}, we find that in correlations functions, the ${U(1)}$ gauge fields are closed forms and have quantized holonomies.\footnote{This can also be deduced from the equations of motion. Indeed, in the ${(a,H)}$ representation, Eq.~\eqref{HiggsModel1}, the ${H}$ equations of motion in the deep IR are
\begin{equation}\label{HiggsEOM}
F_H =  N a.
\end{equation}
Therefore, because of the Bianchi identity ${\dd F_H = 0}$, ${a}$ is a closed ${p}$-form. Furthermore, since $F_H$ satisfies ${\oint F_H = 2\pi\Z}$, the holonomies of ${a}$ are quantized as ${\oint a = 2\pi\Z/N}$.} Therefore, the Wilson operators
\begin{equation}
W_a[C_p] = \ee^{\ii\oint_{C_p}a},\quad\quad W_{\h H}[C_{d-p}] = \ee^{\ii\oint_{C_{d-p}}\hspace{-2pt}\h H}.
\end{equation}
satisfy ${\< (W_a)^N \> = \<(W_{\h H})^N \> = 1}$, and are ${1}$ when ${C_p\in B_p(X)}$ (i.e., there exists an ${O_{p+1}}$ such that ${C_p = \pp O_{p+1}}$). The latter property implies that these Wilson operators are topological. At a fixed time slice, in the deep IR, any contractible Wilson operators can condense into the vacuum, but for non-contractible Wilson operators, only $N$ can condense into the vacuum.

The path integral counts the number of nontrivial Wilson operators ${W(C)}$ which satisfy ${W(C)^N=1}$, and thus the number of configurations ${a\in H^p(X;\Z_N)}$:
\begin{align}
\cZ_{BF} = \sum_{a\in H^{p}(X;\Z_N)}1= |H^{p}(X,\Z_N)|.
\end{align}
The number of ground states is given by the partition function evaluated on ${X = \R\times M}$, where ${M}$ is a space. Therefore, there are $|H^{p}(M;\Z_N)|$ degenerate ground states.

\subsubsection{Symmetries}\label{BFReviewSec}

Having reviewed the basics of ${p}$-form ${BF}$ theory in the previous section, we now turn to identifying the theory's symmetries, showing that there is a ${\ZN{p}\times\ZN{d-p}}$ symmetry~\cite{KS14010740}.

Let's first consider the symmetries manifest in the ${(a, H)}$ representation, Eq.~\eqref{HiggsModel1}. The path integral is invariant under the transformation
\begin{equation}\label{ZNpformSym}
a \to a + \Gamma,\quad\quad\quad\quad F_H \to F_H + N \Gamma,
\end{equation}
with ${\dd\Gamma = 0}$. Since $F_H$ satisfies ${\oint F_H \in 2\pi\Z}$, in order to shift ${F_H \to F_H + N \Gamma}$ we require that ${\oint \Gamma \in 2\pi\Z/N}$. When ${\Gamma = \dd \om}$, the transformation Eq.~\eqref{ZNpformSym} becomes the gauge transformation Eq.~\eqref{ZnGaugeRedundancy}. Therefore, the ${\Ga}$ that correspond to physical transformations are ${ \frac{N}{2\pi}\Gamma \in H^{p}(X;\Z)}$.

The quantization condition of the periods of ${\Ga}$ has a significant consequence. Indeed, note that the Wilson operator ${W_a}$ is charged under this symmetry. Because of the quantization condition ${\oint \Gamma \in 2\pi\Z/N}$, it transforms as
\begin{equation}\label{Zn_pSym}
W_a[C_p]\to \ee^{\ii\oint\Gamma} W_a[C_p],\quad \quad \quad \ee^{\ii\oint\Gamma}\in\Z_N.
\end{equation}
Since the charged operators are ${p}$-dimensional and transform by an element of ${\Z_N}$, this is a ${\Z^{(p)}_N}$ symmetry.

It's tempting to think that an additional symmetry may be associated with the $H$ field. Indeed, Eq.~\eqref{HiggsModel1} is invariant under the transformation ${H\to H + \om}$ for ${\dd\om = 0}$. However, there are no physical observables that transform under this. Indeed, while the Wilson operator ${\exp[\ii\oint H]}$ picks up a phase, it is not a physical operator since it is not invariant under the gauge redundancy Eq.~\eqref{ZnGaugeRedundancy}.

The ${\Z^{(p)}_N}$ symmetry can also be seen in the ${(a,\h H)}$ representation, when the Lagrangian is described by Eq.~\eqref{asdfjlkhlsdfh}. Indeed, under the symmetry transformation, the action amplitude transforms as ${\exp[-S]\to \exp[-S]\exp[-\del S]}$, where
\begin{equation}
\ee^{-\del S} = \ee^{-\frac{\ii N}{2\pi}\int \Ga\wdg F_{\h H}},
\end{equation}
with ${ \frac{N}{2\pi}\Gamma \in H^{p}(X;\Z)}$. Plugging in ${F_{\h H} = \dd \h H + \om_{\h H}}$, the phase factor ${\exp[-\del S]}$ becomes
\begin{align*}
\exp[-\del S] =\ee^{-2\pi \ii \int \frac{N~\Ga}{2\pi}\wdg \frac{\om_{\h H}}{2\pi}},
\end{align*}
where we used integration by parts and that ${\dd\Ga = 0}$. Recall that ${ \frac{N}{2\pi}\Gamma \in H^{p}(X;\Z)}$ and ${ \frac{\om_{\h H}}{2\pi} \in H^{D-p}(X;\Z)}$. Then, since the wedge product preserves integral de Rham cohomology classes, ${ \frac{N~\Ga}{2\pi}\wdg \frac{\om_{\h H}}{2\pi} \in H^{D}(X;\Z)}$, so ${\exp[-\del S]=1}$.

The Lagrangian in the ${(a,\h H)}$ representation can also be written as Eq.~\eqref{asdfjlkhlsdfh2} without changing the partition function. In this form, following the same argument used to show that there is a ${\ZN{p}}$ symmetry, we find there is also ${\ZN{d-p}}$ symmetry. Indeed, the action amplitude is invariant under ${\h H\to \h H + \h \Ga}$ where ${ \frac{N}{2\pi}\h\Gamma \in H^{d-p}(X;\Z)}$. The charged operator of this ${\ZN{d-p}}$ symmetry is the Wilson operator ${W_{\hat{H}} = \exp\left[\ii\oint\hat{H}\right]}$, which transforms as
\begin{equation}
W_{\hat{H}}(C)\to \ee^{\ii\oint\hat{\Gamma}} W_{\hat{H}}(C),\quad \quad  \ee^{\ii\oint\hat{\Gamma}}\in\Z_N.
\end{equation}

The symmetry operator of the ${\ZN{p}}$ symmetry is just ${W_{\h H}}$ and can be written as
\begin{equation}
U(\Si) = \ee^{\ii\oint_\Si \h H}= \exp^{\ii\oint_{M_d} \h H\wdg \Ga},
\end{equation}
where ${\Ga}$ is the Poincaré dual of the ${p}$-cycle ${\Si}$ with respect to space ${M_d}$. Indeed, using the equal-time commutation relation \eqref{BFcomrel}, which form Eq.~\eqref{asdfjlkhlsdfh} is 
\begin{equation}
\hspace{-2pt}[a_{\mu_1\cdots\mu_{p}}(\bm{x}),\h H_{\mu_{p+1}\cdots\mu_{d}}(\bm{y})] = \frac{2\pi \ii}N  \eps_{0\mu_{1}\cdots\mu_{d}}\del^d(\bm{x}-\bm{y}),
\end{equation}
we have that
\begin{equation}
\begin{aligned}
U(\Si)W_a(C)U^\da(\Si) &= \ee^{\frac{2\pi\ii}N\int_C \Ga}W_a(C),\\
&= \ee^{\frac{2\pi\ii}N\#(\Si,C)}W_a(C).
\end{aligned}
\end{equation}
Similarly, the symmetry operator of the ${\ZN{d-p}}$ symmetry is
\begin{equation}
\h U(\h \Si) = \ee^{\ii\oint_{\h \Si}a},
\end{equation}
which is just ${W_a}$.

\subsubsection{Mixed 't Hooft anomaly and anomaly inflow}

In the last section, we reviewed how ${p}$-form ${BF}$ theory has ${\ZN{p}}$ and ${\ZN{d-p}}$ symmetries. However, these symmetries are not independent of one another: the symmetry operator of one symmetry is a charged operator of the other symmetry, thus satisfying the Heisenberg algebra. This is a manifestation of the fact that the ${\ZN{p}\times\ZN{d-p}}$ symmetry is anomalous. In this section, we will turn on a background gauge field for these symmetries to learn more about this mixed 't Hooft anomaly.

Let's first turn on a background gauge field for the ${\ZN{p}}$ symmetry. We introduce the background gauge field ${\cA\in \frac{2\pi}{N}H^{p+1}(X;\Z)}$ and the gauge redundancy
\begin{equation}\label{gaugeTranshatb}
a \to a + \bt,\quad\quad\quad\quad\cA \to \cA + \dd\bt.
\end{equation}
Minimally coupling $\cA$, the ${p}$-form ${BF}$ theory path integral becomes
\begin{equation}\label{Zngaugedtheory}
\cZ[\cA] = \int \cD[a]\cD[\h H]~e^{\frac{\ii N}{2\pi}\int_X \left(a\wdg \dd \h H +(-1)^{p} \cA \wdg\h H\right)}.
\end{equation}
Since ${\cZ[\cA] = \cZ[\cA + \dd\beta]}$, the ${\ZN{p}}$ symmetry is anomaly free.

Second, we next turn off ${\cA}$ and turn on a background gauge field for the ${\ZN{d-p}}$ symmetry, which introduces the background gauge field ${\h\cA\in\frac{2\pi}{N}H^{d-p+1}(X;\Z)}$ and the gauge redundancy
\begin{equation}\label{gaugeTranshatH}
\h H \to \h H + \zeta,\quad\quad\quad\quad\h\cA \to \h\cA+ \dd\zeta.
\end{equation}
Minimally coupling ${\h\cA}$, the path integral becomes
\begin{equation}
\cZ[\h\cA] = \int \cD[a]\cD[\h H]~e^{\frac{\ii N}{2\pi}\int_X a \wdg( F_{\h H} - \h \cA)}.
\end{equation}
Since ${\cZ[\h\cA] = \cZ[\h\cA + \dd\zeta]}$, the ${\ZN{d-p}}$ symmetry is anomaly free.

Now let us turn on both background gauge fields. Coupling them into the action as above yields
\begin{equation}
\hspace{-4pt}\cZ[\cA,\h\cA]\hspace{-2pt} = \hspace{-4pt}\int \hspace{-2pt} \cD[a]\cD[\h H]\ee^{ \frac{\ii N}{2\pi}\hspace{-1pt}\int_X\left(a \wdg( F_{\h H} - \h \cA) +(-1)^{p} \cA \wdg\h H\right)}\hspace{-2pt}.
\end{equation}
While this is invariant under Eq.~\eqref{gaugeTranshatb}, under the gauge transformation Eq.~\eqref{gaugeTranshatH} it transforms as
\begin{equation}\label{ZgaugeTrans}
\cZ[\cA,\h\cA+\dd\zeta]\to \ee^{(-1)^p\frac{\ii N}{2\pi}\int_X\cA\wdg\zeta}\cZ[\cA,\h\cA].
\end{equation}
In fact, no local counterterms can be added such that the path integral is invariant under both gauge transformations. It always gets multiplied by a phase. Thus, we see an obstruction to coupling a background gauge field of both symmetries and hence a mixed 't Hooft anomaly.

A 't Hooft anomaly can be classified by an SPT in one higher dimension whose boundary realizes the symmetry anomalous. Let's now extend the background fields to one higher dimension and have $X$ be the boundary of the new spacetime $Y$. The path integral governing all of ${Y}$ is
\begin{equation}
\cZ_Y[\cA,\h\cA,Y]= \cZ_{\txt{SPT}}[\cA,\h\cA,Y]  \cZ[\cA,\h\cA,\pp Y=X],
\end{equation}
where the SPT path integral ${\cZ_{\txt{SPT}}}$ is defined such that it cancels out the phase in Eq.~\eqref{ZgaugeTrans} such that ${\cZ_Y}$ is gauge invariant. 

Noting that this phase can be written as
\begin{equation}
\ee^{(-1)^p\frac{\ii N}{2\pi}\int_X\cA\wdg\zeta} = \ee^{-\frac{\ii N}{2\pi}\int_Y\cA\wdg \dd \zeta},
\end{equation} 
let's consider the SPT
\begin{equation}\label{ZinvZN}
\cZ_{\txt{SPT}}[\cA,\h\cA] = \ee^{\frac{\ii N}{2\pi}\int_Y \cA\wdg\h\cA}.
\end{equation}
Under the gauge transformations Eq.~\eqref{gaugeTranshatH}, ${\cZ_{\txt{SPT}} }$ transforms as
\begin{equation}
\cZ_{\txt{SPT}}[\cA,\h\cA+\zeta] = \ee^{\frac{\ii N}{2\pi}\int_Y\cA\wdg\dd\zeta} \cZ_{\txt{SPT}}[\cA,\h\cA].
\end{equation}
This phase picked up is the inverse of the phase picked up by ${\cZ}$ in Eq~\eqref{ZgaugeTrans}, and thus ${\cZ_Y}$ is indeed gauge invariant. The mixed anomaly between the ${\ZN{p}}$ and ${\ZN{d-p}}$ symmetry is then said to be classified by the SPT Eq.~\eqref{ZinvZN}.

\section{Continuum of ${U(1)}$ ${p}$-gauge theory}\label{subSec:contLim}

In this appendix, we take the continuum limit of Eq.~\eqref{deepIRU1Ham} in section~\ref{section:SSB} of the main text. Doing so also demonstrates the connection between exact emergent higher-form symmetries in lattice models and exact higher-form symmetries in Lagrangian quantum field theories, where higher-form symmetries are most commonly studied.

The lattice Heisenberg operators ${L^z_{c_p}(t)}$ and ${\Th_{c_p}(t)}$ in the IR are dressed by two local unitary operators ${U^{(1)}_{\txt{LU}}}$ and ${U^{(2)}_{\txt{LU}}}$ and denoted as ${\t{L}^{\prime z}_{c_p}(t)}$ and ${\t\Th'_{c_p}(t)}$. Furthermore, in the mid-IR we defined the variable ${\om_{c_{p+1}}\equiv -2\pi\toZ{(\dd\t\Th)_{c_{p+1}}/(2\pi)}}$, which in the IR was dressed by ${U^{(2)}_{\txt{LU}}}$ and denoted as ${\om'_{c_{p+1}}(t)}$. Therefore, the three elementary operators in the IR are ${\t{L}^{\prime z}_{c_p}(t)}$, ${\t\Th'_{c_p}(t)}$, and ${\om'_{c_{p+1}}(t)}$. We relate these lattice operators to their continuum counterparts ${\t{L}^{\prime z}(t,\bm x)}$, ${\t\Th'(t,\bm x)}$, and ${\om'(t,\bm x)}$ by
\begin{equation}
\t{L}^{\prime z}_{c_p} \hspace{-2pt}= \hspace{-2pt} \int_{c_p} \hspace{-2pt} \t{L}^{\prime z},\hspace{20pt}\t\Th'_{c_p}  \hspace{-2pt}= \hspace{-2pt} \int_{c_p}  \hspace{-2pt} \t\Th',\hspace{20pt}\om'_{c_{p+1}}  \hspace{-2pt}= \hspace{-2pt} \int_{c_{p+1}} \hspace{-5pt} \om',
\end{equation}
where, for instance, ${ \int_{c_p}}$ denotes spatial integral over the ${p}$-cell ${c_p}$. The continuum quantum fields are \txti{globally} differential forms in space ${M}$ (${\t{L}^{\prime z}}$ and ${\t\Th'}$ are ${p}$-forms while ${\om'}$ is a ${(p+1)}$-form), mapping from spacetime ${X}$ to ${\R}$. In the continuum, the lattice differential operators become their continuum versions. For instance, the lattice operator ${F'_{c_{p+1}} = (\dd \t\Th')_{c_{p+1}} + (\om')_{c_{p+1}}}$ becomes ${F'_{c_{p+1}} = \int_{c_{p+1}}F'}$ where ${F' = \dd \t\Th' + \om'}$. So, taking the continuum limit of Eq.~\eqref{deepIRU1Ham}, the deep IR continuum Hamiltonian is
\begin{equation}\label{continuumH}
\hspace{-6pt}H_{\text{deep~IR}} \hspace{-2pt}= \hspace{-4pt}\int\hspace{-4pt} \dd^d\bm x \hspace{-2pt}\left(\hspace{-3pt}\frac{\ka U}{2}\frac{|\t{L}^{\prime z}_{i_{1}\cdots i_p}|^2}{p!} \hspace{-1pt}+\hspace{-1pt} \frac{U}2\frac{|F'_{i_{1}\cdots i_{p+1}}|^2}{(p+1)!}\hspace{-1pt}\right)\hspace{-2pt},
\end{equation}
where, for instance, ${|\t{L}^{\prime z}_{i_{1}\cdots i_p}|^2\equiv\sum_{i_{1},\cdots ,i_p=1}^d(\t{L}^{\prime z}_{i_{1}\cdots i_p})^2}$.

To write down the path integral, we can find the Lorentzian action and then perform a functional integral over field configurations obeying the following constraints:
\begin{enumerate}
\item Since the IR does not include dressed charge excitations, ${\t{L}^{\prime z}}$ must satisfy ${\t\rho~^\prime = 0}$. The expression for ${\t\rho~^\prime_{c_{p-1}}}$ in Eq.~\eqref{2bodyU1Ham} can be rewritten using discrete exterior calculus notation as ${\t\rho~^\prime_{c_{p-1}} \sim (\hstar\dd\hstar \t{L}^{\prime z})_{c_{p-1}}}$. So, ${\t\rho~^\prime = 0}$ in the continuum limit is the Gauss law
\begin{equation}\label{gaussLawCont}
\pp_j\t{L}^{\prime z}_{ji_1\cdots i_{p-1}} = 0.
\end{equation}
Despite there being no dressed charges in the IR, ${\t{L}^{\prime z}}$ can still be sourced along nontrivial ${p}$-cycles. Because ${\t{L}^{\prime z}_{c_p}\in\Z}$ on the lattice, the flux of ${\t{L}^{\prime z}}$ in the continuum is quantized
\begin{equation}\label{fluxCon}
\oint_{C_{d-p}} \hspace{-5pt}\hstar \t{L}^{\prime z}\in \Z,
\end{equation}
where ${C_{d-p}}$ is a nontrivial ${(d-p)}$-cycle in space: ${C_{d-p}\in H_{d-p}(M;\Z)}$.
\item Since the IR does not include dressed topological defects, ${\t\Th'}$ and ${\om'}$ must satisfy ${\h{\rho}' = 0}$. The expression for ${(\hstar \h\rho')_{c_{p+2}} }$ of Eq.~\eqref{topdefectrhoop} in the continuum limit becomes  ${\h{\rho}' = \frac{1}{2\pi}\hstar\dd F'}$, and ${\h{\rho}' = 0}$ becomes the Bianchi identity
\begin{equation}\label{magCon1}
\frac{1}{2\pi}\hstar\dd F' = 0.
\end{equation}
Despite there being no dressed topological defects in the IR, ${\hstar F'}$ can still be sourced along nontrivial ${(d-p)}$-cycles. Indeed, because ${\om'_{c_p}\in 2\pi\Z}$ on the lattice, the flux of ${\hstar F'}$ , in the continuum is quantized
\begin{equation}\label{magCon2}
\oint_{C_{p+1}} \hspace{-5pt}F'\in 2\pi\Z,
\end{equation}
where ${C_{p+1}}$ is a nontrivial ${(p+1)}$-cycle in space. Plugging ${F' = \dd \t\Th' + \om'}$ into Eqs.~\eqref{magCon1} and~\eqref{magCon2}, the ${\t\Th'}$ vanishes and constraints ${\om'}$ as
\begin{equation}\label{omegaConCohom}
\frac{\om'}{2\pi} \in H^{p+1}(M;\Z),
\end{equation}
the ${(p+1)}$th de Rham cohomology group with integral periods.
\end{enumerate}
Enforcing the three constraints Eqs.~\eqref{gaussLawCont},~ \eqref{fluxCon}, and~\eqref{omegaConCohom} by hand, the path integral in Lorentzian signature is
\begin{align}
&\cZ_{\txt{deep~IR}} =\int \cD [\t\Th']~\cD \t{L}^{\prime z} \hspace{-14pt}\sum_{\om' \in 2\pi H^{p+1}\hspace{-1pt}(M;\Z)}\hspace{-14pt} \delta(\pp_{i_{1}} \t{L}^{\prime z}_{i_{1}\cdots i_{p}}) \\
&\hspace{80pt}\times\delta\left(\oint   \hstar \t{L}^{\prime z}\in\Z\right)\ee^{\ii\int_X \dd t  \dd^{d}\bm x  ~\cL_{\txt{deep~IR}} }\nonumber\\
&\cL_{\txt{deep~IR}} = \frac{\t{L}^{\prime z}_{i_1\cdots i_p}\pp_t \t\Th'_{i_1\cdots i_p}}{p!} \nonumber\\
&\hspace{50pt}- \left(\frac{\ka U}2 \frac{|\t{L}^{\prime z}_{i_{1}\cdots i_{p}}|^2}{ p!} + \frac{U}2\frac{|F'_{i_{1}\cdots i_{p+1}}|^2}{(p+1)!}\right).\nonumber
\end{align}
The first term in ${\cL_{\txt{deep~IR}}}$ enforces the equal-time commutation relation
\begin{equation}
\hspace{-4pt}\frac{[\t\Th'_{i_1\cdots i_p}(\bm{x}),\t{L}^{\prime z}_{j_1\cdots j_p}(\bm{y})]}{p!} \hspace{-1pt}=\hspace{-1pt}\ii \del\indices{^{i_1}_{[j_1}}\hspace{-3pt}\cdots\del\indices{^{i_p}_{j_p]}}\del^d(\bm{x}-\bm{y}),
\end{equation}
and the second term in parenthesis is Hamiltonian density from Eq.~\eqref{continuumH}

This expression of the path integral is correct, but let us rewrite this phase space path integral as a coordinate space path integral to get it into a more familiar form. We first rewrite the delta functions by integrating in new fields and modifying the action. The delta function enforcing the ${\t{L}^{\prime z}}$ quantization condition can be represented as
\begin{equation}
\delta\left(\oint \hstar \t{L}^{\prime z}\in\Z\right)  = \hspace{-15pt}\sum_{\eta\in 2\pi H^{p}(M;\Z)}\hspace{-12pt}\ee^{\ii \int_X  \dd t \dd^{d}\bm x  \frac{\t{L}^{\prime z}_{i_1\cdots i_p}\eta_{i_1\cdots i_p}}{p!}}\hspace{-1pt},
\end{equation}
and the delta function enforcing Gauss law can be rewritten using the ${(p-1)}$-form Lagrange multiplier ${\la}$ as
\begin{equation}
\hspace{-2pt}\delta(\pp_{i_{1}} \t{L}^{\prime z}_{i_{1}\cdots i_{p}}) = \hspace{-2pt}\int\cD\la\ee^{\ii\hspace{-1pt}\int_X \la_{i_2\cdots i_{p}}\pp_{i_{1}} \t{L}^{\prime z}_{i_{1}i_2\cdots i_{p}}/(p-1)!}\hspace{-1pt}.
\end{equation}
Plugging these in, the path integral takes the cumbersome form
\begin{align}
\cZ_{\txt{deep~IR}} &= \hspace{-4pt}\int \cD [\t\Th']~\cD \t{L}^{\prime z}~\cD\la\hspace{-10pt}\sum_{\substack{\om' \in 2\pi H^{p+1}(M;\Z)\\ \eta \in  2\pi H^{p}(M;\Z)}}\hspace{-10pt}\ee^{\ii\int_{X}\dd t  \dd^{d}\bm x ~ \cL_{\txt{deep~IR}} },\nonumber\\
\cL_{\txt{deep~IR}} &= \frac{\la_{i_2\cdots i_{p}}\pp_{i_{1}} \t{L}^{\prime z}_{i_{1}i_2\cdots i_{p}}}{(p-1)!}+\frac{\t{L}^{\prime z}_{i_1\cdots i_{p}}\eta_{i_{1}\cdots i_{p}}}{p!}\\
&\hspace{20pt} +\frac{\t{L}^{\prime z}_{i_1\cdots i_p}\pp_t \t\Th'_{i_1\cdots i_p}}{p!} - \frac{\ka U |\t{L}^{\prime z}_{i_{1}\cdots i_{p}}|^2}{2 p!}\nonumber\\
&\hspace{100pt}  - \frac{U|F'_{i_{1}\cdots i_{p+1}}|^2}{2 (p+1)!}.\nonumber
\end{align}
It is now straight forward to integrate out the ${\t{L}^{\prime z}}$ field, after which the path integral becomes
\begin{align}
\hspace{-3pt}\cZ_{\txt{deep~IR}} &= \hspace{-4pt}\int\cD [\t\Th']~\cD\la\hspace{-10pt}\sum_{\substack{\om' \in 2\pi H^{p+1}(M;\Z)\\ \eta \in  2\pi H^{p}(M;\Z)}}\hspace{-10pt}\ee^{\ii\int_{X}\dd t  \dd^{d}\bm x ~ \cL_{\txt{deep~IR}} },\\
\cL_{\txt{deep~IR}} \hspace{-2pt}&=\hspace{-2pt} \frac{|\pp_t \t\Th'_{i_1\cdots i_p} \hspace{-5pt}- p\hspace{1pt}\pp_{[i_{1}}\la_{i_2\cdots i_{p}]}\hspace{-1pt}+\hspace{-1pt}\eta_{ i_1\cdots i_{p}}|^2}{2g^2p!}\hspace{-1pt}-\hspace{-1pt} \frac{|F'_{i_{1}\cdots i_{p+1}}|^2}{2 g^2 (p+1)!}.\nonumber
\end{align}
where we have also rescaled ${t \to t/(U\sqrt{\ka })}$, ${\la \to U\sqrt{\ka }~\la}$, and ${\eta \to U\sqrt{\ka }~\eta}$ and introduced ${g = 1/(\sqrt{U})}$.

Having found the coordinate path integral, let's massage it into a canonical form. Namely, we introduce the ${p}$-form ${a}$ and ${(p+1)}$-form ${\om_a}$ in spacetime whose components are 
\begin{align*}
a_{i_1\cdots i_p}&=\t\Th'_{i_1\cdots i_p} ,\quad a_{0i_2\cdots i_p}=\la_{i_2\cdots i_p} \\
(\om_a)_{i_1\cdots i_{p+1}}&=\om'_{i_1\cdots i_{p+1}},\quad (\om_a)_{0i_1\cdots i_p}=\eta_{i_1\cdots i_p}
\end{align*}
Letting ${F_a = \dd a + \om_a}$, after some simplifying, we then reexpress the path integral as
\begin{align}
\cZ_{\txt{deep~IR}} &= \hspace{-4pt}\int\cD [a]\hspace{-10pt}\sum_{ \om_a \in 2\pi H^{p+1}(X;\Z)}\hspace{-10pt}\ee^{\ii\int_{X}\dd t  \dd^{d}\bm x ~ \cL_{\txt{deep~IR}} },\\
\cL_{\txt{deep~IR}} &= \frac{1}{2g^2}\left(\frac{|(F_a)_{0i_1\cdots i_p}|^2}{p!} - \frac{|(F_a)_{i_{1}\cdots i_{p+1}}|^2}{(p+1)!}\right)\nonumber
\end{align}
Furthermore, working in flat spacetime, ${X}$ is equipped with Minkowski metric ${(-,+,\cdots +)}$. Thus, summing over spacetime indices ${\mu = 0,\cdots,d}$, ${\cL_{\txt{deep~IR}}}$ can be rewritten as
\begin{equation}\label{eqn:SIR}
\cL_{\txt{deep~IR}} = -\frac1{2g^2(p+1)!} (F_a)_{\mu_1\cdots\mu_{p+1}}(F_a)^{\mu_1\cdots\mu_{p+1}}
\end{equation}
which is exactly ${p}$-form Maxwell theory, as stated in the main text.

\subsection{Review of ${p}$-form Maxwell theory}\label{pformMaxReviewSec}

In the remainder of this appendix section, we will review ${p}$-form Maxwell theory, Eq.~\eqref{eqn:SIR}, focusing on its symmetries and anomalies, working in ${D=d+1}$ dimensional spacetime. In particular, we consider the theory where there is no electric nor magnetic matter, thus ${\dd\hstar F_a = 0}$ and ${\dd F_a = 0}$.

The canonical momentum field $\Pi$ is locally a ${(D-p-1)}$-form associated with a codimension-1 submanifold of spacetime, which we choose to be a constant time slice. Then, $\Pi$'s components are defined by varying the action with respect to $\pp_0 a_{\mu_1\cdots\mu_p}$ which yields ${\Pi = \frac{1}{g^2} \hstar F_a}$. Therefore, from canonical quantization, we have the equal-time commutation relation
\begin{equation}\label{commutationRelation}
\hspace{-4pt}\left[a_{\mu_1\cdots\mu_p}\hspace{-2pt}(\bm{x}),\frac{(\hstar F_a)_{\mu_{p+1}\cdots \mu_{d}}\hspace{-1pt}(\bm{y})}{g^2}\right] \hspace{-2pt}=\hspace{-2pt} \ii \eps_{0\mu_1\cdots\mu_{d}}\del^d(\bm{x}-\bm{y}).
\end{equation}

\subsubsection{${\U{p}}$ symmetry}

The action amplitude is only a function of the field strength, and so the path integral is invariant under ${a}$ being shifted by a closed $p$-form: 
\begin{equation}\label{bTransform}
a \to a + \Gamma, \quad \quad \dd \Gamma = 0.
\end{equation}
This is a symmetry because the Wilson operator ${W_a(C_p) = \exp[\ii\int a]}$ transforms nontrivially under Eq.~\eqref{bTransform} as 
\begin{equation}\label{U1pformTransform}
W_{a}\left(C_p\right) \to \ee^{\ii \oint_{C_p} \Gamma}W_{a}\left(C_p\right).
\end{equation}
There are no restrictions on the holonomies of ${a}$ and ${\Gamma}$ satisfies ${\oint\Gamma\in\R}$, so ${ \exp\left[\ii \oint_{C^p} \Gamma\right]\in U(1)}$. Therefore, Eq.~\eqref{U1pformTransform} is the symmetry transformation of a $\U{p}$ symmetry. Of course, for $\Gamma$ that satisfy ${\exp\left[\ii \oint_{C_p} \Gamma\right] = 1}$, the transformation ${a\to a+\Gamma}$ is instead a gauge transformations corresponding to formal redundancies. The physical transformation on ${a}$ requires that ${\Gamma}$ be closed but not exact and have periods not in $2\pi\Z$.

To find the $\U{p}$ symmetry transformation operator, let us turn on the $(p+1)$-form background field $\cA$ and the gauge redundancy
\begin{equation}
a\to a + \bt,\quad\quad\quad\quad\cA \to \cA + \dd\bt.
\end{equation}
Minimally coupling the background field such that the theory is gauge invariant, the action becomes
\begin{equation}
S[\cA] = -\frac{1}{2g^2}\int_X |F_a - \cA|^2.
\end{equation}
The conserved Noether current $J$ (${\dd^\da J = 0}$) of the symmetry will minimally couple to $\cA$ as ${\int \cA\wdg \hstar J}$. We thus find that ${J = \frac{1}{g^2}F_a}$, and so the charge operator is $Q\equiv \frac{1}{g^2}\int \hstar J$ and the symmetry operator is
\begin{equation}\label{UpSymOpbRep}
U_{\al}\left(\Si\right) = \ee^{\ii\al\oint_{\Si} \frac{\hstar F_a}{g^2}},
\end{equation}
where $\al\in [0,2\pi)$ parametrizes the $U(1)$ transformation. Notice that because ${\dd\hstar J = 0}$, $U_\al(\Si)$ is a topological operator, depending only on the homology class of ${\Si}$.

Let's check that $U_\al$ indeed transforms $W_a$ as Eq.~\eqref{U1pformTransform}. Letting the $p$-form $\h \Si$ be the Poincaré dual of $\Si$ with respect to space $M$, we rewrite the symmetry operator as ${U_{\al}\left(\Si\right) = \exp\left[\frac{\ii\al}{g^2} \oint_{M} \hstar F_a \wdg \h \Si\right]}$, and using the Baker–Campbell–Hausdorff formula, the Wilson operator transforms as
\begin{equation*}
U_{\al}\left(\Si\right)W_a\left(C\right)U^\da_{\al}\left(\Si\right)=  \ee^{\frac{\al}{g^2} \left[\int_{M} \hstar F_a \wdg \h \Si, \int_{C}a \right]}W_a\left(C\right).
\end{equation*}
Using the canonical commutation relations Eq.~\eqref{commutationRelation}, this simplifies to 
\begin{align}
U_{\al}\left(\Si\right)W_a\left(C\right)U^\da_{\al}\left(\Si\right) &= \ee^{\ii \al \oint_{C} \h \Si}~W_a\left(C\right),\\
&= \ee^{\ii \al~ \#\left(\Si, C\right)}~W_a\left(C\right),
\end{align}
where ${\oint_{C} \h \Si = \int_{\Si\cap C} 1 = \#\left(\Si, C\right)}$ is the intersection number between $\Si$ and $C$ in $M$.

\subsubsection{Abelian duality}\label{abelianDualAp}

There is another symmetry present in ${p}$-form Maxwell theory, but to make it manifest, we must effectively change the representation of our degrees of freedom using Abelian duality~\cite{W9505186} to dualize the field ${a}$ to the field ${\h a}$. To do so, we can use the following trick. Instead of integrating over the equivalence classes of ${a}$ and summing over ${\om_a}$, we can instead integrate over $F_a$ since the action only depends on $F_a$. However, in doing so, we have to ensure that we only integrate over ${F_a}$ satisfying the Bianchi identity
\begin{equation}
\frac1{2\pi}\hstar\dd F_a  = 0,
\end{equation}
and which obey the quantization condition
\begin{equation}
\oint_{C_{p+1}} F_a \in 2\pi\Z,
\end{equation}
for all $C_{p+1}\in H_{p+1}(X)$. So, with these two constraints in mind, we can change variables and write the Euclidean path integral of ${p}$-form Maxwell theory as
\begin{equation}\label{PIdeltafuncts}
\cZ = \hspace{-4pt}\int \cD F_a ~\del \hspace{-2pt}\left(\frac{\hstar \dd F_a}{2\pi}\right)\del \hspace{-2pt}\left(\oint \frac{F_a}{2\pi} \in \Z\right) \ee^{-\frac1{2g^2} \int_X |F_a|^2 }.
\end{equation}

Let's now rewrite the delta functions by integrating in fields. For the first delta function, introducing the ${(D-p-2)}$-form $\h a$, we can represent it as
\begin{equation}
\del \hspace{-2pt}\left(\frac{\hstar \dd F_a}{2\pi}\right) = \int \cD \h a~\ee^{\frac{\ii}{2\pi}\int_X\dd\h a\wdg F_a }.
\end{equation}
For the second delta function, we can rewrite it as
\begin{equation}
\begin{aligned}
\del \hspace{-2pt}\left(\oint \frac{F_a}{2\pi} \in \Z\right) &= \hspace{-5pt}\sum_{\hat{\om}_{\h a}\in H_{p+1}(X;\Z)}   \hspace{-5pt}\ee^{2\pi\ii\left(\oint_{\hat{\om}_{\h a}} \frac{F_a}{2\pi}\right)},\\
&= \hspace{-5pt} \sum_{\om_{\h a}\in 2\pi H^{D-p-1}(X;\Z)} \hspace{-10pt} \ee^{\frac{\ii}{2\pi}\int_{X} \om_{\h a}\wdg F_a},
\end{aligned}
\end{equation}
where we first sum over all closed ${(p+1)}$-submanifold ${\h\om_{\h a}}$, and then using Poincaré duality instead sum over the dual ${(D-p-1)}$-forms ${\om_{\h a}/(2\pi)}$ satisfying ${\oint\om_{\h a}\in 2\pi\Z}$. Plugging these representations of the delta functions into the path integral Eq.~\eqref{PIdeltafuncts}, it becomes
\begin{equation}
\cZ = \hspace{-4pt}\int \cD F_a \cD \h a\hspace{-20pt}\sum_{\om_{\h a}\in 2\pi H^{D-p-1}(X;\Z)} \hspace{-20pt} \ee^{-\int_X \frac1{2g^2} |F_a|^2 - \frac{\ii}{2\pi} F_{\h a}\wdg F_a },
\end{equation}
where ${F_{\h a} = \dd\h a + \om_{\h a}}$, which satisfies
\begin{equation}
\oint F_{\h a}\in 2\pi\Z,\quad\quad \frac1{2\pi}\hstar\dd F_{\h a}  = 0.
\end{equation}

To integrate out ${F_a}$, we complete the square and introduce ${G = F_a - \ii\frac{g^2}{2\pi}\hstar F_{\h a}}$ so the path integral becomes
\begin{equation}
\cZ = \hspace{-4pt}\int \cD G ~ \cD \h a\hspace{-20pt}\sum_{\om_{\h a}\in 2\pi H^{D-p-1}(X;\Z)} \hspace{-20pt} \ee^{-\int_X \frac1{2g^2} |G|^2 + \frac{g^2}{8\pi^2} |F_{\h a}|^2 }.
\end{equation}
Integrating out $G$, the path integral is only in terms of the dual field $\h a$:
\begin{equation}\label{dualNLSMPI}
\cZ[X,g] = \int\cD [\h a]\hspace{-20pt}\sum_{\om_{\h a}\in 2\pi H^{D-p-1}(X;\Z)} \hspace{-20pt} \ee^{-\frac{g^2}{8\pi^2} \int_X  |F_{\h a}|^2 }.
\end{equation}
Remarkably, this theory has the same form as what we started with, expect now that initial $p$-form ${a}$ is a ${(D-p-2)}$-form $\h a$ and the coupling constant $g$ is now $2\pi/g$. Thus, strongly coupling ($g\gg 1$) in the ${a}$ representation gets mapped to weak coupling in the $\h a$ representation, and vice versa.

Having gone through the process of dualizing ${a}$ to $\h a$, let's now see how operators in terms of ${a}$ transform under dualizing. We introduce the map $\mathbb{S}$ that takes an operator in the ${a}$ representation to the $\h a$ representation.

Let's first check to see what the field strength $F_a$ maps to by inserting $F_a$ into the path integral. When completing the square, we did a change of variables ${F_a = G + \ii\frac{g^2}{2\pi}\hstar F_{\h a}}$ under which the insertion becomes
\begin{equation}
\< F_a\>_a = \< G\>_{\h a} + \left\< \ii\frac{g^2}{2\pi}\hstar F_{\h a}\right\>_{\h a}.
\end{equation}
We use the notation that ${\<\cdot\>_a}$ is the vev evaluated in the ${a}$-representation and ${\<\cdot\>_{\h a}}$ is the vev evaluated in the ${\h a}$-representation. Since the ${G}$ part of the action is Gaussian, ${\< G \> = 0}$ and so ${\< F_a\>=\< \ii\frac{g^2}{2\pi}\hstar F_{\h a}\>}$. Thus, in Euclidean signature
\begin{equation}\label{abelianDualF}
\mathbb{S}:F_a\to\ii\frac{g^2}{2\pi}\hstar F_{\h a}.
\end{equation}
In the Lorentzian signature, this is ${\mathbb{S}:F_a\to\frac{g^2}{2\pi}\hstar F_{\h a}}$. We can dualizing $\h a$ back to ${a}$ by simply repeating the same steps as before to find
\begin{equation}
\mathbb{S}:F_{\h a}\to\ii\frac{2\pi }{g^2}\hstar F_a.
\end{equation}
Therefore, Dualizing ${a}$ twice gives ${\mathbb{S}^2:F_a \to \ii^2\hstar\hstar F_a
}$ in Euclidean spacetime, or equivalently
\begin{equation}
\mathbb{S}^2:F_a \to (-1)^{D(p+1)+p} F_a.
\end{equation}
When both $D$ and $p$ are even, dualizing twice acts as ${\mathbb{S}^2:F_a \to F_a}$. However, if $D$ or $p$ or both are odd, then ${\mathbb{S}^2:F_a \to -F_a}$ and thus one must dualize four times to get the identity map: ${\mathbb{S}^4:F_a \to F_a}$.

Repeating the argument manipulations, we find that ${\dd^\da F_a}$ and ${\hstar \dd F_a}$ get mapped to ${\hstar\dd F_{\h a}}$ and ${\dd^\da F_{\h a}}$, respectively, and vice versa. Therefore, the excitations (topological defects) of ${a}$ are the topological defects (excitations) of $\h a$. Hence, abelian duality is a particle-vortex type duality.

We emphasize that the above mappings do not imply that ${F_a=\ii\frac{g^2}{2\pi}\hstar F_{\h a}}$. Instead, while operators linear in $F_a$ simply have ${F_a}$ replaced with ${\ii\frac{g^2}{2\pi}\hstar F_{\h a}}$, more care is required to find the dual representation of operators nonlinear in $F_a$. For instance, $(F_a)^2$ does \txti{not} become $(\ii\frac{g^2}{2\pi}\hstar F_{\h a})^2$ due to the addition terms pick up when squaring ${F_a = G + \ii\frac{g^2}{2\pi}\hstar F_{\h a}}$.

Next, let us find what the Wilson operator of ${a}$ gets mapped to. We assume that ${W_a}$ is supported on a contractible manifold $C = \pp M$. However, there are infinitely many such ${M}$ whose boundary is ${C}$. To avoid this ambiguity, we sum over all such $M$ and write
\begin{equation}
W_a[C] = \sum_{M:\pp M = C}\exp\left[\ii \int_{M} F_a\right].
\end{equation}
Next we introduce the ${D-p-1}$ form ${\hat{M}}$ dual to $M$, which satisfies ${\oint \h M\in2\pi\Z}$ and is related to the Poincaré dual of $C$ by ${\h C = \dd \h M/(2\pi)}$, and rewrite ${W_a}$ as
\begin{equation*}
W_a[C] = \int\cD\h M ~\del(\dd\h M - 2\pi\h C) \exp\left[\frac{\ii}{2\pi} \int_{X} F_a \wdg \h M\right].
\end{equation*}
Inserting this into the path integral, integrating in $\h a$, and then integrating out ${F_a}$, we find
\begin{align}\label{WilsonopbhatrepZ}
\<W_a[C]\> &= \int\cD\h M \cD [\h a]\hspace{-20pt}\sum_{\om_{\h a}\in 2\pi H^{D-p-1}(X;\Z)} \hspace{-20pt} \del(\dd\h M - 2\pi\h C)~ \ee^{-\int_X\cL },\nonumber\\
\cL &= \frac{g^2}{8\pi^2}  |F_{\h a}-\h M|^2.
\end{align}
Notice how this is the same as the dualized path integral without the Wilson loop insertion but now with the ${p+1}$ form connection ${\h M}$ satisfying ${\dd \h M = 2\pi\h C}$. We can get rid of $\h M$ and have the path integral look similar to Eq.~\eqref{dualNLSMPI} if we let ${\h a}$ be a singular field not defined on ${C}$:
\begin{equation}
\cZ = \int \cD [\h a]\hspace{-20pt}\sum_{\om_{\h a}\in 2\pi H^{D-p-1}(X;\Z)} \hspace{-20pt} \ee^{-\int_{X/C} \frac{g^2}{8\pi^2}  |F_{\h a}|^2 }.
\end{equation}
However, we require that ${\int_{\Si} F_{\hat{a}} = 2\pi}$ for any submanifold ${\Si}$ with a nonzero intersection number with ${C}$. Therefore, the Wilson loop in the ${a}$ representation has become a 't Hooft loop in the ${\h a}$ representation.

\subsubsection{${\U{d-p-1}}$ symmetry}

In the ${a}$ representation, it appears that the model only has a ${\U{p}}$ symmetry. However, upon dualizing ${a}$ to $\h a$, the path integral Eq.~\eqref{dualNLSMPI} took a similar form in terms of ${F_{\hat{a}}}$ but with ${g}$ replaced by ${2\pi/g}$. Thus, we find a new globally defined differential ${(d-p-1)}$-form $\h a$, which shifting by a closed form leaves the path integral invariant. Following the same process as used in investigating the ${\U{p}}$ symmetry, this transformation has a physical part associated with a ${\U{d-p-1}}$ symmetry.

Everything about this ${\U{d-p-1}}$ symmetry follows in a similar fashion from the ${\U{p}}$ case. In particular, the symmetry transformation acts on ${\h a}$ as
\begin{equation}
\h a\to \h a + \h \Gamma, \quad \quad  \dd \h \Gamma = 0
\end{equation}
and the charged operators are Wilson operators in terms of ${\h a}$: ${W_{\h a}\left(C\right)  = \exp\left[\ii \oint_{C} \h a\right]}$. These correspond to the 't Hooft operators in the ${a}$ representation. The Noether's current associated with this $\U{d-p-1}$ symmetry is ${\hat{J} = \frac{g^2}{4\pi^2}F_{\h a}}$, and thus the symmetry operator is
\begin{equation}\label{dualSymOp}
\hat{U}_{\hat{\al}}\left(\Si\right) = \ee^{\ii\hat{\al}~\frac{g^2}{4\pi^2} \oint_{\Si} \hstar F_{\h a} },
\end{equation}
where $\hat{\al}\in [0,2\pi)$ parametrizes the $U(1)$ transformation.

To see what ${\h U}$ is in the ${a}$ representation, let's start with the operator ${\exp\left[\ii\theta \int_{\Si} F_{a} \right]}$ and find its image under ${\mathbb S}$. We've already done this calculation while finding the image of the Wilson operator. This time, we simply do not sum over all ${\Si}$. We, therefore, have that (in Lorentzian signature) 
\begin{equation}
\mathbb{S}:\ee^{\ii\theta \oint_{\Si} F_{a} }\to \ee^{\ii\theta \oint_{\Si} \frac{g^2}{2\pi}\hstar F_{\h a} - \ii \frac{\theta^2 g^2}{2} \int_X |\h\Si|^2}.
\end{equation}
Setting ${\theta = \frac{\h\al}{2\pi}}$, the ${\U{d-p-1}}$ symmetry operator in the ${a}$-representation is
\begin{equation}
\h U_{\h\al}(\Si) = \ee^{\ii \h\al\oint_{\Si} \frac{F_{a}}{2\pi}  + \ii \frac{\h{\al}^2 g^2}{8\pi^2} \oint_X |\h\Si|^2 }
\end{equation}
since under $\mathbb{S}$ it transforms to Eq.~\eqref{dualSymOp}. However, note that the term ${\int_X|\h\Si|^2}$ is an overall phase and therefore does not affect the symmetry transformation. So, we can drop this overall phase and treat the ${\U{d-p-1}}$ symmetry operator in the ${a}$-representation instead as
\begin{equation}\label{dualSymOpbRep}
\h U_{\h\al}(\Si) = \ee^{\ii \h\al\oint_{\Si} \frac{F_{a}}{2\pi} }.
\end{equation}
From this expression, it is easy to see that the Noether current of the ${\U{d-p-1}}$ symmetry in the ${a}$ representation ${\h J}$ satisfies
\begin{equation}
\hstar \h J = \frac{1}{2\pi}  F_{a}.
\end{equation}
The fact that the ${\h J}$ is conserved reflects the Bianchi identity.

\subsubsection{Mixed 't Hooft anomaly and anomaly inflow}

Throughout this subsection thus far, we reviewed that ${p}$-form Maxwell theory has a $\U{p}$ and a $\U{d-p-1}$ symmetry. However, these two symmetries are not fully independent from one another: there is a mixed 't Hooft anomaly preventing us from simultaneously turning on a background gauge field of both symmetries.

Let's first turn on a background field $\cA$ of the ${\U{p}}$ symmetry which includes the gauge redundancy
\begin{equation}\label{gaugeRed10}
a\to a + \bt,\quad\quad\quad\quad \cA \to \cA + \dd\bt.
\end{equation}
Minimally coupling $\cA$ to ${a}$, the path integral becomes
\begin{equation}\label{UpGaugedThy}
\cZ[\cA] = \hspace{-4pt}\int \cD [a]~ \ee^{-\int_X \frac1{2g^2} |F_{a}-\cA|^2 }.
\end{equation}
This is a gauge invariant theory, so the ${\U{p}}$ symmetry is anomaly free.

Let's dualize ${a}$ to ${\h a}$ to see how ${\cA}$ couples to ${\h a}$. Repeating the first few steps of abelian duality reviewed in section~\ref{abelianDualAp}, the path integral becomes
\begin{equation}\label{ZU1pGaugedhalf}
\cZ[\cA] = \hspace{-4pt}\int \hspace{-2pt}\cD K \cD \h a\ee^{-\int_X \frac1{2g^2} |K|^2 - \frac{\ii}{2\pi} F_{\h a}\wdg K- \frac{\ii}{2\pi}  \cA\wdg F_{\h a} }\hspace{-2pt},
\end{equation}
where we made the change of variables ${F_{a} = K + \cA}$. Integrating out ${K}$, this becomes
\begin{equation}\label{UpGaugethyDualRep}
\cZ[\cA]  = \hspace{-4pt}\int \cD [\h a]~ \ee^{-\int_X \frac{g^2}{8\pi^2} |F_{\h a}|^2 - \frac{\ii}{2\pi}  \cA\wdg F_{\h a} }.
\end{equation}
Note that because ${\dd F_{\h a} = 0}$ and spacetime is closed, the path integral in the ${\h a}$-representation is still invariant under the gauge transformation Eq~\eqref{gaugeRed10}.

Eq.~\eqref{UpGaugethyDualRep} reveals that turning on a $\U{p}$ symmetry background gauge field is equivalent to adding a topological term in the $\h a$ representation. This new term has a noticeable effect. The Noether current for the ${\U{D-p-2}}$ symmetry, ${ \hat{J} = \frac{g^2}{4\pi^2}F_{\h a}}$, is no longer conserved:
\begin{equation}
\dd^\da \hat{J} = \frac{1}{2\pi}\hstar\dd\cA.
\end{equation}
This is a manifestation of the mixed 't Hooft anomaly.

Let's now turn off ${\cA}$ and turn on a background gauge field ${\h\cA}$ for the ${\U{d-p-1}}$ symmetry with the gauge redundancy
\begin{equation}\label{gaugeRed11}
\h a\to \h a + \h \bt,\quad\quad\quad\quad\h \cA \to \h \cA +\dd\h\bt.
\end{equation}
Inspired by how $\cA$ coupled to ${\h a}$ in Eq.~\eqref{ZU1pGaugedhalf}, we minimally couple $\h\cA$ to ${a}$ in a similar fashion and consider
\begin{equation}\label{gaugedUdp1brep}
\cZ[\h\cA] = \hspace{-4pt}\int \cD [a]~\ee^{-\int_X \frac1{2g^2} |F_{a}|^2 - \frac{\ii}{2\pi}\h\cA \wdg F_{a} }.
\end{equation}
Because ${F_{a}}$ is closed, shifting ${\h\cA}$ by an exact form does not change the path integral. Thus, the path integral is gauge invariant and ${\U{d-p-1}}$ is anomaly free. To verify this way of coupling ${\h\cA}$ to ${a}$ is correct, let's dualize ${a}$ to $\h a$ in Eq.~\eqref{gaugedUdp1brep}. Doing so, we find
\begin{equation}
\cZ[\h\cA] = \int \cD [\h a]\cD[\h \cA]~ \ee^{-\int_X \frac{g^2}{8\pi^2} |F_{\h a} - \h\cA|^2 },
\end{equation}
as expected. Returning back to Eq.~\eqref{gaugedUdp1brep}, due to the new topological term, the ${\U{p}}$ symmetry Noether current ${J = \frac{1}{g^2}F_a}$ is no longer conserved:
\begin{equation}
\dd^\da J = \frac{1}{2\pi}\hstar\dd\hat{\cA}.
\end{equation}
Once again, this is a manifestation of the mixed 't Hooft anomaly.

Let us now turn on both of the background gauge fields ${\cA}$ and ${\h\cA}$. Working in the ${a}$ representation and using what we just found, the path integral becomes
\begin{equation}
\cZ[\cA,\h\cA] = \hspace{-4pt}\int \cD [a]~\ee^{-\int_X \frac1{2g^2} |F_{a}-\cA|^2 - \frac{\ii}{2\pi}\h\cA \wdg F_{a} }.
\end{equation}
This path integral is invariant under the gauge transformation Eq.~\eqref{gaugeRed11}. However, due to the second term in ${\cL}$, this path integral is no longer invariant under the gauge transformation Eq.~\eqref{gaugeRed10}, and transforms as
\begin{equation}\label{Zgaugeinvbrk}
\cZ[\cA,\h\cA]\to \ee^{\frac{\ii}{2\pi}\int_X \h\cA \wdg \dd\bt}\cZ[\cA,\h\cA].
\end{equation}
No local counter term can remedy this property, and thus the ${\U{p}\times\U{d-p-1}}$ symmetry is anomalous.

We can make the theory gauge invariant by introducing the ${(D+1)}$-dimensional spacetime $Y$ such that $X = \partial Y$ and extending the background gauge fields $\cA$ and $\h\cA$ into $Y$. Indeed, notice how then the phase picked up in Eq.~\eqref{Zgaugeinvbrk} can be rewritten as
\begin{equation}
\int_{X=\pp Y} \hspace{-10pt}\h\cA \wdg \dd\bt = \int_{Y} \dd(\h\cA \wdg \dd\bt)= \int_{Y} \dd\h\cA \wdg \dd\bt.
\end{equation}
This then motivates the new \txti{gauge invariant} partition function
\begin{equation}
\begin{aligned}
\cZ[\cA,\h\cA] &=\exp\left[-\frac{\ii}{2\pi}\int_Y \dd\h\cA \wdg \cA\right]\int \cD [a]~\ee^{-\int_{\pp Y} \cL },\\
\cL &= \frac1{2g^2} |F_{a}-\cA|^2 - \frac{\ii}{2\pi}\h\cA \wdg F_{a}.
\end{aligned}
\end{equation}
Indeed, $\cZ[\cA,\hat{\cA}]$ is invariant under the gauge transformations Eq.~\eqref{Zgaugeinvbrk} since the phase picked up from ${\int \cD [\h a]~\ee^{-\int_{\pp Y} \cL }}$ cancels with the phase we added.

Thus, we see the 't Hooft anomaly through the modern perspective of anomaly inflow. In order to turn on both ${\cA}$ and ${\h\cA}$, we must have the theory resides on the boundary of an SPT, which in this case was~\cite{HT200311550}
\begin{equation}
\label{U1U1}
\cZ_{\txt{SPT}}[\cA,\h\cA] = \ee^{-\frac{\ii}{2\pi}\int_Y \dd\h\cA \wdg \cA}.
\end{equation}

\bibliography{
../../bib/all,
../../bib/allnew,
../../bib/publst,
../../bib/publstnew,
../../bib/salRefs
}

\begin{thebibliography}{130}%
\makeatletter
\providecommand \@ifxundefined [1]{%
 \@ifx{#1\undefined}
}%
\providecommand \@ifnum [1]{%
 \ifnum #1\expandafter \@firstoftwo
 \else \expandafter \@secondoftwo
 \fi
}%
\providecommand \@ifx [1]{%
 \ifx #1\expandafter \@firstoftwo
 \else \expandafter \@secondoftwo
 \fi
}%
\providecommand \natexlab [1]{#1}%
\providecommand \enquote  [1]{``#1''}%
\providecommand \bibnamefont  [1]{#1}%
\providecommand \bibfnamefont [1]{#1}%
\providecommand \citenamefont [1]{#1}%
\providecommand \href@noop [0]{\@secondoftwo}%
\providecommand \href [0]{\begingroup \@sanitize@url \@href}%
\providecommand \@href[1]{\@@startlink{#1}\@@href}%
\providecommand \@@href[1]{\endgroup#1\@@endlink}%
\providecommand \@sanitize@url [0]{\catcode `\\12\catcode `\$12\catcode
  `\&12\catcode `\#12\catcode `\^12\catcode `\_12\catcode `\%12\relax}%
\providecommand \@@startlink[1]{}%
\providecommand \@@endlink[0]{}%
\providecommand \url  [0]{\begingroup\@sanitize@url \@url }%
\providecommand \@url [1]{\endgroup\@href {#1}{\urlprefix }}%
\providecommand \urlprefix  [0]{URL }%
\providecommand \Eprint [0]{\href }%
\providecommand \doibase [0]{https://doi.org/}%
\providecommand \selectlanguage [0]{\@gobble}%
\providecommand \bibinfo  [0]{\@secondoftwo}%
\providecommand \bibfield  [0]{\@secondoftwo}%
\providecommand \translation [1]{[#1]}%
\providecommand \BibitemOpen [0]{}%
\providecommand \bibitemStop [0]{}%
\providecommand \bibitemNoStop [0]{.\EOS\space}%
\providecommand \EOS [0]{\spacefactor3000\relax}%
\providecommand \BibitemShut  [1]{\csname bibitem#1\endcsname}%
\let\auto@bib@innerbib\@empty
\bibitem [{\citenamefont {Nussinov}\ and\ \citenamefont
  {Ortiz}(2009{\natexlab{a}})}]{NOc0605316}%
  \BibitemOpen
  \bibfield  {author} {\bibinfo {author} {\bibfnamefont {Z.}~\bibnamefont
  {Nussinov}}\ and\ \bibinfo {author} {\bibfnamefont {G.}~\bibnamefont
  {Ortiz}},\ }\bibfield  {title} {\bibinfo {title} {Sufficient symmetry
  conditions for topological quantum order},\ }\href
  {https://doi.org/10.1073/pnas.0803726105} {\bibfield  {journal} {\bibinfo
  {journal} {Proc. Natl. Acad. Sci. U.S.A.}\ }\textbf {\bibinfo {volume}
  {106}},\ \bibinfo {pages} {16944} (\bibinfo {year} {2009}{\natexlab{a}})},\
  \Eprint {https://arxiv.org/abs/cond-mat/0605316} {arXiv:cond-mat/0605316}
  \BibitemShut {NoStop}%
\bibitem [{\citenamefont {Nussinov}\ and\ \citenamefont
  {Ortiz}(2009{\natexlab{b}})}]{NOc0702377}%
  \BibitemOpen
  \bibfield  {author} {\bibinfo {author} {\bibfnamefont {Z.}~\bibnamefont
  {Nussinov}}\ and\ \bibinfo {author} {\bibfnamefont {G.}~\bibnamefont
  {Ortiz}},\ }\bibfield  {title} {\bibinfo {title} {A symmetry principle for
  topological quantum order},\ }\href
  {https://doi.org/10.1016/j.aop.2008.11.002} {\bibfield  {journal} {\bibinfo
  {journal} {Ann. Phys.}\ }\textbf {\bibinfo {volume} {324}},\ \bibinfo {pages}
  {977} (\bibinfo {year} {2009}{\natexlab{b}})},\ \Eprint
  {https://arxiv.org/abs/cond-mat/0702377} {arXiv:cond-mat/0702377}
  \BibitemShut {NoStop}%
\bibitem [{\citenamefont {Gaiotto}\ \emph {et~al.}(2015)\citenamefont
  {Gaiotto}, \citenamefont {Kapustin}, \citenamefont {Seiberg},\ and\
  \citenamefont {Willett}}]{GW14125148}%
  \BibitemOpen
  \bibfield  {author} {\bibinfo {author} {\bibfnamefont {D.}~\bibnamefont
  {Gaiotto}}, \bibinfo {author} {\bibfnamefont {A.}~\bibnamefont {Kapustin}},
  \bibinfo {author} {\bibfnamefont {N.}~\bibnamefont {Seiberg}},\ and\ \bibinfo
  {author} {\bibfnamefont {B.}~\bibnamefont {Willett}},\ }\bibfield  {title}
  {\bibinfo {title} {Generalized global symmetries},\ }\href
  {https://doi.org/10.1007/jhep02(2015)172} {\bibfield  {journal} {\bibinfo
  {journal} {J. High Energ. Phys.}\ }\textbf {\bibinfo {volume}
  {2015}}\bibfield  {number} {\bibinfo  {number} { (2)},\ \bibinfo {pages}
  {172}},\ }\Eprint {https://arxiv.org/abs/1412.5148} {arXiv:1412.5148}
  \BibitemShut {NoStop}%
\bibitem [{\citenamefont {{Kapustin}}\ and\ \citenamefont
  {{Thorngren}}(2017)}]{KT13094721}%
  \BibitemOpen
  \bibfield  {author} {\bibinfo {author} {\bibfnamefont {A.}~\bibnamefont
  {{Kapustin}}}\ and\ \bibinfo {author} {\bibfnamefont {R.}~\bibnamefont
  {{Thorngren}}},\ }\bibinfo {title} {{Higher Symmetry and Gapped Phases of
  Gauge Theories}},\ in\ \href {https://doi.org/10.1007/978-3-319-59939-7_5}
  {\emph {\bibinfo {booktitle} {{Algebra, Geometry, and Physics in the 21st
  Century: Kontsevich Festschrift}}}},\ \bibinfo {editor} {edited by\ \bibinfo
  {editor} {\bibfnamefont {D.}~\bibnamefont {Auroux}}, \bibinfo {editor}
  {\bibfnamefont {L.}~\bibnamefont {Katzarkov}}, \bibinfo {editor}
  {\bibfnamefont {T.}~\bibnamefont {Pantev}}, \bibinfo {editor} {\bibfnamefont
  {Y.}~\bibnamefont {Soibelman}},\ and\ \bibinfo {editor} {\bibfnamefont
  {Y.}~\bibnamefont {Tschinkel}}}\ (\bibinfo  {publisher} {{Springer
  International Publishing}},\ \bibinfo {year} {2017})\ pp.\ \bibinfo {pages}
  {177--202},\ \Eprint {https://arxiv.org/abs/1309.4721} {arXiv:1309.4721}
  \BibitemShut {NoStop}%
\bibitem [{\citenamefont {Sharpe}(2015)}]{S150804770}%
  \BibitemOpen
  \bibfield  {author} {\bibinfo {author} {\bibfnamefont {E.}~\bibnamefont
  {Sharpe}},\ }\bibfield  {title} {\bibinfo {title} {Notes on generalized
  global symmetries in {QFT}},\ }\href {https://doi.org/10.1002/prop.201500048}
  {\bibfield  {journal} {\bibinfo  {journal} {Fortschr. Phys.}\ }\textbf
  {\bibinfo {volume} {63}},\ \bibinfo {pages} {659} (\bibinfo {year} {2015})},\
  \Eprint {https://arxiv.org/abs/1508.04770} {arXiv:1508.04770} \BibitemShut
  {NoStop}%
\bibitem [{\citenamefont {Benini}\ \emph {et~al.}(2019)\citenamefont {Benini},
  \citenamefont {C\'ordova},\ and\ \citenamefont {Hsin}}]{BH180309336}%
  \BibitemOpen
  \bibfield  {author} {\bibinfo {author} {\bibfnamefont {F.}~\bibnamefont
  {Benini}}, \bibinfo {author} {\bibfnamefont {C.}~\bibnamefont {C\'ordova}},\
  and\ \bibinfo {author} {\bibfnamefont {P.-S.}\ \bibnamefont {Hsin}},\
  }\bibfield  {title} {\bibinfo {title} {On 2-group global symmetries and their
  anomalies},\ }\href {https://doi.org/10.1007/jhep03(2019)118} {\bibfield
  {journal} {\bibinfo  {journal} {J. High Energ. Phys.}\ }\textbf {\bibinfo
  {volume} {2019}}\bibfield  {number} {\bibinfo  {number} { (3)},\ \bibinfo
  {pages} {118}},\ }\Eprint {https://arxiv.org/abs/1803.09336}
  {arXiv:1803.09336} \BibitemShut {NoStop}%
\bibitem [{\citenamefont {C\'ordova}\ \emph {et~al.}(2019)\citenamefont
  {C\'ordova}, \citenamefont {Dumitrescu},\ and\ \citenamefont
  {Intriligator}}]{CI180204790}%
  \BibitemOpen
  \bibfield  {author} {\bibinfo {author} {\bibfnamefont {C.}~\bibnamefont
  {C\'ordova}}, \bibinfo {author} {\bibfnamefont {T.~T.}\ \bibnamefont
  {Dumitrescu}},\ and\ \bibinfo {author} {\bibfnamefont {K.}~\bibnamefont
  {Intriligator}},\ }\bibfield  {title} {\bibinfo {title} {Exploring 2-group
  global symmetries},\ }\href {https://doi.org/10.1007/jhep02(2019)184}
  {\bibfield  {journal} {\bibinfo  {journal} {J. High Energ. Phys.}\ }\textbf
  {\bibinfo {volume} {2019}}\bibfield  {number} {\bibinfo  {number} { (2)},\
  \bibinfo {pages} {184}},\ }\Eprint {https://arxiv.org/abs/1802.04790}
  {arXiv:1802.04790} \BibitemShut {NoStop}%
\bibitem [{\citenamefont {{Barkeshli}}\ \emph {et~al.}(2023)\citenamefont
  {{Barkeshli}}, \citenamefont {{Chen}}, \citenamefont {{Huang}}, \citenamefont
  {{Kobayashi}}, \citenamefont {{Tantivasadakarn}},\ and\ \citenamefont
  {{Zhu}}}]{BCH220807367}%
  \BibitemOpen
  \bibfield  {author} {\bibinfo {author} {\bibfnamefont {M.}~\bibnamefont
  {{Barkeshli}}}, \bibinfo {author} {\bibfnamefont {Y.-A.}\ \bibnamefont
  {{Chen}}}, \bibinfo {author} {\bibfnamefont {S.-J.}\ \bibnamefont {{Huang}}},
  \bibinfo {author} {\bibfnamefont {R.}~\bibnamefont {{Kobayashi}}}, \bibinfo
  {author} {\bibfnamefont {N.}~\bibnamefont {{Tantivasadakarn}}},\ and\
  \bibinfo {author} {\bibfnamefont {G.}~\bibnamefont {{Zhu}}},\ }\bibfield
  {title} {\bibinfo {title} {{Codimension-2 defects and higher symmetries in
  (3+1)D topological phases}},\ }\href
  {https://doi.org/10.21468/SciPostPhys.14.4.065} {\bibfield  {journal}
  {\bibinfo  {journal} {SciPost Phys.}\ }\textbf {\bibinfo {volume} {14}},\
  \bibinfo {pages} {065} (\bibinfo {year} {2023})},\ \Eprint
  {https://arxiv.org/abs/2208.07367} {arXiv:2208.07367} \BibitemShut {NoStop}%
\bibitem [{\citenamefont {{Barkeshli}}\ \emph {et~al.}(2022)\citenamefont
  {{Barkeshli}}, \citenamefont {{Chen}}, \citenamefont {{Hsin}},\ and\
  \citenamefont {{Kobayashi}}}]{BCH221111764}%
  \BibitemOpen
  \bibfield  {author} {\bibinfo {author} {\bibfnamefont {M.}~\bibnamefont
  {{Barkeshli}}}, \bibinfo {author} {\bibfnamefont {Y.-A.}\ \bibnamefont
  {{Chen}}}, \bibinfo {author} {\bibfnamefont {P.-S.}\ \bibnamefont {{Hsin}}},\
  and\ \bibinfo {author} {\bibfnamefont {R.}~\bibnamefont {{Kobayashi}}},\
  }\href@noop {} {\bibinfo {title} {{Higher-group symmetry in finite gauge
  theory and stabilizer codes}}} (\bibinfo {year} {2022}),\ \Eprint
  {https://arxiv.org/abs/2211.11764} {arXiv:2211.11764} \BibitemShut {NoStop}%
\bibitem [{\citenamefont {{Gorantla}}\ \emph {et~al.}(2022)\citenamefont
  {{Gorantla}}, \citenamefont {{Lam}}, \citenamefont {{Seiberg}},\ and\
  \citenamefont {{Shao}}}]{GLS220110589}%
  \BibitemOpen
  \bibfield  {author} {\bibinfo {author} {\bibfnamefont {P.}~\bibnamefont
  {{Gorantla}}}, \bibinfo {author} {\bibfnamefont {H.~T.}\ \bibnamefont
  {{Lam}}}, \bibinfo {author} {\bibfnamefont {N.}~\bibnamefont {{Seiberg}}},\
  and\ \bibinfo {author} {\bibfnamefont {S.-H.}\ \bibnamefont {{Shao}}},\
  }\bibfield  {title} {\bibinfo {title} {{Global dipole symmetry, compact
  Lifshitz theory, tensor gauge theory, and fractons}},\ }\href
  {https://doi.org/10.1103/PhysRevB.106.045112} {\bibfield  {journal} {\bibinfo
   {journal} {Phys. Rev. B}\ }\textbf {\bibinfo {volume} {106}},\ \bibinfo
  {pages} {045112} (\bibinfo {year} {2022})},\ \Eprint
  {https://arxiv.org/abs/2201.10589} {arXiv:2201.10589} \BibitemShut {NoStop}%
\bibitem [{\citenamefont {{Kapustin}}\ and\ \citenamefont
  {{Spodyneiko}}(2022)}]{KS220809056}%
  \BibitemOpen
  \bibfield  {author} {\bibinfo {author} {\bibfnamefont {A.}~\bibnamefont
  {{Kapustin}}}\ and\ \bibinfo {author} {\bibfnamefont {L.}~\bibnamefont
  {{Spodyneiko}}},\ }\bibfield  {title} {\bibinfo {title}
  {{Hohenberg-Mermin-Wagner-type theorems and dipole symmetry}},\ }\href
  {https://doi.org/10.1103/PhysRevB.106.245125} {\bibfield  {journal} {\bibinfo
   {journal} {Phys. Rev. B}\ }\textbf {\bibinfo {volume} {106}},\ \bibinfo
  {pages} {245125} (\bibinfo {year} {2022})},\ \Eprint
  {https://arxiv.org/abs/2208.09056} {arXiv:2208.09056} \BibitemShut {NoStop}%
\bibitem [{\citenamefont {{Qi}}\ \emph {et~al.}(2021)\citenamefont {{Qi}},
  \citenamefont {{Radzihovsky}},\ and\ \citenamefont
  {{Hermele}}}]{QRH201002254}%
  \BibitemOpen
  \bibfield  {author} {\bibinfo {author} {\bibfnamefont {M.}~\bibnamefont
  {{Qi}}}, \bibinfo {author} {\bibfnamefont {L.}~\bibnamefont
  {{Radzihovsky}}},\ and\ \bibinfo {author} {\bibfnamefont {M.}~\bibnamefont
  {{Hermele}}},\ }\bibfield  {title} {\bibinfo {title} {{Fracton phases via
  exotic higher-form symmetry-breaking}},\ }\href
  {https://doi.org/10.1016/j.aop.2020.168360} {\bibfield  {journal} {\bibinfo
  {journal} {Ann. Phys.}\ }\textbf {\bibinfo {volume} {424}},\ \bibinfo {pages}
  {168360} (\bibinfo {year} {2021})},\ \Eprint
  {https://arxiv.org/abs/2010.02254} {arXiv:2010.02254} \BibitemShut {NoStop}%
\bibitem [{\citenamefont {{Oh}}\ \emph {et~al.}(2023)\citenamefont {{Oh}},
  \citenamefont {{Pace}}, \citenamefont {{Han}}, \citenamefont {{You}},\ and\
  \citenamefont {{Lee}}}]{OPH230104706}%
  \BibitemOpen
  \bibfield  {author} {\bibinfo {author} {\bibfnamefont {Y.-T.}\ \bibnamefont
  {{Oh}}}, \bibinfo {author} {\bibfnamefont {S.~D.}\ \bibnamefont {{Pace}}},
  \bibinfo {author} {\bibfnamefont {J.~H.}\ \bibnamefont {{Han}}}, \bibinfo
  {author} {\bibfnamefont {Y.}~\bibnamefont {{You}}},\ and\ \bibinfo {author}
  {\bibfnamefont {H.-Y.}\ \bibnamefont {{Lee}}},\ }\bibfield  {title} {\bibinfo
  {title} {{Aspects of ${\mathbb{Z}}_{N}$ rank-2 gauge theory in $(2+1)$
  dimensions: Construction schemes, holonomies, and sublattice one-form
  symmetries}},\ }\href {https://doi.org/10.1103/PhysRevB.107.155151}
  {\bibfield  {journal} {\bibinfo  {journal} {Phys. Rev. B}\ }\textbf {\bibinfo
  {volume} {107}},\ \bibinfo {pages} {155151} (\bibinfo {year} {2023})},\
  \Eprint {https://arxiv.org/abs/2301.04706} {arXiv:2301.04706} \BibitemShut
  {NoStop}%
\bibitem [{\citenamefont {{Bhardwaj}}\ and\ \citenamefont
  {{Tachikawa}}(2018)}]{BT170402330}%
  \BibitemOpen
  \bibfield  {author} {\bibinfo {author} {\bibfnamefont {L.}~\bibnamefont
  {{Bhardwaj}}}\ and\ \bibinfo {author} {\bibfnamefont {Y.}~\bibnamefont
  {{Tachikawa}}},\ }\bibfield  {title} {\bibinfo {title} {{On finite symmetries
  and their gauging in two dimensions}},\ }\href
  {https://doi.org/10.1007/JHEP03(2018)189} {\bibfield  {journal} {\bibinfo
  {journal} {J. High Energ. Phys.}\ }\textbf {\bibinfo {volume}
  {2018}}\bibfield  {number} {\bibinfo  {number} { (3)},\ \bibinfo {pages}
  {189}},\ }\Eprint {https://arxiv.org/abs/1704.02330} {arXiv:1704.02330}
  \BibitemShut {NoStop}%
\bibitem [{\citenamefont {{Thorngren}}\ and\ \citenamefont
  {{Wang}}(2019)}]{TW191202817}%
  \BibitemOpen
  \bibfield  {author} {\bibinfo {author} {\bibfnamefont {R.}~\bibnamefont
  {{Thorngren}}}\ and\ \bibinfo {author} {\bibfnamefont {Y.}~\bibnamefont
  {{Wang}}},\ }\href@noop {} {\bibinfo {title} {{Fusion Category Symmetry I:
  Anomaly In-Flow and Gapped Phases}}} (\bibinfo {year} {2019}),\ \Eprint
  {https://arxiv.org/abs/1912.02817} {arXiv:1912.02817} \BibitemShut {NoStop}%
\bibitem [{\citenamefont {{Kong}}\ \emph
  {et~al.}(2020{\natexlab{a}})\citenamefont {{Kong}}, \citenamefont {{Lan}},
  \citenamefont {{Wen}}, \citenamefont {{Zhang}},\ and\ \citenamefont
  {{Zheng}}}]{KZ200514178}%
  \BibitemOpen
  \bibfield  {author} {\bibinfo {author} {\bibfnamefont {L.}~\bibnamefont
  {{Kong}}}, \bibinfo {author} {\bibfnamefont {T.}~\bibnamefont {{Lan}}},
  \bibinfo {author} {\bibfnamefont {X.-G.}\ \bibnamefont {{Wen}}}, \bibinfo
  {author} {\bibfnamefont {Z.-H.}\ \bibnamefont {{Zhang}}},\ and\ \bibinfo
  {author} {\bibfnamefont {H.}~\bibnamefont {{Zheng}}},\ }\bibfield  {title}
  {\bibinfo {title} {{Algebraic higher symmetry and categorical symmetry: A
  holographic and entanglement view of symmetry}},\ }\href
  {https://doi.org/10.1103/PhysRevResearch.2.043086} {\bibfield  {journal}
  {\bibinfo  {journal} {Phys. Rev. Res.}\ }\textbf {\bibinfo {volume} {2}},\
  \bibinfo {pages} {043086} (\bibinfo {year} {2020}{\natexlab{a}})},\ \Eprint
  {https://arxiv.org/abs/2005.14178} {arXiv:2005.14178} \BibitemShut {NoStop}%
\bibitem [{\citenamefont {Lootens}\ \emph {et~al.}(2021)\citenamefont
  {Lootens}, \citenamefont {Fuchs}, \citenamefont {Haegeman}, \citenamefont
  {Schweigert},\ and\ \citenamefont {Verstraete}}]{LV200811187}%
  \BibitemOpen
  \bibfield  {author} {\bibinfo {author} {\bibfnamefont {L.}~\bibnamefont
  {Lootens}}, \bibinfo {author} {\bibfnamefont {J.}~\bibnamefont {Fuchs}},
  \bibinfo {author} {\bibfnamefont {J.}~\bibnamefont {Haegeman}}, \bibinfo
  {author} {\bibfnamefont {C.}~\bibnamefont {Schweigert}},\ and\ \bibinfo
  {author} {\bibfnamefont {F.}~\bibnamefont {Verstraete}},\ }\bibfield  {title}
  {\bibinfo {title} {{Matrix product operator symmetries and intertwiners in
  string-nets with domain walls}},\ }\href
  {https://doi.org/10.21468/SciPostPhys.10.3.053} {\bibfield  {journal}
  {\bibinfo  {journal} {SciPost Phys.}\ }\textbf {\bibinfo {volume} {10}},\
  \bibinfo {pages} {053} (\bibinfo {year} {2021})},\ \Eprint
  {https://arxiv.org/abs/2008.11187} {arXiv:2008.11187} \BibitemShut {NoStop}%
\bibitem [{\citenamefont {{Moradi}}\ \emph {et~al.}(2023)\citenamefont
  {{Moradi}}, \citenamefont {{Faroogh Moosavian}},\ and\ \citenamefont
  {{Tiwari}}}]{MT220710712}%
  \BibitemOpen
  \bibfield  {author} {\bibinfo {author} {\bibfnamefont {H.}~\bibnamefont
  {{Moradi}}}, \bibinfo {author} {\bibfnamefont {S.}~\bibnamefont {{Faroogh
  Moosavian}}},\ and\ \bibinfo {author} {\bibfnamefont {A.}~\bibnamefont
  {{Tiwari}}},\ }\bibfield  {title} {\bibinfo {title} {{Topological holography:
  Towards a unification of Landau and beyond-Landau physics}},\ }\href
  {https://doi.org/10.21468/SciPostPhysCore.6.4.066} {\bibfield  {journal}
  {\bibinfo  {journal} {SciPost Phys. Core}\ }\textbf {\bibinfo {volume} {6}},\
  \bibinfo {pages} {066} (\bibinfo {year} {2023})},\ \Eprint
  {https://arxiv.org/abs/2207.10712} {arXiv:2207.10712} \BibitemShut {NoStop}%
\bibitem [{\citenamefont {{Freed}}\ \emph {et~al.}(2022)\citenamefont
  {{Freed}}, \citenamefont {{Moore}},\ and\ \citenamefont
  {{Teleman}}}]{FT220907471}%
  \BibitemOpen
  \bibfield  {author} {\bibinfo {author} {\bibfnamefont {D.~S.}\ \bibnamefont
  {{Freed}}}, \bibinfo {author} {\bibfnamefont {G.~W.}\ \bibnamefont
  {{Moore}}},\ and\ \bibinfo {author} {\bibfnamefont {C.}~\bibnamefont
  {{Teleman}}},\ }\href@noop {} {\bibinfo {title} {{Topological symmetry in
  quantum field theory}}} (\bibinfo {year} {2022}),\ \Eprint
  {https://arxiv.org/abs/2209.07471} {arXiv:2209.07471} \BibitemShut {NoStop}%
\bibitem [{\citenamefont {{McGreevy}}(2023)}]{M220403045}%
  \BibitemOpen
  \bibfield  {author} {\bibinfo {author} {\bibfnamefont {J.}~\bibnamefont
  {{McGreevy}}},\ }\bibfield  {title} {\bibinfo {title} {{Generalized
  Symmetries in Condensed Matter}},\ }\href
  {https://doi.org/10.1146/annurev-conmatphys-040721-021029} {\bibfield
  {journal} {\bibinfo  {journal} {Annu. Rev. Condens. Matter Phys.}\ }\textbf
  {\bibinfo {volume} {14}},\ \bibinfo {pages} {57} (\bibinfo {year} {2023})},\
  \Eprint {https://arxiv.org/abs/2204.03045} {arXiv:2204.03045} \BibitemShut
  {NoStop}%
\bibitem [{\citenamefont {{C\'ordova}}\ \emph
  {et~al.}(2022{\natexlab{a}})\citenamefont {{C\'ordova}}, \citenamefont
  {Dumitrescu}, \citenamefont {Intriligator},\ and\ \citenamefont
  {Shao}}]{CD220509545}%
  \BibitemOpen
  \bibfield  {author} {\bibinfo {author} {\bibfnamefont {C.}~\bibnamefont
  {{C\'ordova}}}, \bibinfo {author} {\bibfnamefont {T.~T.}\ \bibnamefont
  {Dumitrescu}}, \bibinfo {author} {\bibfnamefont {K.}~\bibnamefont
  {Intriligator}},\ and\ \bibinfo {author} {\bibfnamefont {S.-H.}\ \bibnamefont
  {Shao}},\ }\href@noop {} {\bibinfo {title} {{Snowmass White Paper:
  Generalized Symmetries in Quantum Field Theory and Beyond}}} (\bibinfo {year}
  {2022}{\natexlab{a}}),\ \Eprint {https://arxiv.org/abs/2205.09545}
  {arXiv:2205.09545} \BibitemShut {NoStop}%
\bibitem [{\citenamefont {Wen}(1990)}]{W9039}%
  \BibitemOpen
  \bibfield  {author} {\bibinfo {author} {\bibfnamefont {X.-G.}\ \bibnamefont
  {Wen}},\ }\bibfield  {title} {\bibinfo {title} {Topological orders in rigid
  states},\ }\href {https://doi.org/10.1142/s0217979290000139} {\bibfield
  {journal} {\bibinfo  {journal} {Int. J. Mod. Phys. B}\ }\textbf {\bibinfo
  {volume} {04}},\ \bibinfo {pages} {239} (\bibinfo {year} {1990})}\BibitemShut
  {NoStop}%
\bibitem [{\citenamefont {Landau}(1937)}]{L3726}%
  \BibitemOpen
  \bibfield  {author} {\bibinfo {author} {\bibfnamefont {L.~D.}\ \bibnamefont
  {Landau}},\ }\bibfield  {title} {\bibinfo {title} {Theory of phase
  transformations {I}},\ }\href@noop {} {\bibfield  {journal} {\bibinfo
  {journal} {Phys. Z. Sowjetunion}\ }\textbf {\bibinfo {volume} {11}},\
  \bibinfo {pages} {26} (\bibinfo {year} {1937})}\BibitemShut {NoStop}%
\bibitem [{\citenamefont {Ginzburg}\ and\ \citenamefont
  {Landau}(1950)}]{GL5064}%
  \BibitemOpen
  \bibfield  {author} {\bibinfo {author} {\bibfnamefont {V.~L.}\ \bibnamefont
  {Ginzburg}}\ and\ \bibinfo {author} {\bibfnamefont {L.~D.}\ \bibnamefont
  {Landau}},\ }\bibfield  {title} {\bibinfo {title} {On the theory of
  superconductivity},\ }\href@noop {} {\bibfield  {journal} {\bibinfo
  {journal} {Zh. Eksp. Teor. Fiz.}\ }\textbf {\bibinfo {volume} {20}},\
  \bibinfo {pages} {1064} (\bibinfo {year} {1950})}\BibitemShut {NoStop}%
\bibitem [{\citenamefont {Wen}(2019)}]{W181202517}%
  \BibitemOpen
  \bibfield  {author} {\bibinfo {author} {\bibfnamefont {X.-G.}\ \bibnamefont
  {Wen}},\ }\bibfield  {title} {\bibinfo {title} {Emergent (anomalous) higher
  symmetries from topological order and from dynamical electromagnetic field in
  condensed matter systems},\ }\href
  {https://doi.org/10.1103/physrevb.99.205139} {\bibfield  {journal} {\bibinfo
  {journal} {Phys. Rev. B}\ }\textbf {\bibinfo {volume} {99}},\ \bibinfo
  {pages} {205139} (\bibinfo {year} {2019})},\ \Eprint
  {https://arxiv.org/abs/1812.02517} {arXiv:1812.02517} \BibitemShut {NoStop}%
\bibitem [{\citenamefont {Hsin}\ \emph {et~al.}(2019)\citenamefont {Hsin},
  \citenamefont {Lam},\ and\ \citenamefont {Seiberg}}]{HS181204716}%
  \BibitemOpen
  \bibfield  {author} {\bibinfo {author} {\bibfnamefont {P.-S.}\ \bibnamefont
  {Hsin}}, \bibinfo {author} {\bibfnamefont {H.~T.}\ \bibnamefont {Lam}},\ and\
  \bibinfo {author} {\bibfnamefont {N.}~\bibnamefont {Seiberg}},\ }\bibfield
  {title} {\bibinfo {title} {Comments on one-form global symmetries and their
  gauging in 3d and 4d},\ }\href {https://doi.org/10.21468/scipostphys.6.3.039}
  {\bibfield  {journal} {\bibinfo  {journal} {SciPost Phys.}\ }\textbf
  {\bibinfo {volume} {6}},\ \bibinfo {pages} {039} (\bibinfo {year} {2019})},\
  \Eprint {https://arxiv.org/abs/1812.04716} {arXiv:1812.04716} \BibitemShut
  {NoStop}%
\bibitem [{\citenamefont {{Hsin}}\ and\ \citenamefont
  {{Turzillo}}(2020)}]{HT190411550}%
  \BibitemOpen
  \bibfield  {author} {\bibinfo {author} {\bibfnamefont {P.-S.}\ \bibnamefont
  {{Hsin}}}\ and\ \bibinfo {author} {\bibfnamefont {A.}~\bibnamefont
  {{Turzillo}}},\ }\bibfield  {title} {\bibinfo {title} {{Symmetry-enriched
  quantum spin liquids in ${(3+ 1)d}$}},\ }\href
  {https://doi.org/10.1007/JHEP09(2020)022} {\bibfield  {journal} {\bibinfo
  {journal} {J. High Energ. Phys.}\ }\textbf {\bibinfo {volume}
  {2020}}\bibfield  {number} {\bibinfo  {number} { (9)},\ \bibinfo {pages}
  {22}},\ }\Eprint {https://arxiv.org/abs/1904.11550} {arXiv:1904.11550}
  \BibitemShut {NoStop}%
\bibitem [{\citenamefont {{Kaidi}}\ \emph {et~al.}(2022)\citenamefont
  {{Kaidi}}, \citenamefont {{Komargodski}}, \citenamefont {{Ohmori}},
  \citenamefont {{Seifnashri}},\ and\ \citenamefont {{Shao}}}]{KKO210713091}%
  \BibitemOpen
  \bibfield  {author} {\bibinfo {author} {\bibfnamefont {J.}~\bibnamefont
  {{Kaidi}}}, \bibinfo {author} {\bibfnamefont {Z.}~\bibnamefont
  {{Komargodski}}}, \bibinfo {author} {\bibfnamefont {K.}~\bibnamefont
  {{Ohmori}}}, \bibinfo {author} {\bibfnamefont {S.}~\bibnamefont
  {{Seifnashri}}},\ and\ \bibinfo {author} {\bibfnamefont {S.-H.}\ \bibnamefont
  {{Shao}}},\ }\bibfield  {title} {\bibinfo {title} {{Higher central charges
  and topological boundaries in 2+1-dimensional TQFTs}},\ }\href
  {https://doi.org/10.21468/SciPostPhys.13.3.067} {\bibfield  {journal}
  {\bibinfo  {journal} {SciPost Phys.}\ }\textbf {\bibinfo {volume} {13}},\
  \bibinfo {pages} {067} (\bibinfo {year} {2022})},\ \Eprint
  {https://arxiv.org/abs/2107.13091} {arXiv:2107.13091} \BibitemShut {NoStop}%
\bibitem [{\citenamefont {{Pace}}(2023)}]{P230805730}%
  \BibitemOpen
  \bibfield  {author} {\bibinfo {author} {\bibfnamefont {S.~D.}\ \bibnamefont
  {{Pace}}},\ }\href@noop {} {\bibinfo {title} {{Emergent generalized
  symmetries in ordered phases}}} (\bibinfo {year} {2023}),\ \Eprint
  {https://arxiv.org/abs/2308.05730} {arXiv:2308.05730} \BibitemShut {NoStop}%
\bibitem [{\citenamefont {{Kovner}}\ and\ \citenamefont
  {{Rosenstein}}(1994)}]{KR9210154}%
  \BibitemOpen
  \bibfield  {author} {\bibinfo {author} {\bibfnamefont {A.}~\bibnamefont
  {{Kovner}}}\ and\ \bibinfo {author} {\bibfnamefont {B.}~\bibnamefont
  {{Rosenstein}}},\ }\bibfield  {title} {\bibinfo {title} {{New look at
  ${\mathrm{QED}}_{4}$: the photon as a Goldstone boson and the topological
  interpretation of electric charge}},\ }\href
  {https://doi.org/10.1103/PhysRevD.49.5571} {\bibfield  {journal} {\bibinfo
  {journal} {Phys. Rev. D}\ }\textbf {\bibinfo {volume} {49}},\ \bibinfo
  {pages} {5571} (\bibinfo {year} {1994})},\ \Eprint
  {https://arxiv.org/abs/hep-th/9210154} {arXiv:hep-th/9210154} \BibitemShut
  {NoStop}%
\bibitem [{\citenamefont {{Lake}}(2018)}]{L180207747}%
  \BibitemOpen
  \bibfield  {author} {\bibinfo {author} {\bibfnamefont {E.}~\bibnamefont
  {{Lake}}},\ }\href@noop {} {\bibinfo {title} {{Higher-form symmetries and
  spontaneous symmetry breaking}}} (\bibinfo {year} {2018}),\ \Eprint
  {https://arxiv.org/abs/1802.07747} {arXiv:1802.07747} \BibitemShut {NoStop}%
\bibitem [{\citenamefont {Hofman}\ and\ \citenamefont
  {Iqbal}(2019)}]{HI180209512}%
  \BibitemOpen
  \bibfield  {author} {\bibinfo {author} {\bibfnamefont {D.}~\bibnamefont
  {Hofman}}\ and\ \bibinfo {author} {\bibfnamefont {N.}~\bibnamefont {Iqbal}},\
  }\bibfield  {title} {\bibinfo {title} {Goldstone modes and photonization for
  higher form symmetries},\ }\href
  {https://doi.org/10.21468/scipostphys.6.1.006} {\bibfield  {journal}
  {\bibinfo  {journal} {SciPost Phys.}\ }\textbf {\bibinfo {volume} {6}},\
  \bibinfo {pages} {006} (\bibinfo {year} {2019})},\ \Eprint
  {https://arxiv.org/abs/1802.09512} {arXiv:1802.09512} \BibitemShut {NoStop}%
\bibitem [{\citenamefont {{Kim}}\ and\ \citenamefont
  {{Hirono}}(2019)}]{KH190504617}%
  \BibitemOpen
  \bibfield  {author} {\bibinfo {author} {\bibfnamefont {K.-S.}\ \bibnamefont
  {{Kim}}}\ and\ \bibinfo {author} {\bibfnamefont {Y.}~\bibnamefont
  {{Hirono}}},\ }\bibfield  {title} {\bibinfo {title} {{Higher-form symmetries
  and $d$-wave superconductors from doped Mott insulators}},\ }\href
  {https://doi.org/10.1103/PhysRevB.100.085142} {\bibfield  {journal} {\bibinfo
   {journal} {Phys. Rev. B}\ }\textbf {\bibinfo {volume} {100}},\ \bibinfo
  {pages} {085142} (\bibinfo {year} {2019})},\ \Eprint
  {https://arxiv.org/abs/1905.04617} {arXiv:1905.04617} \BibitemShut {NoStop}%
\bibitem [{\citenamefont {{Hidaka}}\ \emph {et~al.}(2021)\citenamefont
  {{Hidaka}}, \citenamefont {{Hirono}},\ and\ \citenamefont
  {{Yokokura}}}]{HHY200715901}%
  \BibitemOpen
  \bibfield  {author} {\bibinfo {author} {\bibfnamefont {Y.}~\bibnamefont
  {{Hidaka}}}, \bibinfo {author} {\bibfnamefont {Y.}~\bibnamefont {{Hirono}}},\
  and\ \bibinfo {author} {\bibfnamefont {R.}~\bibnamefont {{Yokokura}}},\
  }\bibfield  {title} {\bibinfo {title} {{Counting Nambu-Goldstone Modes of
  Higher-Form Global Symmetries}},\ }\href
  {https://doi.org/10.1103/PhysRevLett.126.071601} {\bibfield  {journal}
  {\bibinfo  {journal} {Phys. Rev. Lett.}\ }\textbf {\bibinfo {volume} {126}},\
  \bibinfo {pages} {071601} (\bibinfo {year} {2021})},\ \Eprint
  {https://arxiv.org/abs/2007.15901} {arXiv:2007.15901} \BibitemShut {NoStop}%
\bibitem [{\citenamefont {{Yamamoto}}\ and\ \citenamefont
  {{Yokokura}}(2021)}]{YY200907621}%
  \BibitemOpen
  \bibfield  {author} {\bibinfo {author} {\bibfnamefont {N.}~\bibnamefont
  {{Yamamoto}}}\ and\ \bibinfo {author} {\bibfnamefont {R.}~\bibnamefont
  {{Yokokura}}},\ }\bibfield  {title} {\bibinfo {title} {{Topological mass
  generation in gapless systems}},\ }\href
  {https://doi.org/10.1103/PhysRevD.104.025010} {\bibfield  {journal} {\bibinfo
   {journal} {Phys. Rev. D}\ }\textbf {\bibinfo {volume} {104}},\ \bibinfo
  {pages} {025010} (\bibinfo {year} {2021})},\ \Eprint
  {https://arxiv.org/abs/2009.07621} {arXiv:2009.07621} \BibitemShut {NoStop}%
\bibitem [{\citenamefont {{Iqbal}}\ and\ \citenamefont
  {{McGreevy}}(2022)}]{IM210612610}%
  \BibitemOpen
  \bibfield  {author} {\bibinfo {author} {\bibfnamefont {N.}~\bibnamefont
  {{Iqbal}}}\ and\ \bibinfo {author} {\bibfnamefont {J.}~\bibnamefont
  {{McGreevy}}},\ }\bibfield  {title} {\bibinfo {title} {{Mean string field
  theory: Landau-Ginzburg theory for 1-form symmetries}},\ }\href
  {https://doi.org/10.21468/SciPostPhys.13.5.114} {\bibfield  {journal}
  {\bibinfo  {journal} {SciPost Phys.}\ }\textbf {\bibinfo {volume} {13}},\
  \bibinfo {pages} {114} (\bibinfo {year} {2022})},\ \Eprint
  {https://arxiv.org/abs/2106.12610} {arXiv:2106.12610} \BibitemShut {NoStop}%
\bibitem [{\citenamefont {{Thorngren}}\ and\ \citenamefont {{von
  Keyserlingk}}(2015)}]{TK151102929}%
  \BibitemOpen
  \bibfield  {author} {\bibinfo {author} {\bibfnamefont {R.}~\bibnamefont
  {{Thorngren}}}\ and\ \bibinfo {author} {\bibfnamefont {C.}~\bibnamefont {{von
  Keyserlingk}}},\ }\href@noop {} {\bibinfo {title} {{Higher {SPT's} and a
  generalization of anomaly in-flow}}} (\bibinfo {year} {2015}),\ \Eprint
  {https://arxiv.org/abs/1511.02929} {arXiv:1511.02929} \BibitemShut {NoStop}%
\bibitem [{\citenamefont {{Yoshida}}(2016)}]{Y150803468}%
  \BibitemOpen
  \bibfield  {author} {\bibinfo {author} {\bibfnamefont {B.}~\bibnamefont
  {{Yoshida}}},\ }\bibfield  {title} {\bibinfo {title} {{Topological phases
  with generalized global symmetries}},\ }\href
  {https://doi.org/10.1103/PhysRevB.93.155131} {\bibfield  {journal} {\bibinfo
  {journal} {Phys. Rev. B}\ }\textbf {\bibinfo {volume} {93}},\ \bibinfo
  {pages} {155131} (\bibinfo {year} {2016})},\ \Eprint
  {https://arxiv.org/abs/1508.03468} {arXiv:1508.03468} \BibitemShut {NoStop}%
\bibitem [{\citenamefont {{Gaiotto}}\ \emph {et~al.}(2017)\citenamefont
  {{Gaiotto}}, \citenamefont {{Kapustin}}, \citenamefont {{Komargodski}},\ and\
  \citenamefont {{Seiberg}}}]{GKK170300501}%
  \BibitemOpen
  \bibfield  {author} {\bibinfo {author} {\bibfnamefont {D.}~\bibnamefont
  {{Gaiotto}}}, \bibinfo {author} {\bibfnamefont {A.}~\bibnamefont
  {{Kapustin}}}, \bibinfo {author} {\bibfnamefont {Z.}~\bibnamefont
  {{Komargodski}}},\ and\ \bibinfo {author} {\bibfnamefont {N.}~\bibnamefont
  {{Seiberg}}},\ }\bibfield  {title} {\bibinfo {title} {{Theta, time reversal
  and temperature}},\ }\href {https://doi.org/10.1007/JHEP05(2017)091}
  {\bibfield  {journal} {\bibinfo  {journal} {J. High Energ. Phys.}\ }\textbf
  {\bibinfo {volume} {2017}}\bibfield  {number} {\bibinfo  {number} { (5)},\
  \bibinfo {pages} {91}},\ }\Eprint {https://arxiv.org/abs/1703.00501}
  {arXiv:1703.00501} \BibitemShut {NoStop}%
\bibitem [{\citenamefont {Kobayashi}\ \emph {et~al.}(2019)\citenamefont
  {Kobayashi}, \citenamefont {Shiozaki}, \citenamefont {Kikuchi},\ and\
  \citenamefont {Ryu}}]{KR180505367}%
  \BibitemOpen
  \bibfield  {author} {\bibinfo {author} {\bibfnamefont {R.}~\bibnamefont
  {Kobayashi}}, \bibinfo {author} {\bibfnamefont {K.}~\bibnamefont {Shiozaki}},
  \bibinfo {author} {\bibfnamefont {Y.}~\bibnamefont {Kikuchi}},\ and\ \bibinfo
  {author} {\bibfnamefont {S.}~\bibnamefont {Ryu}},\ }\bibfield  {title}
  {\bibinfo {title} {Lieb-{Schultz}-mattis type theorem with higher-form
  symmetry and the quantum dimer models},\ }\href
  {https://doi.org/10.1103/physrevb.99.014402} {\bibfield  {journal} {\bibinfo
  {journal} {\prb}\ }\textbf {\bibinfo {volume} {99}},\ \bibinfo {pages}
  {014402} (\bibinfo {year} {2019})},\ \Eprint
  {https://arxiv.org/abs/1805.05367} {arXiv:1805.05367} \BibitemShut {NoStop}%
\bibitem [{\citenamefont {{Wan}}\ and\ \citenamefont
  {{Wang}}(2019)}]{WW181211967}%
  \BibitemOpen
  \bibfield  {author} {\bibinfo {author} {\bibfnamefont {Z.}~\bibnamefont
  {{Wan}}}\ and\ \bibinfo {author} {\bibfnamefont {J.}~\bibnamefont {{Wang}}},\
  }\bibfield  {title} {\bibinfo {title} {{Higher anomalies, higher symmetries,
  and cobordisms I: classification of higher-symmetry-protected topological
  states and their boundary fermionic/bosonic anomalies via a generalized
  cobordism theory}},\ }\href {https://doi.org/10.4310/AMSA.2019.v4.n2.a2}
  {\bibfield  {journal} {\bibinfo  {journal} {Annals of Mathematical Sciences
  and Applications}\ }\textbf {\bibinfo {volume} {4}},\ \bibinfo {pages} {107}
  (\bibinfo {year} {2019})},\ \Eprint {https://arxiv.org/abs/1812.11967}
  {arXiv:1812.11967} \BibitemShut {NoStop}%
\bibitem [{\citenamefont {{Tsui}}\ and\ \citenamefont
  {{Wen}}(2020)}]{TW190802613}%
  \BibitemOpen
  \bibfield  {author} {\bibinfo {author} {\bibfnamefont {L.}~\bibnamefont
  {{Tsui}}}\ and\ \bibinfo {author} {\bibfnamefont {X.-G.}\ \bibnamefont
  {{Wen}}},\ }\bibfield  {title} {\bibinfo {title} {{Lattice models that
  realize ${\mathbb{Z}}_{n}$-1 symmetry-protected topological states for even
  $n$}},\ }\href {https://doi.org/10.1103/PhysRevB.101.035101} {\bibfield
  {journal} {\bibinfo  {journal} {Phys. Rev. B}\ }\textbf {\bibinfo {volume}
  {101}},\ \bibinfo {pages} {035101} (\bibinfo {year} {2020})},\ \Eprint
  {https://arxiv.org/abs/1908.02613} {arXiv:1908.02613} \BibitemShut {NoStop}%
\bibitem [{\citenamefont {Jian}\ \emph {et~al.}(2021)\citenamefont {Jian},
  \citenamefont {Wu}, \citenamefont {Xu},\ and\ \citenamefont
  {Xu}}]{JW200900023}%
  \BibitemOpen
  \bibfield  {author} {\bibinfo {author} {\bibfnamefont {C.-M.}\ \bibnamefont
  {Jian}}, \bibinfo {author} {\bibfnamefont {X.-C.}\ \bibnamefont {Wu}},
  \bibinfo {author} {\bibfnamefont {Y.}~\bibnamefont {Xu}},\ and\ \bibinfo
  {author} {\bibfnamefont {C.}~\bibnamefont {Xu}},\ }\bibfield  {title}
  {\bibinfo {title} {{Physics of Symmetry Protected Topological phases
  involving Higher Symmetries and their Applications}},\ }\href
  {https://doi.org/10.1103/PhysRevB.103.064426} {\bibfield  {journal} {\bibinfo
   {journal} {Phys. Rev. B}\ }\textbf {\bibinfo {volume} {103}},\ \bibinfo
  {pages} {064426} (\bibinfo {year} {2021})},\ \Eprint
  {https://arxiv.org/abs/2009.00023} {arXiv:2009.00023} \BibitemShut {NoStop}%
\bibitem [{\citenamefont {{Moy}}\ \emph {et~al.}(2023)\citenamefont {{Moy}},
  \citenamefont {{Goldman}}, \citenamefont {{Sohal}},\ and\ \citenamefont
  {{Fradkin}}}]{MF220607725}%
  \BibitemOpen
  \bibfield  {author} {\bibinfo {author} {\bibfnamefont {B.}~\bibnamefont
  {{Moy}}}, \bibinfo {author} {\bibfnamefont {H.}~\bibnamefont {{Goldman}}},
  \bibinfo {author} {\bibfnamefont {R.}~\bibnamefont {{Sohal}}},\ and\ \bibinfo
  {author} {\bibfnamefont {E.}~\bibnamefont {{Fradkin}}},\ }\bibfield  {title}
  {\bibinfo {title} {{Theory of oblique topological insulators}},\ }\href
  {https://doi.org/10.21468/SciPostPhys.14.2.023} {\bibfield  {journal}
  {\bibinfo  {journal} {SciPost Phys.}\ }\textbf {\bibinfo {volume} {14}},\
  \bibinfo {pages} {023} (\bibinfo {year} {2023})},\ \Eprint
  {https://arxiv.org/abs/2206.07725} {arXiv:2206.07725} \BibitemShut {NoStop}%
\bibitem [{\citenamefont {{Pace}}\ and\ \citenamefont
  {{Wen}}(2023)}]{PW220703544}%
  \BibitemOpen
  \bibfield  {author} {\bibinfo {author} {\bibfnamefont {S.~D.}\ \bibnamefont
  {{Pace}}}\ and\ \bibinfo {author} {\bibfnamefont {X.-G.}\ \bibnamefont
  {{Wen}}},\ }\bibfield  {title} {\bibinfo {title} {{Emergent higher-symmetry
  protected topological orders in the confined phase of $U(1)$ gauge theory}},\
  }\href {https://doi.org/10.1103/PhysRevB.107.075112} {\bibfield  {journal}
  {\bibinfo  {journal} {Phys. Rev. B}\ }\textbf {\bibinfo {volume} {107}},\
  \bibinfo {pages} {075112} (\bibinfo {year} {2023})},\ \Eprint
  {https://arxiv.org/abs/2207.03544} {arXiv:2207.03544} \BibitemShut {NoStop}%
\bibitem [{\citenamefont {{Verresen}}\ \emph {et~al.}(2022)\citenamefont
  {{Verresen}}, \citenamefont {{Borla}}, \citenamefont {{Vishwanath}},
  \citenamefont {{Moroz}},\ and\ \citenamefont {{Thorngren}}}]{VBV221101376}%
  \BibitemOpen
  \bibfield  {author} {\bibinfo {author} {\bibfnamefont {R.}~\bibnamefont
  {{Verresen}}}, \bibinfo {author} {\bibfnamefont {U.}~\bibnamefont {{Borla}}},
  \bibinfo {author} {\bibfnamefont {A.}~\bibnamefont {{Vishwanath}}}, \bibinfo
  {author} {\bibfnamefont {S.}~\bibnamefont {{Moroz}}},\ and\ \bibinfo {author}
  {\bibfnamefont {R.}~\bibnamefont {{Thorngren}}},\ }\href@noop {} {\bibinfo
  {title} {{Higgs Condensates are Symmetry-Protected Topological Phases: I.
  Discrete Symmetries}}} (\bibinfo {year} {2022}),\ \Eprint
  {https://arxiv.org/abs/2211.01376} {arXiv:2211.01376} \BibitemShut {NoStop}%
\bibitem [{\citenamefont {{Thorngren}}\ \emph {et~al.}(2023)\citenamefont
  {{Thorngren}}, \citenamefont {{Rakovszky}}, \citenamefont {{Verresen}},\ and\
  \citenamefont {{Vishwanath}}}]{TRV230308136}%
  \BibitemOpen
  \bibfield  {author} {\bibinfo {author} {\bibfnamefont {R.}~\bibnamefont
  {{Thorngren}}}, \bibinfo {author} {\bibfnamefont {T.}~\bibnamefont
  {{Rakovszky}}}, \bibinfo {author} {\bibfnamefont {R.}~\bibnamefont
  {{Verresen}}},\ and\ \bibinfo {author} {\bibfnamefont {A.}~\bibnamefont
  {{Vishwanath}}},\ }\href@noop {} {\bibinfo {title} {{Higgs Condensates are
  Symmetry-Protected Topological Phases: II. $U(1)$ Gauge Theory and
  Superconductors}}} (\bibinfo {year} {2023}),\ \Eprint
  {https://arxiv.org/abs/2303.08136} {arXiv:2303.08136} \BibitemShut {NoStop}%
\bibitem [{\citenamefont {{Cheng}}\ and\ \citenamefont
  {{Seiberg}}(2023)}]{CS221112543}%
  \BibitemOpen
  \bibfield  {author} {\bibinfo {author} {\bibfnamefont {M.}~\bibnamefont
  {{Cheng}}}\ and\ \bibinfo {author} {\bibfnamefont {N.}~\bibnamefont
  {{Seiberg}}},\ }\bibfield  {title} {\bibinfo {title} {{Lieb-Schultz-Mattis,
  Luttinger, and 't Hooft - anomaly matching in lattice systems}},\ }\href
  {https://doi.org/10.21468/SciPostPhys.15.2.051} {\bibfield  {journal}
  {\bibinfo  {journal} {SciPost Phys.}\ }\textbf {\bibinfo {volume} {15}},\
  \bibinfo {pages} {051} (\bibinfo {year} {2023})},\ \Eprint
  {https://arxiv.org/abs/2211.12543} {arXiv:2211.12543} \BibitemShut {NoStop}%
\bibitem [{\citenamefont {Foerster}\ \emph {et~al.}(1980)\citenamefont
  {Foerster}, \citenamefont {Nielsen},\ and\ \citenamefont
  {Ninomiya}}]{FNN8035}%
  \BibitemOpen
  \bibfield  {author} {\bibinfo {author} {\bibfnamefont {D.}~\bibnamefont
  {Foerster}}, \bibinfo {author} {\bibfnamefont {H.}~\bibnamefont {Nielsen}},\
  and\ \bibinfo {author} {\bibfnamefont {M.}~\bibnamefont {Ninomiya}},\
  }\bibfield  {title} {\bibinfo {title} {Dynamical stability of local gauge
  symmetry creation of light from chaos},\ }\href
  {https://doi.org/10.1016/0370-2693(80)90842-4} {\bibfield  {journal}
  {\bibinfo  {journal} {Phys. Lett. B}\ }\textbf {\bibinfo {volume} {94}},\
  \bibinfo {pages} {135} (\bibinfo {year} {1980})}\BibitemShut {NoStop}%
\bibitem [{\citenamefont {Hastings}\ and\ \citenamefont {Wen}(2005)}]{HW0541}%
  \BibitemOpen
  \bibfield  {author} {\bibinfo {author} {\bibfnamefont {M.~B.}\ \bibnamefont
  {Hastings}}\ and\ \bibinfo {author} {\bibfnamefont {X.-G.}\ \bibnamefont
  {Wen}},\ }\bibfield  {title} {\bibinfo {title} {Quasiadiabatic continuation
  of quantum states: {The} stability of topological ground-state degeneracy and
  emergent gauge invariance},\ }\href
  {https://doi.org/10.1103/physrevb.72.045141} {\bibfield  {journal} {\bibinfo
  {journal} {Phys. Rev. B}\ }\textbf {\bibinfo {volume} {72}},\ \bibinfo
  {pages} {045141} (\bibinfo {year} {2005})},\ \Eprint
  {https://arxiv.org/abs/cond-mat/0503554} {arXiv:cond-mat/0503554}
  \BibitemShut {NoStop}%
\bibitem [{\citenamefont {{Poppitz}}\ and\ \citenamefont
  {{Shang}}(2008)}]{PS08010587}%
  \BibitemOpen
  \bibfield  {author} {\bibinfo {author} {\bibfnamefont {E.}~\bibnamefont
  {{Poppitz}}}\ and\ \bibinfo {author} {\bibfnamefont {Y.}~\bibnamefont
  {{Shang}}},\ }\bibfield  {title} {\bibinfo {title} {{``Light from chaos'' in
  two dimensions}},\ }\href {https://doi.org/10.1142/S0217751X08041281}
  {\bibfield  {journal} {\bibinfo  {journal} {International Journal of Modern
  Physics A}\ }\textbf {\bibinfo {volume} {23}},\ \bibinfo {pages} {4545}
  (\bibinfo {year} {2008})},\ \Eprint {https://arxiv.org/abs/0801.0587}
  {arXiv:0801.0587} \BibitemShut {NoStop}%
\bibitem [{\citenamefont {{Somoza}}\ \emph {et~al.}(2021)\citenamefont
  {{Somoza}}, \citenamefont {{Serna}},\ and\ \citenamefont
  {{Nahum}}}]{SSN201215845}%
  \BibitemOpen
  \bibfield  {author} {\bibinfo {author} {\bibfnamefont {A.~M.}\ \bibnamefont
  {{Somoza}}}, \bibinfo {author} {\bibfnamefont {P.}~\bibnamefont {{Serna}}},\
  and\ \bibinfo {author} {\bibfnamefont {A.}~\bibnamefont {{Nahum}}},\
  }\bibfield  {title} {\bibinfo {title} {{Self-Dual Criticality in
  Three-Dimensional ${\mathbb{Z}}_{2}$ Gauge Theory with Matter}},\ }\href
  {https://doi.org/10.1103/PhysRevX.11.041008} {\bibfield  {journal} {\bibinfo
  {journal} {Phys. Rev. X}\ }\textbf {\bibinfo {volume} {11}},\ \bibinfo
  {pages} {041008} (\bibinfo {year} {2021})},\ \Eprint
  {https://arxiv.org/abs/2012.15845} {arXiv:2012.15845} \BibitemShut {NoStop}%
\bibitem [{\citenamefont {{C\'ordova}}\ \emph
  {et~al.}(2022{\natexlab{b}})\citenamefont {{C\'ordova}}, \citenamefont
  {{Ohmori}},\ and\ \citenamefont {{Rudelius}}}]{COR220205866}%
  \BibitemOpen
  \bibfield  {author} {\bibinfo {author} {\bibfnamefont {C.}~\bibnamefont
  {{C\'ordova}}}, \bibinfo {author} {\bibfnamefont {K.}~\bibnamefont
  {{Ohmori}}},\ and\ \bibinfo {author} {\bibfnamefont {T.}~\bibnamefont
  {{Rudelius}}},\ }\bibfield  {title} {\bibinfo {title} {{Generalized symmetry
  breaking scales and weak gravity conjectures}},\ }\href
  {https://doi.org/10.1007/JHEP11(2022)154} {\bibfield  {journal} {\bibinfo
  {journal} {J. High Energ. Phys.}\ }\textbf {\bibinfo {volume}
  {2022}}\bibfield  {number} {\bibinfo  {number} { (11)},\ \bibinfo {pages}
  {154}},\ }\Eprint {https://arxiv.org/abs/2202.05866} {arXiv:2202.05866}
  \BibitemShut {NoStop}%
\bibitem [{\citenamefont {{Cian}}\ \emph {et~al.}(2022)\citenamefont {{Cian}},
  \citenamefont {{Hafezi}},\ and\ \citenamefont {{Barkeshli}}}]{CHB220914302}%
  \BibitemOpen
  \bibfield  {author} {\bibinfo {author} {\bibfnamefont {Z.-P.}\ \bibnamefont
  {{Cian}}}, \bibinfo {author} {\bibfnamefont {M.}~\bibnamefont {{Hafezi}}},\
  and\ \bibinfo {author} {\bibfnamefont {M.}~\bibnamefont {{Barkeshli}}},\
  }\href@noop {} {\bibinfo {title} {{Extracting Wilson loop operators and
  fractional statistics from a single bulk ground state}}} (\bibinfo {year}
  {2022}),\ \Eprint {https://arxiv.org/abs/2209.14302} {arXiv:2209.14302}
  \BibitemShut {NoStop}%
\bibitem [{\citenamefont {{Hidaka}}\ and\ \citenamefont
  {{Kondo}}(2022)}]{HK221011492}%
  \BibitemOpen
  \bibfield  {author} {\bibinfo {author} {\bibfnamefont {Y.}~\bibnamefont
  {{Hidaka}}}\ and\ \bibinfo {author} {\bibfnamefont {D.}~\bibnamefont
  {{Kondo}}},\ }\href@noop {} {\bibinfo {title} {{Emergent higher-form symmetry
  in Higgs phases with superfluidity}}} (\bibinfo {year} {2022}),\ \Eprint
  {https://arxiv.org/abs/2210.11492} {arXiv:2210.11492} \BibitemShut {NoStop}%
\bibitem [{\citenamefont {{Cherman}}\ and\ \citenamefont
  {{Jacobson}}(2023)}]{CJ230413751}%
  \BibitemOpen
  \bibfield  {author} {\bibinfo {author} {\bibfnamefont {A.}~\bibnamefont
  {{Cherman}}}\ and\ \bibinfo {author} {\bibfnamefont {T.}~\bibnamefont
  {{Jacobson}}},\ }\href@noop {} {\bibinfo {title} {{Emergent 1-form
  symmetries}}} (\bibinfo {year} {2023}),\ \Eprint
  {https://arxiv.org/abs/2304.13751} {arXiv:2304.13751} \BibitemShut {NoStop}%
\bibitem [{\citenamefont {{Chatterjee}}\ and\ \citenamefont
  {{Wen}}(2023)}]{CW220303596}%
  \BibitemOpen
  \bibfield  {author} {\bibinfo {author} {\bibfnamefont {A.}~\bibnamefont
  {{Chatterjee}}}\ and\ \bibinfo {author} {\bibfnamefont {X.-G.}\ \bibnamefont
  {{Wen}}},\ }\bibfield  {title} {\bibinfo {title} {{Symmetry as a shadow of
  topological order and a derivation of topological holographic principle}},\
  }\href {https://doi.org/10.1103/PhysRevB.107.155136} {\bibfield  {journal}
  {\bibinfo  {journal} {Phys. Rev. B}\ }\textbf {\bibinfo {volume} {107}},\
  \bibinfo {pages} {155136} (\bibinfo {year} {2023})},\ \Eprint
  {https://arxiv.org/abs/2203.03596} {arXiv:2203.03596} \BibitemShut {NoStop}%
\bibitem [{\citenamefont {{Laughlin}}\ and\ \citenamefont
  {{Pines}}(2000)}]{LP0028}%
  \BibitemOpen
  \bibfield  {author} {\bibinfo {author} {\bibfnamefont {R.~B.}\ \bibnamefont
  {{Laughlin}}}\ and\ \bibinfo {author} {\bibfnamefont {D.}~\bibnamefont
  {{Pines}}},\ }\bibfield  {title} {\bibinfo {title} {{The Theory of
  Everything}},\ }\href {https://doi.org/10.1073/pnas.97.1.28} {\bibfield
  {journal} {\bibinfo  {journal} {Proc. Natl. Acad. Sci. U.S.A.}\ }\textbf
  {\bibinfo {volume} {97}},\ \bibinfo {pages} {28} (\bibinfo {year}
  {2000})}\BibitemShut {NoStop}%
\bibitem [{\citenamefont {{Seiberg}}\ and\ \citenamefont
  {{Shao}}(2021)}]{SS200310466}%
  \BibitemOpen
  \bibfield  {author} {\bibinfo {author} {\bibfnamefont {N.}~\bibnamefont
  {{Seiberg}}}\ and\ \bibinfo {author} {\bibfnamefont {S.-H.}\ \bibnamefont
  {{Shao}}},\ }\bibfield  {title} {\bibinfo {title} {{Exotic symmetries,
  duality, and fractons in 2+1-dimensional quantum field theory}},\ }\href
  {https://doi.org/10.21468/SciPostPhys.10.2.027} {\bibfield  {journal}
  {\bibinfo  {journal} {SciPost Phys.}\ }\textbf {\bibinfo {volume} {10}},\
  \bibinfo {pages} {027} (\bibinfo {year} {2021})},\ \Eprint
  {https://arxiv.org/abs/2003.10466} {arXiv:2003.10466} \BibitemShut {NoStop}%
\bibitem [{\citenamefont {{Gorantla}}\ \emph
  {et~al.}(2021{\natexlab{a}})\citenamefont {{Gorantla}}, \citenamefont
  {{Lam}}, \citenamefont {{Seiberg}},\ and\ \citenamefont
  {{Shao}}}]{GLS210800020}%
  \BibitemOpen
  \bibfield  {author} {\bibinfo {author} {\bibfnamefont {P.}~\bibnamefont
  {{Gorantla}}}, \bibinfo {author} {\bibfnamefont {H.~T.}\ \bibnamefont
  {{Lam}}}, \bibinfo {author} {\bibfnamefont {N.}~\bibnamefont {{Seiberg}}},\
  and\ \bibinfo {author} {\bibfnamefont {S.-H.}\ \bibnamefont {{Shao}}},\
  }\bibfield  {title} {\bibinfo {title} {{Low-energy limit of some exotic
  lattice theories and UV/IR mixing}},\ }\href
  {https://doi.org/10.1103/PhysRevB.104.235116} {\bibfield  {journal} {\bibinfo
   {journal} {Phys. Rev. B}\ }\textbf {\bibinfo {volume} {104}},\ \bibinfo
  {pages} {235116} (\bibinfo {year} {2021}{\natexlab{a}})},\ \Eprint
  {https://arxiv.org/abs/2108.00020} {arXiv:2108.00020} \BibitemShut {NoStop}%
\bibitem [{\citenamefont {{Oh}}\ \emph {et~al.}(2022)\citenamefont {{Oh}},
  \citenamefont {{Kim}}, \citenamefont {{Moon}},\ and\ \citenamefont
  {{Han}}}]{OJE211002658}%
  \BibitemOpen
  \bibfield  {author} {\bibinfo {author} {\bibfnamefont {Y.-T.}\ \bibnamefont
  {{Oh}}}, \bibinfo {author} {\bibfnamefont {J.}~\bibnamefont {{Kim}}},
  \bibinfo {author} {\bibfnamefont {E.-G.}\ \bibnamefont {{Moon}}},\ and\
  \bibinfo {author} {\bibfnamefont {J.~H.}\ \bibnamefont {{Han}}},\ }\bibfield
  {title} {\bibinfo {title} {{Rank-2 toric code in two dimensions}},\ }\href
  {https://doi.org/10.1103/PhysRevB.105.045128} {\bibfield  {journal} {\bibinfo
   {journal} {Phys. Rev. B}\ }\textbf {\bibinfo {volume} {105}},\ \bibinfo
  {pages} {045128} (\bibinfo {year} {2022})},\ \Eprint
  {https://arxiv.org/abs/2110.02658} {arXiv:2110.02658} \BibitemShut {NoStop}%
\bibitem [{\citenamefont {{Pace}}\ and\ \citenamefont
  {{Wen}}(2022)}]{PW220407111}%
  \BibitemOpen
  \bibfield  {author} {\bibinfo {author} {\bibfnamefont {S.~D.}\ \bibnamefont
  {{Pace}}}\ and\ \bibinfo {author} {\bibfnamefont {X.-G.}\ \bibnamefont
  {{Wen}}},\ }\bibfield  {title} {\bibinfo {title} {{Position-dependent
  excitations and UV/IR mixing in the ${\mathbb{Z}}_{N}$ rank-2 toric code and
  its low-energy effective field theory}},\ }\href
  {https://doi.org/10.1103/PhysRevB.106.045145} {\bibfield  {journal} {\bibinfo
   {journal} {Phys. Rev. B}\ }\textbf {\bibinfo {volume} {106}},\ \bibinfo
  {pages} {045145} (\bibinfo {year} {2022})},\ \Eprint
  {https://arxiv.org/abs/2204.07111} {arXiv:2204.07111} \BibitemShut {NoStop}%
\bibitem [{\citenamefont {{Bravyi}}\ \emph {et~al.}(2010)\citenamefont
  {{Bravyi}}, \citenamefont {{Hastings}},\ and\ \citenamefont
  {{Michalakis}}}]{BHS10010344}%
  \BibitemOpen
  \bibfield  {author} {\bibinfo {author} {\bibfnamefont {S.}~\bibnamefont
  {{Bravyi}}}, \bibinfo {author} {\bibfnamefont {M.~B.}\ \bibnamefont
  {{Hastings}}},\ and\ \bibinfo {author} {\bibfnamefont {S.}~\bibnamefont
  {{Michalakis}}},\ }\bibfield  {title} {\bibinfo {title} {{Topological quantum
  order: stability under local perturbations}},\ }\href
  {https://doi.org/10.1063/1.3490195} {\bibfield  {journal} {\bibinfo
  {journal} {J. Math. Phys.}\ }\textbf {\bibinfo {volume} {51}},\ \bibinfo
  {pages} {093512} (\bibinfo {year} {2010})},\ \Eprint
  {https://arxiv.org/abs/1001.0344} {arXiv:1001.0344} \BibitemShut {NoStop}%
\bibitem [{\citenamefont {{Yin}}\ and\ \citenamefont
  {{Lucas}}(2023)}]{YL220911242}%
  \BibitemOpen
  \bibfield  {author} {\bibinfo {author} {\bibfnamefont {C.}~\bibnamefont
  {{Yin}}}\ and\ \bibinfo {author} {\bibfnamefont {A.}~\bibnamefont
  {{Lucas}}},\ }\bibfield  {title} {\bibinfo {title} {{Prethermalization and
  the Local Robustness of Gapped Systems}},\ }\href
  {https://doi.org/10.1103/PhysRevLett.131.050402} {\bibfield  {journal}
  {\bibinfo  {journal} {Phys. Rev. Lett.}\ }\textbf {\bibinfo {volume} {131}},\
  \bibinfo {pages} {050402} (\bibinfo {year} {2023})},\ \Eprint
  {https://arxiv.org/abs/2209.11242} {arXiv:2209.11242} \BibitemShut {NoStop}%
\bibitem [{\citenamefont {{Moudgalya}}\ and\ \citenamefont
  {{Motrunich}}(2023{\natexlab{a}})}]{MM220903370}%
  \BibitemOpen
  \bibfield  {author} {\bibinfo {author} {\bibfnamefont {S.}~\bibnamefont
  {{Moudgalya}}}\ and\ \bibinfo {author} {\bibfnamefont {O.~I.}\ \bibnamefont
  {{Motrunich}}},\ }\bibfield  {title} {\bibinfo {title} {{From symmetries to
  commutant algebras in standard Hamiltonians}},\ }\href
  {https://doi.org/10.1016/j.aop.2023.169384} {\bibfield  {journal} {\bibinfo
  {journal} {Ann. Phys.}\ ,\ \bibinfo {pages} {169384}} (\bibinfo {year}
  {2023}{\natexlab{a}})},\ \Eprint {https://arxiv.org/abs/2209.03370}
  {arXiv:2209.03370} \BibitemShut {NoStop}%
\bibitem [{\citenamefont {{Moudgalya}}\ and\ \citenamefont
  {{Motrunich}}(2023{\natexlab{b}})}]{MM230203028}%
  \BibitemOpen
  \bibfield  {author} {\bibinfo {author} {\bibfnamefont {S.}~\bibnamefont
  {{Moudgalya}}}\ and\ \bibinfo {author} {\bibfnamefont {O.~I.}\ \bibnamefont
  {{Motrunich}}},\ }\bibfield  {title} {\bibinfo {title} {{Numerical methods
  for detecting symmetries and commutant algebras}},\ }\href@noop {} {\bibfield
   {journal} {\bibinfo  {journal} {Phys. Rev. B}\ }\textbf {\bibinfo {volume}
  {107}},\ \bibinfo {pages} {224312} (\bibinfo {year} {2023}{\natexlab{b}})},\
  \Eprint {https://arxiv.org/abs/2302.03028} {arXiv:2302.03028} \BibitemShut
  {NoStop}%
\bibitem [{\citenamefont {{Roumpedakis}}\ \emph {et~al.}(2023)\citenamefont
  {{Roumpedakis}}, \citenamefont {{Seifnashri}},\ and\ \citenamefont
  {{Shao}}}]{RSS220402407}%
  \BibitemOpen
  \bibfield  {author} {\bibinfo {author} {\bibfnamefont {K.}~\bibnamefont
  {{Roumpedakis}}}, \bibinfo {author} {\bibfnamefont {S.}~\bibnamefont
  {{Seifnashri}}},\ and\ \bibinfo {author} {\bibfnamefont {S.-H.}\ \bibnamefont
  {{Shao}}},\ }\bibfield  {title} {\bibinfo {title} {{Higher Gauging and
  Non-invertible Condensation Defects}},\ }\href
  {https://doi.org/10.1007/s00220-023-04706-9} {\bibfield  {journal} {\bibinfo
  {journal} {Commun. Math. Phys.}\ }\textbf {\bibinfo {volume} {401}},\
  \bibinfo {pages} {3043} (\bibinfo {year} {2023})},\ \Eprint
  {https://arxiv.org/abs/2204.02407} {arXiv:2204.02407} \BibitemShut {NoStop}%
\bibitem [{\citenamefont {{Choi}}\ \emph {et~al.}(2023)\citenamefont {{Choi}},
  \citenamefont {{C\'ordova}}, \citenamefont {{Hsin}}, \citenamefont {{Lam}},\
  and\ \citenamefont {{Shao}}}]{CCH220409025}%
  \BibitemOpen
  \bibfield  {author} {\bibinfo {author} {\bibfnamefont {Y.}~\bibnamefont
  {{Choi}}}, \bibinfo {author} {\bibfnamefont {C.}~\bibnamefont {{C\'ordova}}},
  \bibinfo {author} {\bibfnamefont {P.-S.}\ \bibnamefont {{Hsin}}}, \bibinfo
  {author} {\bibfnamefont {H.~T.}\ \bibnamefont {{Lam}}},\ and\ \bibinfo
  {author} {\bibfnamefont {S.-H.}\ \bibnamefont {{Shao}}},\ }\bibfield  {title}
  {\bibinfo {title} {{Non-invertible Condensation, Duality, and Triality
  Defects in 3+1 Dimensions}},\ }\href
  {https://doi.org/10.1007/s00220-023-04727-4} {\bibfield  {journal} {\bibinfo
  {journal} {Commun. Math. Phys.}\ }\textbf {\bibinfo {volume} {402}},\
  \bibinfo {pages} {489} (\bibinfo {year} {2023})},\ \Eprint
  {https://arxiv.org/abs/2204.09025} {arXiv:2204.09025} \BibitemShut {NoStop}%
\bibitem [{\citenamefont {{Bhardwaj}}\ \emph {et~al.}(2022)\citenamefont
  {{Bhardwaj}}, \citenamefont {{Sch\"afer-Nameki}},\ and\ \citenamefont
  {{Wu}}}]{BSW220805973}%
  \BibitemOpen
  \bibfield  {author} {\bibinfo {author} {\bibfnamefont {L.}~\bibnamefont
  {{Bhardwaj}}}, \bibinfo {author} {\bibfnamefont {S.}~\bibnamefont
  {{Sch\"afer-Nameki}}},\ and\ \bibinfo {author} {\bibfnamefont
  {J.}~\bibnamefont {{Wu}}},\ }\bibfield  {title} {\bibinfo {title} {{Universal
  Non-Invertible Symmetries}},\ }\href {https://doi.org/10.1002/prop.202200143}
  {\bibfield  {journal} {\bibinfo  {journal} {Fortschr. Phys.}\ }\textbf
  {\bibinfo {volume} {70}},\ \bibinfo {pages} {2200143} (\bibinfo {year}
  {2022})},\ \Eprint {https://arxiv.org/abs/2208.05973} {arXiv:2208.05973}
  \BibitemShut {NoStop}%
\bibitem [{\citenamefont {{Lin}}\ \emph {et~al.}(2022)\citenamefont {{Lin}},
  \citenamefont {{Robbins}},\ and\ \citenamefont {{Sharpe}}}]{LRS220805982}%
  \BibitemOpen
  \bibfield  {author} {\bibinfo {author} {\bibfnamefont {L.}~\bibnamefont
  {{Lin}}}, \bibinfo {author} {\bibfnamefont {D.~G.}\ \bibnamefont
  {{Robbins}}},\ and\ \bibinfo {author} {\bibfnamefont {E.}~\bibnamefont
  {{Sharpe}}},\ }\bibfield  {title} {\bibinfo {title} {{Decomposition,
  Condensation Defects, and Fusion}},\ }\href
  {https://doi.org/10.1002/prop.202200130} {\bibfield  {journal} {\bibinfo
  {journal} {Fortschr. Phys.}\ }\textbf {\bibinfo {volume} {70}},\ \bibinfo
  {pages} {2200130} (\bibinfo {year} {2022})},\ \Eprint
  {https://arxiv.org/abs/2208.05982} {arXiv:2208.05982} \BibitemShut {NoStop}%
\bibitem [{\citenamefont {{Delcamp}}\ and\ \citenamefont
  {{Tiwari}}(2023)}]{DT230101259}%
  \BibitemOpen
  \bibfield  {author} {\bibinfo {author} {\bibfnamefont {C.}~\bibnamefont
  {{Delcamp}}}\ and\ \bibinfo {author} {\bibfnamefont {A.}~\bibnamefont
  {{Tiwari}}},\ }\href@noop {} {\bibinfo {title} {{Higher categorical
  symmetries and gauging in two-dimensional spin systems}}} (\bibinfo {year}
  {2023}),\ \Eprint {https://arxiv.org/abs/2301.01259} {arXiv:2301.01259}
  \BibitemShut {NoStop}%
\bibitem [{\citenamefont {{Bhardwaj}}\ and\ \citenamefont
  {{Sch\"afer-Nameki}}(2023{\natexlab{a}})}]{BS230402660}%
  \BibitemOpen
  \bibfield  {author} {\bibinfo {author} {\bibfnamefont {L.}~\bibnamefont
  {{Bhardwaj}}}\ and\ \bibinfo {author} {\bibfnamefont {S.}~\bibnamefont
  {{Sch\"afer-Nameki}}},\ }\href@noop {} {\bibinfo {title} {{Generalized
  Charges, Part I: Invertible Symmetries and Higher Representations}}}
  (\bibinfo {year} {2023}{\natexlab{a}}),\ \Eprint
  {https://arxiv.org/abs/2304.02660} {arXiv:2304.02660} \BibitemShut {NoStop}%
\bibitem [{\citenamefont {{Bartsch}}\ \emph {et~al.}(2023)\citenamefont
  {{Bartsch}}, \citenamefont {{Bullimore}},\ and\ \citenamefont
  {{Grigoletto}}}]{BBG230403789}%
  \BibitemOpen
  \bibfield  {author} {\bibinfo {author} {\bibfnamefont {T.}~\bibnamefont
  {{Bartsch}}}, \bibinfo {author} {\bibfnamefont {M.}~\bibnamefont
  {{Bullimore}}},\ and\ \bibinfo {author} {\bibfnamefont {A.}~\bibnamefont
  {{Grigoletto}}},\ }\href@noop {} {\bibinfo {title} {{Higher representations
  for extended operators}}} (\bibinfo {year} {2023}),\ \Eprint
  {https://arxiv.org/abs/2304.03789} {arXiv:2304.03789} \BibitemShut {NoStop}%
\bibitem [{\citenamefont {{Bhardwaj}}\ and\ \citenamefont
  {{Sch\"afer-Nameki}}(2023{\natexlab{b}})}]{BS230517159}%
  \BibitemOpen
  \bibfield  {author} {\bibinfo {author} {\bibfnamefont {L.}~\bibnamefont
  {{Bhardwaj}}}\ and\ \bibinfo {author} {\bibfnamefont {S.}~\bibnamefont
  {{Sch\"afer-Nameki}}},\ }\href@noop {} {\bibinfo {title} {{Generalized
  Charges, Part II: Non-Invertible Symmetries and the Symmetry TFT}}} (\bibinfo
  {year} {2023}{\natexlab{b}}),\ \Eprint {https://arxiv.org/abs/2305.17159}
  {arXiv:2305.17159} \BibitemShut {NoStop}%
\bibitem [{\citenamefont {{Lee}}\ and\ \citenamefont
  {{Lee}}(2005)}]{LLc0507191}%
  \BibitemOpen
  \bibfield  {author} {\bibinfo {author} {\bibfnamefont {S.-S.}\ \bibnamefont
  {{Lee}}}\ and\ \bibinfo {author} {\bibfnamefont {P.~A.}\ \bibnamefont
  {{Lee}}},\ }\bibfield  {title} {\bibinfo {title} {{Emergent U(1) gauge theory
  with fractionalized boson/fermion from the Bose condensation of excitons in a
  multiband insulator}},\ }\href {https://doi.org/10.1103/PhysRevB.72.235104}
  {\bibfield  {journal} {\bibinfo  {journal} {\prb}\ }\textbf {\bibinfo
  {volume} {72}},\ \bibinfo {pages} {235104} (\bibinfo {year} {2005})},\
  \Eprint {https://arxiv.org/abs/cond-mat/0507191} {arXiv:cond-mat/0507191}
  \BibitemShut {NoStop}%
\bibitem [{\citenamefont {{Wu}}\ \emph {et~al.}(2012)\citenamefont {{Wu}},
  \citenamefont {{Deng}},\ and\ \citenamefont {{Prokof'ev}}}]{WP12016409}%
  \BibitemOpen
  \bibfield  {author} {\bibinfo {author} {\bibfnamefont {F.}~\bibnamefont
  {{Wu}}}, \bibinfo {author} {\bibfnamefont {Y.}~\bibnamefont {{Deng}}},\ and\
  \bibinfo {author} {\bibfnamefont {N.}~\bibnamefont {{Prokof'ev}}},\
  }\bibfield  {title} {\bibinfo {title} {{Phase diagram of the toric code model
  in a parallel magnetic field}},\ }\href
  {https://doi.org/10.1103/PhysRevB.85.195104} {\bibfield  {journal} {\bibinfo
  {journal} {\prb}\ }\textbf {\bibinfo {volume} {85}},\ \bibinfo {pages}
  {195104} (\bibinfo {year} {2012})},\ \Eprint
  {https://arxiv.org/abs/1201.6409} {arXiv:1201.6409} \BibitemShut {NoStop}%
\bibitem [{\citenamefont {{Huba\v{c}}}\ and\ \citenamefont
  {{Wilson}}(2010)}]{HW1037}%
  \BibitemOpen
  \bibfield  {author} {\bibinfo {author} {\bibfnamefont {I.}~\bibnamefont
  {{Huba\v{c}}}}\ and\ \bibinfo {author} {\bibfnamefont {S.}~\bibnamefont
  {{Wilson}}},\ }\bibfield  {title} {\bibinfo {title} {{Brillouin-Wigner
  Perturbation Theory}},\ }in\ \href
  {https://doi.org/10.1007/978-90-481-3373-4} {\emph {\bibinfo {booktitle}
  {{Brillouin-Wigner Methods for Many-Body Systems}}}}\ (\bibinfo  {publisher}
  {Springer},\ \bibinfo {year} {2010})\ pp.\ \bibinfo {pages}
  {37--68}\BibitemShut {NoStop}%
\bibitem [{\citenamefont {{Bravyi}}\ \emph {et~al.}(2011)\citenamefont
  {{Bravyi}}, \citenamefont {{DiVincenzo}},\ and\ \citenamefont
  {{Loss}}}]{BDL11050675}%
  \BibitemOpen
  \bibfield  {author} {\bibinfo {author} {\bibfnamefont {S.}~\bibnamefont
  {{Bravyi}}}, \bibinfo {author} {\bibfnamefont {D.~P.}\ \bibnamefont
  {{DiVincenzo}}},\ and\ \bibinfo {author} {\bibfnamefont {D.}~\bibnamefont
  {{Loss}}},\ }\bibfield  {title} {\bibinfo {title} {{Schrieffer-Wolff
  transformation for quantum many-body systems}},\ }\href
  {https://doi.org/10.1016/j.aop.2011.06.004} {\bibfield  {journal} {\bibinfo
  {journal} {Ann. Phys.}\ }\textbf {\bibinfo {volume} {326}},\ \bibinfo {pages}
  {2793} (\bibinfo {year} {2011})},\ \Eprint {https://arxiv.org/abs/1105.0675}
  {arXiv:1105.0675} \BibitemShut {NoStop}%
\bibitem [{\citenamefont {{Kong}}\ and\ \citenamefont
  {{Zheng}}(2022)}]{KZ220105726}%
  \BibitemOpen
  \bibfield  {author} {\bibinfo {author} {\bibfnamefont {L.}~\bibnamefont
  {{Kong}}}\ and\ \bibinfo {author} {\bibfnamefont {H.}~\bibnamefont
  {{Zheng}}},\ }\bibfield  {title} {\bibinfo {title} {{Categories of quantum
  liquids III}},\ }\href@noop {} {\  (\bibinfo {year} {2022})},\ \Eprint
  {https://arxiv.org/abs/2201.05726} {arXiv:2201.05726} \BibitemShut {NoStop}%
\bibitem [{\citenamefont {Ji}\ and\ \citenamefont {Wen}(2020)}]{JW191213492}%
  \BibitemOpen
  \bibfield  {author} {\bibinfo {author} {\bibfnamefont {W.}~\bibnamefont
  {Ji}}\ and\ \bibinfo {author} {\bibfnamefont {X.-G.}\ \bibnamefont {Wen}},\
  }\bibfield  {title} {\bibinfo {title} {Categorical symmetry and noninvertible
  anomaly in symmetry-breaking and topological phase transitions},\ }\href
  {https://doi.org/10.1103/PhysRevResearch.2.033417} {\bibfield  {journal}
  {\bibinfo  {journal} {Phys. Rev. Res.}\ }\textbf {\bibinfo {volume} {2}},\
  \bibinfo {pages} {033417} (\bibinfo {year} {2020})},\ \Eprint
  {https://arxiv.org/abs/1912.13492} {arXiv:1912.13492} \BibitemShut {NoStop}%
\bibitem [{\citenamefont {{Kong}}\ \emph
  {et~al.}(2020{\natexlab{b}})\citenamefont {{Kong}}, \citenamefont {{Lan}},
  \citenamefont {{Wen}}, \citenamefont {{Zhang}},\ and\ \citenamefont
  {{Zheng}}}]{KZ200308898}%
  \BibitemOpen
  \bibfield  {author} {\bibinfo {author} {\bibfnamefont {L.}~\bibnamefont
  {{Kong}}}, \bibinfo {author} {\bibfnamefont {T.}~\bibnamefont {{Lan}}},
  \bibinfo {author} {\bibfnamefont {X.-G.}\ \bibnamefont {{Wen}}}, \bibinfo
  {author} {\bibfnamefont {Z.-H.}\ \bibnamefont {{Zhang}}},\ and\ \bibinfo
  {author} {\bibfnamefont {H.}~\bibnamefont {{Zheng}}},\ }\bibfield  {title}
  {\bibinfo {title} {Classification of topological phases with finite internal
  symmetries in all dimensions},\ }\href
  {https://doi.org/doi.org/10.1007/JHEP09(2020)093} {\bibfield  {journal}
  {\bibinfo  {journal} {J. High Energ. Phys.}\ }\textbf {\bibinfo {volume}
  {2020}}\bibfield  {number} {\bibinfo  {number} { (9)},\ \bibinfo {pages}
  {93}},\ }\Eprint {https://arxiv.org/abs/2003.08898} {arXiv:2003.08898}
  \BibitemShut {NoStop}%
\bibitem [{\citenamefont {Apruzzi}\ \emph {et~al.}(2023)\citenamefont
  {Apruzzi}, \citenamefont {Bonetti}, \citenamefont {Etxebarria}, \citenamefont
  {Hosseini},\ and\ \citenamefont {Schäfer-Nameki}}]{AS211202092}%
  \BibitemOpen
  \bibfield  {author} {\bibinfo {author} {\bibfnamefont {F.}~\bibnamefont
  {Apruzzi}}, \bibinfo {author} {\bibfnamefont {F.}~\bibnamefont {Bonetti}},
  \bibinfo {author} {\bibfnamefont {I.~G.}\ \bibnamefont {Etxebarria}},
  \bibinfo {author} {\bibfnamefont {S.~S.}\ \bibnamefont {Hosseini}},\ and\
  \bibinfo {author} {\bibfnamefont {S.}~\bibnamefont {Schäfer-Nameki}},\
  }\bibfield  {title} {\bibinfo {title} {Symmetry {TFTs} from string theory},\
  }\href {https://doi.org/10.1007/s00220-023-04737-2} {\bibfield  {journal}
  {\bibinfo  {journal} {Communications in Mathematical Physics}\ }\textbf
  {\bibinfo {volume} {402}},\ \bibinfo {pages} {895} (\bibinfo {year}
  {2023})},\ \Eprint {https://arxiv.org/abs/2112.02092} {arXiv:2112.02092
  [hep-th]} \BibitemShut {NoStop}%
\bibitem [{\citenamefont {Ji}\ and\ \citenamefont {Wen}(2019)}]{JW190513279}%
  \BibitemOpen
  \bibfield  {author} {\bibinfo {author} {\bibfnamefont {W.}~\bibnamefont
  {Ji}}\ and\ \bibinfo {author} {\bibfnamefont {X.-G.}\ \bibnamefont {Wen}},\
  }\bibfield  {title} {\bibinfo {title} {Non-invertible anomalies and
  mapping-class-group transformation of anomalous partition functions},\ }\href
  {https://doi.org/10.1103/PhysRevResearch.1.033054} {\bibfield  {journal}
  {\bibinfo  {journal} {Phys. Rev. Research}\ }\textbf {\bibinfo {volume}
  {1}},\ \bibinfo {pages} {033054} (\bibinfo {year} {2019})},\ \Eprint
  {https://arxiv.org/abs/1905.13279} {arXiv:1905.13279} \BibitemShut {NoStop}%
\bibitem [{\citenamefont {Wen}(2013)}]{W1313}%
  \BibitemOpen
  \bibfield  {author} {\bibinfo {author} {\bibfnamefont {X.-G.}\ \bibnamefont
  {Wen}},\ }\bibfield  {title} {\bibinfo {title} {Classifying gauge anomalies
  through symmetry-protected trivial orders and classifying gravitational
  anomalies through topological orders},\ }\href
  {https://doi.org/10.1103/physrevd.88.045013} {\bibfield  {journal} {\bibinfo
  {journal} {Phys. Rev. D}\ }\textbf {\bibinfo {volume} {88}},\ \bibinfo
  {pages} {045013} (\bibinfo {year} {2013})},\ \Eprint
  {https://arxiv.org/abs/1303.1803} {arXiv:1303.1803} \BibitemShut {NoStop}%
\bibitem [{\citenamefont {Kong}\ and\ \citenamefont {Wen}(2014)}]{KW1458}%
  \BibitemOpen
  \bibfield  {author} {\bibinfo {author} {\bibfnamefont {L.}~\bibnamefont
  {Kong}}\ and\ \bibinfo {author} {\bibfnamefont {X.-G.}\ \bibnamefont {Wen}},\
  }\bibfield  {title} {\bibinfo {title} {Braided fusion categories,
  gravitational anomalies, and the mathematical framework for topological
  orders in any dimensions},\ }\href@noop {} {\  (\bibinfo {year} {2014})},\
  \Eprint {https://arxiv.org/abs/1405.5858} {arXiv:1405.5858} \BibitemShut
  {NoStop}%
\bibitem [{\citenamefont {Kitaev}\ and\ \citenamefont
  {Kong}(2012)}]{KK11045047}%
  \BibitemOpen
  \bibfield  {author} {\bibinfo {author} {\bibfnamefont {A.}~\bibnamefont
  {Kitaev}}\ and\ \bibinfo {author} {\bibfnamefont {L.}~\bibnamefont {Kong}},\
  }\bibfield  {title} {\bibinfo {title} {Models for gapped boundaries and
  domain walls},\ }\href {https://doi.org/10.1007/s00220-012-1500-5} {\bibfield
   {journal} {\bibinfo  {journal} {Commun. Math. Phys.}\ }\textbf {\bibinfo
  {volume} {313}},\ \bibinfo {pages} {351} (\bibinfo {year} {2012})},\ \Eprint
  {https://arxiv.org/abs/1104.5047} {arXiv:1104.5047} \BibitemShut {NoStop}%
\bibitem [{\citenamefont {Wen}(1991{\natexlab{a}})}]{W9125}%
  \BibitemOpen
  \bibfield  {author} {\bibinfo {author} {\bibfnamefont {X.-G.}\ \bibnamefont
  {Wen}},\ }\bibfield  {title} {\bibinfo {title} {Gapless boundary excitations
  in the quantum hall states and in the chiral spin states},\ }\href
  {https://doi.org/10.1103/physrevb.43.11025} {\bibfield  {journal} {\bibinfo
  {journal} {Phys. Rev. B}\ }\textbf {\bibinfo {volume} {43}},\ \bibinfo
  {pages} {11025} (\bibinfo {year} {1991}{\natexlab{a}})}\BibitemShut {NoStop}%
\bibitem [{\citenamefont {{Delacr{\'e}taz}}\ \emph {et~al.}(2020)\citenamefont
  {{Delacr{\'e}taz}}, \citenamefont {{Hofman}},\ and\ \citenamefont
  {{Mathys}}}]{DM190806977}%
  \BibitemOpen
  \bibfield  {author} {\bibinfo {author} {\bibfnamefont {L.}~\bibnamefont
  {{Delacr{\'e}taz}}}, \bibinfo {author} {\bibfnamefont {D.}~\bibnamefont
  {{Hofman}}},\ and\ \bibinfo {author} {\bibfnamefont {G.}~\bibnamefont
  {{Mathys}}},\ }\bibfield  {title} {\bibinfo {title} {{Superfluids as
  higher-form anomalies}},\ }\href
  {https://doi.org/10.21468/SciPostPhys.8.3.047} {\bibfield  {journal}
  {\bibinfo  {journal} {SciPost Phys.}\ }\textbf {\bibinfo {volume} {8}},\
  \bibinfo {pages} {047} (\bibinfo {year} {2020})},\ \Eprint
  {https://arxiv.org/abs/1908.06977} {arXiv:1908.06977} \BibitemShut {NoStop}%
\bibitem [{\citenamefont {Kapustin}\ and\ \citenamefont
  {Seiberg}(2014)}]{KS14010740}%
  \BibitemOpen
  \bibfield  {author} {\bibinfo {author} {\bibfnamefont {A.}~\bibnamefont
  {Kapustin}}\ and\ \bibinfo {author} {\bibfnamefont {N.}~\bibnamefont
  {Seiberg}},\ }\bibfield  {title} {\bibinfo {title} {Coupling a {QFT} to a
  {TQFT} and duality},\ }\href {https://doi.org/10.1007/jhep04(2014)001}
  {\bibfield  {journal} {\bibinfo  {journal} {J. High Energ. Phys.}\ }\textbf
  {\bibinfo {volume} {2014}}\bibfield  {number} {\bibinfo  {number} { (4)},\
  \bibinfo {pages} {1}},\ }\Eprint {https://arxiv.org/abs/1401.0740}
  {arXiv:1401.0740} \BibitemShut {NoStop}%
\bibitem [{\citenamefont {{Iqbal}}\ and\ \citenamefont
  {{McGreevy}}(2020)}]{IM200304349}%
  \BibitemOpen
  \bibfield  {author} {\bibinfo {author} {\bibfnamefont {N.}~\bibnamefont
  {{Iqbal}}}\ and\ \bibinfo {author} {\bibfnamefont {J.}~\bibnamefont
  {{McGreevy}}},\ }\bibfield  {title} {\bibinfo {title} {{Toward a 3d Ising
  model with a weakly-coupled string theory dual}},\ }\href
  {https://doi.org/10.21468/SciPostPhys.9.2.019} {\bibfield  {journal}
  {\bibinfo  {journal} {SciPost Phys.}\ }\textbf {\bibinfo {volume} {9}},\
  \bibinfo {pages} {019} (\bibinfo {year} {2020})},\ \Eprint
  {https://arxiv.org/abs/2003.04349} {arXiv:2003.04349} \BibitemShut {NoStop}%
\bibitem [{\citenamefont {{Gorantla}}\ \emph
  {et~al.}(2021{\natexlab{b}})\citenamefont {{Gorantla}}, \citenamefont
  {{Lam}}, \citenamefont {{Seiberg}},\ and\ \citenamefont
  {{Shao}}}]{GLS210301257}%
  \BibitemOpen
  \bibfield  {author} {\bibinfo {author} {\bibfnamefont {P.}~\bibnamefont
  {{Gorantla}}}, \bibinfo {author} {\bibfnamefont {H.~T.}\ \bibnamefont
  {{Lam}}}, \bibinfo {author} {\bibfnamefont {N.}~\bibnamefont {{Seiberg}}},\
  and\ \bibinfo {author} {\bibfnamefont {S.-H.}\ \bibnamefont {{Shao}}},\
  }\bibfield  {title} {\bibinfo {title} {{A modified Villain formulation of
  fractons and other exotic theories}},\ }\href
  {https://doi.org/10.1063/5.0060808} {\bibfield  {journal} {\bibinfo
  {journal} {J. Math. Phys.}\ }\textbf {\bibinfo {volume} {62}},\ \bibinfo
  {pages} {102301} (\bibinfo {year} {2021}{\natexlab{b}})},\ \Eprint
  {https://arxiv.org/abs/2103.01257} {arXiv:2103.01257} \BibitemShut {NoStop}%
\bibitem [{\citenamefont {{Fazza}}\ and\ \citenamefont
  {{Sulejmanpasic}}(2023)}]{FS221113047}%
  \BibitemOpen
  \bibfield  {author} {\bibinfo {author} {\bibfnamefont {L.}~\bibnamefont
  {{Fazza}}}\ and\ \bibinfo {author} {\bibfnamefont {T.}~\bibnamefont
  {{Sulejmanpasic}}},\ }\bibfield  {title} {\bibinfo {title} {{Lattice quantum
  Villain Hamiltonians: compact scalars, U(1) gauge theories, fracton models
  and quantum Ising model dualities}},\ }\href
  {https://doi.org/10.1007/JHEP05(2023)017} {\bibfield  {journal} {\bibinfo
  {journal} {J. High Energ. Phys.}\ }\textbf {\bibinfo {volume}
  {2023}}\bibfield  {number} {\bibinfo  {number} { (5)},\ \bibinfo {pages}
  {17}},\ }\Eprint {https://arxiv.org/abs/2211.13047} {arXiv:2211.13047}
  \BibitemShut {NoStop}%
\bibitem [{\citenamefont {Kitaev}(2003)}]{K032}%
  \BibitemOpen
  \bibfield  {author} {\bibinfo {author} {\bibfnamefont {A.}~\bibnamefont
  {Kitaev}},\ }\bibfield  {title} {\bibinfo {title} {Fault-tolerant quantum
  computation by anyons},\ }\href
  {https://doi.org/10.1016/s0003-4916(02)00018-0} {\bibfield  {journal}
  {\bibinfo  {journal} {Ann. Phys.}\ }\textbf {\bibinfo {volume} {303}},\
  \bibinfo {pages} {2} (\bibinfo {year} {2003})},\ \Eprint
  {https://arxiv.org/abs/quant-ph/9707021} {arXiv:quant-ph/9707021}
  \BibitemShut {NoStop}%
\bibitem [{\citenamefont {Fradkin}\ and\ \citenamefont
  {Shenker}(1979)}]{FS7982}%
  \BibitemOpen
  \bibfield  {author} {\bibinfo {author} {\bibfnamefont {E.}~\bibnamefont
  {Fradkin}}\ and\ \bibinfo {author} {\bibfnamefont {S.~H.}\ \bibnamefont
  {Shenker}},\ }\bibfield  {title} {\bibinfo {title} {Phase diagrams of lattice
  gauge theories with higgs fields},\ }\href
  {https://doi.org/10.1103/physrevd.19.3682} {\bibfield  {journal} {\bibinfo
  {journal} {Phys. Rev. D}\ }\textbf {\bibinfo {volume} {19}},\ \bibinfo
  {pages} {3682} (\bibinfo {year} {1979})}\BibitemShut {NoStop}%
\bibitem [{\citenamefont {{Tupitsyn}}\ \emph {et~al.}(2010)\citenamefont
  {{Tupitsyn}}, \citenamefont {{Kitaev}}, \citenamefont {{Prokof'ev}},\ and\
  \citenamefont {{Stamp}}}]{TKP08043175}%
  \BibitemOpen
  \bibfield  {author} {\bibinfo {author} {\bibfnamefont {I.~S.}\ \bibnamefont
  {{Tupitsyn}}}, \bibinfo {author} {\bibfnamefont {A.}~\bibnamefont
  {{Kitaev}}}, \bibinfo {author} {\bibfnamefont {N.~V.}\ \bibnamefont
  {{Prokof'ev}}},\ and\ \bibinfo {author} {\bibfnamefont {P.~C.~E.}\
  \bibnamefont {{Stamp}}},\ }\bibfield  {title} {\bibinfo {title} {{Topological
  multicritical point in the phase diagram of the toric code model and
  three-dimensional lattice gauge Higgs model}},\ }\href
  {https://doi.org/10.1103/PhysRevB.82.085114} {\bibfield  {journal} {\bibinfo
  {journal} {Phys. Rev. B}\ }\textbf {\bibinfo {volume} {82}},\ \bibinfo
  {pages} {085114} (\bibinfo {year} {2010})},\ \Eprint
  {https://arxiv.org/abs/0804.3175} {arXiv:0804.3175} \BibitemShut {NoStop}%
\bibitem [{\citenamefont {{Vidal}}\ \emph {et~al.}(2009)\citenamefont
  {{Vidal}}, \citenamefont {{Dusuel}},\ and\ \citenamefont
  {{Schmidt}}}]{VDS08070487}%
  \BibitemOpen
  \bibfield  {author} {\bibinfo {author} {\bibfnamefont {J.}~\bibnamefont
  {{Vidal}}}, \bibinfo {author} {\bibfnamefont {S.}~\bibnamefont {{Dusuel}}},\
  and\ \bibinfo {author} {\bibfnamefont {K.~P.}\ \bibnamefont {{Schmidt}}},\
  }\bibfield  {title} {\bibinfo {title} {{Low-energy effective theory of the
  toric code model in a parallel magnetic field}},\ }\href
  {https://doi.org/10.1103/PhysRevB.79.033109} {\bibfield  {journal} {\bibinfo
  {journal} {Phys. Rev. B}\ }\textbf {\bibinfo {volume} {79}},\ \bibinfo
  {pages} {033109} (\bibinfo {year} {2009})},\ \Eprint
  {https://arxiv.org/abs/0807.0487} {arXiv:0807.0487} \BibitemShut {NoStop}%
\bibitem [{\citenamefont {{Dusuel}}\ \emph {et~al.}(2011)\citenamefont
  {{Dusuel}}, \citenamefont {{Kamfor}}, \citenamefont {{Or\'us}}, \citenamefont
  {{Schmidt}},\ and\ \citenamefont {{Vidal}}}]{DKO10121740}%
  \BibitemOpen
  \bibfield  {author} {\bibinfo {author} {\bibfnamefont {S.}~\bibnamefont
  {{Dusuel}}}, \bibinfo {author} {\bibfnamefont {M.}~\bibnamefont {{Kamfor}}},
  \bibinfo {author} {\bibfnamefont {R.}~\bibnamefont {{Or\'us}}}, \bibinfo
  {author} {\bibfnamefont {K.~P.}\ \bibnamefont {{Schmidt}}},\ and\ \bibinfo
  {author} {\bibfnamefont {J.}~\bibnamefont {{Vidal}}},\ }\bibfield  {title}
  {\bibinfo {title} {{Robustness of a Perturbed Topological Phase}},\ }\href
  {https://doi.org/10.1103/PhysRevLett.106.107203} {\bibfield  {journal}
  {\bibinfo  {journal} {Phys. Rev. Lett.}\ }\textbf {\bibinfo {volume} {106}},\
  \bibinfo {pages} {107203} (\bibinfo {year} {2011})},\ \Eprint
  {https://arxiv.org/abs/1012.1740} {arXiv:1012.1740} \BibitemShut {NoStop}%
\bibitem [{\citenamefont {{Iqbal}}\ and\ \citenamefont
  {{Schuch}}(2021)}]{IS201106611}%
  \BibitemOpen
  \bibfield  {author} {\bibinfo {author} {\bibfnamefont {M.}~\bibnamefont
  {{Iqbal}}}\ and\ \bibinfo {author} {\bibfnamefont {N.}~\bibnamefont
  {{Schuch}}},\ }\bibfield  {title} {\bibinfo {title} {{Entanglement Order
  Parameters and Critical Behavior for Topological Phase Transitions and
  Beyond}},\ }\href {https://doi.org/10.1103/PhysRevX.11.041014} {\bibfield
  {journal} {\bibinfo  {journal} {Phys. Rev. X}\ }\textbf {\bibinfo {volume}
  {11}},\ \bibinfo {pages} {041014} (\bibinfo {year} {2021})},\ \Eprint
  {https://arxiv.org/abs/2011.06611} {arXiv:2011.06611} \BibitemShut {NoStop}%
\bibitem [{\citenamefont {{Bonati}}\ \emph {et~al.}(2022)\citenamefont
  {{Bonati}}, \citenamefont {{Pelissetto}},\ and\ \citenamefont
  {{Vicari}}}]{BPV211201824}%
  \BibitemOpen
  \bibfield  {author} {\bibinfo {author} {\bibfnamefont {C.}~\bibnamefont
  {{Bonati}}}, \bibinfo {author} {\bibfnamefont {A.}~\bibnamefont
  {{Pelissetto}}},\ and\ \bibinfo {author} {\bibfnamefont {E.}~\bibnamefont
  {{Vicari}}},\ }\bibfield  {title} {\bibinfo {title} {{Multicritical point of
  the three-dimensional ${\mathbb{Z}}_{2}$ gauge Higgs model}},\ }\href
  {https://doi.org/10.1103/PhysRevB.105.165138} {\bibfield  {journal} {\bibinfo
   {journal} {Phys. Rev. B}\ }\textbf {\bibinfo {volume} {105}},\ \bibinfo
  {pages} {165138} (\bibinfo {year} {2022})},\ \Eprint
  {https://arxiv.org/abs/2112.01824} {arXiv:2112.01824} \BibitemShut {NoStop}%
\bibitem [{\citenamefont {Read}\ and\ \citenamefont {Sachdev}(1991)}]{RS9173}%
  \BibitemOpen
  \bibfield  {author} {\bibinfo {author} {\bibfnamefont {N.}~\bibnamefont
  {Read}}\ and\ \bibinfo {author} {\bibfnamefont {S.}~\bibnamefont {Sachdev}},\
  }\bibfield  {title} {\bibinfo {title} {{Large-$N$ expansion for frustrated
  quantum antiferromagnets}},\ }\href
  {https://doi.org/10.1103/PhysRevLett.66.1773} {\bibfield  {journal} {\bibinfo
   {journal} {\prl}\ }\textbf {\bibinfo {volume} {66}},\ \bibinfo {pages}
  {1773} (\bibinfo {year} {1991})}\BibitemShut {NoStop}%
\bibitem [{\citenamefont {Wen}(1991{\natexlab{b}})}]{W9164}%
  \BibitemOpen
  \bibfield  {author} {\bibinfo {author} {\bibfnamefont {X.-G.}\ \bibnamefont
  {Wen}},\ }\bibfield  {title} {\bibinfo {title} {Mean-field theory of
  spin-liquid states with finite energy gap and topological orders},\ }\href
  {https://doi.org/10.1103/physrevb.44.2664} {\bibfield  {journal} {\bibinfo
  {journal} {Phys. Rev. B}\ }\textbf {\bibinfo {volume} {44}},\ \bibinfo
  {pages} {2664} (\bibinfo {year} {1991}{\natexlab{b}})}\BibitemShut {NoStop}%
\bibitem [{\citenamefont {{C\'ordova}}\ and\ \citenamefont
  {{Ohmori}}(2020)}]{CO191004962}%
  \BibitemOpen
  \bibfield  {author} {\bibinfo {author} {\bibfnamefont {C.}~\bibnamefont
  {{C\'ordova}}}\ and\ \bibinfo {author} {\bibfnamefont {K.}~\bibnamefont
  {{Ohmori}}},\ }\href@noop {} {\bibinfo {title} {{Anomaly Obstructions to
  Symmetry Preserving Gapped Phases}}} (\bibinfo {year} {2020}),\ \Eprint
  {https://arxiv.org/abs/1910.04962} {arXiv:1910.04962} \BibitemShut {NoStop}%
\bibitem [{\citenamefont {{Banks}}\ and\ \citenamefont
  {{Rabinovici}}(1979)}]{BR1979349}%
  \BibitemOpen
  \bibfield  {author} {\bibinfo {author} {\bibfnamefont {T.}~\bibnamefont
  {{Banks}}}\ and\ \bibinfo {author} {\bibfnamefont {E.}~\bibnamefont
  {{Rabinovici}}},\ }\bibfield  {title} {\bibinfo {title} {{Finite-temperature
  behavior of the lattice abelian Higgs model}},\ }\href
  {https://doi.org/10.1016/0550-3213(79)90064-6} {\bibfield  {journal}
  {\bibinfo  {journal} {Nucl. Phys. B.}\ }\textbf {\bibinfo {volume} {160}},\
  \bibinfo {pages} {349} (\bibinfo {year} {1979})}\BibitemShut {NoStop}%
\bibitem [{\citenamefont {{Cherman}}\ \emph {et~al.}(2020)\citenamefont
  {{Cherman}}, \citenamefont {{Jacobson}}, \citenamefont {{Sen}},\ and\
  \citenamefont {{Yaffe}}}]{CJS200708539}%
  \BibitemOpen
  \bibfield  {author} {\bibinfo {author} {\bibfnamefont {A.}~\bibnamefont
  {{Cherman}}}, \bibinfo {author} {\bibfnamefont {T.}~\bibnamefont
  {{Jacobson}}}, \bibinfo {author} {\bibfnamefont {S.}~\bibnamefont {{Sen}}},\
  and\ \bibinfo {author} {\bibfnamefont {L.~G.}\ \bibnamefont {{Yaffe}}},\
  }\bibfield  {title} {\bibinfo {title} {{Higgs-confinement phase transitions
  with fundamental representation matter}},\ }\href
  {https://doi.org/10.1103/PhysRevD.102.105021} {\bibfield  {journal} {\bibinfo
   {journal} {Phys. Rev. D}\ }\textbf {\bibinfo {volume} {102}},\ \bibinfo
  {pages} {105021} (\bibinfo {year} {2020})},\ \Eprint
  {https://arxiv.org/abs/2007.08539} {arXiv:2007.08539} \BibitemShut {NoStop}%
\bibitem [{\citenamefont {{Weinberg}}(1972)}]{W19721698}%
  \BibitemOpen
  \bibfield  {author} {\bibinfo {author} {\bibfnamefont {S.}~\bibnamefont
  {{Weinberg}}},\ }\bibfield  {title} {\bibinfo {title} {{Approximate
  Symmetries and Pseudo-Goldstone Bosons}},\ }\href
  {https://doi.org/10.1103/PhysRevLett.29.1698} {\bibfield  {journal} {\bibinfo
   {journal} {Phys. Rev. Lett.}\ }\textbf {\bibinfo {volume} {29}},\ \bibinfo
  {pages} {1698} (\bibinfo {year} {1972})}\BibitemShut {NoStop}%
\bibitem [{\citenamefont {{Pace}}\ and\ \citenamefont
  {{Liu}}(2023)}]{PL231109293}%
  \BibitemOpen
  \bibfield  {author} {\bibinfo {author} {\bibfnamefont {S.~D.}\ \bibnamefont
  {{Pace}}}\ and\ \bibinfo {author} {\bibfnamefont {Y.~L.}\ \bibnamefont
  {{Liu}}},\ }\href@noop {} {\bibinfo {title} {Topological aspects of brane
  fields: solitons and higher-form symmetries}} (\bibinfo {year} {2023}),\
  \Eprint {https://arxiv.org/abs/2311.09293} {arXiv:2311.09293} \BibitemShut
  {NoStop}%
\bibitem [{\citenamefont {{Else}}\ and\ \citenamefont
  {{Senthil}}(2021)}]{ES210615623}%
  \BibitemOpen
  \bibfield  {author} {\bibinfo {author} {\bibfnamefont {D.~V.}\ \bibnamefont
  {{Else}}}\ and\ \bibinfo {author} {\bibfnamefont {T.}~\bibnamefont
  {{Senthil}}},\ }\bibfield  {title} {\bibinfo {title} {{Critical drag as a
  mechanism for resistivity}},\ }\href
  {https://doi.org/10.1103/PhysRevB.104.205132} {\bibfield  {journal} {\bibinfo
   {journal} {Phys. Rev. B}\ }\textbf {\bibinfo {volume} {104}},\ \bibinfo
  {pages} {205132} (\bibinfo {year} {2021})},\ \Eprint
  {https://arxiv.org/abs/2106.15623} {arXiv:2106.15623} \BibitemShut {NoStop}%
\bibitem [{\citenamefont {{Callan Jr.}}\ and\ \citenamefont
  {{Harvey}}(1985)}]{CH8536}%
  \BibitemOpen
  \bibfield  {author} {\bibinfo {author} {\bibfnamefont {C.~G.}\ \bibnamefont
  {{Callan Jr.}}}\ and\ \bibinfo {author} {\bibfnamefont {J.~A.}\ \bibnamefont
  {{Harvey}}},\ }\bibfield  {title} {\bibinfo {title} {{Anomalies and fermion
  zero modes on strings and domain walls}},\ }\href
  {https://doi.org/10.1016/0550-3213(85)90489-4} {\bibfield  {journal}
  {\bibinfo  {journal} {Nucl. Phys. B}\ }\textbf {\bibinfo {volume} {250}},\
  \bibinfo {pages} {427} (\bibinfo {year} {1985})}\BibitemShut {NoStop}%
\bibitem [{\citenamefont {Kapustin}(2014)}]{K1459}%
  \BibitemOpen
  \bibfield  {author} {\bibinfo {author} {\bibfnamefont {A.}~\bibnamefont
  {Kapustin}},\ }\bibfield  {title} {\bibinfo {title} {{Bosonic Topological
  Insulators and Paramagnets: {A} view from cobordisms}},\ }\href@noop {} {\
  (\bibinfo {year} {2014})},\ \Eprint {https://arxiv.org/abs/1404.6659}
  {arXiv:1404.6659} \BibitemShut {NoStop}%
\bibitem [{\citenamefont {{Wang}}\ \emph {et~al.}(2015)\citenamefont {{Wang}},
  \citenamefont {{Gu}},\ and\ \citenamefont {{Wen}}}]{WGW1489}%
  \BibitemOpen
  \bibfield  {author} {\bibinfo {author} {\bibfnamefont {J.}~\bibnamefont
  {{Wang}}}, \bibinfo {author} {\bibfnamefont {Z.-C.}\ \bibnamefont {{Gu}}},\
  and\ \bibinfo {author} {\bibfnamefont {X.-G.}\ \bibnamefont {{Wen}}},\
  }\bibfield  {title} {\bibinfo {title} {{Field-Theory Representation of
  Gauge-Gravity Symmetry-Protected Topological Invariants, Group Cohomology,
  and Beyond}},\ }\href {https://doi.org/10.1103/PhysRevLett.114.031601}
  {\bibfield  {journal} {\bibinfo  {journal} {Phys. Rev. Lett.}\ }\textbf
  {\bibinfo {volume} {114}},\ \bibinfo {pages} {031601} (\bibinfo {year}
  {2015})},\ \Eprint {https://arxiv.org/abs/1405.7689} {arXiv:1405.7689}
  \BibitemShut {NoStop}%
\bibitem [{\citenamefont {Wen}(2015)}]{W150605768}%
  \BibitemOpen
  \bibfield  {author} {\bibinfo {author} {\bibfnamefont {X.-G.}\ \bibnamefont
  {Wen}},\ }\bibfield  {title} {\bibinfo {title} {A theory of {2+1D} bosonic
  topological orders},\ }\href {https://doi.org/10.1093/nsr/nwv077} {\bibfield
  {journal} {\bibinfo  {journal} {Nat. Sci. Rev.}\ }\textbf {\bibinfo {volume}
  {3}},\ \bibinfo {pages} {68} (\bibinfo {year} {2015})},\ \Eprint
  {https://arxiv.org/abs/1506.05768} {arXiv:1506.05768} \BibitemShut {NoStop}%
\bibitem [{\citenamefont {{Serbyn}}\ \emph {et~al.}(2021)\citenamefont
  {{Serbyn}}, \citenamefont {{Abanin}},\ and\ \citenamefont
  {{Papi\'c}}}]{SAP201109486}%
  \BibitemOpen
  \bibfield  {author} {\bibinfo {author} {\bibfnamefont {M.}~\bibnamefont
  {{Serbyn}}}, \bibinfo {author} {\bibfnamefont {D.~A.}\ \bibnamefont
  {{Abanin}}},\ and\ \bibinfo {author} {\bibfnamefont {Z.}~\bibnamefont
  {{Papi\'c}}},\ }\bibfield  {title} {\bibinfo {title} {{Quantum many-body
  scars and weak breaking of ergodicity}},\ }\href
  {https://doi.org/10.1038/s41567-021-01230-2} {\bibfield  {journal} {\bibinfo
  {journal} {Nat. Phys.}\ }\textbf {\bibinfo {volume} {17}},\ \bibinfo {pages}
  {675} (\bibinfo {year} {2021})},\ \Eprint {https://arxiv.org/abs/2011.09486}
  {arXiv:2011.09486} \BibitemShut {NoStop}%
\bibitem [{\citenamefont {Dennis}\ \emph {et~al.}(2002)\citenamefont {Dennis},
  \citenamefont {Kitaev}, \citenamefont {Landahl},\ and\ \citenamefont
  {Preskill}}]{DKL0252}%
  \BibitemOpen
  \bibfield  {author} {\bibinfo {author} {\bibfnamefont {E.}~\bibnamefont
  {Dennis}}, \bibinfo {author} {\bibfnamefont {A.}~\bibnamefont {Kitaev}},
  \bibinfo {author} {\bibfnamefont {A.}~\bibnamefont {Landahl}},\ and\ \bibinfo
  {author} {\bibfnamefont {J.}~\bibnamefont {Preskill}},\ }\bibfield  {title}
  {\bibinfo {title} {Topological quantum memory},\ }\href
  {https://doi.org/10.1063/1.1499754} {\bibfield  {journal} {\bibinfo
  {journal} {J. Math. Phys.}\ }\textbf {\bibinfo {volume} {43}},\ \bibinfo
  {pages} {4452} (\bibinfo {year} {2002})}\BibitemShut {NoStop}%
\bibitem [{\citenamefont {{Castelnovo}}\ and\ \citenamefont
  {{Chamon}}(2008)}]{CC08043591}%
  \BibitemOpen
  \bibfield  {author} {\bibinfo {author} {\bibfnamefont {C.}~\bibnamefont
  {{Castelnovo}}}\ and\ \bibinfo {author} {\bibfnamefont {C.}~\bibnamefont
  {{Chamon}}},\ }\bibfield  {title} {\bibinfo {title} {{Topological order in a
  three-dimensional toric code at finite temperature}},\ }\href
  {https://doi.org/10.1103/PhysRevB.78.155120} {\bibfield  {journal} {\bibinfo
  {journal} {Phys. Rev. B}\ }\textbf {\bibinfo {volume} {78}},\ \bibinfo
  {pages} {155120} (\bibinfo {year} {2008})},\ \Eprint
  {https://arxiv.org/abs/0804.3591} {arXiv:0804.3591} \BibitemShut {NoStop}%
\bibitem [{\citenamefont {{Alicki}}\ \emph {et~al.}(2009)\citenamefont
  {{Alicki}}, \citenamefont {{Fannes}},\ and\ \citenamefont
  {{Horodecki}}}]{AFH08104584}%
  \BibitemOpen
  \bibfield  {author} {\bibinfo {author} {\bibfnamefont {R.}~\bibnamefont
  {{Alicki}}}, \bibinfo {author} {\bibfnamefont {M.}~\bibnamefont {{Fannes}}},\
  and\ \bibinfo {author} {\bibfnamefont {M.}~\bibnamefont {{Horodecki}}},\
  }\bibfield  {title} {\bibinfo {title} {{On thermalization in Kitaev's 2D
  model}},\ }\href {https://doi.org/10.1088/1751-8113/42/6/065303} {\bibfield
  {journal} {\bibinfo  {journal} {Journal of Physics A: Mathematical and
  Theoretical}\ }\textbf {\bibinfo {volume} {42}},\ \bibinfo {pages} {065303}
  (\bibinfo {year} {2009})},\ \Eprint {https://arxiv.org/abs/0810.4584}
  {arXiv:0810.4584} \BibitemShut {NoStop}%
\bibitem [{\citenamefont {{Alicki}}\ \emph {et~al.}(2010)\citenamefont
  {{Alicki}}, \citenamefont {{Horodecki}}, \citenamefont {{Horodecki}},\ and\
  \citenamefont {{Horodecki}}}]{AHH08110033}%
  \BibitemOpen
  \bibfield  {author} {\bibinfo {author} {\bibfnamefont {R.}~\bibnamefont
  {{Alicki}}}, \bibinfo {author} {\bibfnamefont {M.}~\bibnamefont
  {{Horodecki}}}, \bibinfo {author} {\bibfnamefont {P.}~\bibnamefont
  {{Horodecki}}},\ and\ \bibinfo {author} {\bibfnamefont {R.}~\bibnamefont
  {{Horodecki}}},\ }\bibfield  {title} {\bibinfo {title} {{On thermal stability
  of topological qubit in Kitaev's 4D model}},\ }\href
  {https://doi.org/10.1142/S1230161210000023} {\bibfield  {journal} {\bibinfo
  {journal} {{Open Systems \& Information Dynamics}}\ }\textbf {\bibinfo
  {volume} {17}},\ \bibinfo {pages} {1} (\bibinfo {year} {2010})},\ \Eprint
  {https://arxiv.org/abs/0811.0033} {arXiv:0811.0033} \BibitemShut {NoStop}%
\bibitem [{\citenamefont {{Savary}}\ and\ \citenamefont
  {{Balents}}(2016)}]{SB160103742}%
  \BibitemOpen
  \bibfield  {author} {\bibinfo {author} {\bibfnamefont {L.}~\bibnamefont
  {{Savary}}}\ and\ \bibinfo {author} {\bibfnamefont {L.}~\bibnamefont
  {{Balents}}},\ }\bibfield  {title} {\bibinfo {title} {{Quantum spin liquids:
  a review}},\ }\href {https://doi.org/10.1088/0034-4885/80/1/016502}
  {\bibfield  {journal} {\bibinfo  {journal} {Reports on Progress in Physics}\
  }\textbf {\bibinfo {volume} {80}},\ \bibinfo {pages} {016502} (\bibinfo
  {year} {2016})},\ \Eprint {https://arxiv.org/abs/1601.03742}
  {arXiv:1601.03742} \BibitemShut {NoStop}%
\bibitem [{\citenamefont {{Gingras}}\ and\ \citenamefont
  {{McClarty}}(2014)}]{GM13111817}%
  \BibitemOpen
  \bibfield  {author} {\bibinfo {author} {\bibfnamefont {M.~J.}\ \bibnamefont
  {{Gingras}}}\ and\ \bibinfo {author} {\bibfnamefont {P.~A.}\ \bibnamefont
  {{McClarty}}},\ }\bibfield  {title} {\bibinfo {title} {{Quantum spin ice: a
  search for gapless quantum spin liquids in pyrochlore magnets}},\ }\href
  {https://doi.org/10.1088/0034-4885/77/5/056501} {\bibfield  {journal}
  {\bibinfo  {journal} {Reports on Progress in Physics}\ }\textbf {\bibinfo
  {volume} {77}},\ \bibinfo {pages} {056501} (\bibinfo {year} {2014})},\
  \Eprint {https://arxiv.org/abs/1311.1817} {arXiv:1311.1817} \BibitemShut
  {NoStop}%
\bibitem [{\citenamefont {Hermele}\ \emph {et~al.}(2004)\citenamefont
  {Hermele}, \citenamefont {Fisher},\ and\ \citenamefont {Balents}}]{HFB0404}%
  \BibitemOpen
  \bibfield  {author} {\bibinfo {author} {\bibfnamefont {M.}~\bibnamefont
  {Hermele}}, \bibinfo {author} {\bibfnamefont {M.~P.~A.}\ \bibnamefont
  {Fisher}},\ and\ \bibinfo {author} {\bibfnamefont {L.}~\bibnamefont
  {Balents}},\ }\bibfield  {title} {\bibinfo {title} {Pyrochlore photons: The
  {U(1)} spin liquid in a {S=1/2} three-dimensional frustrated magnet},\ }\href
  {https://doi.org/10.1103/physrevb.69.064404} {\bibfield  {journal} {\bibinfo
  {journal} {\prb}\ }\textbf {\bibinfo {volume} {69}},\ \bibinfo {pages}
  {064404} (\bibinfo {year} {2004})},\ \Eprint
  {https://arxiv.org/abs/cond-mat/0305401} {arXiv:cond-mat/0305401}
  \BibitemShut {NoStop}%
\bibitem [{\citenamefont {{Benton}}\ \emph {et~al.}(2012)\citenamefont
  {{Benton}}, \citenamefont {{Sikora}},\ and\ \citenamefont
  {{Shannon}}}]{BSS12041325}%
  \BibitemOpen
  \bibfield  {author} {\bibinfo {author} {\bibfnamefont {O.}~\bibnamefont
  {{Benton}}}, \bibinfo {author} {\bibfnamefont {O.}~\bibnamefont {{Sikora}}},\
  and\ \bibinfo {author} {\bibfnamefont {N.}~\bibnamefont {{Shannon}}},\
  }\bibfield  {title} {\bibinfo {title} {{Seeing the light: Experimental
  signatures of emergent electromagnetism in a quantum spin ice}},\ }\href
  {https://doi.org/10.1103/PhysRevB.86.075154} {\bibfield  {journal} {\bibinfo
  {journal} {Phys. Rev. B}\ }\textbf {\bibinfo {volume} {86}},\ \bibinfo
  {pages} {075154} (\bibinfo {year} {2012})},\ \Eprint
  {https://arxiv.org/abs/1204.1325} {arXiv:1204.1325} \BibitemShut {NoStop}%
\bibitem [{\citenamefont {{Pace}}\ \emph {et~al.}(2021)\citenamefont {{Pace}},
  \citenamefont {{Morampudi}}, \citenamefont {{Moessner}},\ and\ \citenamefont
  {{Laumann}}}]{PMM200904499}%
  \BibitemOpen
  \bibfield  {author} {\bibinfo {author} {\bibfnamefont {S.~D.}\ \bibnamefont
  {{Pace}}}, \bibinfo {author} {\bibfnamefont {S.~C.}\ \bibnamefont
  {{Morampudi}}}, \bibinfo {author} {\bibfnamefont {R.}~\bibnamefont
  {{Moessner}}},\ and\ \bibinfo {author} {\bibfnamefont {C.~R.}\ \bibnamefont
  {{Laumann}}},\ }\bibfield  {title} {\bibinfo {title} {{Emergent Fine
  Structure Constant of Quantum Spin Ice Is Large}},\ }\href
  {https://doi.org/10.1103/PhysRevLett.127.117205} {\bibfield  {journal}
  {\bibinfo  {journal} {Phys. Rev. Lett.}\ }\textbf {\bibinfo {volume} {127}},\
  \bibinfo {pages} {117205} (\bibinfo {year} {2021})},\ \Eprint
  {https://arxiv.org/abs/2009.04499} {arXiv:2009.04499} \BibitemShut {NoStop}%
\bibitem [{\citenamefont {{Sulejmanpasic}}\ and\ \citenamefont
  {{Gattringer}}(2019)}]{SG190102637}%
  \BibitemOpen
  \bibfield  {author} {\bibinfo {author} {\bibfnamefont {T.}~\bibnamefont
  {{Sulejmanpasic}}}\ and\ \bibinfo {author} {\bibfnamefont {C.}~\bibnamefont
  {{Gattringer}}},\ }\bibfield  {title} {\bibinfo {title} {{Abelian gauge
  theories on the lattice: $\theta$-Terms and compact gauge theory with(out)
  monopoles}},\ }\href {https://doi.org/10.1016/j.nuclphysb.2019.114616}
  {\bibfield  {journal} {\bibinfo  {journal} {Nucl. Phys. B}\ }\textbf
  {\bibinfo {volume} {943}},\ \bibinfo {pages} {114616} (\bibinfo {year}
  {2019})},\ \Eprint {https://arxiv.org/abs/1901.02637} {arXiv:1901.02637}
  \BibitemShut {NoStop}%
\bibitem [{\citenamefont {{Wan}}\ \emph {et~al.}(2022)\citenamefont {{Wan}},
  \citenamefont {{Wang}},\ and\ \citenamefont {{Wen}}}]{WWW211212148}%
  \BibitemOpen
  \bibfield  {author} {\bibinfo {author} {\bibfnamefont {Z.}~\bibnamefont
  {{Wan}}}, \bibinfo {author} {\bibfnamefont {J.}~\bibnamefont {{Wang}}},\ and\
  \bibinfo {author} {\bibfnamefont {X.-G.}\ \bibnamefont {{Wen}}},\ }\bibfield
  {title} {\bibinfo {title} {{$(3+1)\mathrm{d}$ boundaries with gravitational
  anomaly of $(4+1)\mathrm{d}$ invertible topological order for
  branch-independent bosonic systems}},\ }\href
  {https://doi.org/10.1103/PhysRevB.106.045127} {\bibfield  {journal} {\bibinfo
   {journal} {Phys. Rev. B}\ }\textbf {\bibinfo {volume} {106}},\ \bibinfo
  {pages} {045127} (\bibinfo {year} {2022})},\ \Eprint
  {https://arxiv.org/abs/2112.12148} {arXiv:2112.12148} \BibitemShut {NoStop}%
\bibitem [{\citenamefont {{Slagle}}\ and\ \citenamefont
  {{Kim}}(2017)}]{SK170804619}%
  \BibitemOpen
  \bibfield  {author} {\bibinfo {author} {\bibfnamefont {K.}~\bibnamefont
  {{Slagle}}}\ and\ \bibinfo {author} {\bibfnamefont {Y.~B.}\ \bibnamefont
  {{Kim}}},\ }\bibfield  {title} {\bibinfo {title} {{Quantum field theory of
  X-cube fracton topological order and robust degeneracy from geometry}},\
  }\href {https://doi.org/10.1103/PhysRevB.96.195139} {\bibfield  {journal}
  {\bibinfo  {journal} {Phys. Rev. B}\ }\textbf {\bibinfo {volume} {96}},\
  \bibinfo {pages} {195139} (\bibinfo {year} {2017})},\ \Eprint
  {https://arxiv.org/abs/1708.04619} {arXiv:1708.04619} \BibitemShut {NoStop}%
\bibitem [{\citenamefont {Maldacena}\ \emph {et~al.}(2001)\citenamefont
  {Maldacena}, \citenamefont {Seiberg},\ and\ \citenamefont {Moore}}]{MMS0105}%
  \BibitemOpen
  \bibfield  {author} {\bibinfo {author} {\bibfnamefont {J.}~\bibnamefont
  {Maldacena}}, \bibinfo {author} {\bibfnamefont {N.}~\bibnamefont {Seiberg}},\
  and\ \bibinfo {author} {\bibfnamefont {G.}~\bibnamefont {Moore}},\ }\bibfield
   {title} {\bibinfo {title} {D-brane charges in five-brane backgrounds},\
  }\href {https://doi.org/10.1088/1126-6708/2001/10/005} {\bibfield  {journal}
  {\bibinfo  {journal} {J. High Energy Phys.}\ }\textbf {\bibinfo {volume}
  {2001}}\bibfield  {number} {\bibinfo  {number} { (10)},\ \bibinfo {pages}
  {005}},\ }\Eprint {https://arxiv.org/abs/hep-th/0108152}
  {arXiv:hep-th/0108152} \BibitemShut {NoStop}%
\bibitem [{\citenamefont {Hansson}\ \emph {et~al.}(2004)\citenamefont
  {Hansson}, \citenamefont {Oganesyan},\ and\ \citenamefont
  {Sondhi}}]{HOS0497}%
  \BibitemOpen
  \bibfield  {author} {\bibinfo {author} {\bibfnamefont {T.}~\bibnamefont
  {Hansson}}, \bibinfo {author} {\bibfnamefont {V.}~\bibnamefont {Oganesyan}},\
  and\ \bibinfo {author} {\bibfnamefont {S.}~\bibnamefont {Sondhi}},\
  }\bibfield  {title} {\bibinfo {title} {Superconductors are topologically
  ordered},\ }\href {https://doi.org/10.1016/j.aop.2004.05.006} {\bibfield
  {journal} {\bibinfo  {journal} {Ann. Phys.}\ }\textbf {\bibinfo {volume}
  {313}},\ \bibinfo {pages} {497} (\bibinfo {year} {2004})},\ \Eprint
  {https://arxiv.org/abs/cond-mat/0404327} {arXiv:cond-mat/0404327}
  \BibitemShut {NoStop}%
\bibitem [{\citenamefont {Banks}\ and\ \citenamefont {Seiberg}(2011)}]{BS1119}%
  \BibitemOpen
  \bibfield  {author} {\bibinfo {author} {\bibfnamefont {T.}~\bibnamefont
  {Banks}}\ and\ \bibinfo {author} {\bibfnamefont {N.}~\bibnamefont
  {Seiberg}},\ }\bibfield  {title} {\bibinfo {title} {Symmetries and strings in
  field theory and gravity},\ }\href
  {https://doi.org/10.1103/physrevd.83.084019} {\bibfield  {journal} {\bibinfo
  {journal} {Phys. Rev. D}\ }\textbf {\bibinfo {volume} {83}},\ \bibinfo
  {pages} {084019} (\bibinfo {year} {2011})},\ \Eprint
  {https://arxiv.org/abs/1011.5120} {arXiv:1011.5120} \BibitemShut {NoStop}%
\bibitem [{\citenamefont {{Mathieu}}\ and\ \citenamefont
  {{Thuillier}}(2016)}]{MT150904236}%
  \BibitemOpen
  \bibfield  {author} {\bibinfo {author} {\bibfnamefont {P.}~\bibnamefont
  {{Mathieu}}}\ and\ \bibinfo {author} {\bibfnamefont {F.}~\bibnamefont
  {{Thuillier}}},\ }\bibfield  {title} {\bibinfo {title} {{Abelian BF theory
  and Turaev-Viro invariant}},\ }\href {https://doi.org/10.1063/1.4942046}
  {\bibfield  {journal} {\bibinfo  {journal} {J. Math. Phys.}\ }\textbf
  {\bibinfo {volume} {57}},\ \bibinfo {pages} {022306} (\bibinfo {year}
  {2016})},\ \Eprint {https://arxiv.org/abs/1509.04236} {arXiv:1509.04236}
  \BibitemShut {NoStop}%
\bibitem [{\citenamefont {{Witten}}(1995)}]{W9505186}%
  \BibitemOpen
  \bibfield  {author} {\bibinfo {author} {\bibfnamefont {E.}~\bibnamefont
  {{Witten}}},\ }\bibfield  {title} {\bibinfo {title} {{On $S$-duality in
  Abelian Gauge Theory}},\ }\href {https://doi.org/10.1007/BF01671570}
  {\bibfield  {journal} {\bibinfo  {journal} {Selecta Mathematica}\ }\textbf
  {\bibinfo {volume} {1}},\ \bibinfo {pages} {383} (\bibinfo {year} {1995})},\
  \Eprint {https://arxiv.org/abs/hep-th/9505186} {arXiv:hep-th/9505186}
  \BibitemShut {NoStop}%
\bibitem [{\citenamefont {Hsieh}\ \emph {et~al.}(2022)\citenamefont {Hsieh},
  \citenamefont {Tachikawa},\ and\ \citenamefont {Yonekura}}]{HT200311550}%
  \BibitemOpen
  \bibfield  {author} {\bibinfo {author} {\bibfnamefont {C.-T.}\ \bibnamefont
  {Hsieh}}, \bibinfo {author} {\bibfnamefont {Y.}~\bibnamefont {Tachikawa}},\
  and\ \bibinfo {author} {\bibfnamefont {K.}~\bibnamefont {Yonekura}},\
  }\bibfield  {title} {\bibinfo {title} {{Anomaly inflow and $p$-form gauge
  theories}},\ }\href {https://doi.org/10.1007/s00220-022-04333-w} {\bibfield
  {journal} {\bibinfo  {journal} {Commun. Math. Phys.}\ }\textbf {\bibinfo
  {volume} {391}},\ \bibinfo {pages} {495} (\bibinfo {year} {2022})},\ \Eprint
  {https://arxiv.org/abs/2003.11550} {arXiv:2003.11550} \BibitemShut {NoStop}%
\end{thebibliography}%

\end{document}